\documentclass[12pt]{article}
\usepackage{amsmath,amssymb,amsfonts,color,graphicx,cite,color,feynarts,soul}
\input paperdef

\graphicspath{{figsmini/}}

\evensidemargin \oddsidemargin
\allowdisplaybreaks

\begin{document}
\thispagestyle{empty}

\def\thefootnote{\fnsymbol{footnote}}

\begin{flushright}
KA-TP--38--2011 \\
arXiv:1112.0760 [hep-ph]
\end{flushright}

\vspace{0.5cm}

\begin{center}

{\large\sc {\bf Chargino Decays in the Complex MSSM:}}

\vspace{0.4cm}

{\large\sc {\bf A Full One-Loop Analysis}}

\vspace{1cm}

{\sc
S.~Heinemeyer$^{1}$%
\footnote{email: Sven.Heinemeyer@cern.ch}%
, F.~von der Pahlen$^{1}$%
\footnote{email: pahlen@ifca.unican.es}%
~and C.~Schappacher$^{2}$%
\footnote{email: cs@particle.uni-karlsruhe.de}%
}

\vspace*{.7cm}

{\sl
$^1$Instituto de F\'isica de Cantabria (CSIC-UC), Santander, Spain

\vspace*{0.1cm}

$^2$Institut f\"ur Theoretische Physik, Karlsruhe Institute of Technology, \\
D--76128 Karlsruhe, Germany

}

\end{center}

\vspace*{0.1cm}

\begin{abstract}
\noindent
We evaluate two-body decay modes of charginos in 
the Minimal Supersymmetric Standard Model with complex parameters
(cMSSM). Assuming heavy scalar quarks we take into account all decay channels
involving charginos, neutralinos, (scalar) leptons, 
Higgs bosons and Standard Model gauge bosons.
The evaluation of the decay widths is based on a full one-loop calculation 
including hard and soft QED radiation. 
Special attention is paid to decays involving the 
Lightest Supersymmetric Particle (LSP), 
i.e.\ the lightest neutralino, or a neutral or charged Higgs boson.
The higher-order corrections of the chargino decay widths
involving the LSP can easily reach 
a level of about $\pm 10\%$, while the corrections to the
decays to Higgs bosons are slightly smaller, translating into
corrections of similar size in the respective branching ratios.
These corrections are important for the correct interpretation of
LSP and Higgs production at the LHC and at a future linear $e^+e^-$
collider.
The results will be implemented into the Fortran code {\tt FeynHiggs}.
\end{abstract}

\def\thefootnote{\arabic{footnote}}
\setcounter{page}{0}
\setcounter{footnote}{0}

\newpage

\newcommand{\DecayNNh}[3]{\neu{#1} \to \neu{#2} h_{#3}}
\newcommand{\DecayNNZ}[2]{\neu{#1} \to \neu{#2} Z}
\newcommand{\DecayCCh}[1]{\cham{2} \to \cham{1} h_{#1}}
\newcommand{\DecayCCZ}{\cham{2} \to \cham{1} Z}
\newcommand{\DecayCNH}[2]{\cham{#1} \to \neu{#2} H^-}
\newcommand{\DecayCNW}[2]{\cham{#1} \to \neu{#2} W^-}
\newcommand{\DecayCnSl}[3]{\cham{#1} \to \nu_{#2}\, \tilde{#2}_{#3}^{-}}
\newcommand{\DecayClSn}[2]{\cham{#1} \to {#2}^{-}\, \tilde{\nu_{#2}}}
\newcommand{\DecayCxy}[1]{\cha{#1} \to {\rm xy}}

\newcommand{\decayCCh}{\DecayCCh{k}}
\newcommand{\decayCCZ}{\DecayCCZ}
\newcommand{\decayCNH}{\DecayCNH{i}{j}}
\newcommand{\decayCNW}{\DecayCNW{i}{j}}
\newcommand{\decayCnSl}{\DecayCnSl{i}{l}{k}}
\newcommand{\decayClSn}{\DecayClSn{i}{l}}
\newcommand{\decayCxy}{\DecayCxy{i}}

\newcommand{\DecayCmCh}[1]{\cham{2} \to \cham{1} h_{#1}}
\newcommand{\DecayCmCZ}{\cham{2} \to \cham{1} Z}
\newcommand{\DecayCmNH}[2]{\cham{#1} \to \neu{#2} H^-}
\newcommand{\DecayCmNW}[2]{\cham{#1} \to \neu{#2} W^-}
\newcommand{\DecayCmnSl}[3]{\cham{#1} \to \bar{\nu}_{#2}\, \tilde{#2}_{#3}^-}
\newcommand{\DecayCmlSn}[2]{\cham{#1} \to {#2}^-\, \tilde{\nu}_{#2}^\dagger}
\newcommand{\DecayCmxy}[1]{\cham{#1} \to {\rm xy}}

\newcommand{\decayCmCh}{\DecayCmCh{k}}
\newcommand{\decayCmCZ}{\DecayCmCZ}
\newcommand{\decayCmNH}{\DecayCmNH{i}{j}}
\newcommand{\decayCmNW}{\DecayCmNW{i}{j}}
\newcommand{\decayCmnSl}{\DecayCmnSl{i}{l}{k}}
\newcommand{\decayCmlSn}{\DecayCmlSn{i}{l}}
\newcommand{\decayCmxy}{\DecayCmxy{i}}


\section{Introduction}

One of the important tasks at the LHC is to search for 
physics beyond the Standard Model (SM), where the 
Minimal Supersymmetric Standard Model (MSSM)~\cite{mssm} is one of the 
leading candidates. 
Two related important tasks are investigating the mechanism of electroweak
symmetry breaking, as well as the production and measurement of 
the properties of Cold Dark Matter (CDM).
The most frequently investigated models for electroweak symmetry
breaking are the Higgs mechanism within the SM and within the MSSM.
The latter also offers a natural candidate for CDM, the
Lightest Supersymmetric Particle (LSP), i.e.\ the lightest neutralino,~$\neu{1}$~\cite{EHNOS}.
Supersymmetry (SUSY) predicts two scalar partners for all SM fermions as well
as fermionic partners to all SM bosons.
Contrary to the case of the SM, in the MSSM 
two Higgs doublets are required.
This results in five physical Higgs bosons instead of the single Higgs
boson in the SM. These are the light and heavy $\cp$-even Higgs bosons, $h$
and $H$, the $\cp$-odd Higgs boson, $A$, and the charged Higgs bosons,
$H^\pm$.
In the MSSM with complex parameters (cMSSM) the three neutral Higgs
bosons mix~\cite{mhiggsCPXgen,mhiggsCPXRG1,mhiggsCPXFD1}, 
giving rise to the states $\He, \Hz, \Hd$.

If SUSY is realized in nature and the scalar quarks and/or the gluino
are in the kinematic reach of the LHC, it is expected that these
strongly interacting particles are copiously produced. 
The primarily produced strongly interacting particles subsequently
decay via cascades to 
SM particles and (if $R$-parity conservation is assumed, as we do) the
LSP. One step in these decay chains is often the decay of a chargino, 
$\cha{1,2}$,  
to a SM particle and the LSP, or as a competing process the
chargino decay to another SUSY particle accompanied by a
SM particle. Also neutral and charged Higgs bosons can be produced this way.
Via these decays some characteristics of the LSP and/or Higgs bosons can be
measured, see, e.g., \citeres{atlas,cms} and references therein. 
At any future $e^+e^-$ collider (such as ILC or CLIC)
a precision determination of the properties of the observed particles is
expected~\cite{teslatdr,ilc}. (For combined LHC/ILC analyses and further
prospects see \citere{lhcilc}.) 
Thus, if kinematically accessible, the pair production of charginos 
with a subsequent decay to the LSP and/or Higgs bosons 
can yield important information about the lightest neutralino and 
the Higgs sector of the model.

In order to yield a sufficient accuracy, one-loop corrections to
the various chargino decay modes have to be considered.
In this paper we evaluate full one-loop corrections to chargino decays
in the cMSSM.
If scalar quarks are sufficiently heavy (as in many GUT based models
such as CMSSM, GMSB or AMSB, see for instance
\citere{newbenchmark}) a chargino decay to a quark and a scalar 
quark is kinematically forbidden. Assuming heavy squarks 
we calculate the full one-loop correction to
all two body decay modes (which are non-zero at the tree-level),
\begin{align}
\label{CNH}
&\Ga(\decayCmNH) \qquad (i = 1,2,\; j = 1,2,3,4)~, \\
\label{CNW}
&\Ga(\decayCmNW) \qquad (i = 1,2,\; j = 1,2,3,4)~, \\
\label{CCh}
&\Ga(\decayCmCh) \qquad (k = 1,2,3)~, \\
\label{CCZ}
&\Ga(\decayCmCZ) ~, \\
\label{CSln}
&\Ga(\decayCmnSl) \qquad (i = 1,2,\; l = e, \mu, \tau,\; k = 1,2)~, \\
\label{CSnl}
&\Ga(\decayCmlSn) \qquad (i = 1,2,\; l = e, \mu, \tau)~.
\end{align}
The total width is defined as the sum of the channels (\ref{CNH}) to
(\ref{CSnl}), where for a given parameter point several channels may
be kinematically forbidden.

As explained above, 
we are especially interested in the branching ratios (BR) of the
decays involving a Higgs boson, \refeqs{CNH}, (\ref{CCh})
as part of an evaluation of a Higgs production cross section, and/or
involving the LSP, \refeqs{CNH}, (\ref{CNW}) as part of the measurement
of CDM properties at the LHC.
Consequently, it is not necessary to investigate three- or four-body decay
modes. These only play a significant role once the two-body modes
are kinematically forbidden, and thus the relevant BR's are zero. 
The same applies to two-body decay modes that exist only at the
one-loop level, such as $\cha{2} \to \cha{1} \ga$ (see, for instance,
\citere{chachaga}). While this channel is 
of \order{\al^2}, the size of the one-loop corrections to \refeqs{CNH}
to (\ref{CSnl}) is of \order{\al}. We have numerically verified that the
contribution of $\Ga(\cha{2} \to \cha{1} \ga)$ to the total width is
completely negligible.

Tree-level results for the decays of charginos in the MSSM
were presented in \citeres{chadectree,chachaga,haber3}.
Higher-order corrections to chargino decays have been evaluated in
various analyses over the last
decade.
However, they were 
either restricted to one specific channel 
or only a very restricted set of parameters were analyzed,
and in many cases only parts of the one-loop calculation have been performed. 
More specifically, the available literature comprises the following.
First order electroweak corrections to the partial decay widths of charginos 
in the MSSM with real parameters (rMSSM) where derived:
for the two-body decays into a neutralino/chargino and $W/Z$~boson 
including only third generation quark-squark exchange
diagrams~\cite{ChaDecCorr1},  
for the three-body decays into the LSP and quarks, 
including corrections to the masses of third generation fermions and SUSY
particles~\cite{ChaDecCPC}, and for the three-body leptonic decays at full
one-loop order in \citere{ChaDec3Body}.
The one-loop electroweak corrections to all two-body decay channels 
of charginos, evaluated in  an on-shell renormalization scheme, 
have been implemented in the code SloopS~\cite{BaroII}. 
In \citere{GRACESUSY} a large set of two-body and three-body decay channels of
charginos including full one-loop corrections has been calculated using the
code GRACE/SUSY-loop, but only a very limited set of numerical results have
been published. However, \citere{BaroII} has compared its results on the
partial widths with those of \citere{GRACESUSY} and concluded that the latter
uses a renormalization scheme which leads to too large corrections. 
The code SDECAY~\cite{SDECAY} also includes all two-body decays of
charginos. However, no radiative corrections to these decay channels have been
included so far. A full one-loop  calculation of the electroweak
corrections to the partial width of the decay of a chargino into a neutralino
and a $W$~boson in the MSSM and NMSSM is presented in \citere{liebler},  
and made available with the code CNNDecays.
A brief comparison with this calculation can be found in
\refse{sec:calc}.  
In the cMSSM only the decay of charginos into a neutralino and a $W$ boson has
been studied. In \citere{ChaDecCPVYang} a partial one-loop calculation of 
rate asymmetries of $\champ{i} \to \neu{1} W^\mp$ has been performed, 
including contributions from the third generation quarks, while 
\citere{ChaDecCPVEberl} evaluated this $\cp$-violating asymmetry 
at the full one-loop level, highlighting the relevance of the contribution
from the chargino wave function corrections. 
However, a complete one-loop result for the two-body total decay width in the cMSSM 
is missing so far.

In this paper we present
for the first time a full one-loop calculation for all non-hadronic 
two-body decay channels of a chargino, taking into
account soft and hard QED radiation, simultaneously and
consistently evaluated in the cMSSM.
In \refse{sec:cMSSM} we review the relevant sectors of the cMSSM
and their renormalization.
Details about the calculation can be
found in \refse{sec:calc}, and the numerical results for all decay
channels are presented in \refse{sec:numeval}. The conclusions can be
found in \refse{sec:conclusions}.
The evaluation of the branching ratios of the charginos will be 
implemented into the Fortran code 
{\tt FeynHiggs}~\cite{feynhiggs,mhiggslong,mhiggsAEC,mhcMSSMlong}.


\section{The relevant sectors of the complex MSSM}
\label{sec:cMSSM}

All the channels (\ref{CNH}) -- (\ref{CSnl}) are calculated at the
one-loop level, including real QED radiation. This requires the
simultaneous renormalization of several sectors of the cMSSM. In
the following 
subsections we briefly review these sectors.
Details about the renormalization of most of the sectors can be found in
\citere{Stop2decay}. Here we only review the renormalization that can
not be found explicitly in \citere{Stop2decay}.


\subsection{The lepton/slepton sector of the cMSSM}
\label{sec:slepton}

For the evaluation of the one-loop contributions to the decay channels 
in \refeqs{CSln}, (\ref{CSnl}) a renormalization of the scalar lepton 
($\Sl$) and neutrino ($\Sn$) sector is needed (we assume no generation 
mixing and discuss the case for one generation only).
The bilinear part of the $\Sl$ and $\Sn$ Lagrangian,
\begin{align}
\cL_{\Sl/\Sn}^{\text{mass}} &= - \begin{pmatrix}
\sll^{\dagger}, \slr^{\dagger} \end{pmatrix}
\matr{M}_{\Sl} \begin{pmatrix} \sll \\ \slr
\end{pmatrix} 
- \begin{pmatrix} \Sn^{\dagger} \end{pmatrix}
\matr{M}_{\Sn}\begin{pmatrix} \Sn \end{pmatrix}~,
\end{align}
contains the slepton and sneutrino mass matrices
$\matr{M}_{\Sl}$ and $\matr{M}_{\Sn}$,
given by 
\begin{align}\label{Sfermionmassenmatrix}
\matr{M}_{\Sl} &= \begin{pmatrix} 
\MslL^2 + \ml^2 + M_Z^2 c_{2 \beta} (I_l^3 - Q_l \sw^2) & 
 \ml \Xl^* \\[.2em]
 \ml \Xl &
\MslR^2 + \ml^2 +M_Z^2 c_{2 \beta} Q_l \sw^2
\end{pmatrix}~, \\[.5em]
\matr{M}_{\Sn} &= \MslL^2 
+ I_\nu^3 c_{2\be} M_Z^2
\end{align}
with
\begin{align}
\Xl &= \Al - \mu^* \tb~.
\end{align}
$\MslL$ and $\MslR$ are the soft SUSY-breaking mass
parameters, where $\MslL$ is equal for all members of an
$SU(2)_L$ doublet.
$\ml$ and $Q_{l}$ are, respectively, the mass and the charge of the
corresponding lepton, $I_{l/\nu}^3$ denotes the isospin of $l/\nu$,  
and $A_l$ is the trilinear soft-breaking parameter.
$\MZ$ and $\MW$ are the masses of the $Z$~and $W$~boson, 
$\cw = \MW/\MZ$, and $\sw = \sqrt{1 - \cw^2}$. Finally we use the
short-hand notations $c_{x} = \cos(x)$, $s_x = \sin(x)$.
The mass matrix $\matr{M}_{\Sl}$ can be diagonalized with the help of a unitary
transformation ${\matr{U}}_{\Sl}$,
\begin{align}\label{transformationkompl}
\matr{D}_{\Sl} &= 
\matr{U}_{\Sl}\, \matr{M}_{\Sl} \, {\matr{U}}_{\Sl}^\dagger = 
\begin{pmatrix} \msle^2 & 0 \\ 0 & \mslz^2 \end{pmatrix}~, \qquad
{\matr{U}}_{\Sl}= 
\begin{pmatrix} U_{\Sl_{11}} & U_{\Sl_{12}} \\  
                U_{\Sl_{21}} & U_{\Sl_{22}} \end{pmatrix}
~.
\end{align}
The mass eigenvalues depend only on $|\Xl|$. 
The scalar lepton masses will always be mass ordered, i.e.\
$m_{\sle} \le m_{\slz}$:
\begin{align}
\label{MSlep}
m_{\Sl_{1,2}}^2 &= \edz \KL M_{\Sl_L}^2 + M_{\Sl_R}^2 \KR
       + \ml^2 + \edz I_l^3 c_{2\be} \MZ^2 \\
&\quad \mp \edz \sqrt{\KKL \MslL^2 - \MslR^2
       + \MZ^2 c_{2\be} (I_l^3 - 2 Q_l \sw^2) \KKR^2 + 4 \ml^2 |\Xl|^2}~, 
\non\\[.5em]
m_{\Sn}^2 &= \MslL^2 + I_\nu^3 c_{2\be} M_Z^2~.
\end{align}


\subsubsection{Renormalization}

The parameter renormalization can be performed as follows, 
\begin{align}
\matr{M}_{\Sl} &\to \matr{M}_{\Sl} + \de\matr{M}_{\Sl}~, \\
\matr{M}_{\Sn} &\to \matr{M}_{\Sn} + \de\matr{M}_{\Sn} 
\end{align}
which means that the parameters in the mass matrix $\matr{M}_{\Sl}$ 
are replaced by the renormalized parameters and a counterterm. After the
expansion $\de\matr{M}_{\Sl}$ contains the counterterm part,
\begin{align}
\label{proc1a}
\de\matr{M}_{\Sl_{11}} &= \de M_{\Sl_L}^2 + 2 \ml \de \ml 
- M_Z^2 c_{2 \beta}\, Q_l \, \de \sw^2 + (I_l^3 - Q_l \sw^2) 
  ( c_{2 \beta}\, \de M_Z^2 + M_Z^2\, \de c_{2\beta})~, \\
\label{proc1b}
\de\matr{M}_{\Sl_{12}} &= (\Al^*  - \mu \tb)\, \de \ml 
+ \ml (\de \Al^* - \mu\, \de \tb - \tb \, \de \mu)~, \\
\label{proc1c}
\de\matr{M}_{\Sl_{21}} &=\de\matr{M}_{\Sl_{12}}^*~, \\
\label{proc1d}
\de\matr{M}_{\Sl_{22}} &= \de M_{\Sl_R}^2 
+ 2 \ml \de \ml +  M_Z^2 c_{2 \beta}\, Q_l \, \de \sw^2
+ Q_l \sw^2 ( c_{2 \beta}\, \de M_Z^2+ M_Z^2\, \de c_{2 \beta})~, \\
\label{proc1e}
\de\matr{M}_{\Sn} &= \de M_{\Sl_L}^2 + I_\nu^3
(c_{2 \beta}\, \de M_Z^2 + M_Z^2\, \de c_{2\beta})~. 
\end{align}

Another possibility for the parameter renormalization of the sleptons is
to start out with the physical parameters which corresponds to
the replacement:
\begin{align} \label{proc2}
\matr{U}_{\Sl}\, \matr{M}_{\Sl} \, 
{\matr{U}}_{\Sl}^\dagger &\to\matr{U}_{\Sl}\, \matr{M}_{\Sl} \, 
{\matr{U}}_{\Sl}^\dagger + \matr{U}_{\Sl}\, \de \matr{M}_{\Sl} \, 
{\matr{U}}_{\Sl}^\dagger =
\begin{pmatrix} \msle^2 & Y_l \\ Y_l^* & \mslz^2 \end{pmatrix} +
\begin{pmatrix}
\de \msle^2 & \de Y_l \\ \de Y_l^* & \de \mslz^2
\end{pmatrix}
\end{align}
where $\de \msle$ and $\de \mslz$ are the counterterms of the 
slepton masses. $\de Y_l$ is the counterterm%
\footnote{The unitary matrix $\matr{U}_{\Sl}$ can be expressed by a
  mixing angle and a corresponding phase. Then the counterterm  $\de
  Y_l$ can be related to the counterterms of the mixing angle and the
  phase (see \citere{mhcMSSM2L}).}%
~to the slepton mixing parameter $Y_l$ (which vanishes
at tree-level, $Y_l = 0$, and corresponds to the 
off-diagonal entries in $\matr{D}_{\Sl} =\matr{U}_{\Sl}\,
 \matr{M}_{\Sl} \,  
{\matr{U}}_{\Sl}^\dagger$, \refeq{transformationkompl}). Using
\refeq{proc2} 
 one can express $\de\matr{M}_{\Sl}$ by the counterterms $\de \msle^2$,
 $\de \mslz^2$ and $\de Y_l$. Especially for $\de\matr{M}_{\Sl_{12}}$
 one finds
\begin{align}\label{dMsq12physpar}
\de\matr{M}_{{\Sl}_{12}} &=
U^*_{\Sl_{11}} U_{\Sl_{12}}
(\de \msle^2 - \de \mslz^2) +
U^*_{\Sl_{11}} U_{\Sl_{22}} \de Y_l + U_{\Sl_{12}}
U^*_{\Sl_{21}} \de Y_l^*~.
\end{align}
In the following the relation given by \refeqs{proc1b} and
\eqref{dMsq12physpar} will be used to express either $\de Y_l$, $\de
A_l$ or $\de \ml$ by the other counterterms.

For the field renormalization the following procedure is applied,
\begin{align}
\begin{pmatrix} \sle \\ \slz \end{pmatrix} &\to 
  \KL \id + \edz \de\matr{Z}_{\Sl} \KR 
  \begin{pmatrix} \sle \\ \slz \end{pmatrix} 
~~{\rm with}~~
\de\matr{Z}_{\Sl} = \begin{pmatrix} 
                   \dZ{\Sl_{11}} & \dZ{\Sl_{12}} \\
                   \dZ{\Sl_{21}} & \dZ{\Sl_{22}} 
                   \end{pmatrix}~, \\
\Sn &\to \KL 1 + \tedz \dZ{\Sn} \KR \Sn~.
\end{align}

This yields for the renormalized self-energies
\begin{align}
\hSi_{\Sl_{11}}(k^2) &= \Si_{\Sl_{11}}(k^2) 
  + \tedz (k^2 - \msle^2) (\dZ{\Sl_{11}} + \dZ{\Sl_{11}}^*)
  - \de\msle^2~, \\
\hSi_{\Sl_{12}}(k^2) &= \Si_{\Sl_{12}}(k^2)
  + \tedz (k^2 - \msle^2) \dZ{\Sl_{12}}
  + \tedz (k^2 - \mslz^2) \dZ{\Sl_{21}}^* 
  - \de Y_l~, \\
\hSi_{\Sl_{21}}(k^2) &= \Si_{\Sl_{21}}(k^2)
  + \tedz (k^2 - \msle^2) \dZ{\Sl_{12}}^*
  + \tedz (k^2 - \mslz^2) \dZ{\Sl_{21}} 
  - \de Y_l^*~, \\
\hSi_{\Sl_{22}}(k^2) &= \Si_{\Sl_{22}}(k^2) 
  + \tedz (k^2 - \mslz^2) (\dZ{\Sl_{22}} + \dZ{\Sl_{22}}^*)
  - \de\mslz^2~, \\
\hSi_{\Sn}(k^2) &= \Si_{\Sn}(k^2) 
  + \tedz (k^2 - \msn^2) (\dZ{\Sn} + \dZ{\Sn}^*)
  - \de\msn^2~.
\end{align}
In order to complete the lepton/slepton sector renormalization also for the
corresponding lepton (i.e. its mass, $\ml$, and the lepton
fields $l_L$, $l_R$, $\nu_L$) renormalization constants have to be introduced:
\begin{align}
\ml &\to \ml + \de \ml~,\\
l_{L/R} &\to (1 + \tedz \dZ{l}^{L/R})\, l_{L/R}~, \\
\nu_L &\to (1 + \tedz \dZ{\nu})\, \nu_L~,
\end{align}
with $\de \ml$ being the lepton mass counterterm and $\dZ{l}^L$ and
$\dZ{l}^R$ being the $Z$~factors of the left-handed and the right-handed
charged lepton fields, respectively; $\dZ{\nu}$ is the neutrino
field renormalization.
Then the renormalized self energy $\hSi_{l}$ 
can be decomposed 
into left/right-handed and scalar left/right-handed parts, 
${\Si}_{l}^{L/R}$ and ${\Si}_{l}^{SL/SR}$, 
respectively,
while only the left-handed part exists for the self energy $\hSi_\nu$ of the
massless neutrino
\begin{align}\label{decomposition}
\hSi_{l} (k) &= \not\! k\, {\omega}_{-} \hSi_l^L (k^2)
                   + \not\! k\, {\omega}_{+} \hSi_l^R (k^2)
                   + {\omega}_{-} \hSi_l^{SL} (k^2) 
                   + {\omega}_{+} \hSi_l^{SR} (k^2)~, \\[.3em]
\hSi_{\nu} (k) &= \not\! k\, {\omega}_{-} \hSi_\nu^L (k^2)
~,
\end{align}
where the components are given by
\begin{align}
\hSi_l^{L/R} (k^2) &= {\Si}_l^{L/R} (k^2) 
   + \frac{1}{2} (\dZ{l}^{L/R} + {\dZ{l}^{L/R}}^*)~, \\
\hSi_l^{SL} (k^2) &=  {\Si}_l^{SL} (k^2) 
   - \frac{\ml}{2} (\dZ{l}^L + {\dZ{l}^R}^*) - \de \ml~,  \\
\hSi_l^{SR} (k^2) &=  {\Si}_l^{SR} (k^2) 
   - \frac{\ml}{2} (\dZ{l}^R + {\dZ{l}^L}^*) - \de \ml~, \\[.3em]
\hSi_\nu^{L} (k^2) &= {\Si}_\nu^{L} (k^2) 
   + \frac{1}{2} (\dZ{\nu}^{L} + {\dZ{\nu}^{L}}^*)~, 
\end{align}
and ${\omega}_{\pm} = \frac{1}{2}(\id \pm \gamma_5)$ 
are the right- and left-handed projectors, respectively.
Note that 
$\wtre\hSi_{l}^{SR} (k^2) = (\wtre\hSi_{l}^{SL} (k^2))^*$ 
holds due to ${\cal CPT}$ invariance.


\subsubsection{The neutrino/sneutrino sector}
\label{sec:sneutrino}

We follow closely the renormalization presented in
\citere{SbotRen,Stop2decay}, slightly modified to be applicable to the
lepton/slepton sector. 

\begin{itemize}

\item[(i)] The neutrino is defined on-shell (OS), yielding the one-loop
field renormalization
\begin{align}
\label{RedZnu}
\re \dZ{\nu} &= - \wtre \Si_\nu(0)~, \\
\label{ImdZnu}
\im \dZ{\nu} &= 0~.
\end{align}
$\wtre$ denotes the real part with respect to
contributions from the loop integral, but leaves the complex
couplings unaffected.

\item[(ii)]
The $\Sn$ mass is defined OS,
\begin{align}
\wtre\hSi_{\Sn}(\msn^2) = 0~.
\end{align}
This yields for the sneutrino mass counter terms
\begin{align}
\de\msn^2 = \wtre\Si_{\Sn}(\msn^2)~.
\end{align}

\item[(iii)]
Due to $m_\nu \equiv 0$ no off-diagonal parameters in the sneutrino mass
matrix have to be renormalized.

\item[(iv)]
We now determine the $Z$~factors in the sneutrino sector in the OS scheme.
The diagonal $Z$~factor is determined such that the real part of the
residua of the propagator are set to unity, 
\begin{align}
\label{residuumSneutOS}
\wtre \hSi'_{\Sn}(k^2)\big|_{k^2 = \msn^2} = 0~.
\end{align}
with $\Si'(k^2) \equiv \frac{\partial \Si(k^2)}{\partial k^2}$.
This condition fixes the real parts of the diagonal $Z$~factor to
\begin{align}
\re\,\dZ{\Sn} = - \wtre \Si'_{\Sn}(k^2)\big|_{k^2 = \msn^2}~,
\end{align}
which is correct, since the imaginary parts of the diagonal 
$Z$~factor does not contain any divergences and can be 
(implicitly) set to zero, 
\begin{align}
\im \dZ{\Sn} &= 0~.
\end{align}

\item[(v)]
Due to $m_\nu \equiv 0$ no off-diagonal field renormalization for the
sneutrinos has to be performed.

\end{itemize}


\subsubsection{The charged lepton/slepton sector}

We choose the slepton masses $\msle$, $\mslz$ and the 
lepton mass $\ml$ as independent parameters.
Since we also require an independent renormalization of the scalar
neutrino, this requires an explicit restoration of the $SU(2)_L$
relation, achieved via a shift in the $M_{\Sl_L}$ parameter entering 
the $\Sl$~mass matrix (see also \citeres{stopsbot_phi_als,dr2lA}).
Requiring the $SU(2)_L$ relation
to be valid at the loop level induces the following shift in 
$M^2_{\Sl_L}(\Sl)$ 
\begin{align}
M_{\Sl_L}^2(\Sl) = M_{\Sl_L}^2(\Sn) 
   + \de M_{\Sl_L}^2(\Sn) - \de M_{\Sl_L}^2(\Sl)
\label{MSnushift}
\end{align}
with
\begin{align}
\de M_{\Sl_L}^2(\Sl) &= |U_{\Sl_{11}}|^2 \de\msle^2
   + |U_{\Sl_{12}}|^2 \de\mslz^2
   - U_{\Sl_{22}} U_{\Sl_{12}}^* \de Y_l
   - U_{\Sl_{12}} U_{\Sl_{22}}^* \de Y_l^* - 2 \ml \de\ml \non \\
&\quad  + \MZ^2\, c_{2\be}\, Q_l\, \de \sw^2 
        - (I_l^3 - Q_l \sw^2) (c_{2\be}\, \de \MZ^2 + \MZ^2\, \de c_{2\be})~, 
\\[.5em]
\de M_{\Sl_L}^2(\Sn) &= \de\msn^2 
   - I_\nu^3(c_{2\be}\, \de \MZ^2 + \MZ^2\, \de c_{2\be})~.
\label{MSnushift-detail}
\end{align}
This choice avoids problems concerning UV- and IR-finiteness as
discussed in detail in \citere{SbotRen}, but also leads to shifts 
in both slepton masses, which are therefore slightly shifted away 
from their on-shell values.
An additional shift in $M_{\Sl_R}$ recovers at least one on-shell 
slepton mass. 
\begin{align}
M_{\Sl_R}^2(\Sl_i) = \frac{\ml^2\, |\Al^* - \mu \tb|^2}
  {M_{\Sl_L}^2(\tilde{l}) + \ml^2 
   + \MZ^2\, c_{2\be} (I_l^3 - Q_l \sw^2) - \msli^2} 
  - \ml^2 - \MZ^2\, c_{2\be}\, Q_l\, \sw^2+ \msli^2~.
\label{backshift}
\end{align}
The choice of slepton for this additional shift, which relates its mass to
the slepton parameter $M_{\Sl_R}$, also represents a choice of scenario, 
with the chosen slepton having a dominantly right-handed character.
A ``natural'' choice is to preserve the character of the sleptons in the
renormalization process.
With our choice of mass ordering, $\msle \le \mslz$ (see
above), this suggests to recover 
$\msle$ for $M_{\Sl_L}^2 > M_{\Sl_R}^2$, and to recover
$\mslz$ for the other mass hierarchy. Consequently, for our numerical
choice given below in \refta{tab:para}, we insert $\mslz$ into
\refeq{backshift} and recover its original value from the
re-diagonalization after applying this shift.

\bigskip

For the scalar lepton sector we can now employ a ``full'' on-shell
scheme, where the following renormalization conditions are imposed:
\begin{itemize}

\item[(i)] The lepton mass is defined on-shell, yielding the one-loop
  counterterm $\de \ml$:
\begin{align}\label{dmt}
\de \ml &= \tedz \wtre \KKKL 
    \ml \KKL\Si_l^L (\ml^2) + \Si_l^R (\ml^2) \KKR  
  + \KKL \Si_l^{SL} (\ml^2) + \Si_l^{SR} (\ml^2) \KKR \KKKR~,
\end{align}
referring to the Lorentz decomposition of the self energy 
${\hSi}_{l}(k)$, see \refeq{decomposition}.\\
The field renormalization constants are given by
\begin{align}
\label{RedZl}
\re \dZ{l}^{L/R} &= - \wtre \Big\{ {\Si}_l^{L/R} (\ml^2)  \non \\
&\qquad + \ml^2 \KKL {{\Si}_l^{L}}'(\ml^2) + {{\Si}_l^{R}}'(\ml^2) \KKR
             + \ml \KKL {{\Si}_l^{SL}}'(\ml^2) + {{\Si}_l^{SR}}'(\ml^2) \KKR
                                 \Big\}~,  \\
\label{ImdZl}
\im \dZ{l}^{L/R} &= \pm \frac{i}{2\, \ml} 
        \wtre \KKKL {\Si}_l^{SR}(\ml^2) - {\Si}_l^{SL}(\ml^2) \KKKR
     = \pm \frac{1}{\ml} \im \KKKL \wtre {\Si}_l^{SL}(\ml^2) \KKKR~. 
\end{align}
with 
$\Si'(m^2) \equiv \frac{\partial \Si(k^2)}{\partial k^2}\big|_{k^2 = m^2}$.

\item[(ii)]
The slepton masses are also determined via on-shell
conditions~\cite{mhiggslong,hr}, yielding  
\begin{align}
\label{dmsl}
\de\msli^2 &= \wtre\Si_{\Sl_{ii}}(\msli^2) \qquad (i = 1,2)~.
\end{align}

\item[(iii)]
The non-diagonal entry of \refeq{proc2} is fixed
as~\cite{mhiggsFDalbals,hr,SbotRen} 
\begin{align}
\de Y_l =  \tedz \wtre 
    \big\{ \Si_{\Sl_{12}}(\msle^2) + \Si_{\Sl_{12}}(\mslz^2) \big\}~, 
\end{align}
which corresponds to two separate conditions in the case of a complex
$\de Y_l$.
The counterterm of the trilinear coupling $\de\Al$ can be obtained from the
relation of \refeqs{proc1b} and~\eqref{dMsq12physpar},
\begin{align}
\de \Al &= \frac{1}{\ml}\bigl[U_{\Sl_{11}} U_{\Sl_{12}}^*
           (\de \msle^2 - \de \mslz^2)
        +  U_{\Sl_{11}} U_{\Sl_{22}}^{*} \de Y_l^*
        + U_{\Sl_{12}}^{*} U_{\Sl_{21}} \de Y_l  
        - (\Al - \mu^* \tb)\, \de\ml  \bigr]  \non \\
&\quad  + (\de\mu^* \tb + \mu^* \dtanb)~.
\end{align}
So far undetermined are $\dtanb$ and $\de\mu$, which are defined 
via the Higgs sector and the chargino/neutralino sector, see
\citere{Stop2decay} for details.

\item[(iv)]
We now determine the $Z$~factors of the scalar lepton sector in 
the OS scheme.
The diagonal $Z$~factors are determined such that the real part of the
residua of the propagators is set to unity, 
\begin{align}
\label{residuumSlepOS}
\wtre \hSi'_{\Sl_{ii}}(k^2) \big|_{k^2 = \msli^2} = 0 \qquad (i = 1,2)~.
\end{align}
This condition fixes the real parts of the diagonal $Z$~factors to
\begin{align}
\re\,\dZ{\Sl_{ii}} = 
- \wtre \Si'_{\Sl_{ii}}(k^2)\big|_{k^2 = \msli^2} \qquad (i = 1,2)~,
\end{align}
which is correct, since the
imaginary parts of the diagonal $Z$~factors 
does not contain any divergences and can be 
(implicitly) set to zero,
\begin{align}
\im \dZ{\Sl_{ii}} &= 0 \qquad (i = 1,2)~.
\end{align}

\item[(v)]
For the non-diagonal $Z$~factors we impose the condition that for
on-shell sleptons no transition from one slepton to the other occurs, 
\begin{align}
\wtre\hSi_{\Sl_{12}}(\msli^2)  = 0~, \qquad 
\wtre\hSi_{\Sl_{21}}(\msli^2) = 0 \qquad (i = 1,2)~.
\end{align}
This yields
\begin{align}
\dZ{\Sl_{12}} = + 2 \frac{\wtre\Si_{\Sl_{12}}(\mslz^2) - \de Y_l}
                       {(\msle^2 - \mslz^2)}~, \qquad
\dZ{\Sl_{21}} = - 2 \frac{\wtre\Si_{\Sl_{21}}(\msle^2) - \de Y_l^*}
                       {(\msle^2 - \mslz^2)}~.
\label{dZslepoffdiagOS}
\end{align}

\end{itemize}

\bigskip
Alternative field renormalizations can be constructed if absorptive 
parts of self-energy type corrections are included into them, 
see \citere{Stop2decay} for more details.
These new combined factors ${\mathcal Z}$ are (in general) different 
for incoming particles/outgoing antiparticles (unbarred) and 
outgoing particles/incoming antiparticles (barred).

\begin{itemize}

\item[(a)]
The alternative diagonal slepton and sneutrino $Z$~factors read
\begin{align}
\de{\mathcal Z}_{\Sl} &= - \Si'_{\Sl}(k^2)\big|_{k^2 = \msl^2}~,\hspace{-2.5cm}
& \de \bar{\mathcal Z}_{\Sl} &= \de{\mathcal Z}_{\Sl}~, \\ 
\de{\mathcal Z}_{\Sn} &= - \Si'_{\Sn}(k^2)\big|_{k^2 = \msn^2}~,\hspace{-2.5cm}
& \de \bar{\mathcal Z}_{\Sn} &= \de{\mathcal Z}_{\Sn}~.
\end{align}

\item[(b)]
For the non-diagonal $Z$~factors we impose the condition that for
on-shell sleptons no transition from one slepton to the other occurs, 
\begin{align}
\hSi_{\Sl_{12}}(\msli^2) = 0~, \qquad
\hSi_{\Sl_{21}}(\msli^2) = 0 \qquad (i = 1,2)~.
\end{align}
This yields the following alternative field renormalization constants,
\begin{align}
\de{\mathcal Z}_{\Sl_{12}}&= 
+ 2 \frac{\Si_{\Sl_{12}}(\mslz^2) - \de Y_l}{(\msle^2 - \mslz^2)}~,
& \de\bar{\mathcal Z}_{\Sl_{12}} &= 
+ 2 \frac{\Si_{\Sl_{21}}(\mslz^2) - \de Y_l^*}{(\msle^2 - \mslz^2)}~, \\
\de{\mathcal Z}_{\Sl_{21}} &= 
- 2 \frac{\Si_{\Sl_{21}}(\msle^2) - \de Y_l^*}{(\msle^2 - \mslz^2)}~,
& \de\bar{\mathcal Z}_{\Sl_{21}} &= 
- 2 \frac{\Si_{\Sl_{12}}(\msle^2) - \de Y_l}{(\msle^2 - \mslz^2)}~.
\end{align}
\end{itemize}


\subsection{The Higgs and gauge boson sector of the cMSSM}
\label{sec:higgs}

The two Higgs doublets of the cMSSM are decomposed in the following way,
\begin{align}
\label{eq:higgsdoublets}
\cHe = \begin{pmatrix} H_{11} \\ H_{12} \end{pmatrix} &=
\begin{pmatrix} v_1 + \tfrac{1}{\sqrt{2}} (\phi_1-i \chi_1) \\
  -\phi^-_1 \end{pmatrix}, \notag \\ 
\cHz = \begin{pmatrix} H_{21} \\ H_{22} \end{pmatrix} &= e^{i \xi}
\begin{pmatrix} \phi^+_2 \\ v_2 + \tfrac{1}{\sqrt{2}} (\phi_2+i
  \chi_2) \end{pmatrix}. 
\end{align}
Besides the vacuum expectation values $v_1$ and $v_2$, in 
\refeq{eq:higgsdoublets} a possible new phase $\xi$ between the two
Higgs doublets is introduced. 
The Higgs potential $\VHiggs$ can be written in powers of the Higgs fields,
\begin{align}
\VHiggs &=  \ldots + T_{\phi_1}\, \phi_1  +T_{\phi_2}\, \phi_2 +
        T_{\chi_1}\, \chi_1 + T_{\chi_2}\, \chi_2 \non \\ 
&\quad - \edz \begin{pmatrix} \phi_1,\phi_2,\chi_1,\chi_2
        \end{pmatrix} 
\matr{M}_{\phi\phi\chi\chi}
\begin{pmatrix} \phi_1 \\ \phi_2 \\ \chi_1 \\ \chi_2 \end{pmatrix} -
\begin{pmatrix} \phi^{+}_1,\phi^{+}_2  \end{pmatrix}
\matr{M}^{\top}_{\phi^\pm\phi^\pm}
\begin{pmatrix} \phi^{-}_1 \\ \phi^{-}_2  \end{pmatrix} + \ldots~,
\end{align}
where the coefficients of the linear terms are called tadpoles and
those of the bilinear terms are the mass matrices
$\matr{M}_{\phi\phi\chi\chi}$ and $\matr{M}_{\phi^\pm\phi^\pm}$. 
After a rotation to the physical fields 
one obtains
\begin{align}
\label{VHiggs}
\VHiggs &=  \ldots + T_{h}\, h + T_{H}\, H + T_{A}\, A \non \\ 
&\quad  - \edz \begin{pmatrix} h, H, A, G 
        \end{pmatrix} 
\matr{M}_{hHAG}^{\rm diag}
\begin{pmatrix} h \\ H \\ A \\ G  \end{pmatrix} -
\begin{pmatrix} H^{+}, G^{+}  \end{pmatrix}
\matr{M}_{H^\pm G^\pm}^{\rm diag}
\begin{pmatrix} H^{-} \\ G^{-} \end{pmatrix} + \ldots~,
\end{align}
where the tree-level masses are denoted as
$\mh$, $\mH$, $\mA$, $\mG$, $\MHp$, $\mGp$.
With the help of a Peccei-Quinn
transformation~\cite{Peccei} $\mu$ and the complex soft SUSY-breaking
parameters in the Higgs sector can be 
redefined~\cite{MSSMcomplphasen} such that the complex phases
vanish at tree-level.

Concerning the renormalization we follow the usual approach where the
gauge-fixing term does not receive a net contribution from the
renormalization transformations. 
As input parameter we choose the mass of the charged Higgs boson, $\MHp$.
All details can be found in \citeres{Stop2decay,mhcMSSMlong}%
\footnote{
Corresponding to the convention used in \fa/\fc, we exchanged in
the charged part the positive Higgs fields with the negative ones, 
which is in contrast to \cite{mhcMSSMlong}. 
As we keep the definition of the matrix 
$\matr{M}_{\phi^\pm\phi^\pm}$ used in \cite{mhcMSSMlong} the
transposed matrix will appear in  
the expression for $\matr{M}_{H^\pm G^\pm}^{\rm diag}$.
}
~(see also \citere{Demir} for the alternative effective potential
approach and \citere{mhcMSSMother} for the renormalization group improved
effective potential approach including Higgs pole mass effects).

Including higher-order corrections the three neutral Higgs bosons can
mix~\cite{mhiggsCPXgen,mhiggsCPXRG1,mhiggsCPXFD1,mhcMSSMlong}, 
\begin{align}
\KL h, H, A \KR \quad \longrightarrow \quad \KL \He, \Hz, \Hd \KR~,
\end{align} 
where we define the loop corrected masses according to
\begin{align}
\MHe \le \MHz \le \MHd~.
\end{align}
A vertex with an external on-shell Higgs boson $h_{k}$ (${k} = 1,2,3$)
is obtained from the decay widths to the tree-level Higgs bosons via the
complex matrix $\matr{Z}$~\cite{mhcMSSMlong},
\begin{align}
\Ga_{h_{k}} &=
[\matr{Z}]_{i1} \Ga_h +
[\matr{Z}]_{i2} \Ga_H +
[\matr{Z}]_{i3} \Ga_A + \ldots ~,
\label{eq:zfactors123}
\end{align}
where the ellipsis represents contributions from the mixing with the
Goldstone boson and the $Z$~boson, see \refse{sec:calc}.
It should be noted that the `rotation' with $\matr{Z}$ is not a
unitary transformation, see \citere{mhcMSSMlong} for details.

Also the charged Higgs boson appearing as an external particle in a
chargino decay has to obey the proper on-shell conditions. This leads to
an extra $Z$~factor,
\begin{align}
\hat Z_{H^-H^+} = 
   \KKL 1 + \re \hSip_{H^-H^+}(p^2)\big|_{p^2 = \MHp^2} \KKR^{-1}~.
\end{align}
Expanding to one-loop order yields the $Z$~factor that has to be applied
to the process with external charged Higgs boson,
\begin{align}
\sqrt{\hat Z_{H^-H^+}} = 1 + \frac{1}{2} \de\hat Z_{H^-H^+} 
\end{align}
with 
\begin{align}
\label{dhZHpHm}
\de\hat Z_{H^-H^+} = - \re\hSip_{H^-H^+}(p^2)\big|_{p^2 = \MHp^2} = 
- \re\Sip_{H^-H^+}(\MHp^2) - \dZ{H^-H^+}~.
\end{align}
As for the neutral Higgs bosons, there are contributions from the 
mixing with the Goldstone boson and the $W$~boson.
This $Z$~factor is by definition UV-finite. However, it contains
IR-divergences that cancel with the soft photon contributions from 
the loop diagrams, see \refse{sec:calc}.

For the renormalization of $\tb$ and the Higgs field
renormalization the \DRbar\ scheme is
chosen~\cite{mhcMSSMlong,Stop2decay}. This leads to the introduction
of the scale $\mu_R$, which will be fixed later to the
mass of the decaying particle.


\subsection{The chargino/neutralino sector of the cMSSM}
\label{sec:chaneu}

The mass eigenstates of the charginos can be determined from the matrix
\begin{align}
  \matr{X} =
  \begin{pmatrix}
    \MTwo & \sqrt{2} \sinb \MW \\
    \sqrt{2} \cosb \MW & \mu
  \end{pmatrix}.
\end{align}
In addition to the higgsino mass parameter $\mu$ it 
contains the soft breaking term $\MTwo$, 
which can also be complex in the cMSSM.
The rotation to the chargino mass eigenstates is done by transforming
the original wino and higgsino fields with the help of two unitary 2$\times$2
matrices $\matr{U}$ and $\matr{V}$,
\begin{align}
\label{eq:charginotransform}
\tilde{\chi}^-_i = 
\begin{pmatrix} \psi^L_i
   \\ \overline{\psi}^R_i \end{pmatrix}
\quad \text{with} \quad \psi^L_{i} = U_{ij} \begin{pmatrix} \tilde{W}^-
  \\ \tilde{H}^-_1 \end{pmatrix}_{j} \quad \text{and} \quad
 \psi^R_{i} = V_{ij} \begin{pmatrix} \tilde{W}^+
  \\ \tilde{H}^+_2 \end{pmatrix}_{j}~,
\end{align}
where the $i$th mass eigenstate can be expressed in terms of either the Weyl
spinors $\psi^L_i$ and $\psi^R_i$ or the Dirac spinor $\tilde{\chi}^-_i$.
These rotations lead to the diagonal mass matrix
\begin{align}
\matr{M}_{\cham{}} = 
  \matr{V}^* \, \matr{X}^{\top} \, \matr{U}^{\dagger} =
  \matr{diag}(m_{\tilde{\chi}^\pm_1}, m_{\tilde{\chi}^\pm_2})~.
\end{align}
{}From this relation, it becomes clear that the mass ordered
chargino masses $\mcha{1} < \mcha{2}$ can be determined as the (real and
positive) singular values of $\matr{X}$,
\begin{align}
\mcha{1,2}^2 &= \edz \KL |\MTwo|^2 + |\mu|^2\KR + \MW^2 \\[.2em] 
& \mp \edz\sqrt{\KL |\MTwo|^2 - |\mu|^2 \KR^2
         + 4 \MW^2 \KL |\MTwo|^2 +|\mu|^2 
         + 2 |\MTwo| |\mu| s_{2 \be} \cos(\phimu + \phiMz)  
         + \MW^2 c_{2 \be}^2 \KR }~. \non
\label{eq:mcha}
\end{align}
The singular value decomposition of $\matr{X}$
also yields results for $\matr{U}$ and~$\matr{V}$.

A similar procedure is used for the determination of the neutralino masses and
mixing matrix, which can both be calculated from the mass matrix
\begin{align}
  \matr{Y} =
  \begin{pmatrix}
    \MOne                  & 0                & -\MZ \, \sw \cosb
    & \MZ \, \sw \sinb \\
    0                      & \MTwo            & \quad \MZ \, \cw \cosb
    & -\MZ \, \cw \sinb \\
    -\MZ \, \sw \cosb      & \MZ \, \cw \cosb & 0
    & -\mu             \\
    \quad \MZ \, \sw \sinb & -\MZ \, \cw \sinb & -\mu                   & 0
  \end{pmatrix}.
\end{align}
This symmetric matrix contains the additional complex soft-breaking
parameter $\MOne$. 
The diagonalization of the matrix
is achieved by a transformation starting from the original
bino/wino/higgsino basis,
\begin{align}
\tilde{\chi}^0_i = \begin{pmatrix} \psi^0_i \\ \overline{\psi}^0_i 
                   \end{pmatrix} \qquad \text{with} \qquad 
\psi^0_{i} = N_{ij}\, 
         (\tilde{B}^0, \tilde{W}^0, \tilde{H}^0_1,\tilde{H}^0_2)_{j}^{\top}~, 
\\
\matr{M}_{\neu{}} = \matr{N}^* \, \matr{Y} \, \matr{N}^{\dagger} =
\matr{diag}(\mneu{1}, \mneu{2}, \mneu{3}, \mneu{4})~,
\end{align}
where $\psi^0_i$ denotes the two component Weyl spinor and $\tilde{\chi}^0_i$
the four component Majorana spinor of the $i$th neutralino field.
The unitary 4$\times$4 matrix $\matr{N}$ and the physical neutralino
masses again result 
from a numerical singular value decomposition of $\matr{Y}$.
The symmetry of $\matr{Y}$ permits the non-trivial condition of
using only one 
matrix $\matr{N}$ for its diagonalization, in contrast to the chargino
case shown above.

Concerning the renormalization we use the results of
\citere{Stop2decay,dissAF,diplTF,dissTF}.  
This includes the contributions from absorptive parts of self-energy 
type corrections into `combined' ${\mathcal Z}$~factors (which in general 
can be different for incoming and outgoing particles). The explicit
expressions can be found in the Appendix of \citere{Stop2decay}.
Since in our renormalization the chargino masses $\mcha{1}, \mcha{2}$
and the lightest neutralino 
mass $\mneu{1}$ have been chosen as independent parameters the one-loop masses
of the heavier neutralinos 
are obtained from the tree-level ones with the shifts
\begin{align}
\De \mneu{i} = -\re \KKKL \mneu{i} \hat\Si_{\neu{i}}^{L}(\mneu{i}^2) 
               + \hat\Si_{\neu{i}}^{SL}(\mneu{i}^2) \KKKR 
               \qquad (i = 2,3,4)~,
\label{Deltamneu}
\end{align}
where the renormalized self energies of the neutralino have been decomposed 
into their left/right-handed and scalar left/right-handed parts as in 
Eq.~(\ref{decomposition}).
$\De \mneu{1} = 0$ is just the real part of one of 
our renormalization conditions.
Special care has to be taken in the regions of the cMSSM parameter space where 
the gaugino-higgsino mixing in the chargino sector is maximal,
i.e.\ where $\mu \approx M_2$. 
Here $\delta M_2$ (see Eq.~(180) in \cite{Stop2decay})
 and $\delta \mu$ (see Eq.~(181) in \cite{Stop2decay}) diverge as 
$(U^*_{11}U^*_{22}V^*_{11}V^*_{22} - U^*_{12}U^*_{21}V^*_{12}V^*_{21})^{-1}$
and the loop calculation does not yield a reliable result.
An analysis of various renormalization schemes was recently
published in \citere{onshellCNmasses}, where this kind of divergences
were discussed.%
\footnote{Similar divergences appearing in the on-shell
renormalization in the sbottom sector, occurring for ``maximal sbottom
mixing'', have been observed and discussed in 
\citeres{SbotRen,Stop2decay}.}%
~In \citere{onshellCNmasses} it was furthermore emphasized that 
in the case of the renormalization of two chargino and one neutralino
mass always the most bino-like neutralino has to be renormalized in order
to find a numerically stable result (see also \citere{BaroII}).
In our numerical set-up, see
\refse{sec:numeval}, the lightest neutralino is nearly always rather 
bino-like. If required, however, it would be trivial to change our 
prescription from the lightest neutralino to any other neutralino. 

As will be outlined in \refse{sec:parameter}, we choose the two 
chargino masses as independent numerical input, which fixes also their 
mass difference. 
In the case of maximal mixing in the chargino sector this mass
difference tends to reach a minimum, where the counterterms depend on
the variation of this difference.
Consequently, our results will be less reliable where this minimum is
reached, and we will exclude a small range of parameters, about 
$\sim 5 \gev$ in mass, from our analysis, see below.
In \citere{onshellCNmasses} it was also suggested that the
numerically most stable result is obtained via the renormalization of
one chargino and two neutralinos. 
This choice is well suited for tree level masses.
However, in our approach to calculate chargino decays, including their
renormalization, this choice leads to IR divergences, 
since an electrically charged particle (the chargino) 
changes its mass by the renormalization procedure 
via an analogous shift to \refeq{Deltamneu}.
Using the shifted mass for the external particle, but the
tree-level mass for internal particles results in the IR divergence.
On the other hand,
inserting the shifted chargino mass everywhere yields an UV divergence,
see the corresponding discussion in 
\citere{Stop2decay}.
Consequently, we choose to stick to our choice of imposing 
on-shell conditions for the two charginos and one neutralino.


\section{Calculation of loop diagrams}
\label{sec:calc}

\begin{figure}[ht!]
\begin{center}
\includegraphics[width=0.90\textwidth]{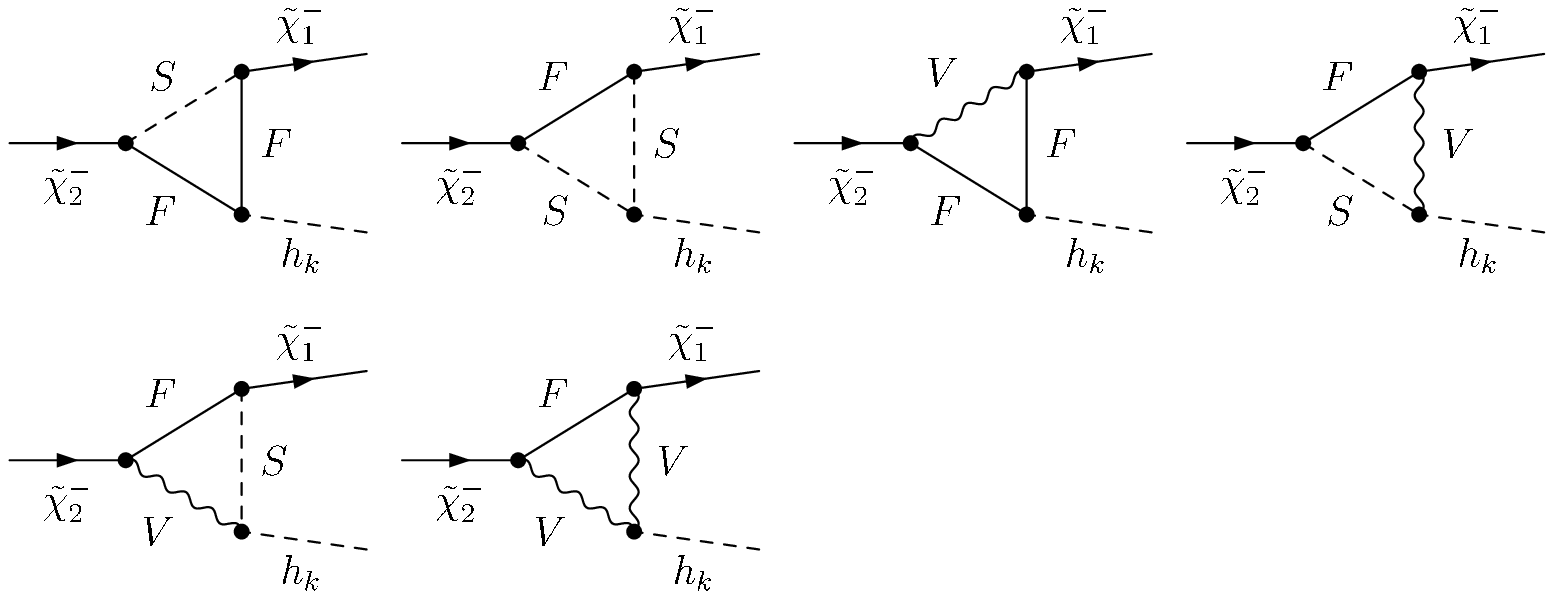}
\caption{
Generic Feynman diagrams for the decay 
$\decayCmCh$ ($k = 1,2,3$).
$F$ can be a SM fermion, chargino or neutralino, $S$ can be a
sfermion or a Higgs boson, $V$ can be a $\ga$, $Z$ or $W^\pm$. 
Not shown are the diagrams with a $Z$--$h_{k}$ or $G$--$h_{k}$ 
transition contribution on the external Higgs boson leg. 
}
\label{fig:fdCCh}
\end{center}
\end{figure}

\begin{figure}[ht!]
\vspace{2em}
\begin{center}
\includegraphics[width=0.90\textwidth]{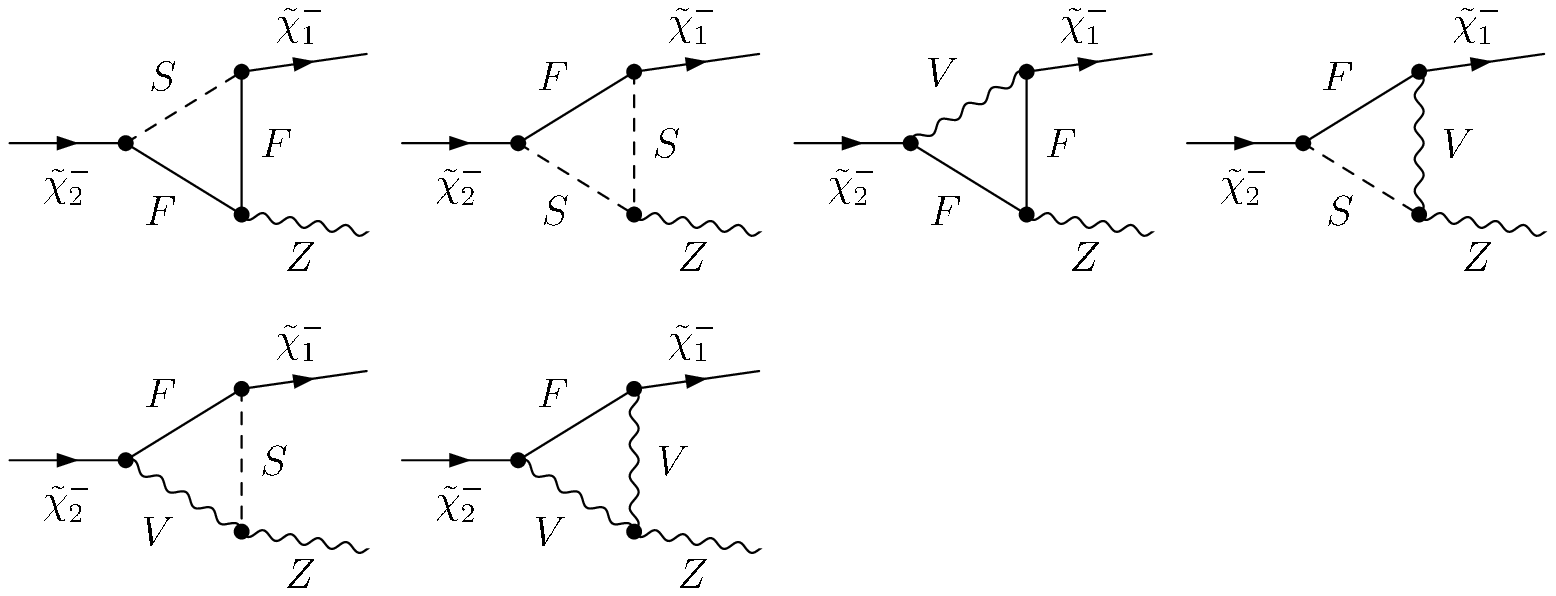}
\caption{
Generic Feynman diagrams for the decay 
$\decayCmCZ$.
$F$ can be a SM fermion, chargino or neutralino, $S$ can be a
sfermion or a Higgs boson, $V$ can be a $\ga$, $Z$ or $W^\pm$. 
}
\label{fig:fdCCZ}
\end{center}
\end{figure}

\begin{figure}[ht!]
\begin{center}
\includegraphics[width=0.90\textwidth]{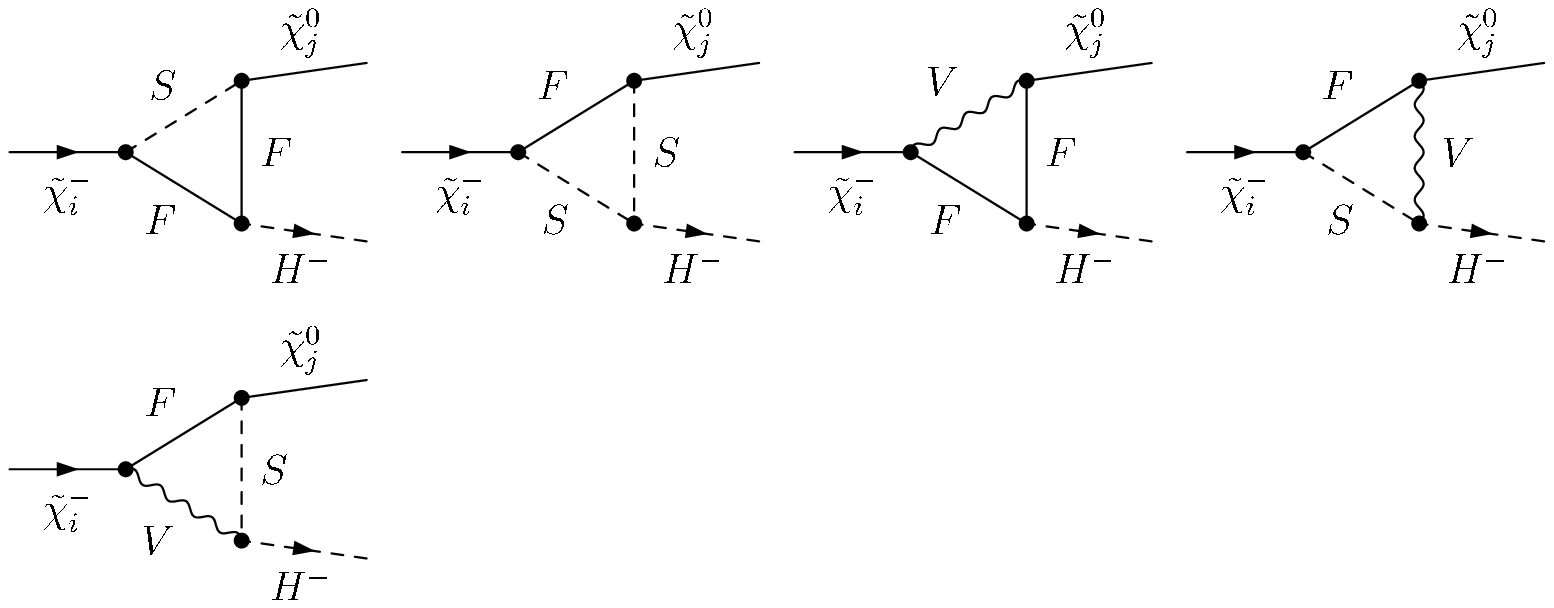}
\caption{
Generic Feynman diagrams for the decay 
$\decayCmNH$ ($i = 1,2, \; j = 1,2,3,4$).
$F$ can be a SM fermion, chargino or neutralino, $S$ can be a
sfermion or a Higgs boson, $V$ can be a $\ga$, $Z$ or $W^\pm$. 
Not shown are the diagrams with a $W^-$--$H^-$ or 
$G^-$--$H^-$ transition 
contribution on the external Higgs boson leg. 
}
\label{fig:fdCNH}
\end{center}
\vspace{-2em}
\end{figure}

\begin{figure}[ht!]
\begin{center}
\includegraphics[width=0.90\textwidth]{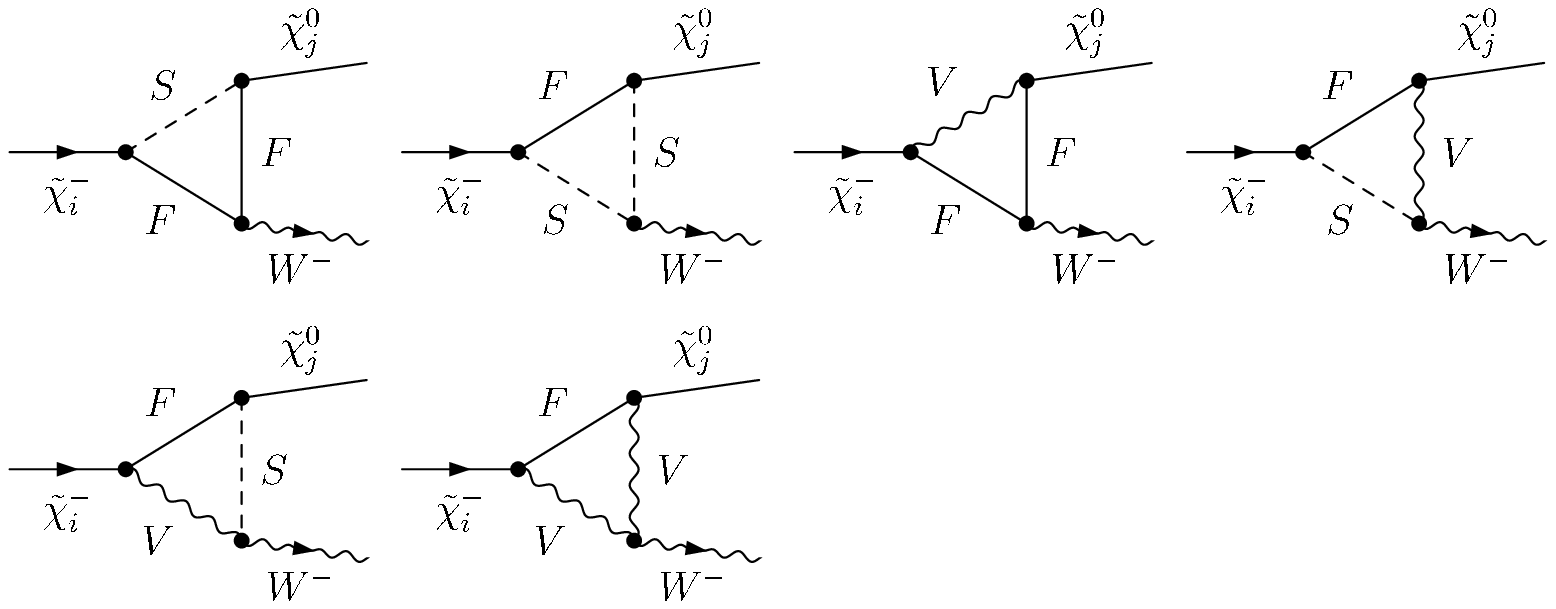}
\caption{
Generic Feynman diagrams for the decay 
$\decayCmNW$ ($i = 1,2, \; j = 1,2,3,4$).
$F$ can be a SM fermion, chargino or neutralino, $S$ can be a
sfermion or a Higgs boson, $V$ can be a $\ga$, $Z$ or $W^\pm$. 
}
\label{fig:fdCNW}
\end{center}
\vspace{-2em}
\end{figure}

\begin{figure}[ht!]
\begin{center}
\includegraphics[width=0.90\textwidth]{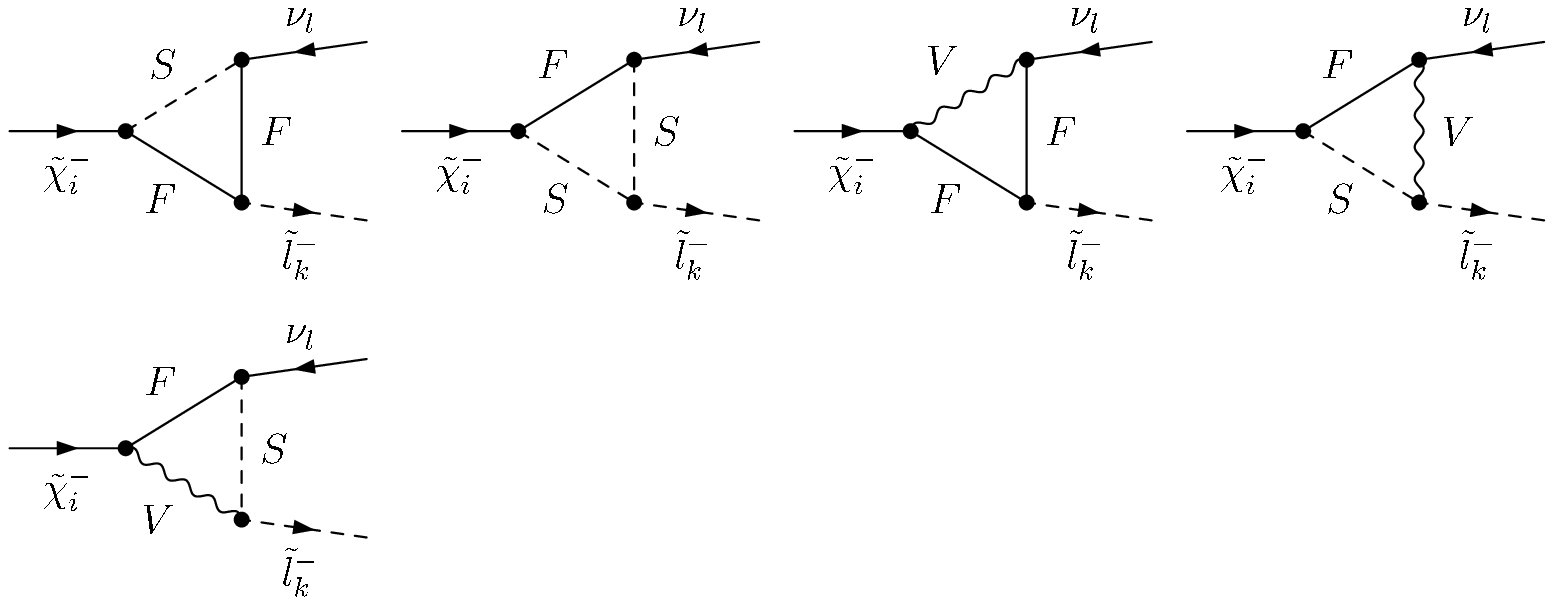}
\caption{
Generic Feynman diagrams for the decay 
$\decayCmnSl$ ($i = 1, 2, \; l = e, \mu, \tau, \; k = 1,2$).
$F$ can be a SM fermion, chargino or neutralino, $S$ can be a
sfermion or a Higgs boson, $V$ can be a $\ga$, $Z$ or $W^\pm$. 
}
\label{fig:fdCSln}
\end{center}
\vspace{-2em}
\end{figure}

\begin{figure}[ht!]
\begin{center}
\includegraphics[width=0.90\textwidth]{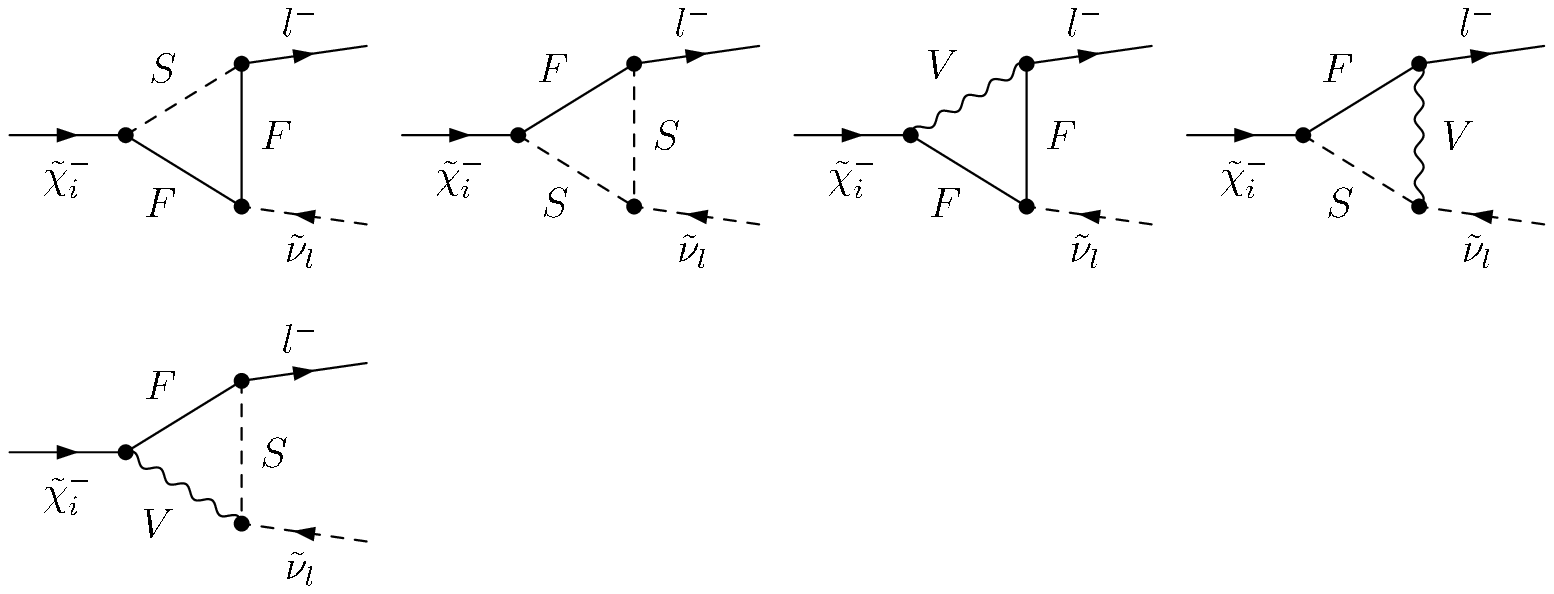}
\caption{
Generic Feynman diagrams for the decay 
$\decayCmlSn$ ($i = 1, 2, \; l = e, \mu, \tau$).
$F$ can be a SM fermion, chargino or neutralino, $S$ can be a
sfermion or a Higgs boson, $V$ can be a $\ga$, $Z$ or $W^\pm$. 
}
\label{fig:fdCSnl}
\end{center}
\vspace{-2em}
\end{figure}

In this section we give some details about the calculation of the
higher-order corrections to the chargino decays. Sample diagrams are
shown in \reffis{fig:fdCCh} -- \ref{fig:fdCSnl}. 
We only show the diagrams for the $\cham{i}$~decays, where the same
set of diagrams exist for the decays of~$\chap{i}$.
Not shown are the diagrams for real (hard or soft) photon 
radiation. They are obtained from the corresponding tree-level diagrams
by attaching a photon to the electrically charged
particles. The internal generically depicted particles in
\reffis{fig:fdCCh} -- \ref{fig:fdCSnl} are labeled as follows:
$F$ can be a SM fermion, chargino or neutralino,
$S$ can be a sfermion or a Higgs, $V$ can be a $\ga$, 
$Z$ or $W^\pm$.
Internally appearing Higgs bosons do not receive higher-order
corrections in their masses or couplings, which would correspond to
effects beyond one-loop. Furthermore, we found that using loop corrected
Higgs boson masses and couplings for the internal Higgs bosons leads to
a divergent result.
For external Higgs bosons, as described in
\refse{sec:higgs}, the appropriate $\matr{Z}$~factors are applied.

Not shown are the diagrams with a gauge boson (Goldstone)--Higgs
self-energy contribution on the external Higgs boson leg. They appear in 
the decay $\cham{2} \to \cham{1} h_{k}$ (${k} = 1, 2, 3$), 
\reffi{fig:fdCCh}, 
with a $Z/G$--$h_{k}$ transition, and in 
the decay {$\cham{i} \to \neu{j} H^-$}, ($i = 1,2, \; j = 1,2,3,4$),
\reffi{fig:fdCNH}, with a $W^-$/$G^-$--$H^-$ transition.%
\footnote{From a technical point of view, the $W^-$/$G^-$--$H^-$
transitions have been absorbed into the respective counterterms, 
while the $Z/G$--$h_k$ transition has been calculated explicitly.}%

On the other hand, the self-energy correction for the chargino decay to
a chargino/neu-tralino and a gauge boson, $\cham{2} \to \cham{1} Z$ or 
$\cham{i} \to \neu{j} W^-$ ($i = 1,2, \; j = 1,2,3,4$), vanish on mass shell, 
i.e.\ for $p^2 = \MZ^2$ ($p^2 = \MW^2$) due to $\eps \cdot p = 0$, 
where $p$ denotes the external momentum and $\eps$ the polarization
vector of the gauge boson.

In the figures we have furthermore omitted in general diagrams of
self-energy type 
of external (on-shell) particles. While the real part of 
such a loop
does not contribute to the decay width due to the on-shell
renormalization, the imaginary part, in product with an imaginary part
of a complex coupling (such as $\Al$) can give a real contribution to
the decay width. While these diagrams are not shown explicitly, they
have been taken into account in the analytical and numerical
evaluation.
The impact of those contributions will be discussed in 
\refse{sec:numeval}.

The diagrams and corresponding amplitudes have been obtained with 
\fa~\cite{feynarts}. The model file, including the MSSM counter
terms, is discussed in more detail in \citere{Stop2decay}. 
The further evaluation has been performed with 
\fc\ (and \looptools)~\cite{formcalc}. 
As regularization scheme for the UV-divergences we
have used constrained differential renormalization~\cite{cdr}, 
which has been shown to be equivalent to 
dimensional reduction~\cite{dred} at the \onel\ level~\cite{formcalc}. 
Thus the employed regularization preserves SUSY~\cite{dredDS,dredDS2}. 
All UV-divergences cancel in the final result.

The IR-divergences from diagrams with an internal photon have
to cancel with the ones from the corresponding real soft radiation, 
where we have included the soft photon contribution
following the description given in \citere{denner}. 
The IR-divergences arising from the diagrams involving a $\ga$ 
are regularized by introducing a finite photon mass,
$\lambda$. 
All IR-divergences, i.e.\ all divergences in the limit
$\lambda \to 0$, cancel to all orders
once virtual and real diagrams for one decay
channel are added.%
\footnote{
The only exception are the decays $\cham{i} \to \neu{2,3,4} W^-$. 
The shift to the neutralino on-shell masses via \refeq{Deltamneu}
results in an IR divergence at the two-loop level, i.e.\ here we find a
cancellation of the divergences ``only'' at the one-loop level, as
required for our one-loop calculation. 
The remaining IR divergences could be eliminated by a symmetry restoring
counterterm in the $\cha{i}\neu{2,3,4}W^\mp$ vertex, similar to the 
evaluation of the decay $\Stopz \to \Sbot_{1,2} W^+$ in \citere{Stop2decay}.
}%
~We have furthermore checked that our result does not depend on $\De E$
defining the energy cut that separates the soft from the hard
radiation. Our numerical results have been obtained for 
$\De E = 10^{-5} \times \mcha{i}$ 
for all channels except for $\DecayCmlSn{2}{e}$,
for which $\De E = 10^{-3} \times \mcha{2}$ has been used.%
\footnote{
The larger cut is necessary to obtain a better convergence of the 
integration over the three body phase space.
The contribution from nearly collinear photons 
(along the direction of the electron) leads to numerical instabilities 
in the integration. 
This problem is more acute for the heavier chargino decay, with a larger 
phase space and thus a larger electron energy.
}

\clearpage
\subsubsection*{Tree-level results}

For completeness we show here also the formulas that have been
used to calculate the tree-level decay widths:
\begin{align}
\label{CNHtree}
\Gtree(\decayCmNH) =& 
\KKL 
\KL |C(\cham{i},\neu{j},H^+)_L|^2 + |C(\cham{i},\neu{j},H^+)_R|^2 \KR
(\mcha{i}^2+\mneu{j}^2-\MHp^2) 
\right.
\non \\ & 
\left.
+ 4 \re \KKKL C(\cham{i},\neu{j},H^+)_L^* C(\cham{i},\neu{j},H^+)_R \KKKR
\mcha{i}\mneu{j}
\KKR
\non \\ &
\times \frac{\la^{1/2}(\mcha{i}^2,\mneu{j}^2,\MHp^2)}{32\pi \mcha{i}^3}
\quad (i = 1,2, \; j = 1,2,3,4)~, \\
\label{CNWtree}
\Gtree(\decayCmNW) =&
\KKL 
\KL |C(\cham{i},\neu{j},W^{+})_L|^2 + |C(\cham{i},\neu{j},W^{+})_R|^2 \KR
\right.
\non \\ &
\left.
\times 
\KL \mcha{i}^2+\mneu{j}^2-2\MW^2+\frac{(\mcha{i}^2-\mneu{j}^2)^2}{\MW^2} \KR
\right.
\non \\ & 
\left.
- 12\re \KKKL C(\cham{i},\neu{j},W^{+})_L^* C(\cham{i},\neu{j},W^{+})_R \KKKR
\mcha{i}\mneu{j}
\KKR
\non \\ &
\times \frac{\la^{1/2}(\mcha{i}^2,\mneu{j}^2,\MW^2)}{32\pi \mcha{i}^3}
\quad (i = 1,2, \; j = 1,2,3,4)~, \\
\label{CCHtree}
\Gtree(\decayCmCh) =& 
\KKL 
\KL |C(\cham{2},\chap{1},h_k)_L|^2 + |C(\cham{2},\chap{1},h_k)_R|^2 \KR
(\mcha{2}^2+\mcha{1}^2-m_{h_k}^2) 
\right.
\non \\ & 
\left.
+ 4\re \KKKL C(\cham{2},\chap{1},h_k)_L^* C(\cham{2},\chap{1},h_k)_R \KKKR
\mcha{2}\mcha{1}
\KKR
\non \\ &
\times \frac{\la^{1/2}(\mcha{2}^2,\mcha{1}^2,m_{h_k}^2)}{32\pi \mcha{2}^3}
\quad (k = 1,2,3)~,  \\
\label{CCZtree}
\Gtree(\decayCmCZ) =& 
\KKL 
\KL |C(\cham{2},\chap{1},Z)_L|^2 + |C(\cham{2},\chap{1},Z)_R|^2 \KR
\right.
\non \\ &
\left.
\times 
\KL \mcha{2}^2+\mcha{1}^2-2\MZ^2+\frac{(\mcha{2}^2-\mcha{1}^2)^2}{\MZ^2} \KR
\right.
\non \\ & 
\left.
- 12\re \KKKL C(\cham{2},\chap{1},Z)_L^* C(\cham{2},\chap{1},Z)_R \KKKR
\mcha{2}\mcha{1}
\KKR
\non \\ &
\times \frac{\la^{1/2}(\mcha{2}^2,\mcha{1}^2,\MZ^2)}{32\pi \mcha{2}^3}
~, \\
\label{CSlntree}
\Gtree(\decayCmnSl) =&  
 |C(\nu_l,\cham{i},\slk^\dagger)_L|^2 
(\mcha{i}^2 - \mslk^2)
\non \\ & 
\times \frac{\la^{1/2}(\mcha{i}^2,0,\mslk^2)}{32\pi \mcha{i}^3}
\quad (i = 1,2, \; l = e, \mu, \tau, \; k = 1,2)~, \\
\Gtree(\decayCmlSn) =&  
\KKL 
\KL |C(\cham{i},\bar l,\Sn)_L|^2 + |C(\cham{i},\bar l,\Sn)_R|^2 \KR
(\mcha{i}^2 + \ml^2 - \msn^2)
\right.
\non \\ &
\left.
+ 4\re \KKKL C(\cham{i},\bar l,\Sn)_L^* C(\cham{i},\bar l,\Sn)_R \KKKR \mcha{i} \ml
\KKR
\non \\ & 
\times \frac{\la^{1/2}(\mcha{i}^2,\ml^2,\msn^2)}{32\pi \mcha{i}^3}
\quad (i = 1,2, \; l = e, \mu, \tau)~,  
\end{align}
where $\lambda(x,y,z) = (x - y - z)^2 - 4yz$ and the couplings 
$C(a, b, c)$ can be found in the \fa~model files~\cite{feynarts-mf}.
$C(a, b, c)_{L,R}$ denote the part of the coupling which
is proportional to $\om_\mp = \ed{2}(\id \mp \ga_5)$.


\subsubsection*{Comparison with other calculations}

We have performed a detailed comparison with \citere{liebler} for
the decay $\DecayCmNW{2}{1}$, where the chargino/neutralino sector
is renormalized differently to our prescription. 
After a correction of the charge
renormalization in \citere{liebler} we found good agreement at the
level expected for different renormalization schemes in the
chargino/neutralino sector, see \citeres{Stop2decay,liebler} for details.


\newcommand{\SN}{${\cal S}$}
\newcommand{\SE}{\ensuremath{{\cal S}_>}}
\newcommand{\SZ}{\ensuremath{{\cal S}_<}}

\section{Numerical analysis}
\label{sec:numeval}

In this section we present a numerical analysis of all decay
channels ($\decayCmxy$). 
We restrict ourselves here to the decay of the charginos with 
{\em negative} charge. Small differences with respect to $\chap{i}$
decays occur for complex parameters~\cite{ChaDecCPVYang,ChaDecCPVEberl}. 
These effects are not
the scope of this paper, and we have checked that for the parameter
choices in this paper these effects are small.
In the various figures we show the decay width and its
relative correction at the tree-level (``tree'') and at the one-loop
level (``full''), 
\begin{align}
\Gtree &\equiv \Gtree(\decayCmxy)~, \\
\Gfull &\equiv \Ga^{\rm full}(\decayCmxy)~, \\
\De\Ga/\Ga &\equiv \frac{\Gfull - \Gtree}{\Gtree}~.
\end{align}
The total decay width is defined as the sum of all 38 decay widths, 
\begin{align}
\Ga_{\rm tot}^{\rm tree} &\equiv \sum_{{\rm xy}} \Gtree(\decayCmxy)~, \\
\Ga_{\rm tot}^{\rm full} &\equiv \sum_{{\rm xy}} \Gfull(\decayCmxy)~.
\end{align}
We also show the absolute and relative changes of the branching ratios,
\begin{align}
\br^{\rm tree} &\equiv 
   \frac{\Ga^{\rm tree}(\decayCmxy)}{\Ga_{\rm tot}^{\rm tree}}~, \\
\br^{\rm full} &\equiv 
   \frac{\Ga^{\rm full}(\decayCmxy)}{\Ga_{\rm tot}^{\rm full}}~, \\
\De\br/\br &\equiv \frac{\br^{\rm full} - \br^{\rm tree}}{\br^{\rm full}} 
\label{brrel}
\end{align}
The last quantity is crucial to analyze the impact of the one-loop
corrections on the phenomenology at the LHC and the ILC, 
see below.


\subsection{Parameter settings}
\label{sec:parameter}

The renormalization scale, $\mu_R$, has been set to the mass of the
decaying particle, i.e.\ $\mu_R~=~\mcha{i}$.
The SM parameters are chosen as follows, see also \cite{pdg}:
\begin{itemize}

\item Fermion masses\index{leptonmasses}:
\begin{align}
m_e    &= 0.51099891\mev~, & m_{\nu_e}     &= 0\mev~, \non \\
m_\mu  &= 105.658367\mev~, & m_{\nu_{\mu}}  &= 0\mev~, \non \\
m_\tau &= 1776.84\mev~,    & m_{\nu_{\tau}} &= 0\mev~, \non \\
m_u &= 53.8\mev~,          & m_d &= 53.8\mev~, \non \\ 
m_c &= 1.27\gev~,          & m_s &= 104\mev~, \non \\
m_t &= 172.0\gev~,         & m_b(m_b) &= 4.25\gev~.
\end{align}
$m_u$ and $m_d$ are effective parameters, calculated through the hadronic
contributions to:
\begin{align}
\Delta\alpha_{\text{had}}^{(5)}(M_Z) &= 
      \frac{\alpha}{\pi}\sum_{f = u,c,d,s,b}
      Q_f^2 \Bigl(\ln\frac{M_Z^2}{m_f^2} - \frac 53\Bigr)~.
\end{align}

\item The CKM matrix has been set to unity.

\item Gauge boson masses:
\begin{align}  
  \MZ & = 91.1876 \gev~, \quad \MW =  80.399 \gev~,
\end{align}  
\item Coupling constants:
\begin{align}  
  \al & = \frac{e^2}{4\pi} = 1/137.035999679 ~.
\end{align}  
\end{itemize}

The Higgs sector quantities (masses, mixings, etc.) have been
evaluated using {\tt FeynHiggs} version\,2.7.4
\cite{feynhiggs,mhiggslong,mhiggsAEC,mhcMSSMlong}, where we used
the running top mass for the evaluation.

When performing an analysis involving complex parameters
it should be noted that the results for physical observables are
affected only 
by certain combinations of the complex phases of the 
parameters $\mu$, the trilinear couplings $\At$, $\Ab$, $\Atau$, \ldots, and
the gaugino mass parameters $\MOne$, $\MTwo$,
$M_3$~\cite{MSSMcomplphasen,SUSYphases}.
It is possible, for instance, to rotate the phase $\phiMz$ away.
Experimental constraints on the (combinations of) complex phases 
arise in particular from their contributions to electric dipole moments of
heavy quarks~\cite{EDMDoink}, of the electron and 
the neutron, see \citeres{EDMrev2,EDMPilaftsis} and references therein, 
and of the deuteron~\cite{EDMRitz}. 
A recent review can be found in \citere{EDMrev3}.
Using the convention that $\phiMz = 0$ (i.e.\ $\MTwo$ real
and positive) as done in this paper, in particular 
the phase $\phimu$ is tightly constrained~\cite{plehnix} to be close to
zero or~$\pi$. Accordingly, we also choose $\mu$ to be real. To be in
agreement with the anomalous magnetic moment of the muon, $(g-2)_\mu$,
we furthermore choose $\mu$ to be positive~\cite{newBNL,g-2,newDavier}.
On the other hand, the bounds on the phases of the third generation
trilinear couplings are much weaker. 
The phases of $\mu$ and $\Atau$ (the scalar top and bottom sector as
well as the gluino enter only as virtual particles,
i.e.\ subleading, in the decays evaluated here) appear 
only in the combination $(\phiatau + \phimu)$
(or in different combinations together with phases of $M_1$ or $M_3$). 
Setting $\phimu = 0$ (see above) as well as $\phigl = 0$ 
(we do not consider the gluino phase in this paper) leaves us with 
$\phiatau$ and $\phiMe$ as the only complex valued
parameters. (The dependence on $\phiat$ and $\phiab$ on decays involving 
SUSY particles has recently been analyzed in detail in
\citeres{SbotRen,Stop2decay}.)

We will show the results for some representative numerical examples. 
The SUSY parameters are chosen according to the scenario, \SN,
shown in \refta{tab:para}, but with one of the parameters varied.

\begin{table}[t!]
\renewcommand{\arraystretch}{1.5}
\BC
\begin{tabular}{|c||c|c|c|c|c|c|c|c|}
\hline
Scen.\ & $\tb$ & $\MHp$ & $\mcha{2}$ & $\mcha{1}$ 
       & $\MslL$ & $\MslR$ & $\Al$ 
\\ \hline\hline
\SN & 20 & 160 & 600 & 350 & 300 & 310 & 400 
\\ \hline
\end{tabular}
\caption{MSSM parameters for the initial numerical
  investigation; all 
  masses are in GeV. $\MOne$, $\MTwo$ and $\mu$ are chosen such that the
  values for $\mcha{1}$ and $\mcha{2}$ and \refeq{M1M2} are fulfilled
(see text).
The diagonal soft SUSY-breaking parameters in the squark sector 
are set to $1200 \gev$
and the corresponding trilinear couplings to $2400 \gev$.
}
\label{tab:para}
\EC
\renewcommand{\arraystretch}{1.0}
\end{table}

\noindent
The absolute value of $\MOne$ (see above) is fixed via the GUT
relation (with $|\MTwo| \equiv \MTwo$)
\begin{align}
|\MOne| &= \frac{5}{3} \tan^2 \thw \MTwo \approx \edz \MTwo~.
\label{M1M2}
\end{align}
For the numerical analysis we fix $\mcha{1,2}$. From the two chargino
masses and \refeq{M1M2} the numerical values for $|\MOne|$, $\MTwo$ and $\mu$
can be evaluated avoiding any ambiguity (leaving $\phiMe$ as a free
parameter), see below.
As default we use $\phiMe = 0$.
This ensures that parameter variations keep the
variation of the phase space at a minimum level and the numerical
results mainly show the effects from the higher-order corrections
to the decay widths. 

We invert the mass relations of \refeq{eq:mcha} 
in order to express the parameters
$\mu$ and $\MTwo$ (which are taken to be real, see above) as a
function of chargino masses.
The resulting quartic equation leads to two sets of solutions. 
Each set of solutions satisfies the relations
\begin{align}
|\MTwo \mu  - \MW^2 \SZb| &= 
\eta_{\chi}(\MTwo \mu  - \MW^2 \SZb) =   \mcha{1}\mcha{2}, 
\label{eq:mcha1mcha2}
\\
\MTwo^2 + \mu^2 + 2 \MW^2 &= \mcha{1}^2 + \mcha{2}^2
=:  2 \bmcha{}^2~,
\end{align}
where $\eta_\chi=\pm 1$ is defined by Eq.~(\ref{eq:mcha1mcha2}).
Choosing $\mu$ and $\MTwo$ real and positive and a lower experimental
bound on $\mcha{1}$ of $\sim 100 \gev$ (see below) yields $\eta_\chi = +1$.
The above two relations are symmetric under an exchange of $\MTwo$
and $\mu$. One finds two solutions, 
\begin{align}
\label{mu-gt-M2}
\{\mu, \MTwo\}  &= \{x_+,x_- \}~, \\
\label{mu-lt-M2}
\{\mu, \MTwo\}  &= \{x_-,x_+ \}~,
\end{align}
with
\begin{align}
x^2_{\pm} & =  
\bmcha{}^2 - \MW^2 
\pm \KKL 
\KL \bmcha{}^2 - \MW^2 \KR^2 
- \KL  \mcha{1}\mcha{2} + \MW^2 s_{2 \be} \KR^2
\KKR^{\edz}\, .
\label{eq.x2pm}
\end{align}
The two choices (\ref{mu-gt-M2}) and (\ref{mu-lt-M2}) correspond to a
more higgsino- or gaugino-like heavy chargino, respectively (and the
reverse for the lighter chargino). 
While the phase space of a chargino decay is not
affected by this choice, the various branching ratios are. Consequently,
for our numerical analysis we define two scenarios,
\begin{align}
\label{eq.SE}
\SE &: \mu > \MTwo \quad (\cha{2} \mbox{~more higgsino-like})~, \\
\label{eq.SZ}
\SZ &: \mu < \MTwo \quad (\cha{2} \mbox{~more gaugino-like})~.
\end{align}

The numerical scenarios are defined such that many decay modes are open
simultaneously to permit an analysis of as many channels as possible.
Only the channels $\cha{2} \to \neu{4} H^\pm/W^\pm$ and 
$\cha{1} \to \neu{2,3,4} H^\pm/W^\pm$ are closed, mostly due to
\refeq{M1M2}. 
We will start with a variation of $\mcha{2}$, and 
analyze later the results for varying $\phiMe$.
The scenarios are in agreement with the 
MSSM Higgs boson searches at LEP~\cite{LEPHiggsSM,LEPHiggsMSSM}. 
Too small values of the lightest Higgs boson mass would be reached for 
$\tb \lsim 5$ within \SN\ as given in \refta{tab:para}.
Furthermore, 
the following exclusion limits for neutralinos~\cite{pdg} hold in
our numerical scenarios: 
\begin{align}
\mneu{1} &> 46 \gev, \;
\mneu{2} > 62 \gev, \;
\mneu{3} > 100 \gev, \;
\mneu{4} > 116 \gev~.
\end{align}
It should be noted that the limit for $\mneu{1}$ arises solely from
\refeq{M1M2}. In the absence of this condition, no limit on a light
neutralino mass exists, see \citere{masslessx} and references therein.

A few examples of the chargino and neutralino masses are shown in
\refta{tab:chaneu}, while Higgs and slepton masses are shown in
\refta{tab:higgsslep}. The values of $\mcha{1,2}$ allow copious 
production of the charginos in SUSY cascades at the LHC.
Furthermore, the production of $\cha{1}\champ{2}$ or $\chap{1}\cham{1}$
at the ILC(1000), i.e.\ with $\sqrt{s} = 1000 \gev$, via 
$e^+e^- \to \cha{1}\champ{1,2}$ will be possible,
with all the subsequent decay modes (\ref{CNH}) -- (\ref{CSnl})
being (in principle) open. The clean environment of the ILC would
permit a detailed study of the chargino decays~\cite{ilc,lhcilc}.
For the parameters of scenarios $\SE$ and $\SZ$~, see \refta{tab:para}
and \refeqs{eq.SE}, (\ref{eq.SZ}), we show the cross sections for 
chargino pair production at the ILC(1000), varying the chargino
masses. The calculation has been performed at the tree-level, which
is sufficient to get an overview about the expected number of events.
Higher-order corrections could change these numbers by
\order{10\%}~\cite{ChaProd,BaroII}. 
For the values in \refta{tab:para} and unpolarized beams
we find, for $\SE$ ($\SZ$),
$\si(e^+e^- \to \cha{1}\champ{2}) \approx 4\, (12)~{\rm fb}$, and
$\si(e^+e^- \to \chap{1}\cham{1}) \approx 55\, (80)~{\rm fb}$. 
Choosing appropriate polarized beams these cross sections can be
enhanced by a factor of approximately $2$ to $3$.
An integrated luminosity of $\sim 1\, \iab$ would yield about 
$4-12 \times 10^3$ $\cha{1}\champ{2}$ events and about
$55 - 80 \times 10^3$ $\chap{1}\cham{1}$ events, with appropriate
enhancements in the case of polarized beams.
The ILC environment would result in an accuracy of
the relative branching ratio \refeq{brrel} close to the statistical
uncertainty. The statistical precisions for the various mass and
polarization assumptions, assuming a (hypothetical) 10\%~BR and $1\, \iab$, 
are shown in the two right-most columns in \refta{tab:ILCproduction}.  
Depending on the combination of allowed decay channels a determination of 
the branching ratios at the per-cent level might be achievable in the 
high-luminosity running of the ILC(1000).

\begin{table}[htb!]
\renewcommand{\arraystretch}{1.6}
\BC
\begin{tabular}{|c||c|r|r|r|r|r|r||c|c|c|}
\hline
Scenario
& 
$\tb$ & 
$\mcha{2}$ & $\mcha{1}$
      & $\mneu{4}$ & $\mneu{3}$ & $\mneu{2}$ & $\mneu{1}$ 
  & $\mu$      & $\MTwo$    & $\MOne$  
\\ \hline\hline
\SE & 20 & 600.0 & 350.0 & 599.4 & 586.0 & 350.1 & 171.4 & 
             581.8 & 362.1 & 172.8 %
 \\ \hline
\SZ & 20 & 600.0 & 350.0 & 600.1 & 366.5 & 358.7 & 267.2 & 
             362.1 & 581.8 & 277.7
 \\ \hline
\end{tabular}
\caption{The chargino and neutralino masses in \SE\ and \SZ.
We also show the values for the ``derived'' parameters $\MOne$, 
$\MTwo$ and $\mu$. 
All masses are in GeV, rounded to $0.1 \gev$ to show the 
  size of small mass
  differences, which can determine whether a certain decay channel is
  kinematically closed or open.
}
\label{tab:chaneu}
\EC
\renewcommand{\arraystretch}{1.0}
\end{table}
%
\begin{table}[htb!]
\renewcommand{\arraystretch}{1.6}
\BC
\begin{tabular}{|c||c|r|r|r|r|r||r|r|r|r|}
\hline
Scenario
& 
$\ms{\mu}{1} $ & 
$\ms{\mu}{2} $ & 
$\ms{\tau}{1}$ & 
$\ms{\tau}{2}$ & 
${m_{\tilde{\nu}_\mu}} $ & 
${m_{\tilde{\nu}_\tau}} $ & 
$\MHp$ & $\mHe$  & $\mHz$ & $\mHd$ 
\\ \hline\hline
\SE & 303.4 & 313.3 & 273.8 & 339.5 & 293.0 & 293.0 & 160.0 & 127.4 & 
        137.7 & 140.0
 \\ \hline
\SZ & 303.7 & 313.1 & 287.5 & 328.0 & 293.0 & 293.0 & 160.0 & 127.2 & 
        137.5 & 140.4
 \\ \hline
\end{tabular}
\caption{The slepton and Higgs masses in \SE\ and \SZ.
The selectron and electron sneutrino masses 
are equal to those of the corresponding smuon and muon sneutrino
up to a few tenths of $\gev$. 
  All masses are in GeV, rounded to $0.1 \gev$.
}
\label{tab:higgsslep}
\EC
\renewcommand{\arraystretch}{1.0}
\end{table}
%
\begin{table}[htb!]
\renewcommand{\arraystretch}{1.6}
\BC
\begin{tabular}{|c||c|c|c|c|c|c|c|c||c|c|}
\hline
Scen. & $\mcha{1},\mcha{2}[{\rm GeV}]$ & process & $\si_{0,0}[{\rm fb}]$
& $\si_{\rm pol}[{\rm fb}]$ & 
stat.\ prec.$_{0,0}$ & stat.\ prec$_{\rm pol}$ 
\\ \hline\hline
\SE & 350, 600 & $e^+e^- \to \chap{1}\cham{1}$  & 58.3 & 167.7 & 1\% & 1\%
\\ \hline
\SE & 450, 600 & $e^+e^- \to \chap{1}\cham{1}$  & 19.8 & 56.0 & 2\% & 1\%
\\ \hline
\SZ & 350, 600 & $e^+e^- \to \chap{1}\cham{1}$  & 77.7 & 185.0 & 1\% & 1\%
\\ \hline
\SZ & 450, 600 & $e^+e^- \to \chap{1}\cham{1}$  & 29.1 & 64.2 & 2\% & 1\%
\\ \hline\hline
\SE & 350, 500 & $e^+e^- \to \cha{1}\champ{2}$  & 21.5 & 56.5 & 2\% & 1\%
\\ \hline
\SE & 350, 600 & $e^+e^- \to \cha{1}\champ{2}$  & 4.1 & 10.5 & 5\% & 3\%
\\ \hline
\SZ & 350, 500 & $e^+e^- \to \cha{1}\champ{2}$  & 34.2 & 93.1 & 2\% & 1\%
\\ \hline
\SZ & 350, 600 & $e^+e^- \to \cha{1}\champ{2}$  & 11.5 & 31.9 & 3\% & 2\%
\\ \hline
\end{tabular}
\caption{
Chargino production cross sections at the ILC 1000. 
Here $\si_{0,0}$ denotes the cross section for
unpolarized beams, while $\si_{\rm pol}$ 
denotes that with electron and positron polarization $-80\%$ and
$+60\%$, respectively. 
The two right-most columns show the statistical precision for a
  (hypothetical) branching ratio of 10\% assuming an integrated
  luminosity of $1\, \iab$, rounded to 1\%.
}
\label{tab:ILCproduction}
\EC
\renewcommand{\arraystretch}{1.0}
\end{table}

The numerical results we show in the next subsections are of
course dependent on the choice of the SUSY parameters. Nevertheless, they
give an idea of the relevance of the full one-loop corrections.
Channels (and their respective one-loop corrections) that may look 
unobservable due to the smallness of their BR in the plots shown below, 
could become important if other channels are kinematically forbidden.
Consequently, the one-loop corrections to {\em all} channels are
evaluated analytically, but in the numerical analysis we only show the
channels that are kinematically open in our numerical scenarios.


The results shown in this and the following subsections consist of 
``tree'', which denotes the tree-level 
value and of ``full'', which is the decay width including {\em all} one-loop 
corrections as described in \refse{sec:calc}.
We start the numerical analysis with $\cham{2}$~decay widths evaluated as
a function of $\mcha{2}$.
For the ``tree'' contributions, 
we start at $\mcha{2} = 469.3 \gev$,  
its lowest value (for fixed $\mcha{1} = 350 \gev$), 
up to $\mcha{2} = 1000\gev$. 
For the ``full'' results we start at $\mcha{2} = 475 \gev$.
For lower values of $\mcha{2}$ the on-shell renormalization scheme 
adopted here leads to insufficient results,
as $\MTwo$ approaches $\mu$, and the potential problems described in
\refse{sec:chaneu} start to take effect. However, this affects only a
parameter range of $\sim 5 \gev$.

In the figures below
the upper panels contain the results for the absolute
value of the various decay widths, $\Ga(\decayCmxy)$ (left) and
the relative correction from the full one-loop contributions
(right). The lower panels show the same results for 
$\br(\decayCmxy)$.
The vertical lines indicate where $\mcha{1} + \mcha{2} = 1000 \gev$, 
i.e.\ the maximum reach of the ILC(1000).

Since all parameters are chosen real no contributions from
absorptive parts of self-energy type corrections on external legs can
contribute at the one-loop level. This will be different in
\refse{sec:1LphiMe}.

In order to understand the qualitative behavior of the various decay widths
we first briefly summarize the composition of the relevant charginos and
neutralinos in the two numerical scenarios and their couplings to other
particles. In our notation the charginos are a mixture of gaugino
($\tilde G$) and higgsino ($\tilde H$), while the neutralinos are
mixtures of bino ($\tilde B$), wino ($\tilde W$), and higgsino,
\begin{align}
\cha{i} = [\tilde{H}^\pm + \tilde{W}^\pm]_i, \qquad 
\neu{j}= [\tilde{H}^0 + \tilde{W}^3 + \tilde{B} ]_j~.
\end{align}
For the two numerical scenarios and depending on the relative size of
$\mcha{2}$ or $\mcha{1}$ we show the decomposition in
\refta{tab:chaneu_character}.

\begin{table}[htb!]
\renewcommand{\arraystretch}{1.5}
\BC
\begin{tabular}{|c||c||c|c|c|c|c||c|c|}
\hline
Scenario & $\mu,M_2$ &\  $\quad\neu{1}\quad$  &  $\quad\neu{2}\quad$  & $\quad\neu{3}\quad$  &  $\quad\neu{4}\quad$ 
       &  $\quad\cha{1}\quad$  &  $\quad\cha{2}\quad$ 
\\ \hline\hline
\SE, ``low $\mcha{2}$''  & $\mu \gsim M_2$ & $\tilde{B}$ & $\tilde{W}|\tilde{H}$ &  $\tilde{H}$ &  $\tilde{H}|\tilde{W}$
& $\tilde{G}|\tilde{H}$ & $\tilde{G}|\tilde{H}$
\\ \hline
\SE, ``high $\mcha{2}$'' & $\mu \gg M_2$  & $\tilde{B}$ & $\tilde{W}$ & $\tilde{H}$ & $\tilde{H}$ 
& $\tilde{G}$ & $\tilde{H}$ 
\\ \hline
\SZ, ``low $\mcha{2}$''  & $\mu \lsim M_2$ & $\tilde{B}$ & $\tilde{H}|\tilde{W}$ & $\tilde{H}$ & $\tilde{W}|\tilde{H}$
& $\tilde{G}|\tilde{H}$ & $\tilde{G}|\tilde{H}$
\\ \hline
\SZ, ``high $\mcha{2}$'' & $\mu \ll M_2$ & $\tilde{H}$ & $\tilde{H}$ & $\tilde{B}$ & $\tilde{W}$  
& $\tilde{H}$ & $\tilde{G}$  
\\ \hline\hline
\SE, ``low $\mcha{1}$''  & $\mu \gg M_2$  & $\tilde{B}$  &  $\tilde{W}$ &  $\tilde{H}$ &   $\tilde{H}$
& $\tilde{G}$ & $\tilde{H}$ 
\\ \hline
\SE, ``high $\mcha{1}$'' & $\mu \gsim M_2$  & $\tilde{B}$ & $\tilde{W}|\tilde{H}$ &  $\tilde{H}$ &  $\tilde{H}|\tilde{W}$
& $\tilde{G}|\tilde{H}$ & $\tilde{G}|\tilde{H}$
\\ \hline
\SZ, ``low $\mcha{1}$''  & $\mu \ll M_2$  &  $\tilde{B}$  &  $\tilde{H}$  &   $\tilde{H}|\tilde{B}$ & $\tilde{W}$
& $\tilde{H}$ & $\tilde{G}$  
\\ \hline
\SZ, ``high $\mcha{1}$'' & $\mu \lsim M_2$  & $\tilde{B}$ & $\tilde{H}|\tilde{W}$ & $\tilde{H}$ & $\tilde{W}|\tilde{H}$
& $\tilde{G}|\tilde{H}$ & $\tilde{G}|\tilde{H}$
\\ \hline
\end{tabular}
\caption{
Character of the charginos and neutralinos in the analyzed regions of parameter space,
 indicating their main electroweak eigenstate component(s).
We introduce the short-hand notation: 
$\tilde{B}$ = bino,  
$\tilde{W}$ = wino,  
$\tilde{H}$ = higgsino, 
$\tilde{G}$ = gaugino, 
$\tilde{G}|\tilde{H}$ = mixed gaugino-higgsino (for charginos), 
and 
$\tilde{W}|\tilde{H}$, $\tilde{H}|\tilde{W}$, $\tilde{H}|\tilde{B}$ = mixed wino-higgsino, mixed higgsino-wino, mixed higgsino-bino (for neutralinos).
}
\label{tab:chaneu_character}
\EC
\renewcommand{\arraystretch}{1.0}
\end{table}

The coupling structure relevant in the chargino decays can be read off
from the interaction Lagrangians, which is symbolically given by 
\begin{align}
{\cal L}_{\cha{}\neu{}W^\pm} = ~&
{\cal L}_{\tilde{H}^\pm\tilde{H}^0 W^\pm}
 + 
{\cal L}_{\tilde{W}^\pm\tilde{W}^3W^\pm }
\label{eq:Lagr.CNW} 
~,\\ 
{\cal L}_{\cha{}\neu{}H^\pm} = ~&
{\cal L}_{\tilde{W}^\pm\tilde{H}^0 H^\pm }
 + 
{\cal L}_{\tilde{H}^\pm \tilde{W}^3 H^\pm} 
+ {\cal L}_{\tilde{H}^\pm\tilde{B}H^\pm }
~,
\label{eq:Lagr.CNH}
\\ 
{\cal L}_{\cha{i}\cha{j}Z} = ~&
{\cal L}_{\tilde{W}^\pm \tilde{W}^\pm Z} + 
{\cal L}_{\tilde{H}^\pm \tilde{H}^\pm Z}
+ \delta_{ij} {\cal L}_{\ldots}
\label{eq:Lagr.CCZ}  ~,
\\ 
{\cal L}_{\cha{i}\cha{j}h_k} = ~&
{\cal L}_{\tilde{W}^\pm \tilde{H}^\pm h_k } 
\label{eq:Lagr.CCH} ~,\\
{\cal L}_{\cha{} \nu_l \slk} = ~&
{\cal L}_{\tilde{W}^\pm \nu_l \tilde{l}_L} 
+ {\cal L}_{\tilde{H}^\pm \nu_l \tilde{l}_R} (\propto \ml\tb)
\label{eq:Lagr.CSln} ~,\\
{\cal L}_{\cha{} l \tilde{\nu}_l} = ~&
{\cal L}_{\tilde{W}^\pm l \tilde{\nu}_l} 
+ {\cal L}_{\tilde{H}^\pm l \tilde{\nu}_l} (\propto \ml\tb)~,
\label{eq:Lagr.CSnl}
\end{align}
where all other field combinations correspond to ``forbidden''
interactions. The allowed combinations can be summarized as follows, 
\begin{itemize}
\item Decay into $W^\pm$: only gaugino-gaugino and
      higgsino-higgsino interaction, but no bino-gaugino-$W$. 
\item Decay into $H^\pm$: only gaugino-higgsino interaction.
\item Decay into $Z$: only gaugino-gaugino and higgsino-higgsino
      interaction. 
\item Decay into $h_k$: only gaugino-higgsino interaction.
\item Decay into $\nu_l\, \Sl$: gaugino-$\tilde{l}_L$
      (EW), higgsino-$\tilde{l}_R$ (Yukawa, suppressed with $\ml$). 
\item Decay into $l\, \tilde{\nu_{l}}$: gaugino (EW), higgsino
      (Yukawa, suppressed with $\ml$). 
\end{itemize}


\subsection{Full one-loop results for varying \boldmath{$\mcha{2}$}}
\label{sec:1Lmcha2}

\begin{figure}[htb!]
\begin{center}
\begin{tabular}{c}
\includegraphics[width=0.49\textwidth,height=7.5cm]{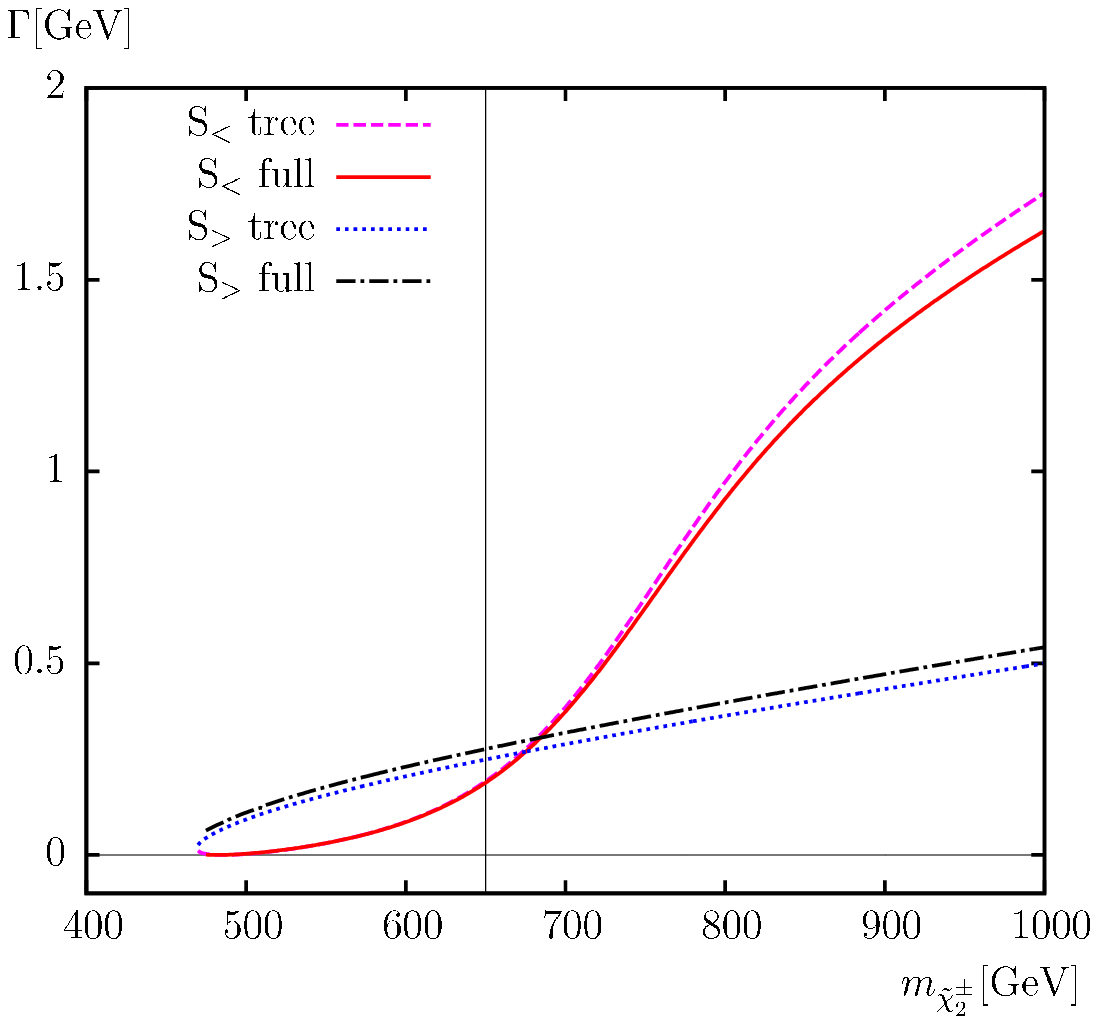}
\hspace{-4mm}
\includegraphics[width=0.49\textwidth,height=7.5cm]{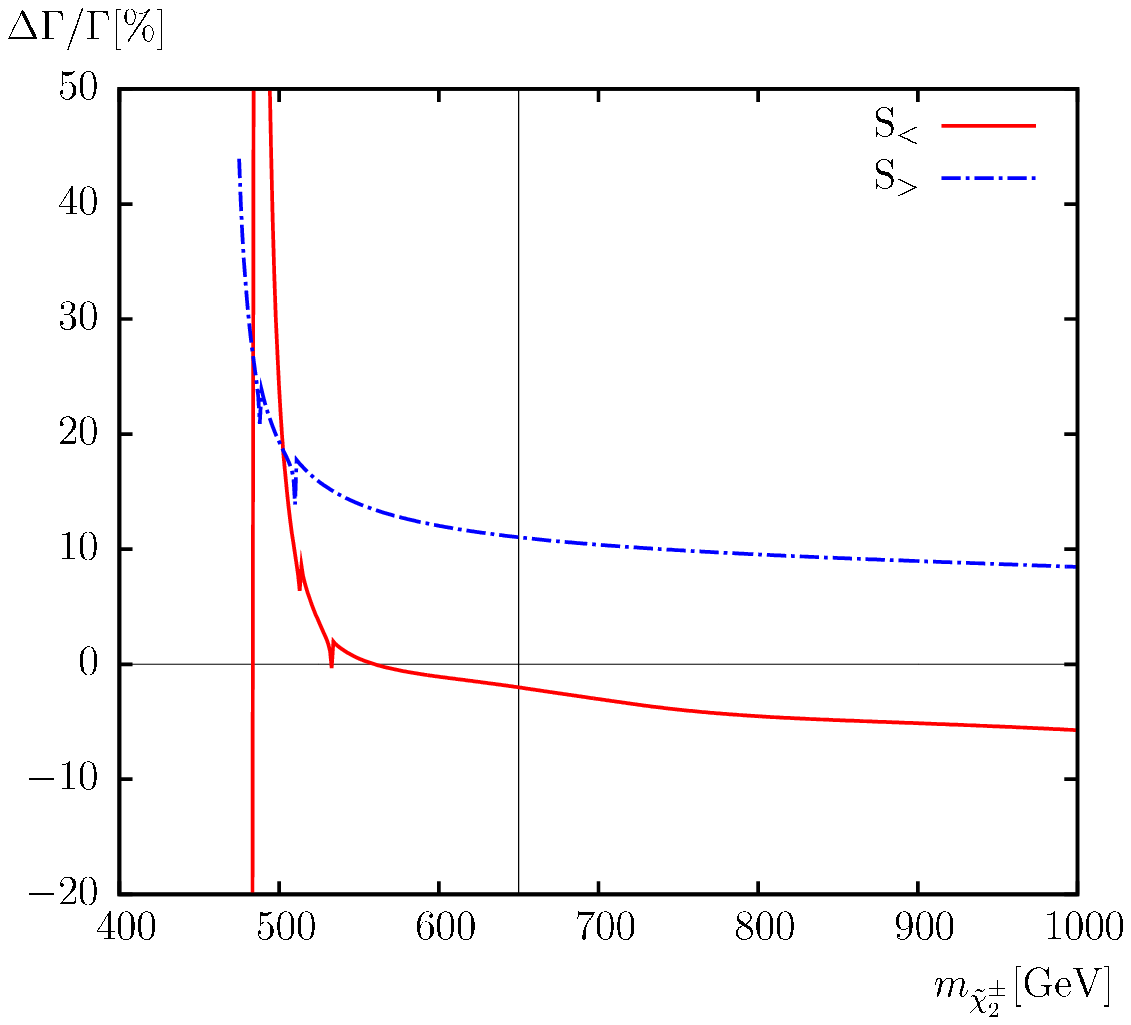} 
\\[4em]
\includegraphics[width=0.49\textwidth,height=7.5cm]{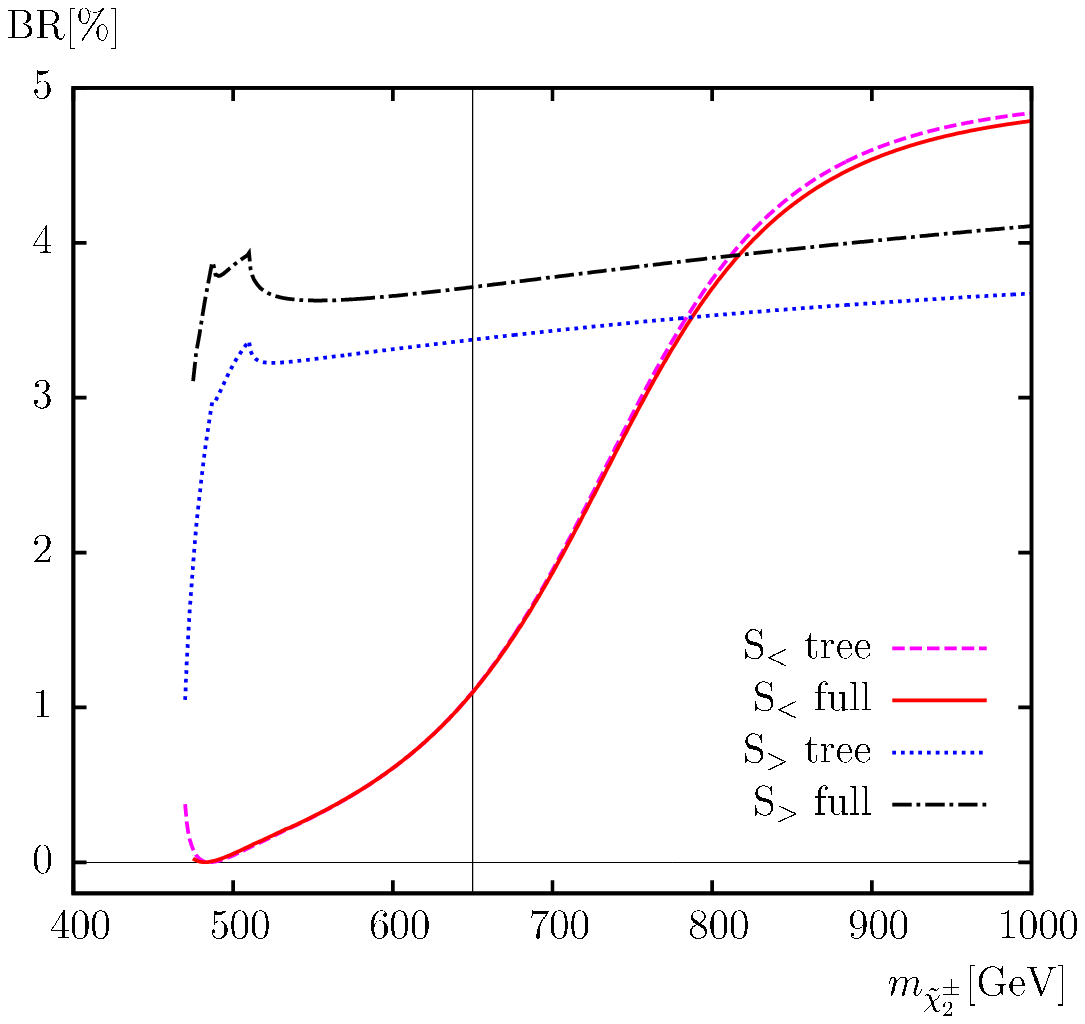}
\hspace{-4mm}
\includegraphics[width=0.49\textwidth,height=7.5cm]{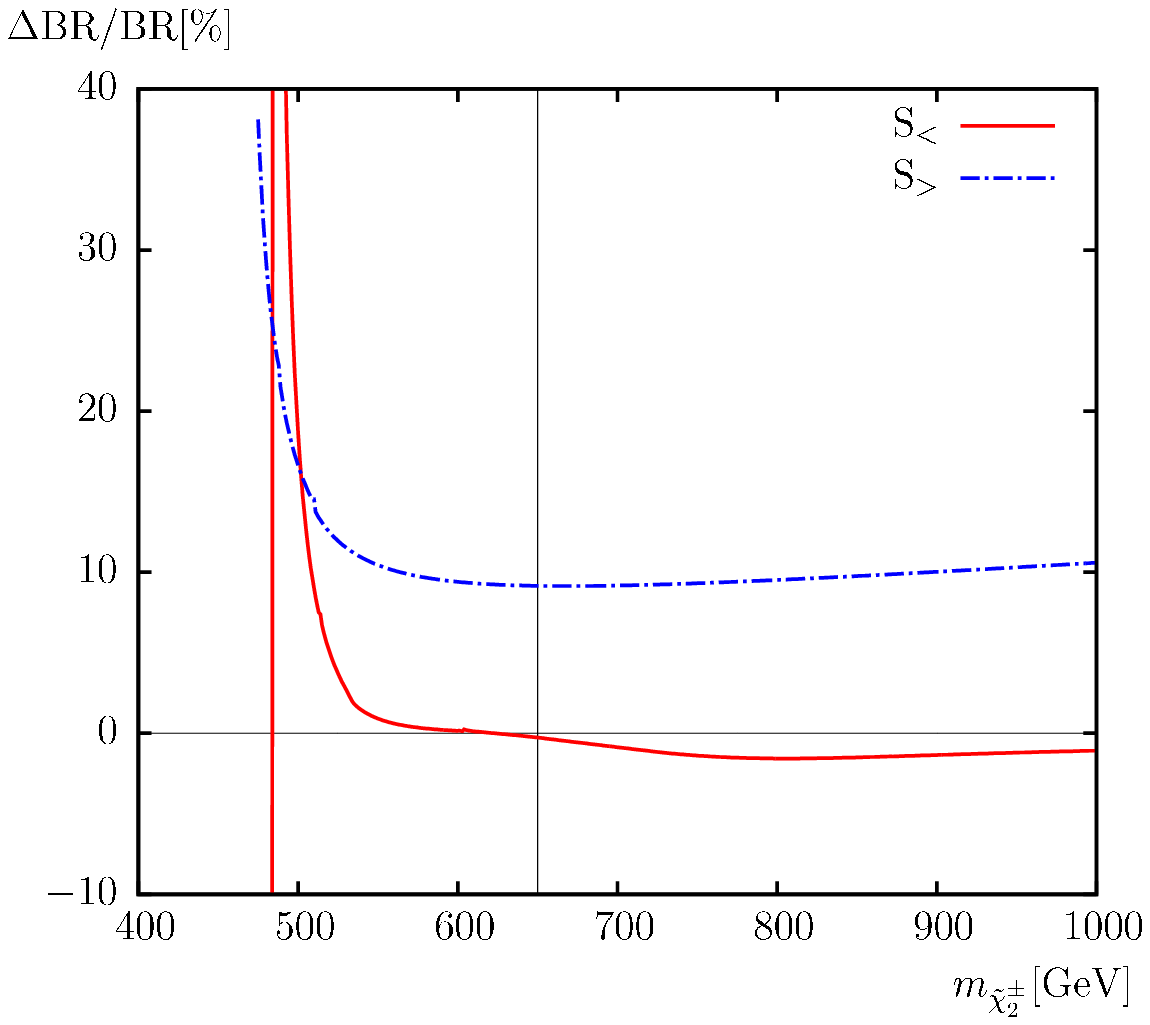}
\end{tabular}
\vspace{2em}
\caption{$\Ga(\DecayCmNH{2}{1})$. 
  Tree-level (``tree'') and full one-loop (``full'') corrected 
  decay widths are shown with the parameters chosen according to \SN\
  (see \refta{tab:para}), with $\mcha{2}$ varied.
  The upper left plot shows the decay width, the upper right plot shows 
  the relative size of the corrections.
  The lower left plot shows the BR, the lower right plot shows 
  the relative size of the BR.
  The vertical lines indicate where $\mcha{1} + \mcha{2} = 1000 \gev$, 
  i.e.\ the maximum reach of the ILC(1000).
}
\label{fig:mC2.cha2neu1hp}
\end{center}
\end{figure}

We start our numerical analysis with the decays $\DecayCmNH{2}{i}$ 
($i = 1, 2, 3$, where $\DecayCmNH{2}{4}$ is kinematically forbidden).
The results for $\DecayCmNH{2}{1}$ are presented in
\reffi{fig:mC2.cha2neu1hp}. 
The dips, visible best in the upper right plots are due to thresholds in
the vertex corrections.
At $\mcha{2} = 488 \gev$ the $\cha{1} \Hz$ 
threshold can be seen in \SE\ (with the $\cha{1} \Hd$ threshold only $\sim 0.5 \gev$
above remains nearly invisible). A second dip can be seen at 
$\mcha{2} = 510\ (513) \gev$ in \SE\ (\SZ) corresponding to the 
$\neu{2} H^-$ threshold, and a third dip in \SZ\ at $\mcha{2} = 533 \gev$ 
corresponding to the $\neu{3} H^-$ threshold. 
In \SE\ and \SZ\ for low $\mcha{2}$ the higgsino
component of the heavy chargino allows the decay to the bino-like
$\neu{1}$ and the charged Higgs. At large $\mcha{2}$ within \SZ\
the lightest neutralino changes from bino to wino dominated and
the decay proceeds via the ``fully allowed'' higgsino-wino interaction with a
corresponding increase in the decay width. 
Within \SE, on the other
hand, we find a pure higgsino-bino induced decay. 
The loop corrections
can be larger than $20\%$ 
at the smallest $\mcha{2}$ in both scenarios,
see the discussion in \refse{sec:chaneu} on the $\mu\simeq M_2$ region.
At large $\mcha{2}$ the size of the loop corrections levels out at
$\sim -6\%\ (+9\%)$ in \SZ\ (\SE). The BR in \SZ\ rises from zero 
at $\mcha{2}\simeq 485\gev$
to above $4.5\%$, whereas in \SE\ it is found around $4\%$ for
most of the parameter space. 
The corrections to the BR's reach the level of $30\%$
at the smallest $\mcha{2}$. In \SE\ they are found to be around
$\sim -1\%$ for $\mcha{2} \gsim 600 \gev$, whereas in \SZ\ they remain
at the $10\%$ level. In view of the ILC precision at the per-cent
level, at least in \SZ\ the loop corrections are highly relevant.

\begin{figure}[htb!]
\begin{center}
\begin{tabular}{c}
\includegraphics[width=0.49\textwidth,height=7.5cm]{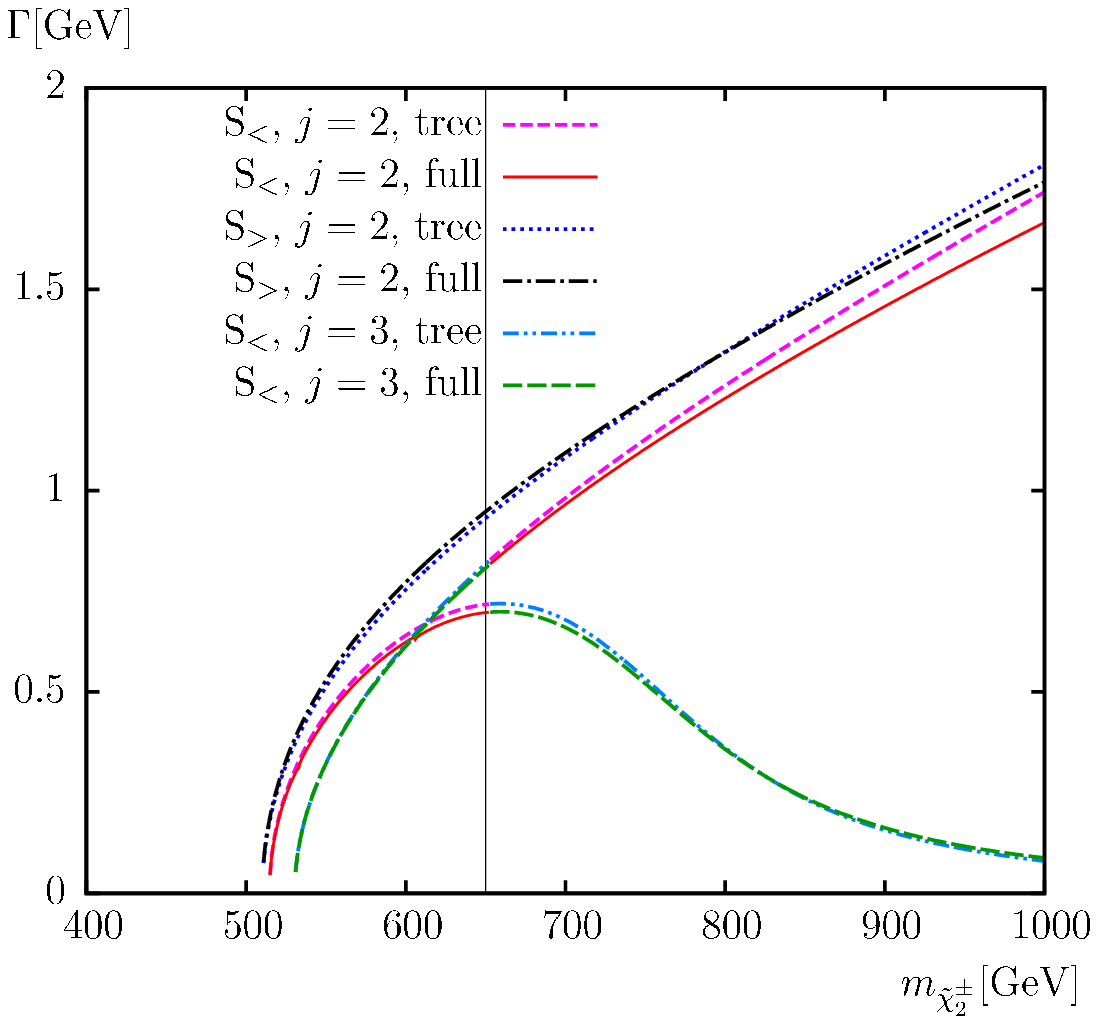}
\hspace{-4mm}
\includegraphics[width=0.49\textwidth,height=7.5cm]{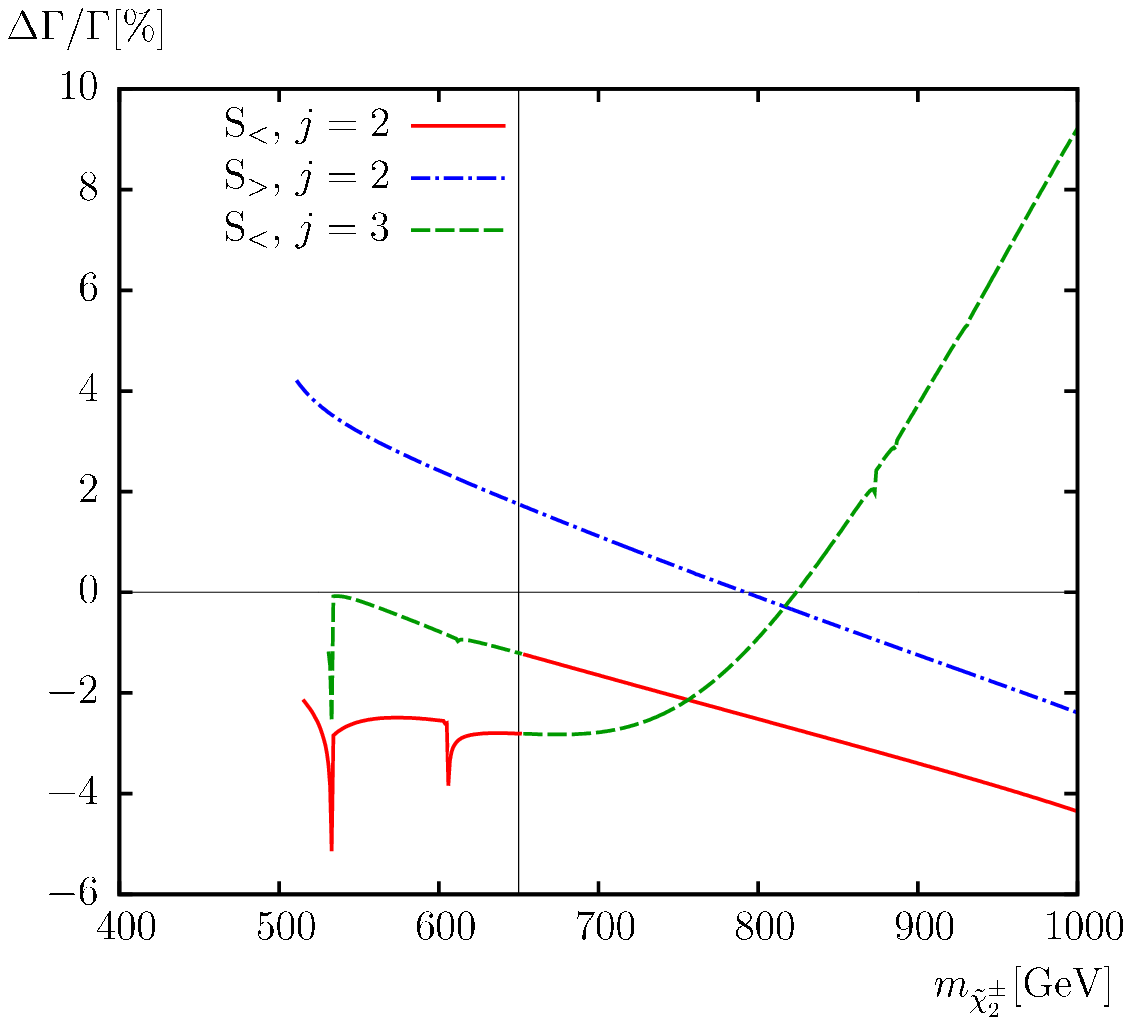} 
\\[4em]
\includegraphics[width=0.49\textwidth,height=7.5cm]{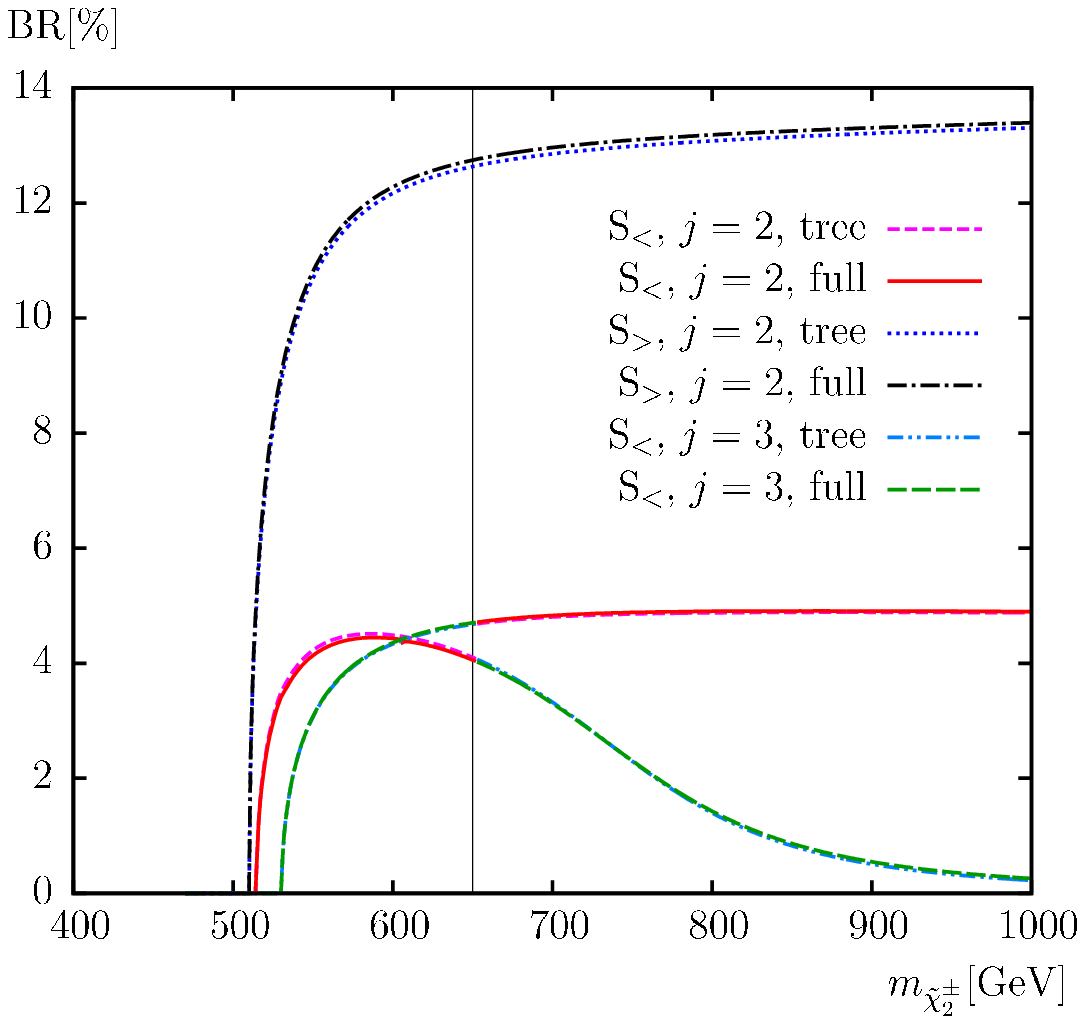}
\hspace{-4mm}
\includegraphics[width=0.49\textwidth,height=7.5cm]{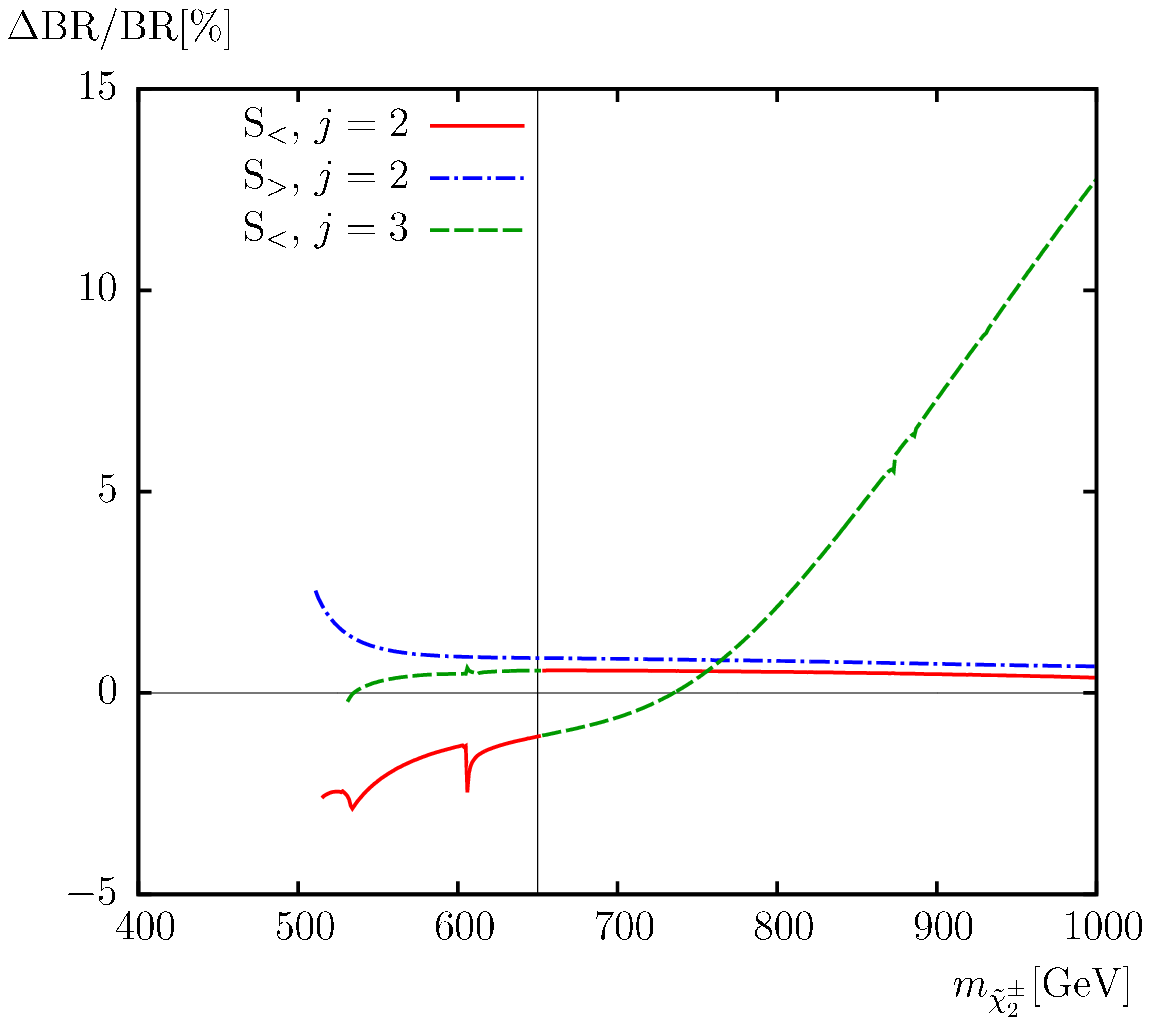}
\end{tabular}
\vspace{2em}
\caption{
  $\Ga(\DecayCmNH{2}{j})$ for $j = 2,3$.
  Tree-level (``tree'') and full one-loop (``full'') corrected 
  decay widths are shown with the parameters chosen according to \SN\
  (see \refta{tab:para}), with $\mcha{2}$ varied.
  The upper left plot shows the decay width, the upper right plot shows 
  the relative size of the corrections.
  The lower left plot shows the BR, the lower right plot shows 
  the relative size of the BR.
  The vertical lines indicate where $\mcha{1} + \mcha{2} = 1000 \gev$, 
  i.e.\ the maximum reach of the ILC(1000).
}
\label{fig:mC2.cha2neujhp}
\end{center}
\end{figure}

The results for $\DecayCmNH{2}{j}$ ($j = 2,3$) are shown in
\reffi{fig:mC2.cha2neujhp}. 
The dip in \SZ\ at $\mcha{2} = 533 \gev$ visible in $\DecayCmNH{2}{2}$
stems from the $\neu{3} H^-$ threshold.
Finally dips in $\DecayCmNH{2}{2(3)}$ can be observed at 
$\mcha{2} \sim 604, 606, (873, 886) \gev$, 
which correspond, respectively, to the thresholds 
$\neu{2}\to\neu{1}Z $, $\neu{2}\to\neu{1}h $
 ($\neu{3}\to \neu{1}h$, $\neu{3}\to \cha{1} W^{\mp}$).
As before the general behavior of the decay
widths can be understood from the decomposition of the heavy chargino
and the neutralinos. In \SE\ only the decay $\DecayCmNH{2}{2}$ is
kinematically allowed, and the width rises up to $\sim 1.7 \gev$ for
large $\mcha{2}$. In \SZ\ two observations can be made. 
At $\mcha{2} = 652 \gev$ the $\neu{2}$ and $\neu{3}$ ``switch
character''.%
\footnote{
The fact that $\mcha{2} + \mcha{1} = 652 + 350 \approx 1000 \gev$ (and
consequently, the ``character switch'' occurs nearly at the vertical
line) is a pure numerical coincidence.
}%
~For this chargino mass one finds $\mneu{2} = \mneu{3}$, and
the neutralino mixing matrix exhibits a discontinuity. As a consequence,
as can be observed in \reffi{fig:mC2.cha2neujhp}, the $\neu{2}$-curves
continue for $\neu{3}$ and vice versa. 
At low masses we find ``partially''
allowed wino/higgsino interaction for $\DecayCmNH{2}{2,3}$. At high
masses the $\neu{2}$ couples as a higgsino to the charged wino, whereas
the $\neu{3}$ becomes a bino and decouples from the wino-like $\cha{2}$,
and the 
decay width goes to zero. The relative size of the one-loop corrections
to the decay width vary roughly between $-4\%$ and $+4\%$. 
The $\br(\DecayCmNH{2}{2})$ in \SE\ reaches $\sim 12\%$ already close to
threshold, whereas in \SZ\ it rises only up to $\sim 5\%$
due to the
larger total width, see below. The $\br(\DecayCmNH{2}{3})$ in \SZ\ 
reaches a
maximum of $4\%$ around $\mcha{2} \approx 600 \gev$ and goes to zero for
larger chargino masses. The relative size of the one-loop effects on the
BR's remains below $3\%$ in the ILC(1000) relevant mass region,
which, however, can still be relevant, see \refta{tab:ILCproduction}.

\begin{figure}[htb!]
\begin{center}
\begin{tabular}{c}
\includegraphics[width=0.49\textwidth,height=7.5cm]{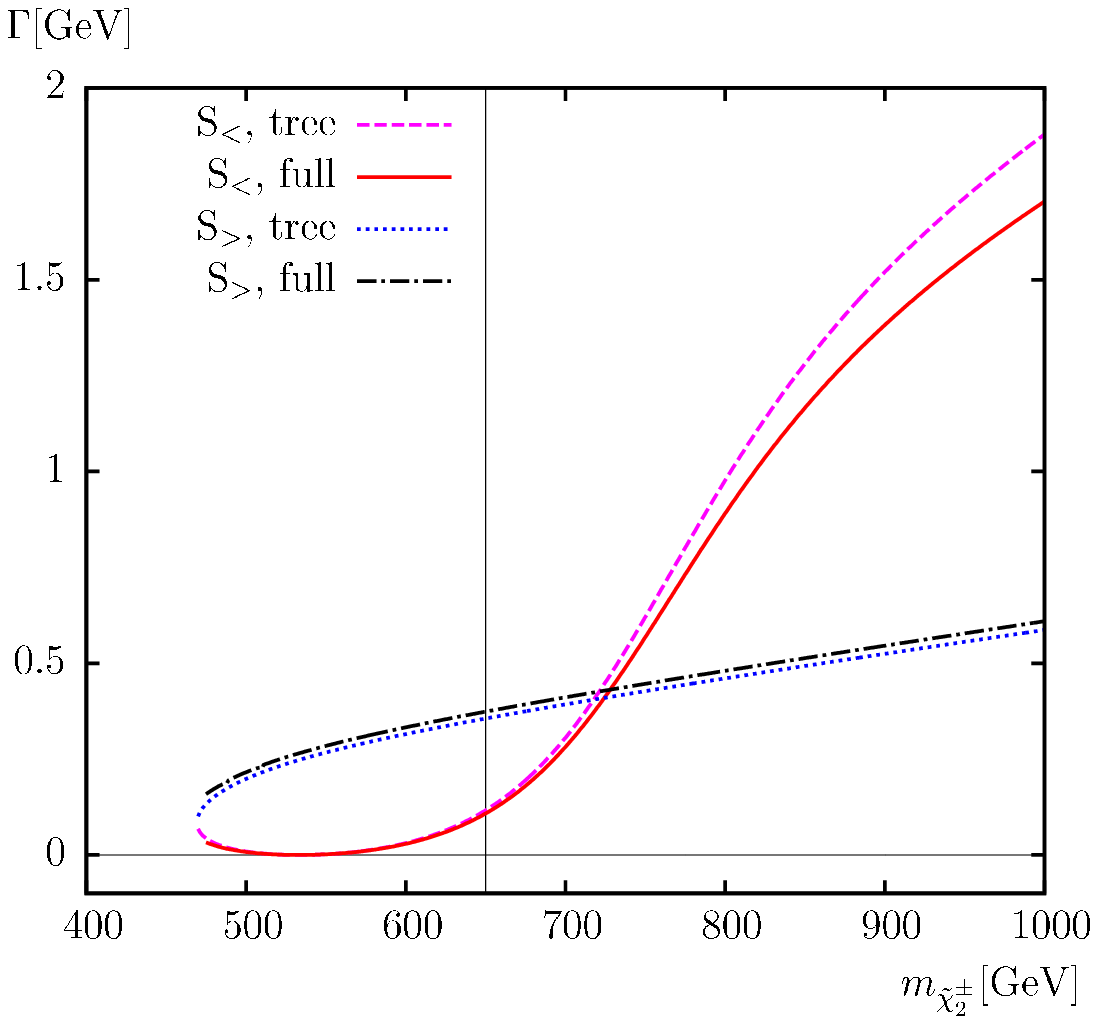}
\hspace{-4mm}
\includegraphics[width=0.49\textwidth,height=7.5cm]{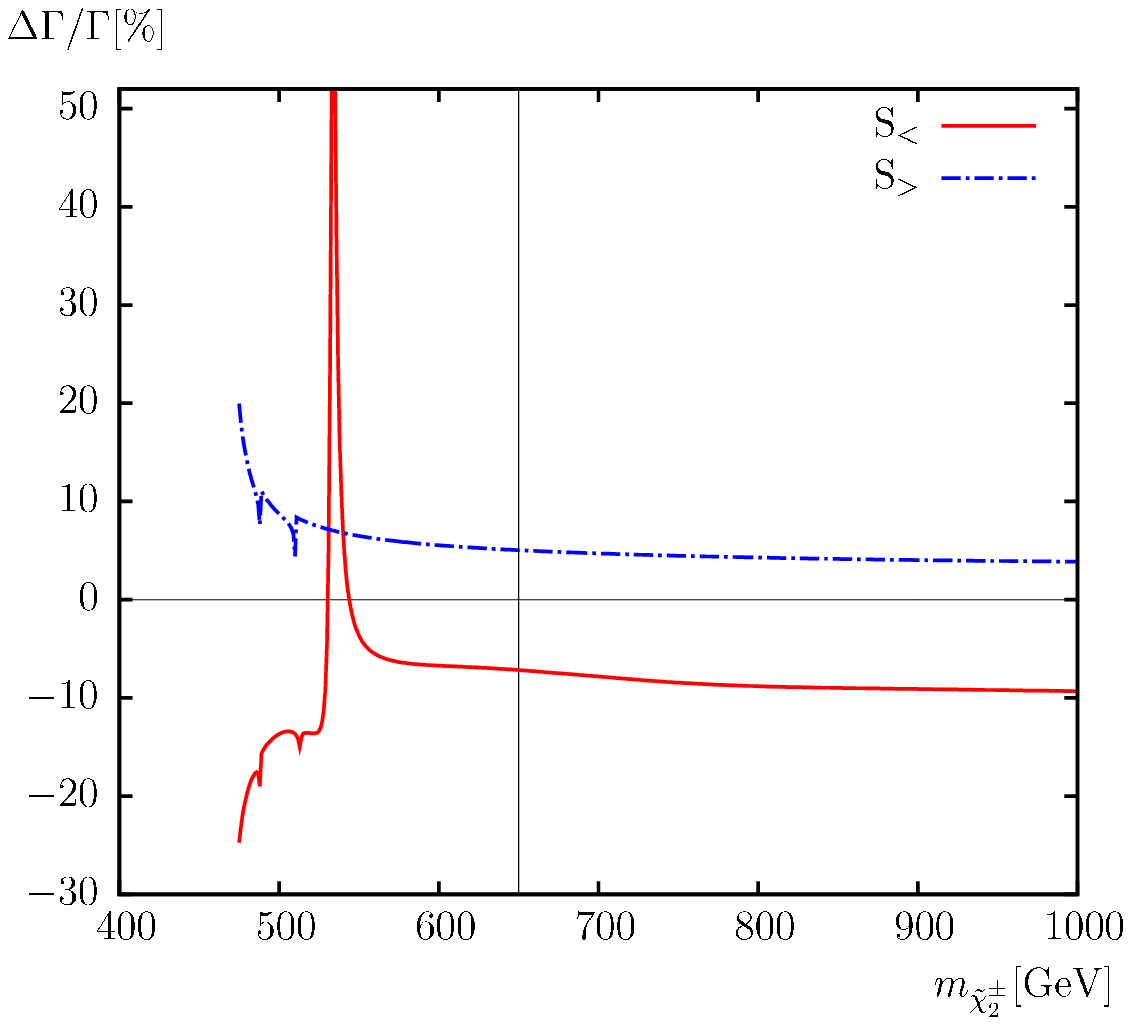} 
\\[4em]
\includegraphics[width=0.49\textwidth,height=7.5cm]{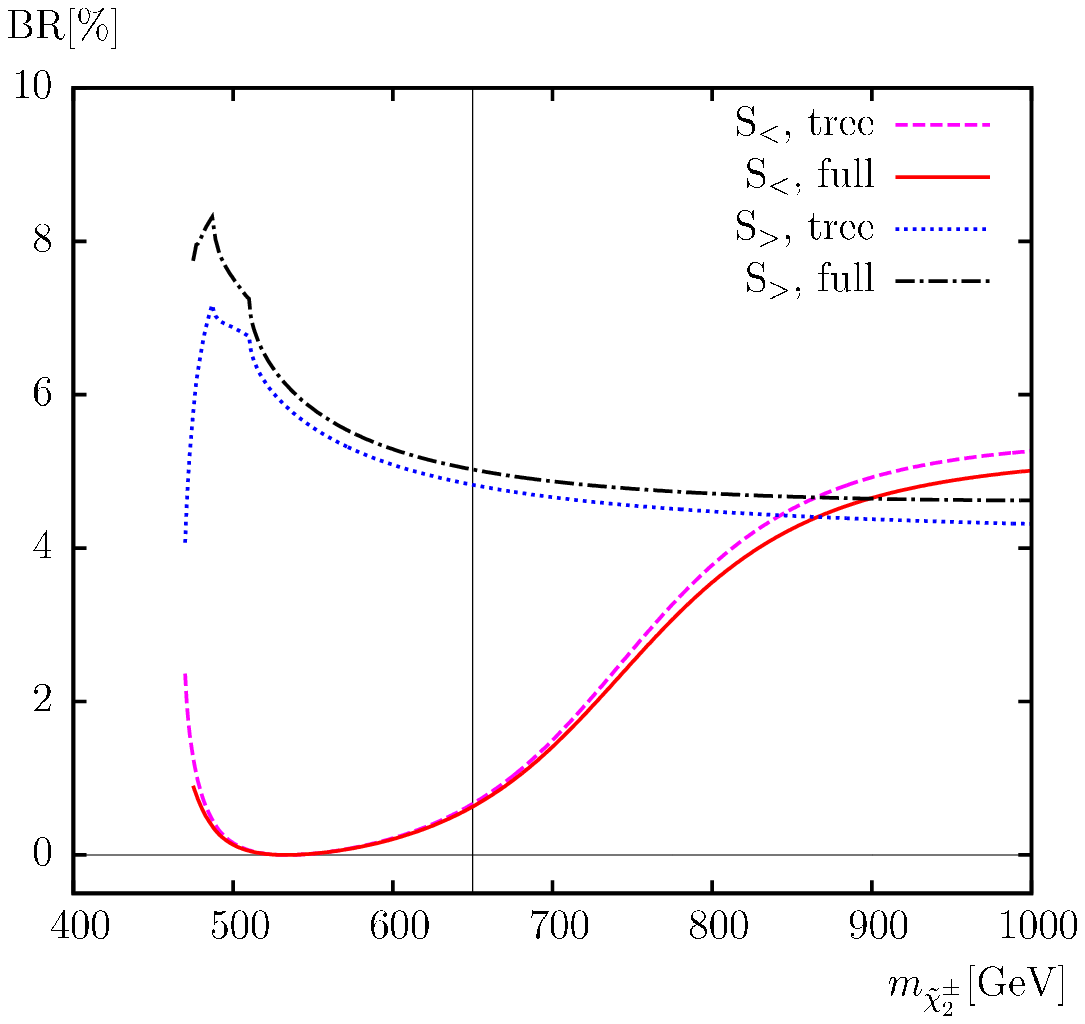}
\hspace{-4mm}
\includegraphics[width=0.49\textwidth,height=7.5cm]{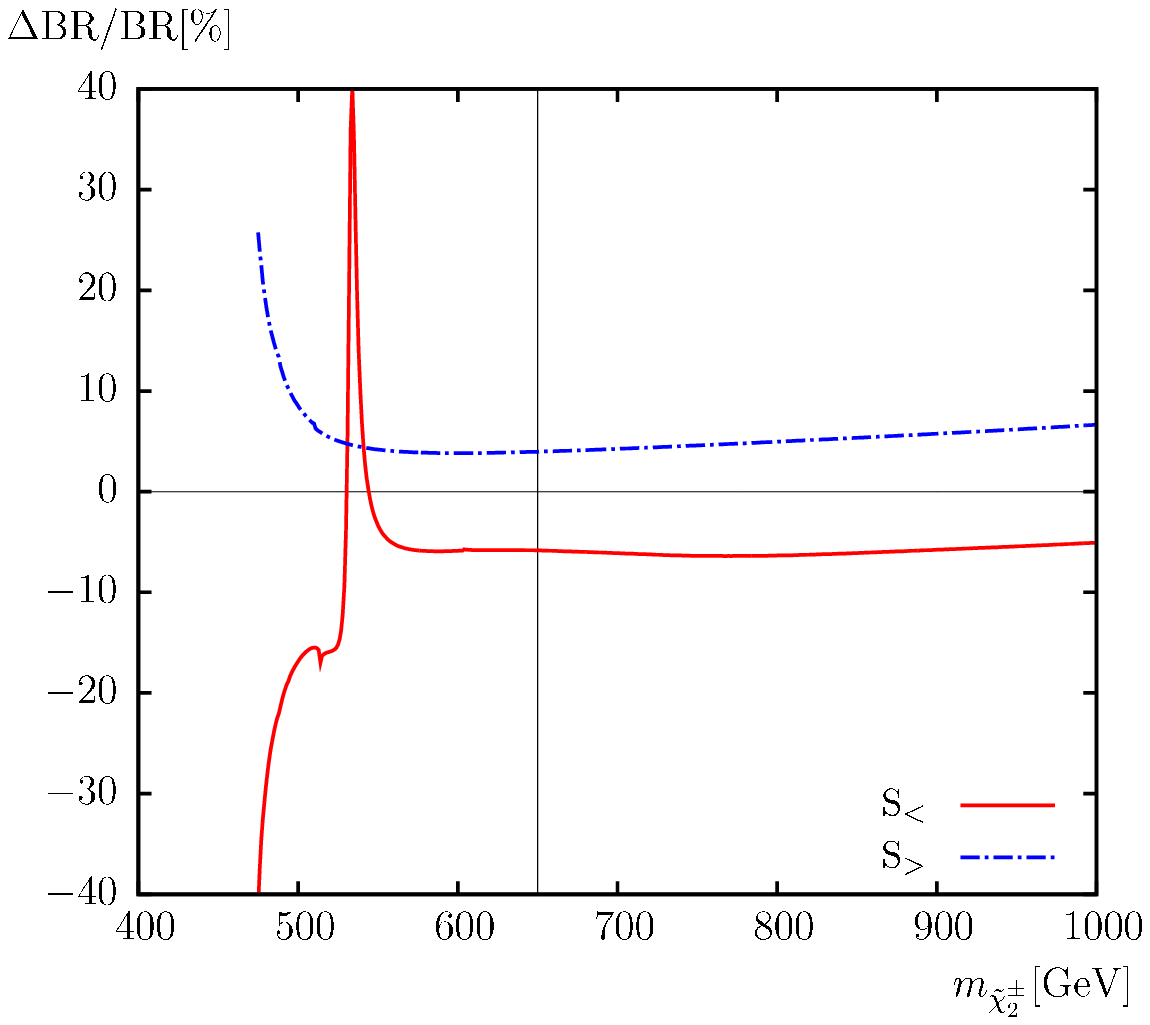}
\end{tabular}
\vspace{2em}
\caption{
  $\Ga(\DecayCmNW{2}{1})$. 
  Tree-level (``tree'') and full one-loop (``full'') corrected 
  decay widths are shown with the parameters chosen according to \SN\
  (see \refta{tab:chaneu}), with $\mcha{2}$ varied.
  The upper left plot shows the decay width, the upper right plot shows 
  the relative size of the corrections.
  The lower left plot shows the BR, the lower right plot shows 
  the relative size of the BR.
  The vertical lines indicate where $\mcha{1} + \mcha{2} = 1000 \gev$, 
  i.e.\ the maximum reach of the ILC(1000).
}
\label{fig:mC2.cha2neu1w}
\end{center}
\end{figure}

\begin{figure}[htb!]
\begin{center}
\begin{tabular}{c}
\includegraphics[width=0.49\textwidth,height=7.5cm]{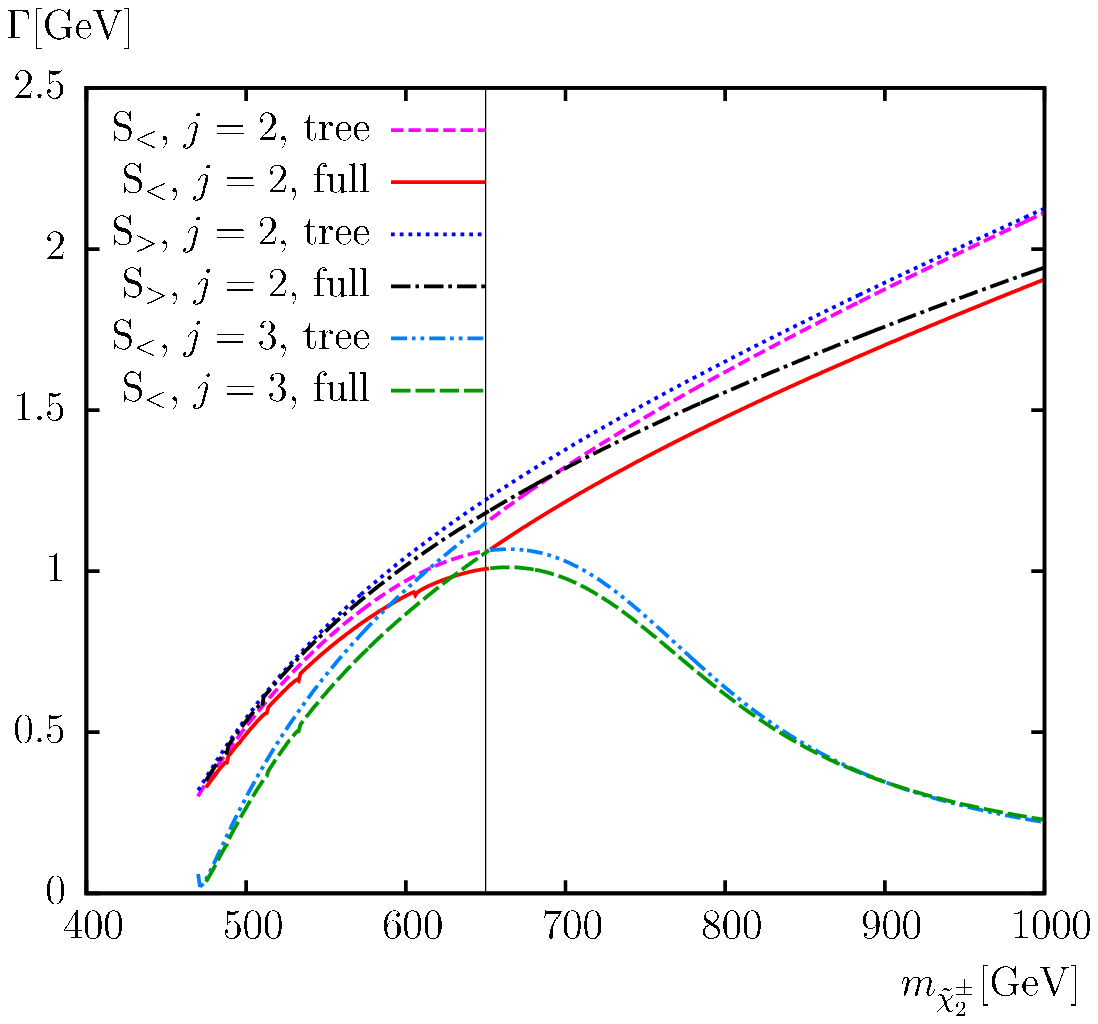}
\hspace{-4mm}
\includegraphics[width=0.49\textwidth,height=7.5cm]{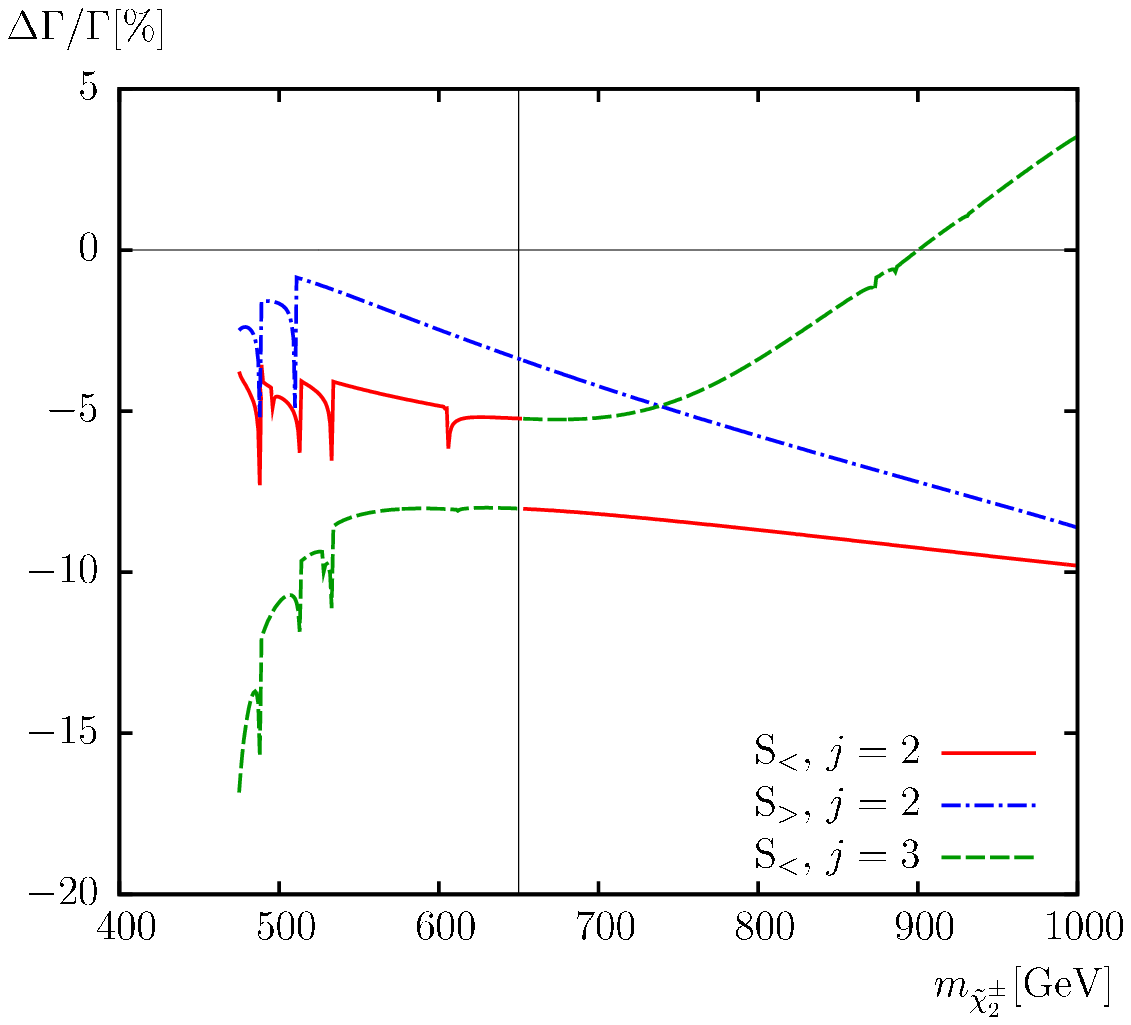} 
\\[4em]
\includegraphics[width=0.49\textwidth,height=7.5cm]{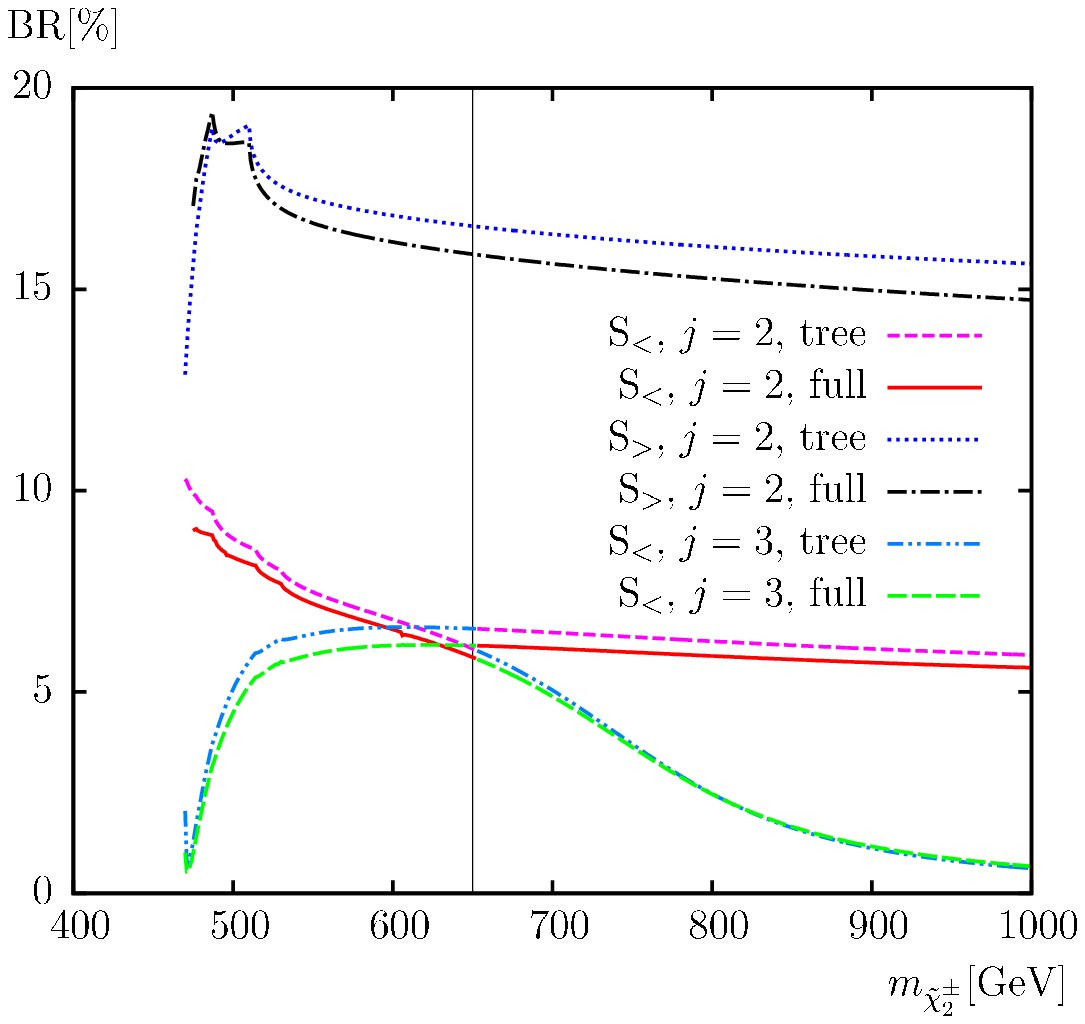}
\hspace{-4mm}
\includegraphics[width=0.49\textwidth,height=7.5cm]{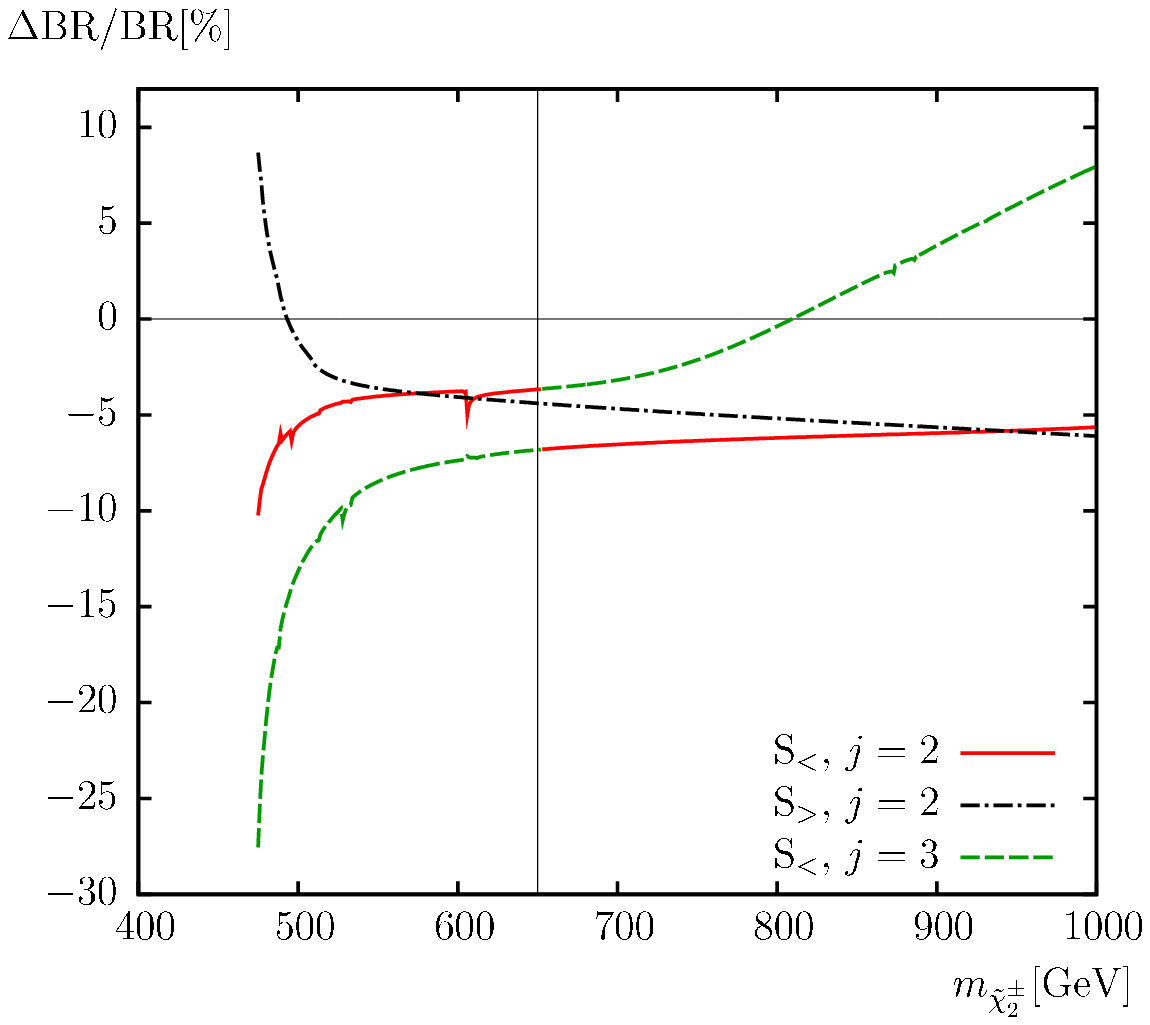}
\end{tabular}
\vspace{2em}
\caption{
  $\Ga(\DecayCmNW{2}{j})$ for $j = 2,3$. 
  Tree-level (``tree'') and full one-loop (``full'') corrected 
  decay widths are shown with the parameters chosen according to \SN\
  (see \refta{tab:para}), with $\mcha{2}$ varied.
  The upper left plot shows the decay width, the upper right plot shows 
  the relative size of the corrections.
  The lower left plot shows the BR, the lower right plot shows 
  the relative size of the BR.
  The vertical lines indicate where $\mcha{1} + \mcha{2} = 1000 \gev$, 
  i.e.\ the maximum reach of the ILC(1000).
}
\label{fig:mC2.cha2neujw}
\end{center}
\end{figure}

\medskip
Next we analyze the decays $\DecayCmNW{2}{1,2,3}$ shown in
\reffis{fig:mC2.cha2neu1w}, \ref{fig:mC2.cha2neujw}. The general
behavior of the decays to $W^-$ is very similar to the decays to $H^-$
discussed above.  
The lightest neutralino is presented in \reffi{fig:mC2.cha2neu1w}. 
The dip at $\mcha{2} = 488 \gev$ corresponds to the $\cha{1} \Hz$ 
threshold.
A second dip can be seen at 
$\mcha{2} = 510\ (513) \gev$ in \SE\ (\SZ) to the 
$\neu{2} H^-$ threshold, as in $\DecayCmNH{2}{1}$,
and a third dip at $\mcha{2} = 533 \gev$ in \SZ.
The size of the decay widths for the decay $\DecayCmNW{2}{1}$ can be
understood as follows. The dominant components of $\cha{2}$ and
$\neu{1}$ have a vanishing coupling according to \refeq{eq:Lagr.CNW}. 
The size of the width is then driven by the ``small'' components with a
generic size of $\sim \MW/\MTwo$. However, their smallness is
compensated by factors of $\MTwo/\MW$, as can be seen in the tree-level
expression, \refeq{CNWtree}. In \SE\ the width rises from 
$0.1 \gev$ at the lowest $\mcha{2}$ up to $\sim 0.6 \gev$ at high $\mcha{2}$ values. 
In \SZ\ we find a minimum (reaching zero) around $\mcha{2} = 540 \gev$, 
and rising up to $\sim 1.6 \gev$ including one-loop corrections. The
relative size of the corrections varies between $\sim +10\%$ and $+4\%$
in \SE\ and around $-10\%$ in \SZ\ (apart from the region where the
width is negligible small). The BR's behave accordingly, reaching $\sim 8\%$
in \SE\ for $\mcha{2} \approx 500 \gev$, going down to $\sim 4.5 \gev$
for large $\mcha{2}$, where in \SZ\ values up to $5\%$ are found. The
relative size of the one-loop effects on the BR's in the ILC(1000)
relevant mass region varies between $+10\%$ and $+3\%$ in \SE\ and 
between $\sim -20\%$ and $-7\%$ in \SZ. This corresponds to several
  times the anticipated ILC(1000) precision.

The decays involving the heavier neutralinos, $\DecayCmNW{2}{2,3}$
(where the decay $\DecayCmNW{2}{4}$ is kinematically forbidden) are
presented in \reffi{fig:mC2.cha2neujw}. 
The two dips in \SE\ for 
$\DecayCmNW{2}{2}$ 
stem from the $\cha{1} \Hz$ and the $\neu{2} H^-$ threshold, which
can also be observed in the two other curves in \SZ. 
In \SZ\ an additional dip from the $\neu{3} H^-$ threshold can be seen
at $\mcha{2} = 533 \gev$ for $\DecayCmNW{2}{2,3}$. 
Finally the dips at
$\mcha{2} \sim 604, 606, 873, 886 \gev$
have been described for $\DecayCmNH{2}{2,3}$.
The behavior of the decays $\DecayCmNW{2}{2,3}$, as stated above, is
similar to the one of $\DecayCmNH{2}{2,3}$, where again the small
chargino/neutralino components determine the size of the widths. As for
$\DecayCmNH{2}{1}$ the smallness of the components is compensated by
factors $\sim \MTwo/\MW$ in the tree-level expressions, see
\refeq{CNWtree}. Also the ``character switch'' between $\neu{2}$ and
$\neu{3}$ in \SZ\ appears as in \reffi{fig:mC2.cha2neujhp}. The relative
size of the one-loop corrections, shown in the upper right plot, are
found around $\sim -10\%$ and $\sim -5\%$ in \SZ, except where the width
becomes very small. In \SE\ the one-loop corrections are small close to
threshold and grow to $\sim -9\%$ at large $\mcha{2}$. 
The $\br(\DecayCmNW{2}{2})$ in \SE\ reaches a maximum of $19\%$ around
$\mcha{2} = 500 \gev$ and settles at $\sim 15\%$ at large $\mcha{2}$. In
\SZ\ the BR's are between $\sim 5\%$ and $10\%$ in the ILC(1000)
relevant region, where $\sim 5\%$ are also found for
$\br(\DecayCmNW{2}{2})$ for large $\mcha{2}$. The relative size of the
one-loop effects on the BR's in the ILC(1000) region (and not directly
at the production threshold) is found between $\sim -10\%$ and $\sim -5\%$.
Again, this corresponds to several times the anticipated ILC(1000)
  precision. 

\begin{figure}[htb!]
\begin{center}
\begin{tabular}{c}
\includegraphics[width=0.49\textwidth,height=7.5cm]{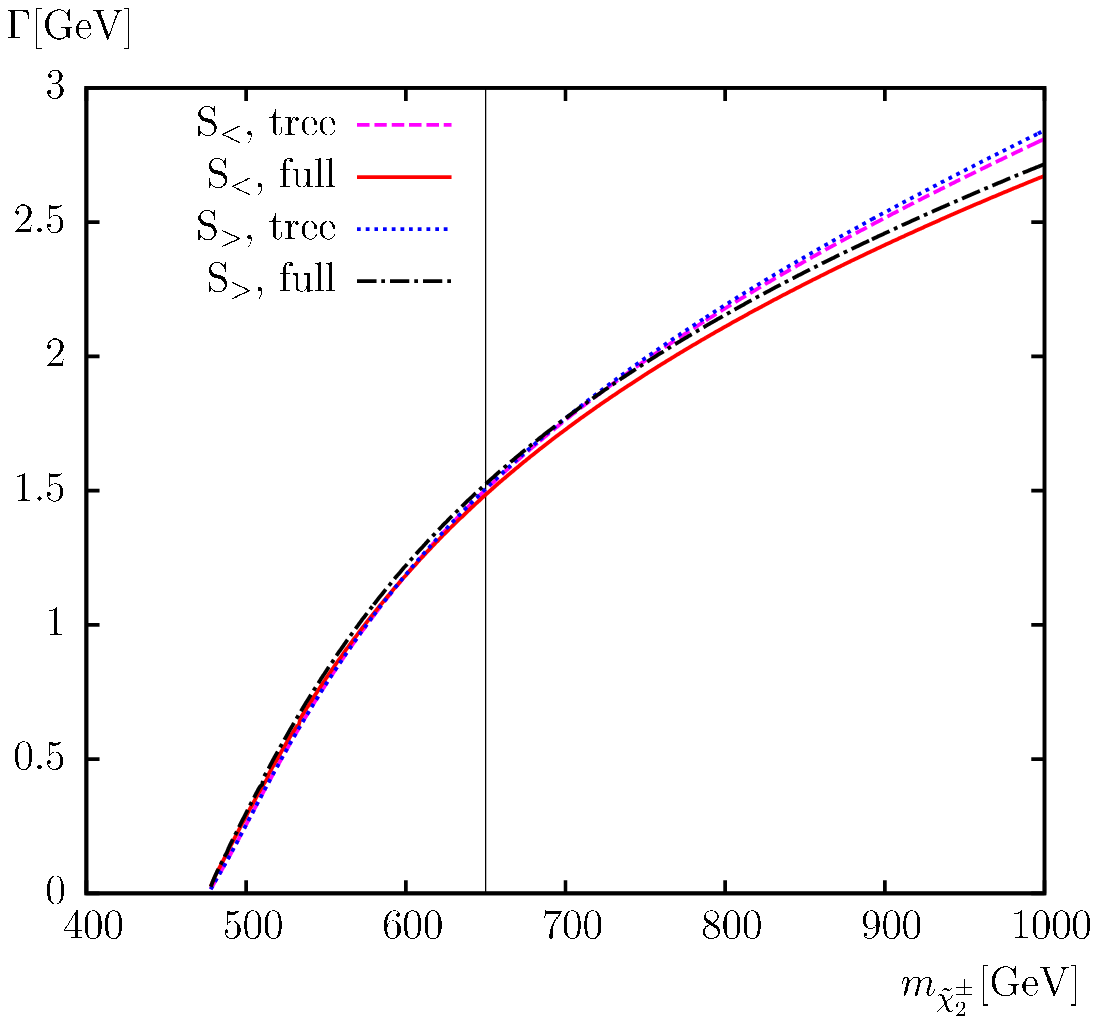}
\hspace{-4mm}
\includegraphics[width=0.49\textwidth,height=7.5cm]{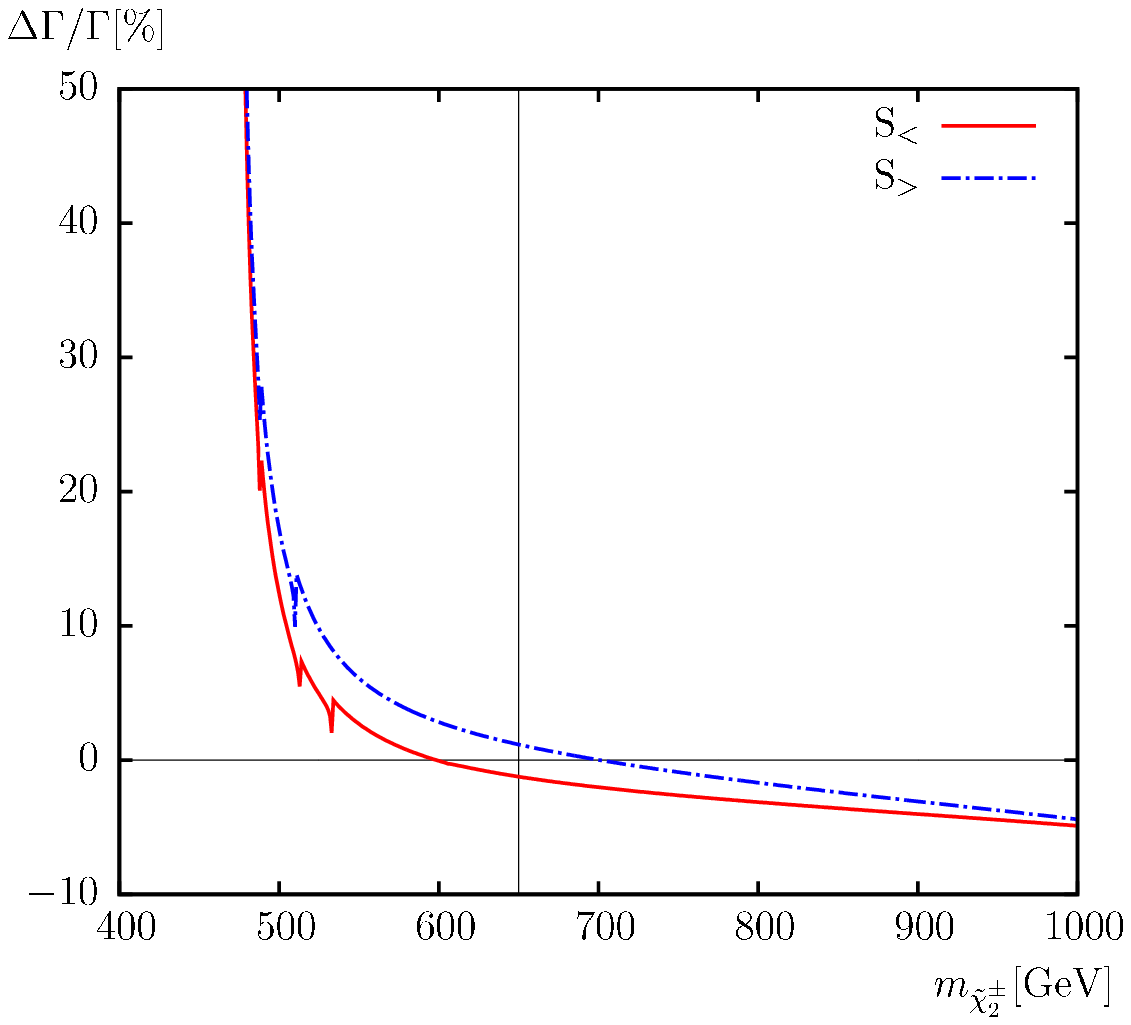} 
\\[4em]
\includegraphics[width=0.49\textwidth,height=7.5cm]{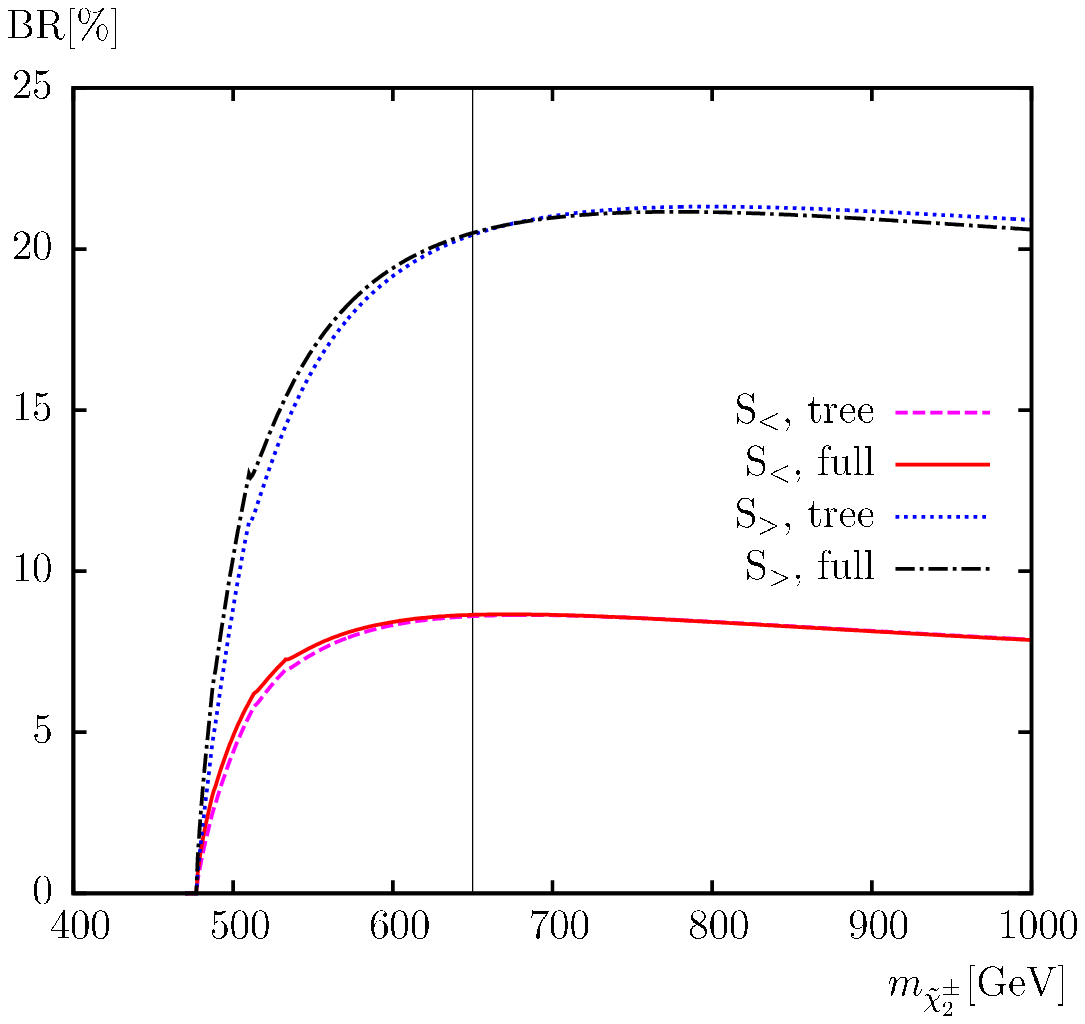}
\hspace{-4mm}
\includegraphics[width=0.49\textwidth,height=7.5cm]{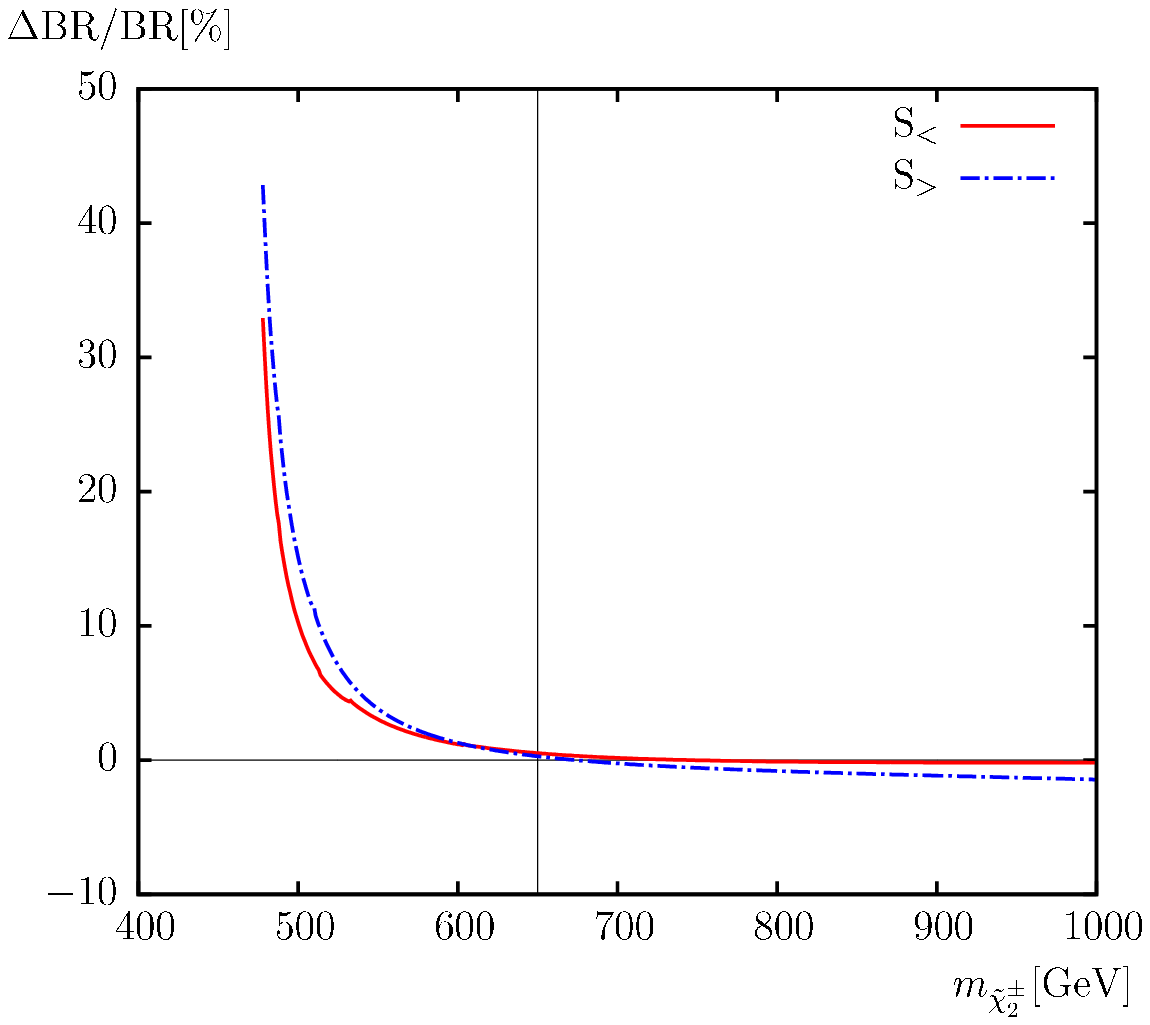}
\end{tabular}
\vspace{2em}
\caption{
  $\Ga(\DecayCmCh{1})$. 
  Tree-level (``tree'') and full one-loop (``full'') corrected 
  decay widths are shown with the parameters chosen according to \SN\
  (see \refta{tab:para}), with $\mcha{2}$ varied.
  The upper left plot shows the decay width, the upper right plot shows 
  the relative size of the corrections.
  The lower left plot shows the BR, the lower right plot shows 
  the relative size of the BR.
  The vertical lines indicate where $\mcha{1} + \mcha{2} = 1000 \gev$, 
  i.e.\ the maximum reach of the ILC(1000).
}
\label{fig:mC2.cha2cha1h1}
\end{center}
\end{figure}

\medskip
Now we turn to the decays involving neutral Higgs bosons. 
The channels $\DecayCmCh{k}$ ($k = 1,2,3$) can serve as source for Higgs
production from SUSY cascades at the LHC, and are therefore of
particular interest. 
The decay $\DecayCmCh{1}$ is shown in \reffi{fig:mC2.cha2cha1h1}. 
The dips are due to the $\cha{1} \Hz$, $\neu{2} H^-$ and $\neu{3} H^-$ thresholds 
and have been described in $\DecayCmNW{2}{2,3}$, see above. 
The tree-level results show a very small difference between \SE\ and \SZ.
This holds also after the full one-loop corrections are included. 
The widths rise from zero at threshold to $\sim 2.6 \gev$. 
The soft rise at threshold is due to the p-wave suppression 
of the decay into a $\cp$-even scalar.
As for the decays
to a charged Higgs boson the admixture of the charginos is crucial for
the size of the decay widths: from the wino/higgsino mixtures at low
$\mcha{2}$ nearly pure wino and higgsino states are reached at large
$\mcha{2}$, corresponding to an ``allowed'' coupling in
\refeq{eq:Lagr.CCH}. The size of the one-loop corrections slightly above
the production threshold is relatively large, 
$\gsim +10\%$, and dominated by their s-wave contribution%
\footnote{\label{threshold-loop}
It should be noted that a calculation very close to threshold requires
the inclusion of additional (non-relativistic) contributions, which is
far beyond the scope of this paper. Consequently, very close to threshold
our calculation (at the tree- or at the loop-level) does not provide a
very accurate description of the decay width.}%
, %
and reaches $\sim -5\%$ at large $\mcha{2}$. 
The BR's reach $20\%$ in \SE\ and $9\%$
in \SZ\ at the highest $\mcha{2}$ values in the ILC(1000) relevant
region, and remain nearly flat for higher $\mcha{2}$ values. The
relative size of the one-loop effects on the BR's is only sizable close
to threshold, and is found at the $\sim 1\%$ level for 
$\mcha{2} \gsim 600 \gev$, which corresponds roughly to the
anticipated ILC(1000) precision, see \refta{tab:ILCproduction}.

\begin{figure}[htb!]
\begin{center}
\begin{tabular}{c}
\includegraphics[width=0.49\textwidth,height=7.5cm]{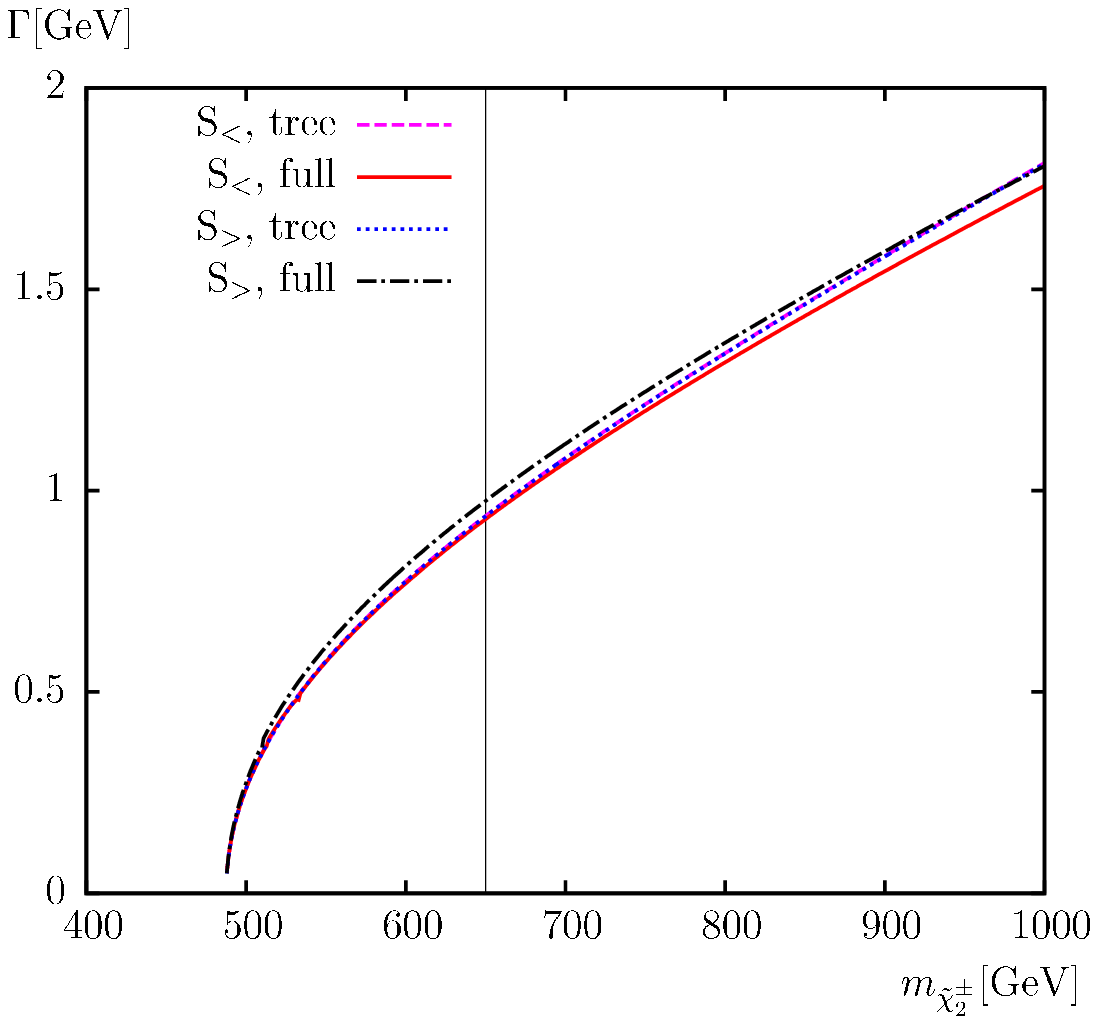}
\hspace{-4mm}
\includegraphics[width=0.49\textwidth,height=7.5cm]{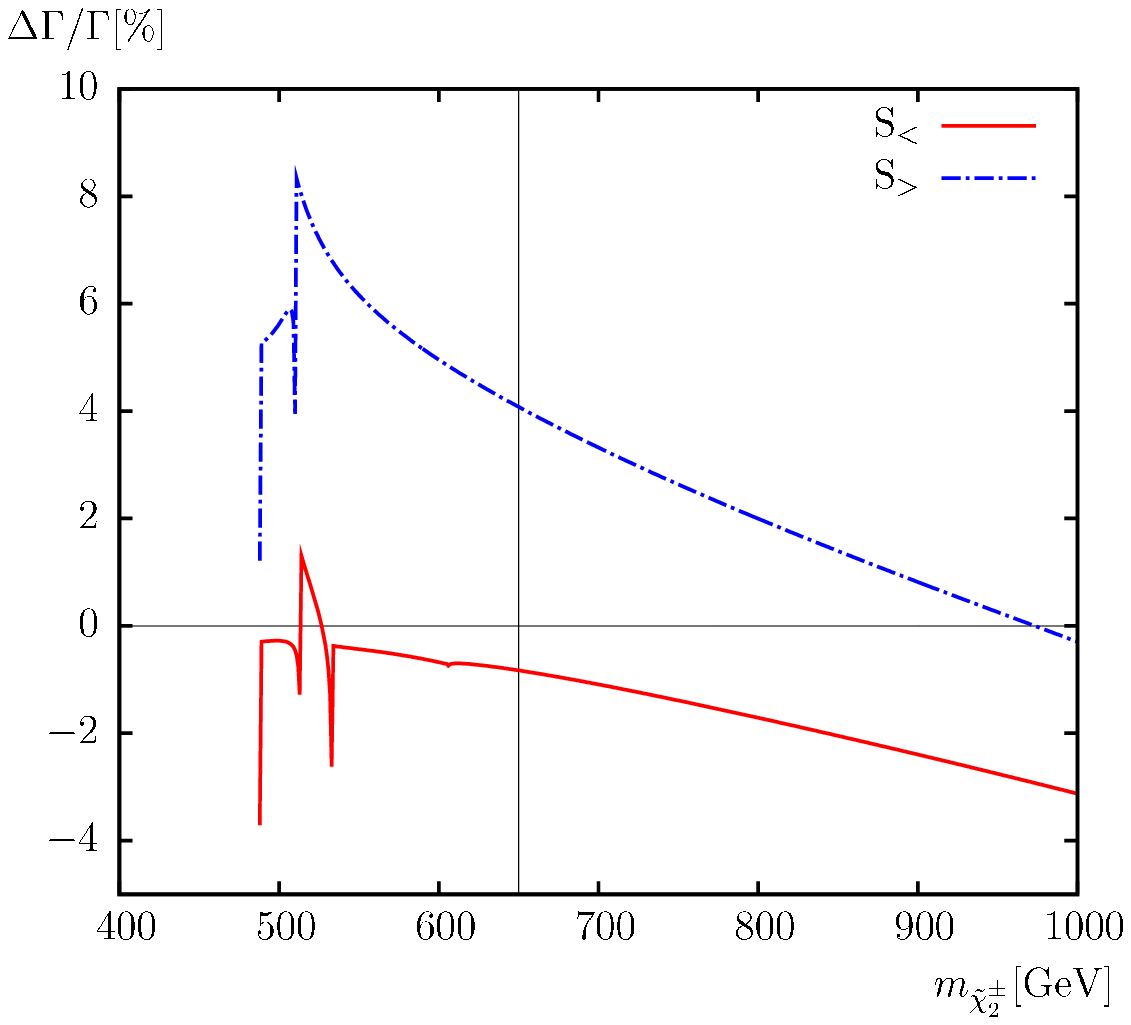} 
\\[4em]
\includegraphics[width=0.49\textwidth,height=7.5cm]{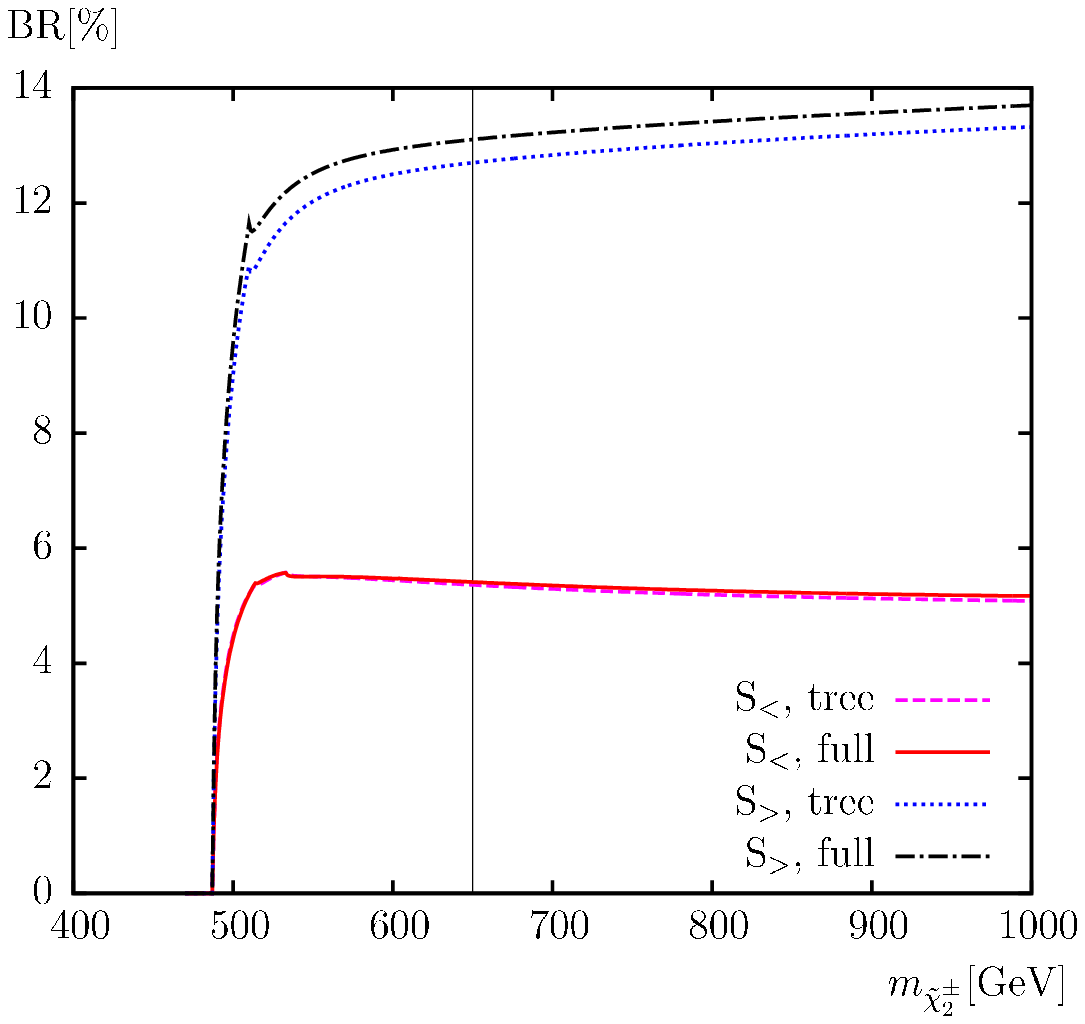}
\hspace{-4mm}
\includegraphics[width=0.49\textwidth,height=7.5cm]{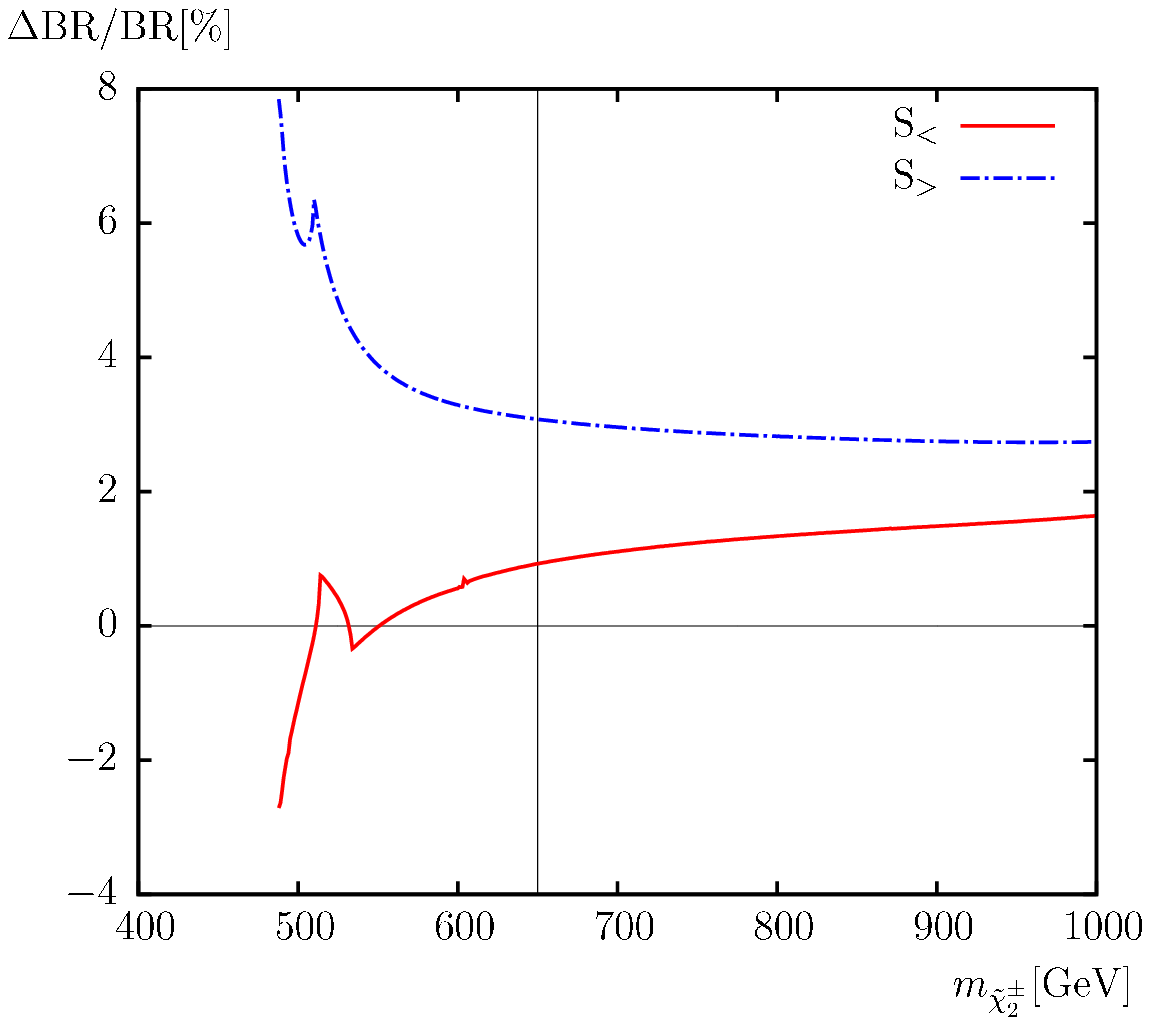}
\end{tabular}
\vspace{2em}
\caption{
  $\Ga(\DecayCmCh{2})$. 
  Tree-level (``tree'') and full one-loop (``full'') corrected 
  decay widths are shown with the parameters chosen according to \SN\
  (see \refta{tab:para}), with $\mcha{2}$ varied.
  The upper left plot shows the decay width, the upper right plot shows 
  the relative size of the corrections.
  The lower left plot shows the BR, the lower right plot shows 
  the relative size of the BR.
  The vertical lines indicate where $\mcha{1} + \mcha{2} = 1000 \gev$, 
  i.e.\ the maximum reach of the ILC(1000).
}
\label{fig:mC2.cha2cha1h2}
\end{center}
\end{figure}

The decay $\DecayCmCh{2}$ are shown in \reffi{fig:mC2.cha2cha1h2}, where
a qualitatively similar result to $\DecayCmCh{1}$ can be found, with the
main difference of somewhat smaller decay width and branching ratio,
mostly due to the smaller and negative contribution of the chirally
violating part in the corresponding coupling, see \refeq{CCHtree}. 
The steep rise at threshold is due to the unsuppressed s-wave contribution of 
this channel, where $h_2$ is the $\cp$-odd Higgs boson.
The dips at $\mcha{2}=510 \gev$ in \SE\ and $\mcha{2}=513, 533 \gev$  
correspond to the thresholds described above for $\DecayCmCh{1}$.
Finally, the dip at $\mcha{2}=606\gev$ corresponds to the
$\DecayCNW{1}{1}$ threshold. 
It should be noticed that the dips correspond 
to the thresholds for tree-level processes while the widths are
evaluated with one-loop masses, 
leading to the effect of the $\cha{1} \Hz$ threshold in this decay, 
(see, however, footnote~\ref{threshold-loop}).
The relative size of the one-loop corrections on the decay width varies
in \SZ\ between $+8\%$ at $\mcha{2} = 500 \gev$ and very small negative
values for large $\mcha{2}$ (where possible larger negative values can
be reached for even larger $\mcha{2}$). 
In \SZ\ the corrections vary around $\sim -3\%$. 
Concerning the one-loop effects on the BR's, in
\SE\ at $\mcha{2} = 500 \gev$ around $+6\%$ are found, going down to
$\sim +3\%$ at large $\mcha{2}$. In \SZ\ the corrections vary between
$-2\%$ at low $\mcha{2}$ and $+2\%$ at high $\mcha{2}$. For small
$\mcha{2}$ values the corrections can be substantially larger than the
ILC(1000) precision.

\begin{figure}[htb!]
\begin{center}
\begin{tabular}{c}
\includegraphics[width=0.49\textwidth,height=7.5cm]{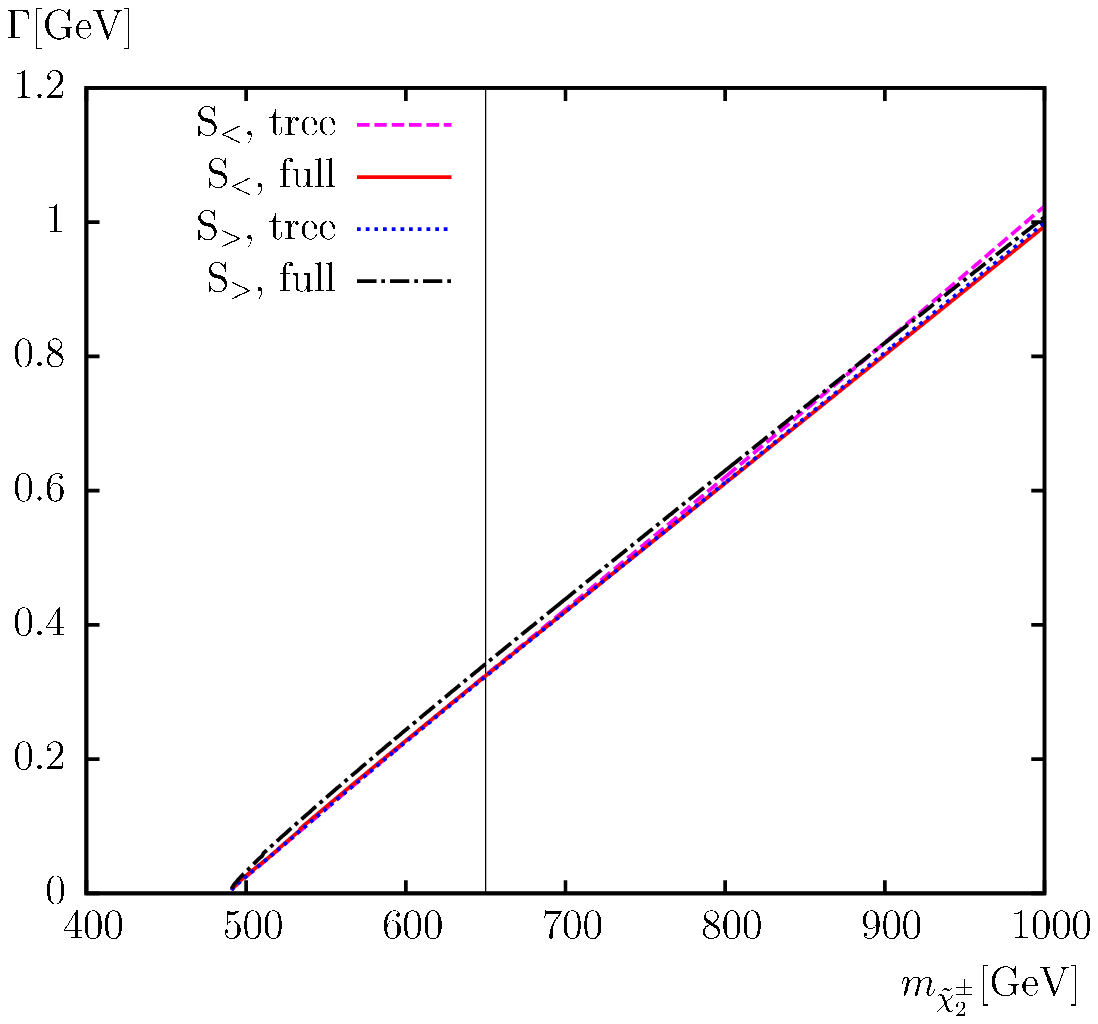}
\hspace{-4mm}
\includegraphics[width=0.49\textwidth,height=7.5cm]{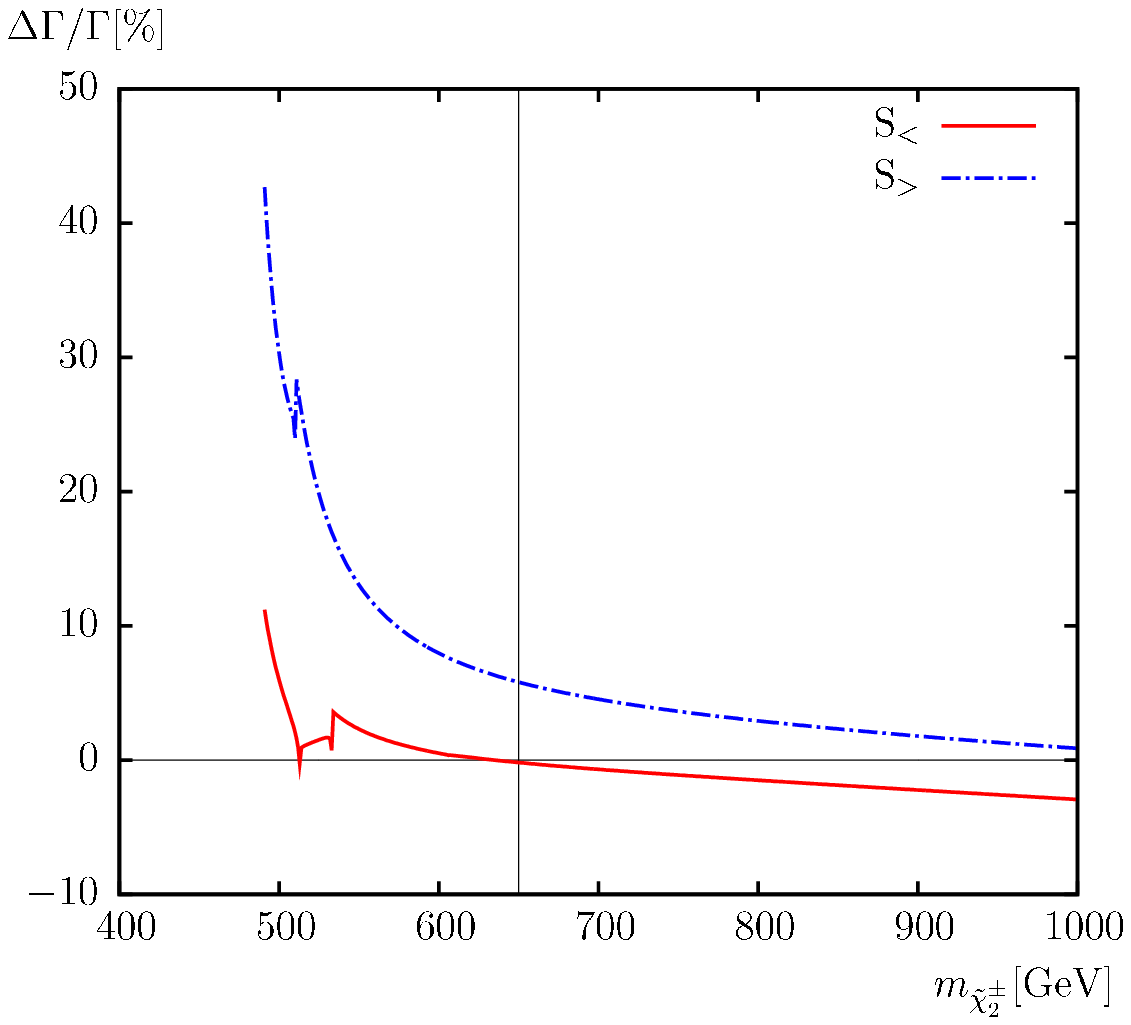} 
\\[4em]
\includegraphics[width=0.49\textwidth,height=7.5cm]{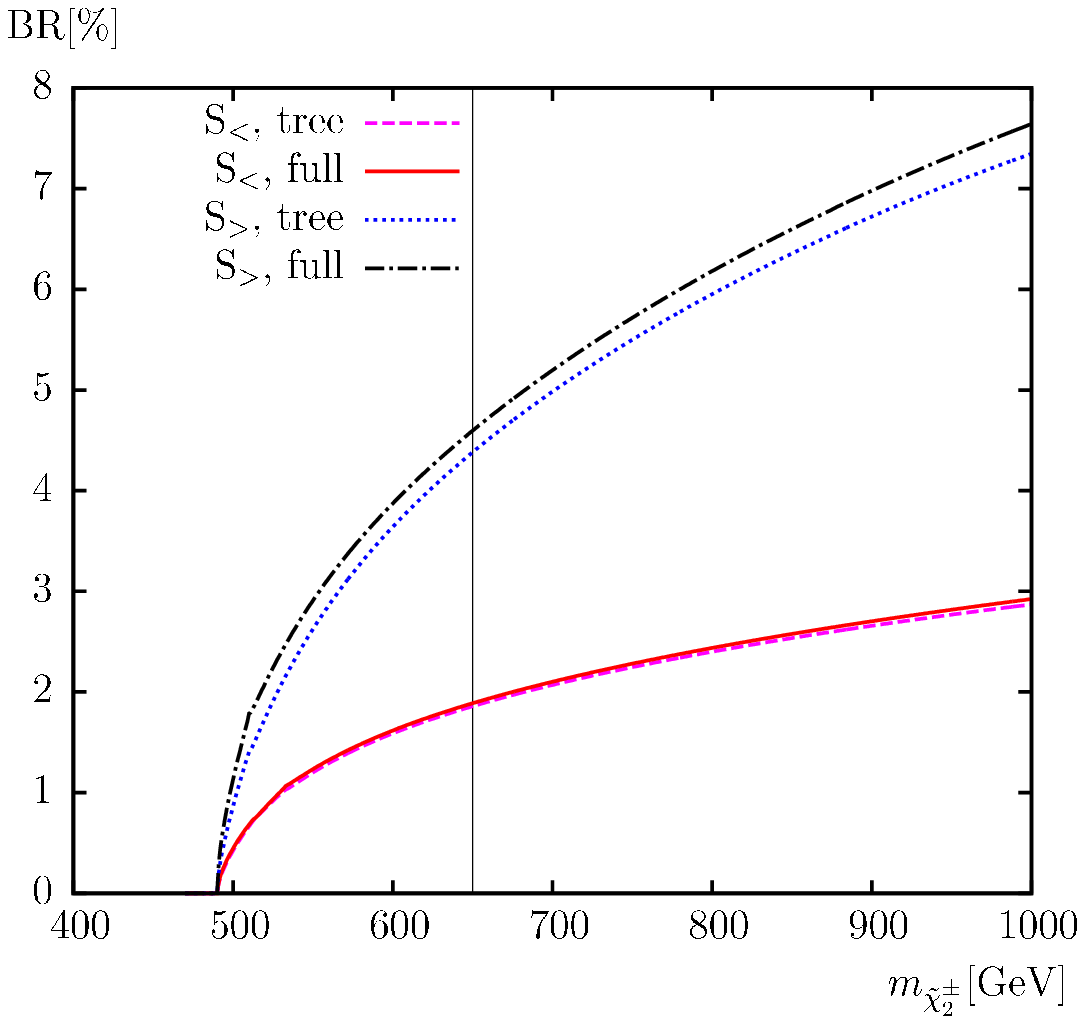}
\hspace{-4mm}
\includegraphics[width=0.49\textwidth,height=7.5cm]{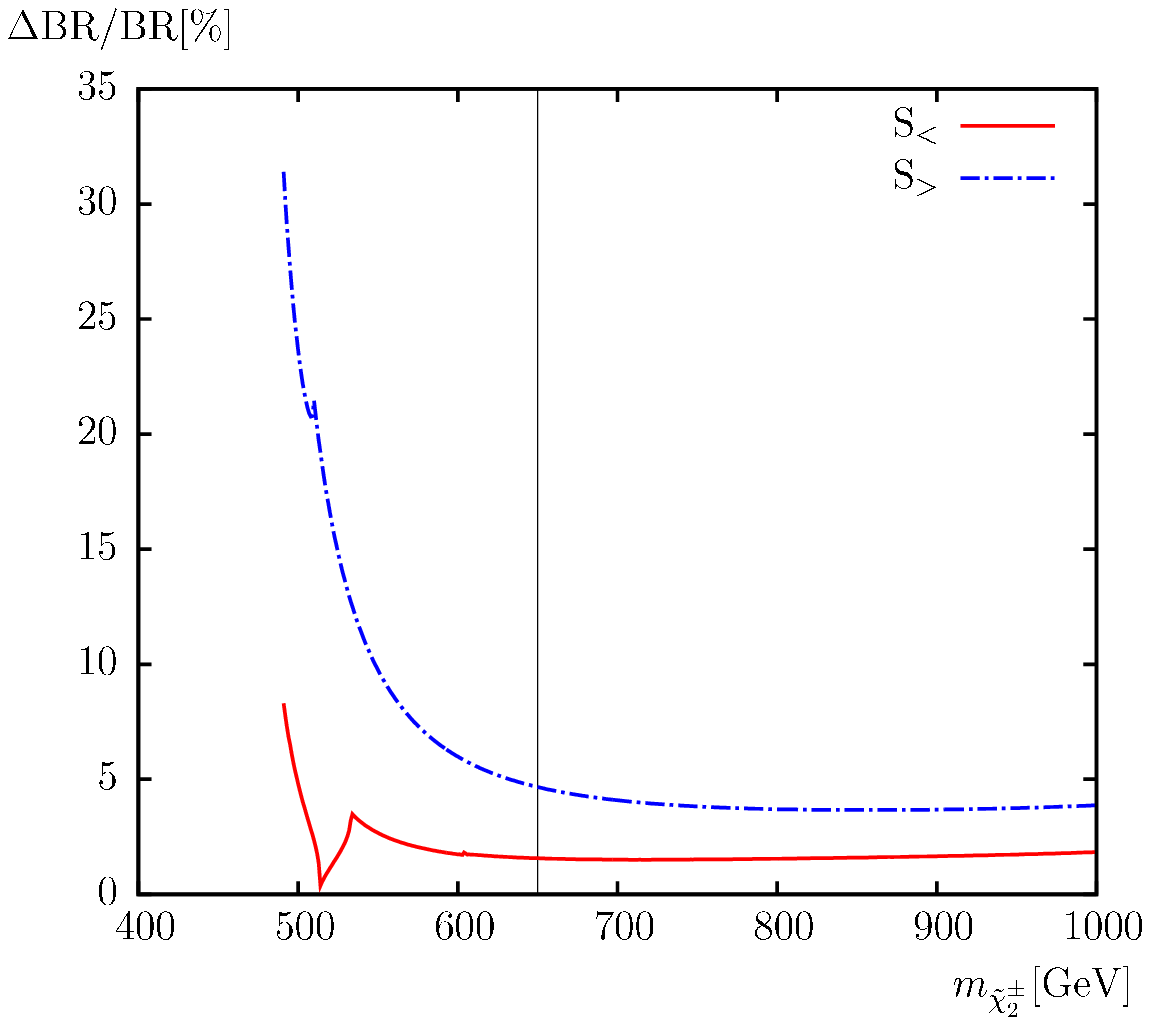}
\end{tabular}
\vspace{2em}
\caption{
  $\Ga(\DecayCmCh{3})$. 
  Tree-level (``tree'') and full one-loop (``full'') corrected 
  decay widths are shown with the parameters chosen according to \SN\
  (see \refta{tab:para}), with $\mcha{2}$ varied.
  The upper left plot shows the decay width, the upper right plot shows 
  the relative size of the corrections.
  The lower left plot shows the BR, the lower right plot shows 
  the relative size of the BR.
  The vertical lines indicate where $\mcha{1} + \mcha{2} = 1000 \gev$, 
  i.e.\ the maximum reach of the ILC(1000).
}
\label{fig:mC2.cha2cha1h3}
\end{center}
\end{figure}

The last channel involving a neutral Higgs, $\DecayCmCh{3}$ is shown in
\reffi{fig:mC2.cha2cha1h3}. 
We observe the same dips as for $\DecayCmCh{2}$.
A straight rise of the decay width in
\SE\ and \SZ\ is found, reaching $\sim 1 \gev$ at $\mcha{2} = 1000 \gev$.
The smaller results with respect to $\DecayCmCh{1}$ are due to the
opposite sign of the chirally violating contributions in the
corresponding coupling, see \refeq{CCHtree}.
As in the decay into $h_1$, the threshold behavior is p-wave suppressed
at tree-level with a small s-wave dominated contribution to the loop
corrections. 
A corresponding rise in the BR up to $7.5\%$ ($3\%$) is found in
\SE\ (\SZ). Within \SE\ the relative size of the one-loop corrections is
small above threshold, reaching $\sim -5\%$ at large $\mcha{2}$. Within
\SZ\ the corrections can be very large slightly above threshold,
reaching $\sim 30\%$ at $\mcha{2} \gsim 500 \gev$, going down to zero
for large $\mcha{2}$. The relative effect on the BR's is similar, where
values around $5\%$--$10\%$ ($\sim 2.5\%$) are found in \SE\ (\SZ),
exceeding by far the anticipated ILC(1000) precision.

\begin{figure}[htb!]
\begin{center}
\begin{tabular}{c}
\includegraphics[width=0.49\textwidth,height=7.5cm]{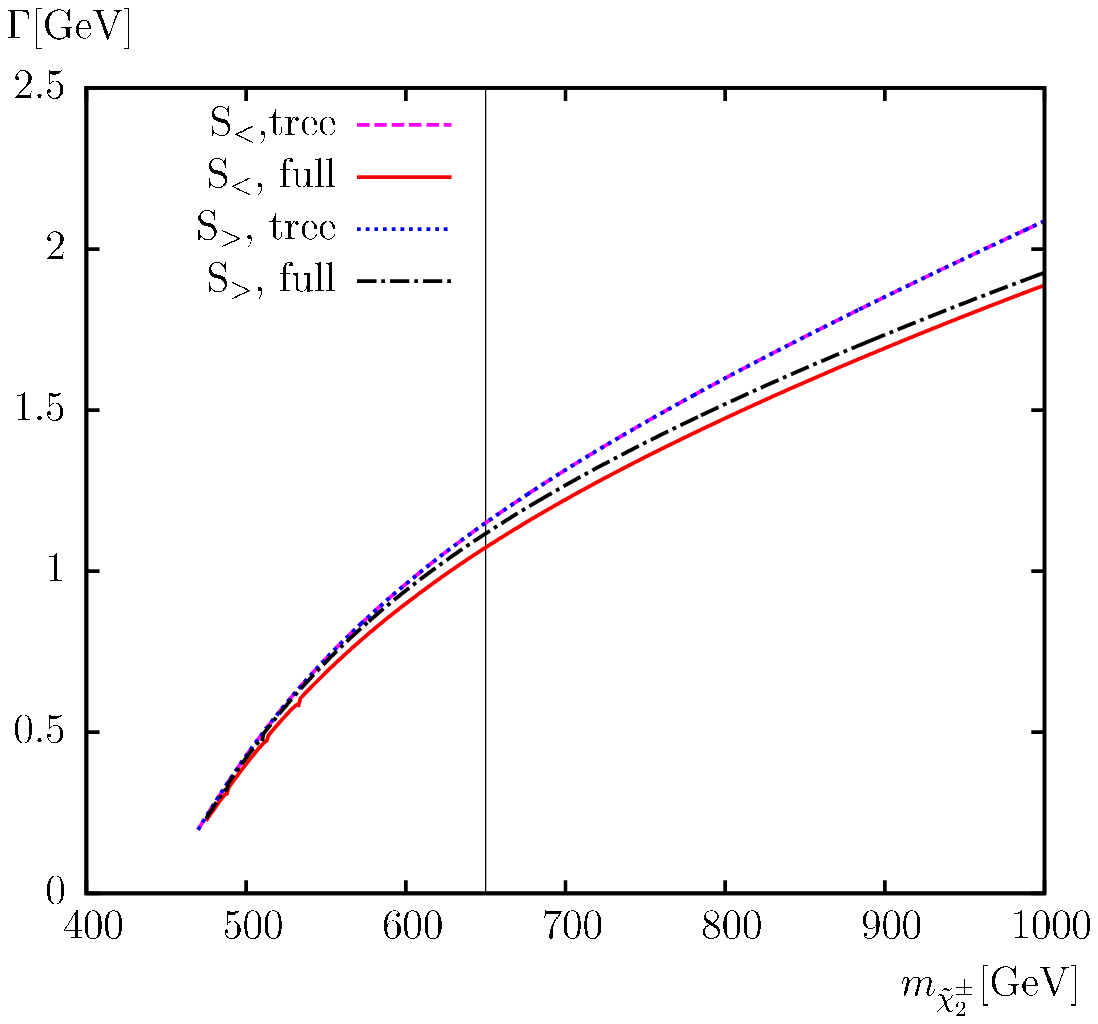}
\hspace{-4mm}
\includegraphics[width=0.49\textwidth,height=7.5cm]{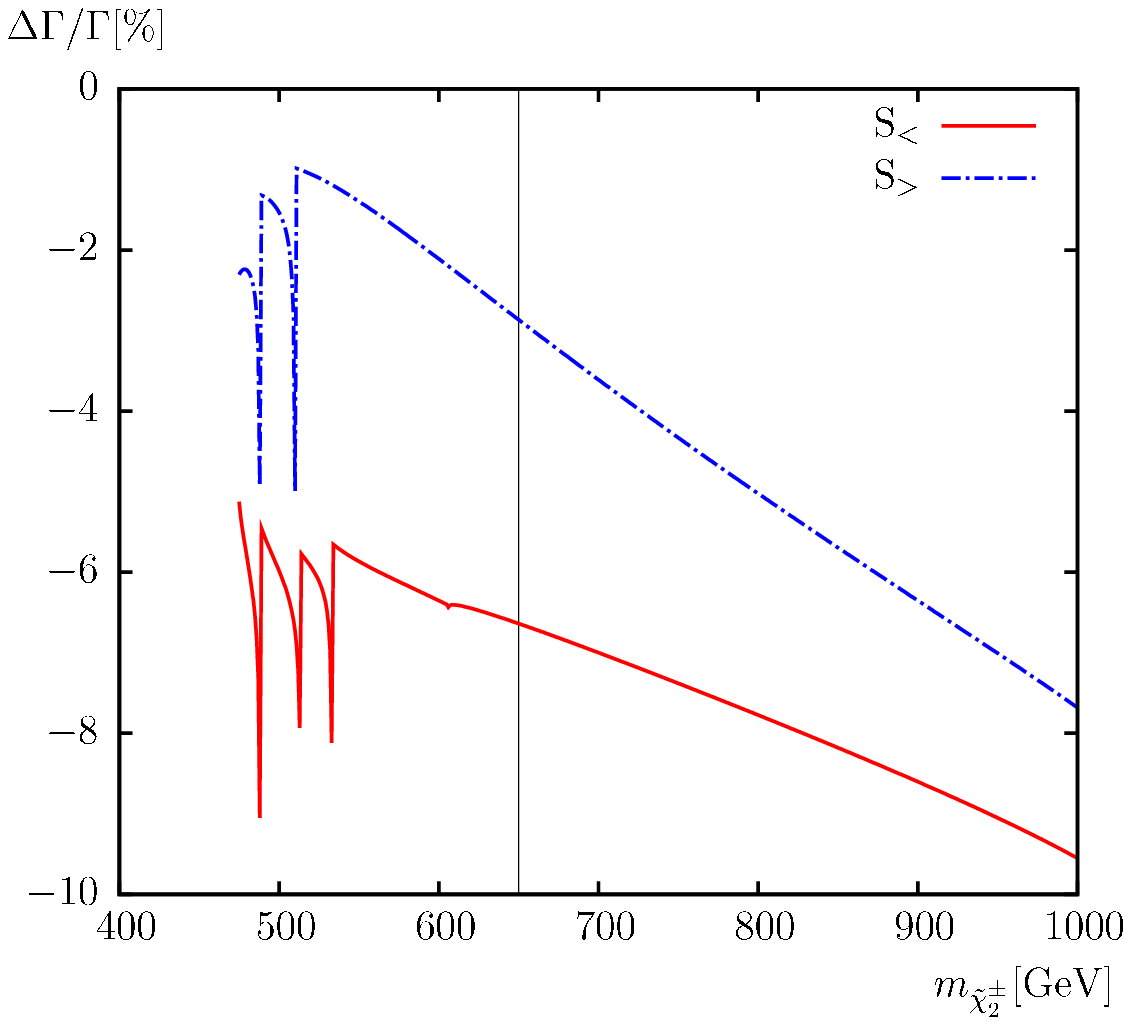} 
\\[4em]
\includegraphics[width=0.49\textwidth,height=7.5cm]{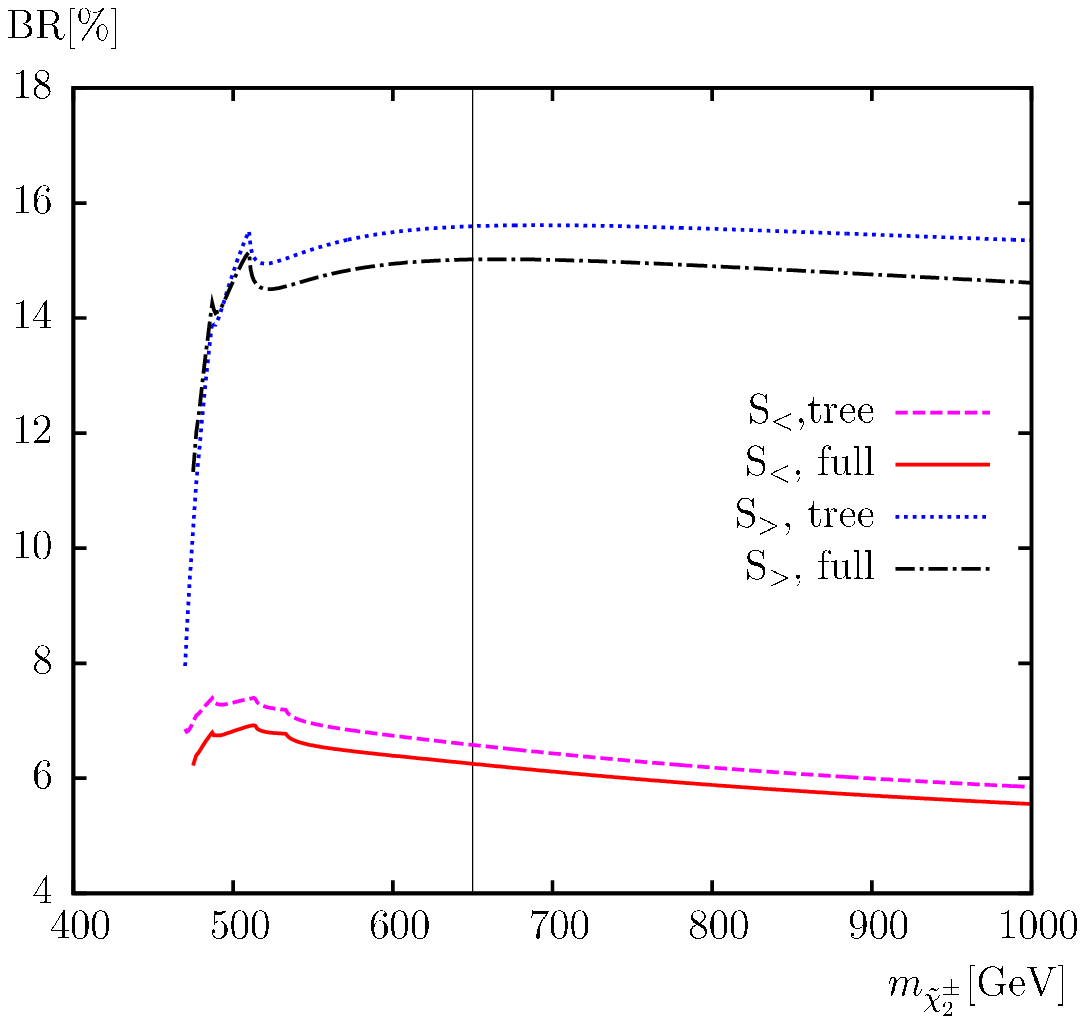}
\hspace{-4mm}
\includegraphics[width=0.49\textwidth,height=7.5cm]{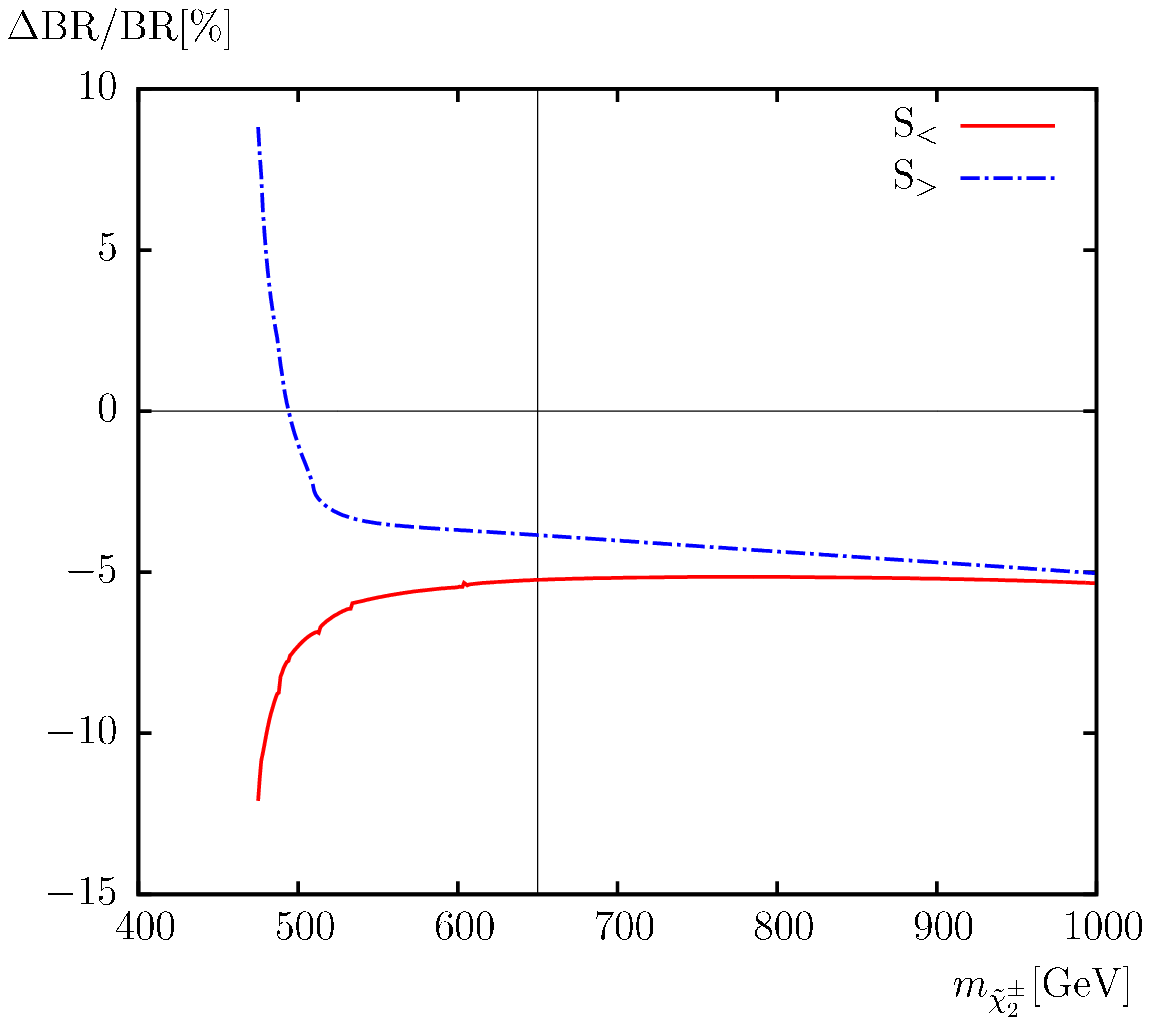}
\end{tabular}
\vspace{2em}
\caption{
  $\Ga(\DecayCmCZ)$. 
  Tree-level (``tree'') and full one-loop (``full'') corrected 
  decay widths are shown with the parameters chosen according to \SN\
  (see \refta{tab:para}), with $\mcha{2}$ varied.
  The upper left plot shows the decay width, the upper right plot shows 
  the relative size of the corrections.
  The lower left plot shows the BR, the lower right plot shows 
  the relative size of the BR.
  The vertical lines indicate where $\mcha{1} + \mcha{2} = 1000 \gev$, 
  i.e.\ the maximum reach of the ILC(1000).
}
\label{fig:mC2.cha2cha1z}
\end{center}
\end{figure}

\medskip
The last channel involving SM gauge bosons, $\DecayCmCZ$, is presented
in \reffi{fig:mC2.cha2cha1z}. 
The same dips as in the decays into neutral Higgs bosons can be observed.
The tree-level widths are equal in the two scenarios \SE\ and \SZ, since
both the $Z\cha{1}\cha{2}$~coupling as well as the two chargino masses
are symmetric under the exchange of $\mu \leftrightarrow \MTwo$.
As for the decays $\DecayCmNW{2}{1,2,3}$
the large admixtures of the charginos yield mostly a vanishing decay
width. However, as above, the smallness of the ``allowed couplings'' is
compensated by factors in the tree-level expression $\sim \MTwo/\MZ$,
see \refeq{CCZtree}.
Including one-loop corrections the decay widths rise up to 
$\sim 1.9 \gev$ 
for $\mcha{2} = 1000 \gev$, where, contrary to the tree-level widths, a
small difference between \SE\ and \SZ\ can be observed. The branching
ratios are, except at the smallest $\mcha{2}$, 
relatively flat around 
$14.5\%$ $(6\%)$ in \SE\ (\SZ). 
The size of the one-loop corrections to
the decay width grows from $\sim -2\%$ $(-6\%)$ at small $\mcha{2}$ to 
$-7.5\%$ $(-9.5\%)$ at large $\mcha{2}$ in \SE\ (\SZ). The effects on
the branching ratios is mostly at the $-5\%$ level, which is larger
than the ILC(1000) precision.

\begin{figure}[htb!]
\begin{center}
\begin{tabular}{c}
\includegraphics[width=0.49\textwidth,height=7.5cm]{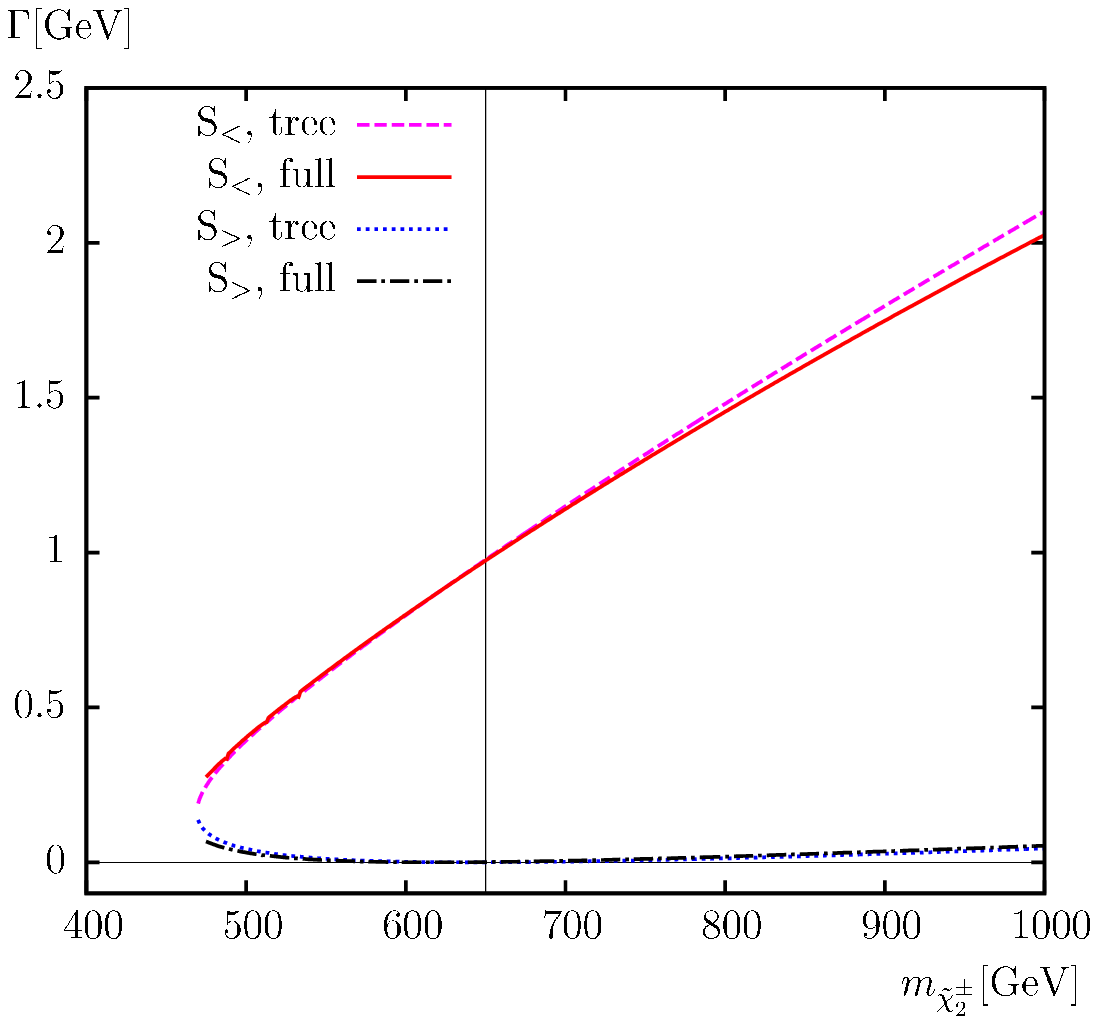}
\hspace{-4mm}
\includegraphics[width=0.49\textwidth,height=7.5cm]{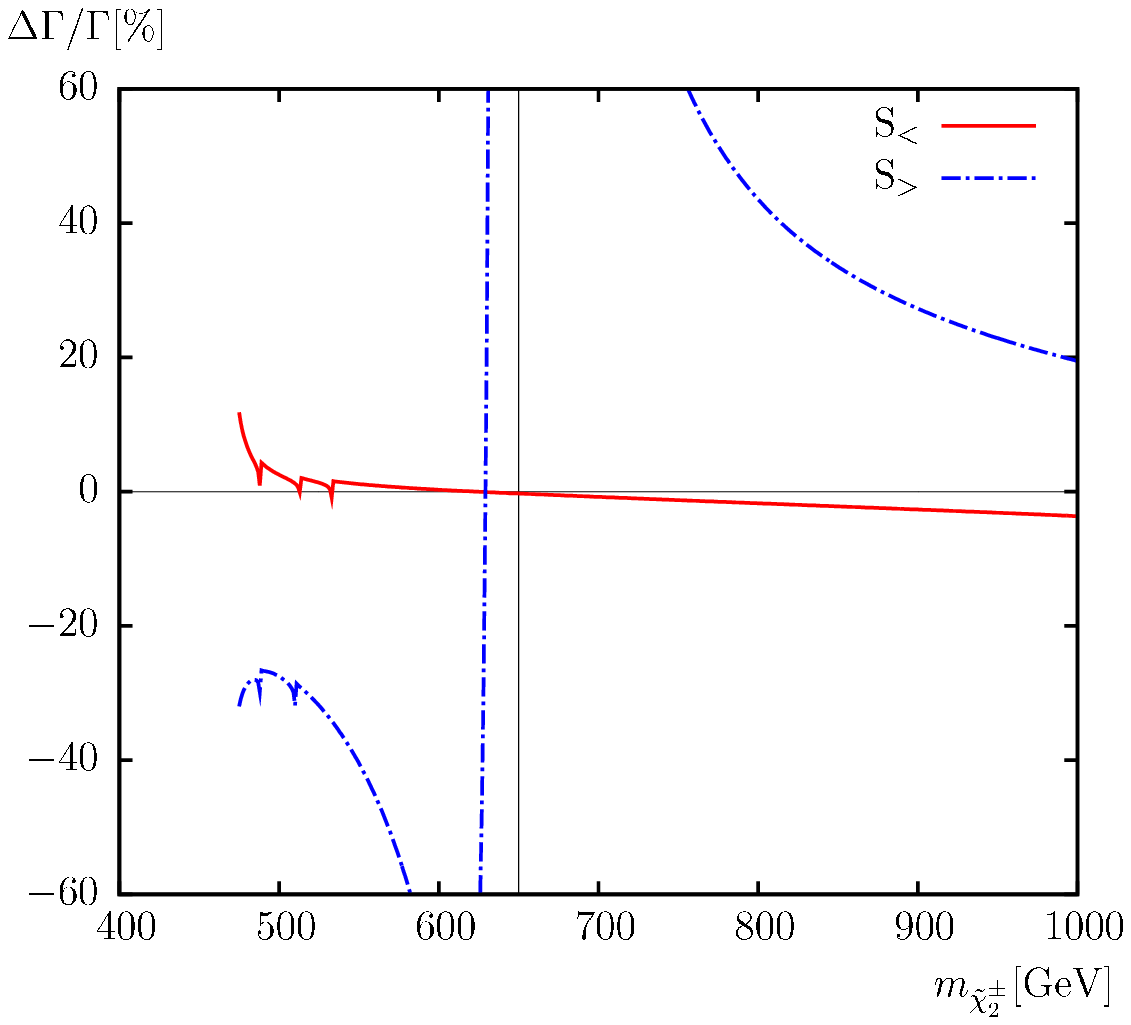} 
\\[4em]
\includegraphics[width=0.49\textwidth,height=7.5cm]{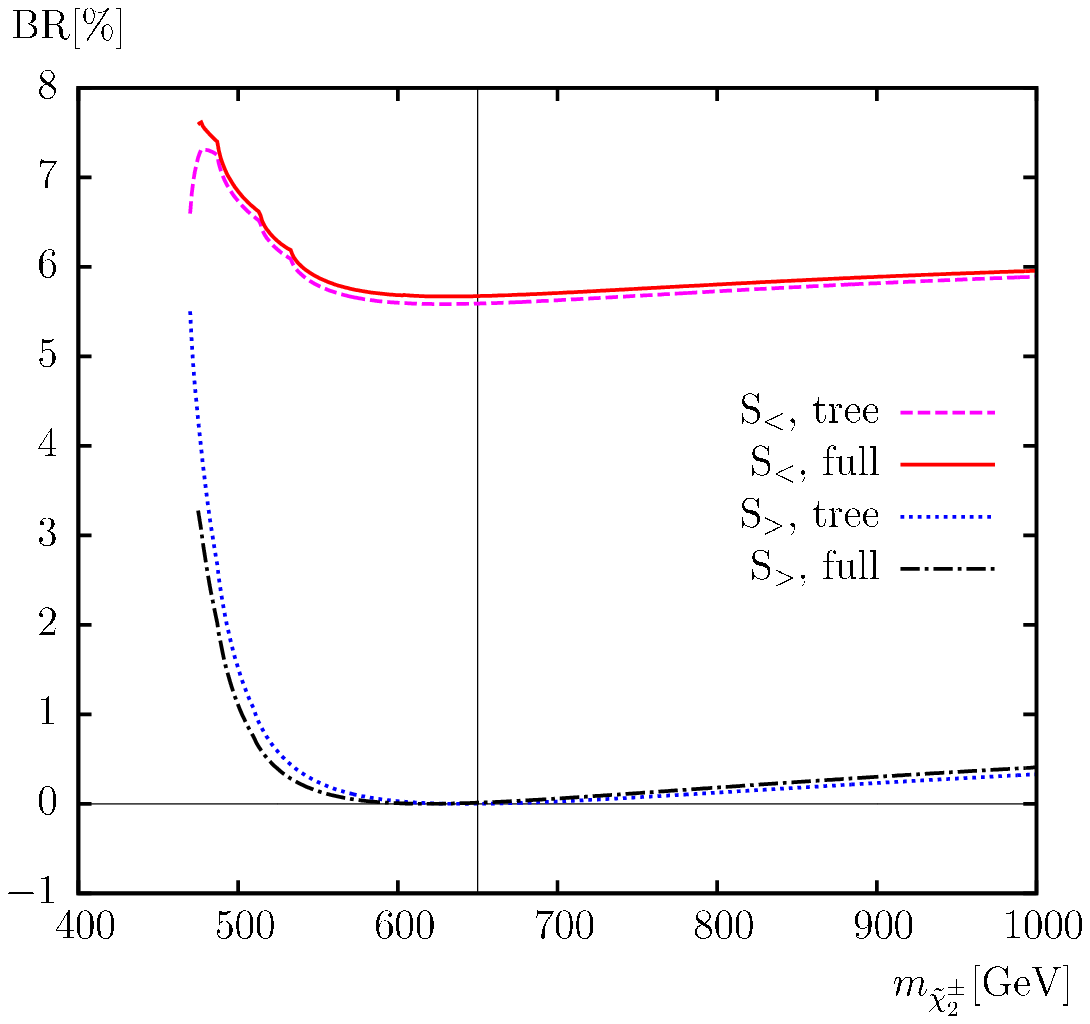}
\hspace{-4mm}
\includegraphics[width=0.49\textwidth,height=7.5cm]{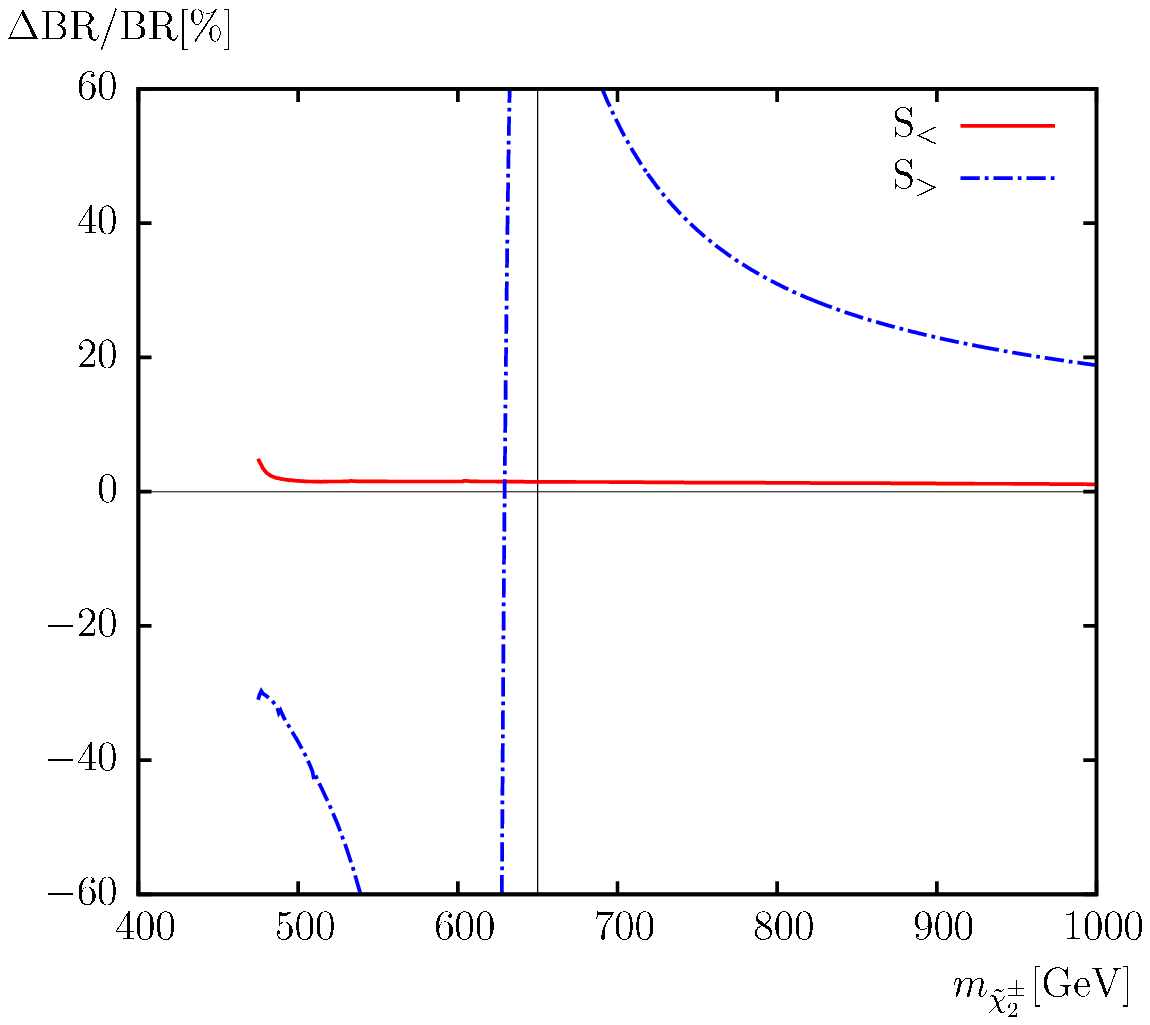}
\end{tabular}
\vspace{2em}
\caption{
  $\Ga(\DecayCmnSl{2}{\tau}{1})$. 
  Tree-level (``tree'') and full one-loop (``full'') corrected 
  decay widths are shown with the parameters chosen according to \SN\
  (see \refta{tab:para}), with $\mcha{2}$ varied.
  The upper left plot shows the decay width, the upper right plot shows 
  the relative size of the corrections.
  The lower left plot shows the BR, the lower right plot shows 
  the relative size of the BR.
  The vertical lines indicate where $\mcha{1} + \mcha{2} = 1000 \gev$, 
  i.e.\ the maximum reach of the ILC(1000).
}
\label{fig:mC2.cha2stau1nu}
\end{center}
\end{figure}

\begin{figure}[htb!]
\begin{center}
\begin{tabular}{c}
\includegraphics[width=0.49\textwidth,height=7.5cm]{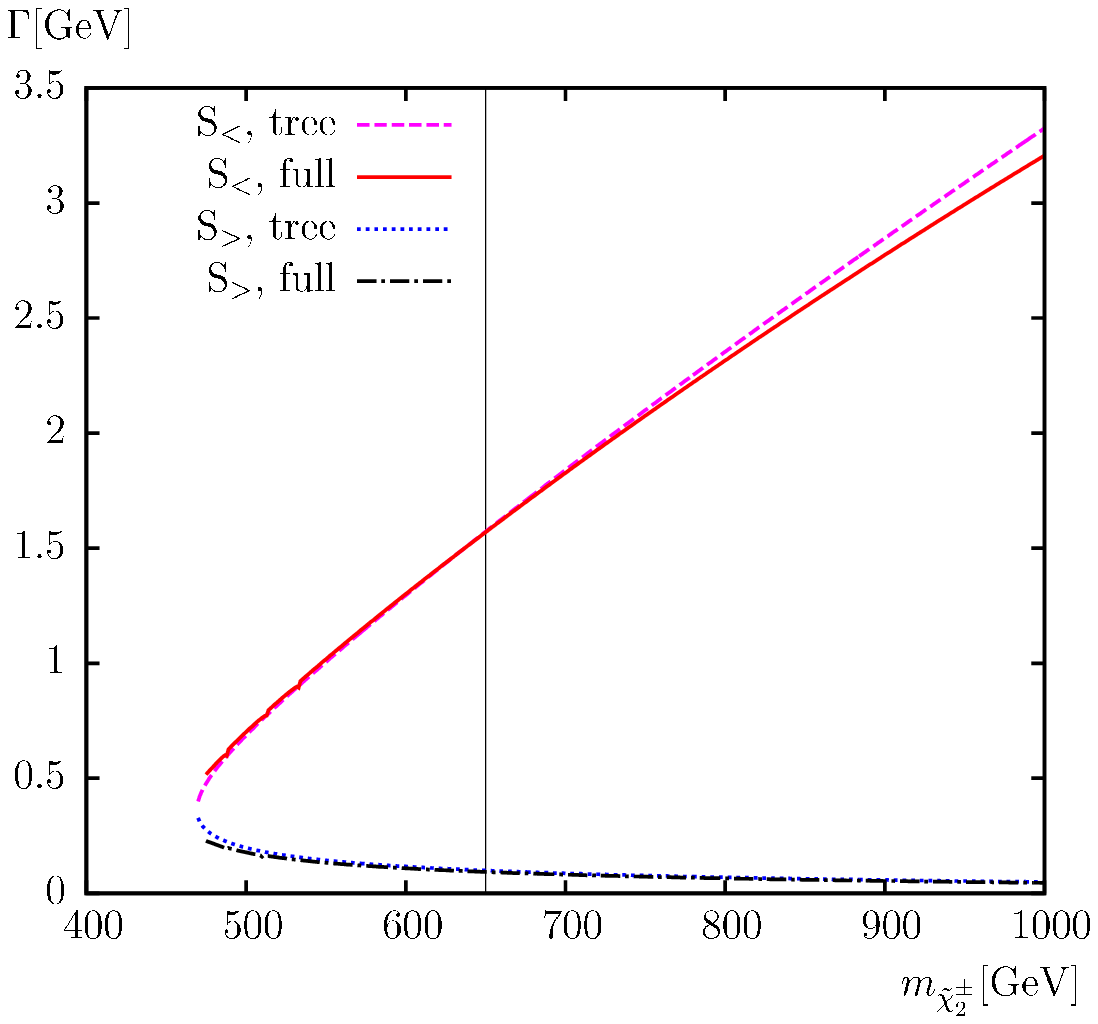}
\hspace{-4mm}
\includegraphics[width=0.49\textwidth,height=7.5cm]{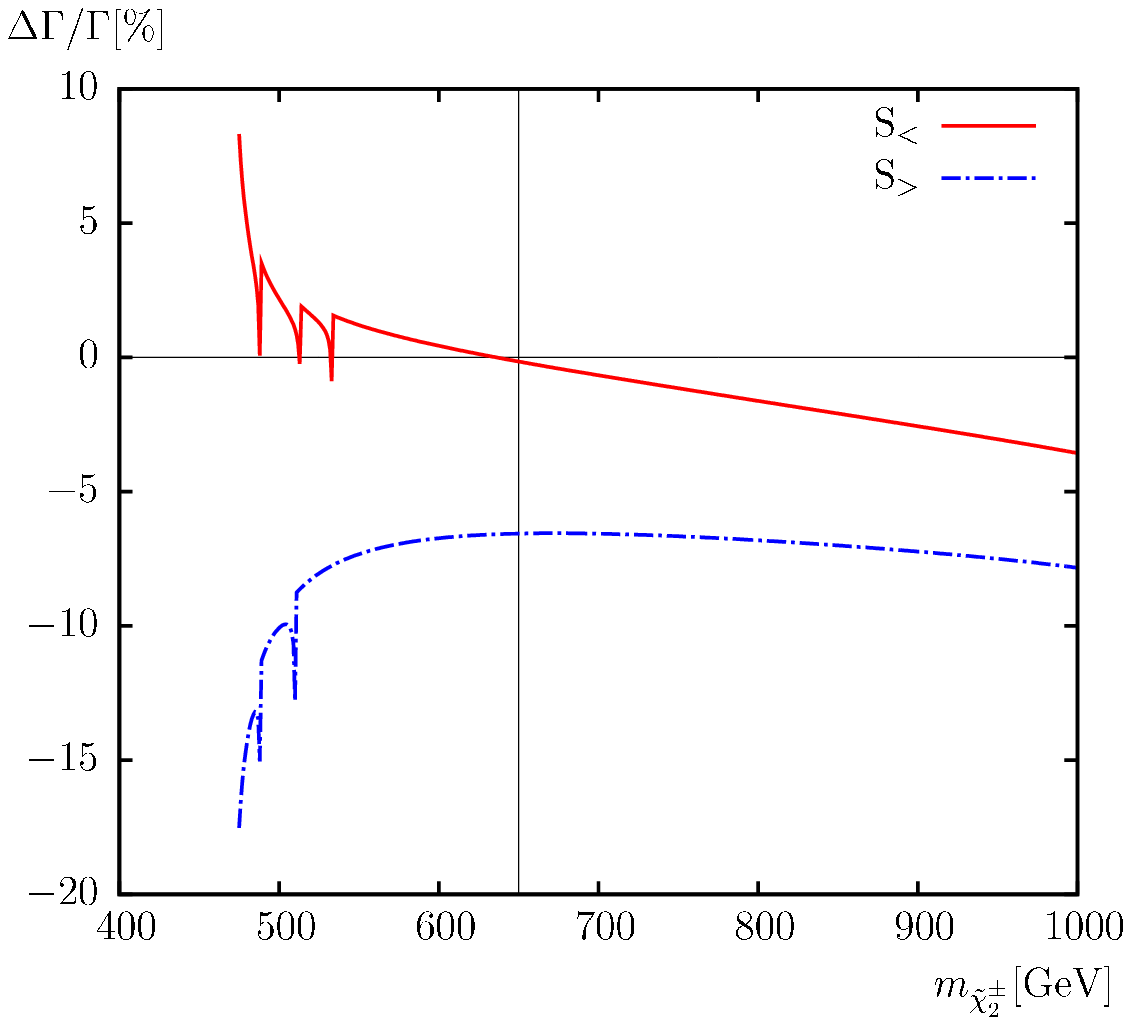} 
\\[4em]
\includegraphics[width=0.49\textwidth,height=7.5cm]{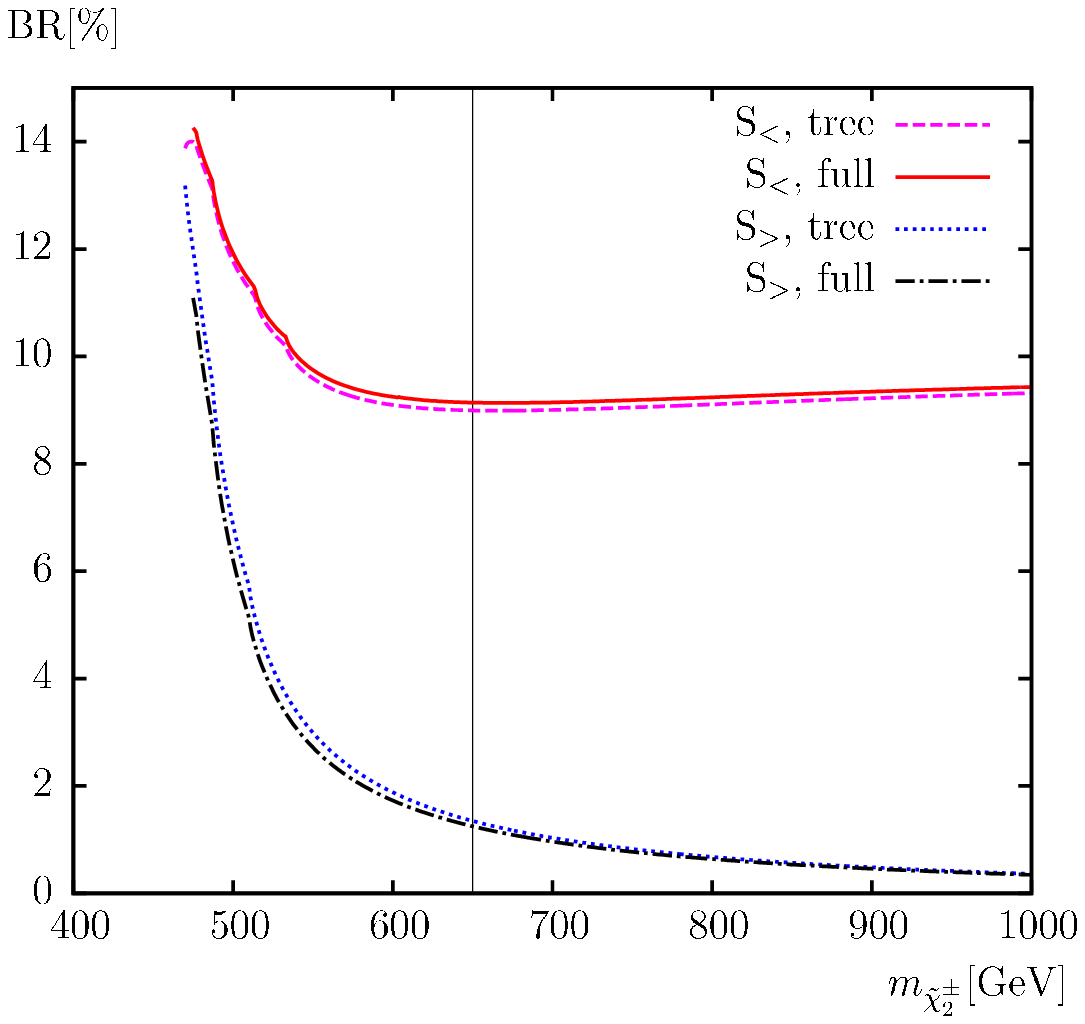}
\hspace{-4mm}
\includegraphics[width=0.49\textwidth,height=7.5cm]{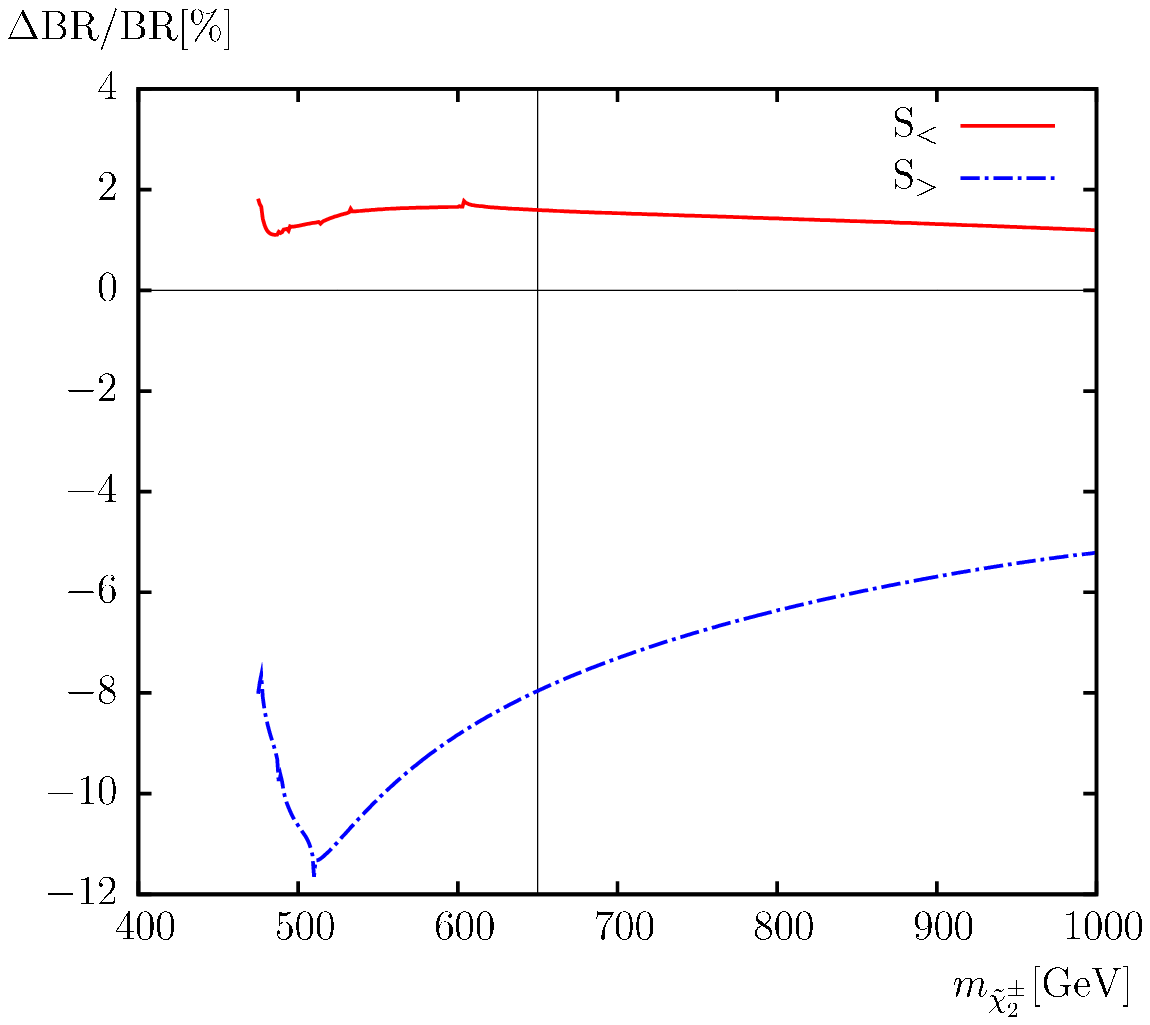}
\end{tabular}
\vspace{2em}
\caption{
  $\Ga(\DecayCmnSl{2}{\mu}{1})$. 
  Tree-level (``tree'') and full one-loop (``full'') corrected 
  decay widths are shown with the parameters chosen according to \SN\
  (see \refta{tab:para}), with $\mcha{2}$ varied.
  The upper left plot shows the decay width, the upper right plot shows 
  the relative size of the corrections.
  The lower left plot shows the BR, the lower right plot shows 
  the relative size of the BR.
  The vertical lines indicate where $\mcha{1} + \mcha{2} = 1000 \gev$, 
  i.e.\ the maximum reach of the ILC(1000).
}
\label{fig:mC2.cha2smu1nu}
\end{center}
\end{figure}

\begin{figure}[htb!]
\begin{center}
\begin{tabular}{c}
\includegraphics[width=0.49\textwidth,height=7.5cm]{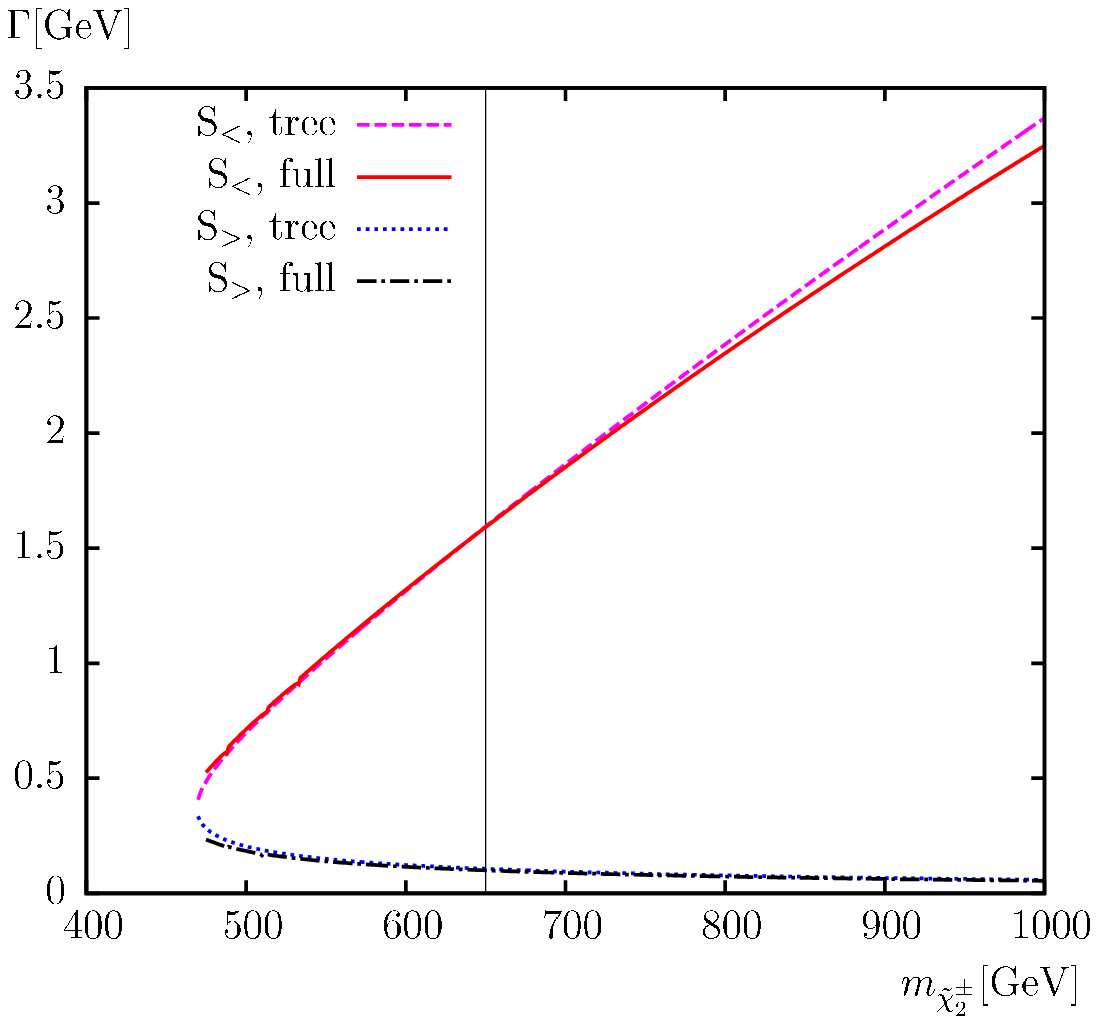}
\hspace{-4mm}
\includegraphics[width=0.49\textwidth,height=7.5cm]{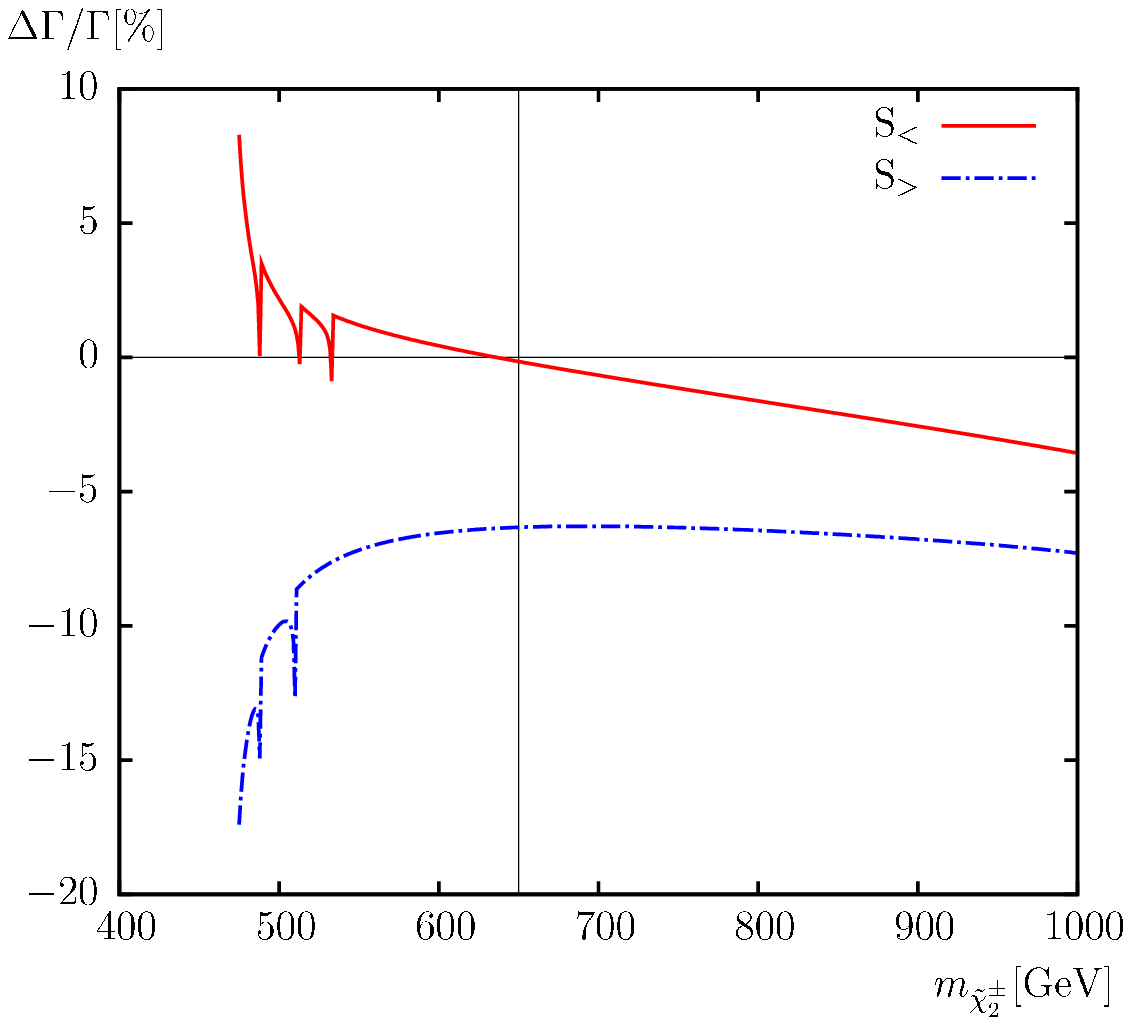} 
\\[4em]
\includegraphics[width=0.49\textwidth,height=7.5cm]{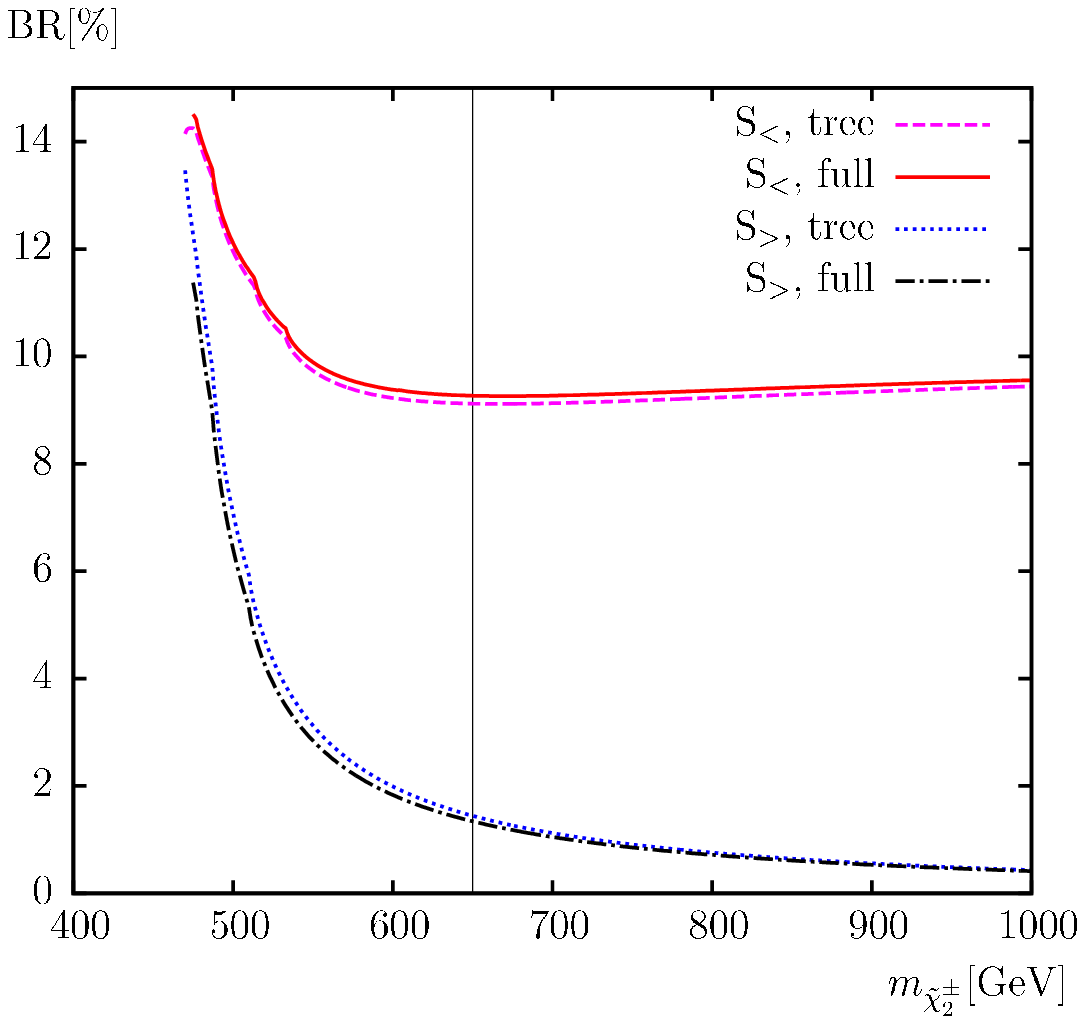}
\hspace{-4mm}
\includegraphics[width=0.49\textwidth,height=7.5cm]{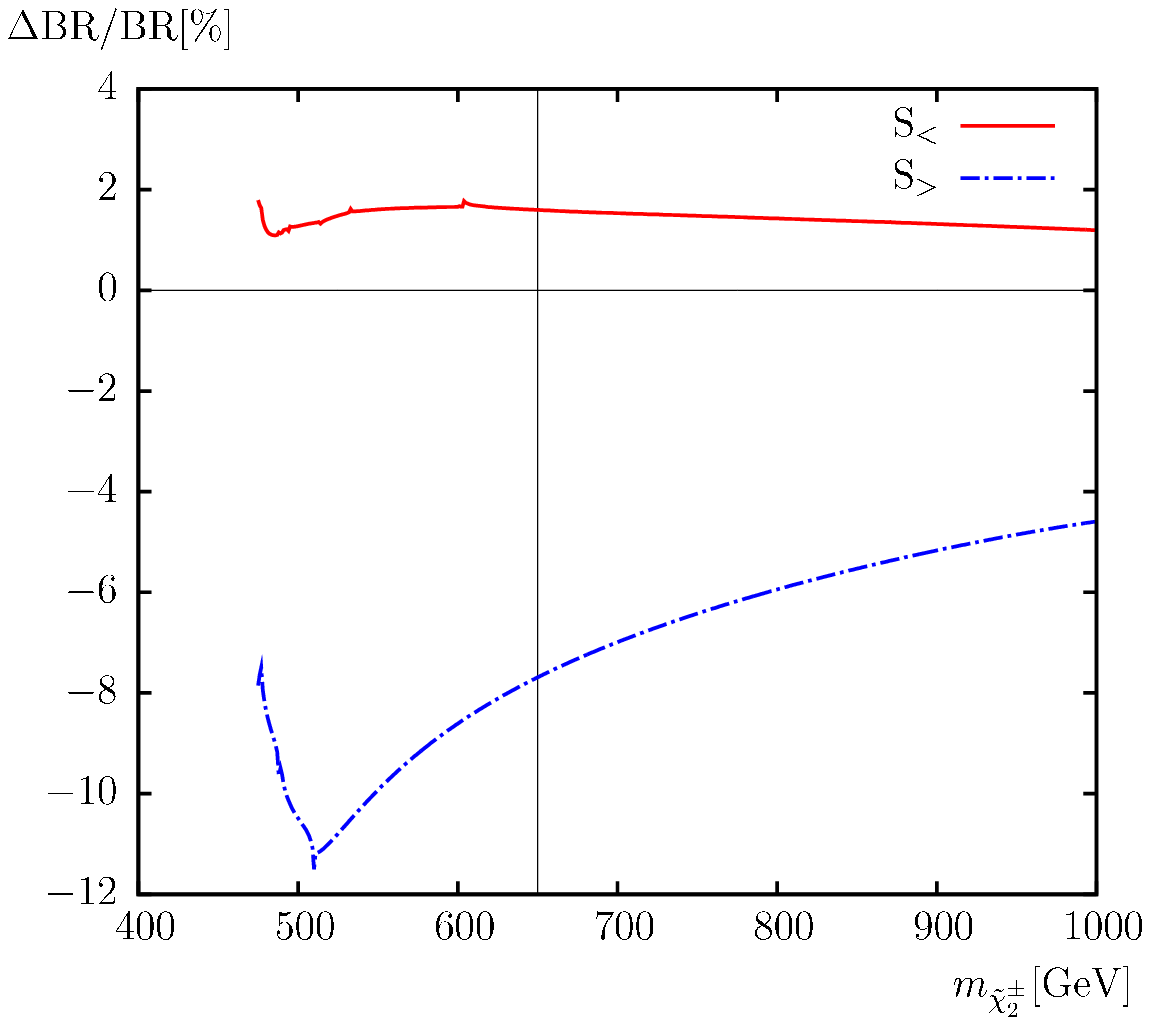}
\end{tabular}
\vspace{2em}
\caption{
  $\Ga(\DecayCmnSl{2}{e}{1})$. 
  Tree-level (``tree'') and full one-loop (``full'') corrected 
  decay widths are shown with the parameters chosen according to \SN\
  (see \refta{tab:para}), with $\mcha{2}$ varied.
  The upper left plot shows the decay width, the upper right plot shows 
  the relative size of the corrections.
  The lower left plot shows the BR, the lower right plot shows 
  the relative size of the BR.
  The vertical lines indicate where $\mcha{1} + \mcha{2} = 1000 \gev$, 
  i.e.\ the maximum reach of the ILC(1000).
}
\label{fig:mC2.cha2sel1nu}
\end{center}
\end{figure}

\medskip
Now we turn to the decays involving (scalar) leptons. 
All these decay widths follow the same pattern that can be understood
from \refeqs{eq:Lagr.CSln}, (\ref{eq:Lagr.CSnl}).
We have chosen $\MslL < \MslR$, leading to a left-handed lighter and a
right-handed heavier slepton, where significant mixing can
be found in the scalar tau sector. 
In \reffis{fig:mC2.cha2stau1nu}, \ref{fig:mC2.cha2smu1nu},
\ref{fig:mC2.cha2sel1nu} we show the decays
$\DecayCmnSl{2}{\tau}{1}, \DecayCmnSl{2}{\mu}{1}, \DecayCmnSl{2}{e}{1}$
respectively. 
The dips, best visible in the upper right panels, are due to 
the $\cha{1} \Hz$, $\neu{2} H^-$ and $\neu{3} H^-$ thresholds
and have been already described for $\DecayCmNW{2}{2,3}$. 
Within \SE\ the $\cha{2}$
turns from a mixed higgsino/wino state at low $\mcha{2}$ to a pure
higgsino state at large $\mcha{2}$ with a vanishing coupling to the
left-handed slepton. 
Consequently, these widths are very small, and only in the case of
$\Staue$, due to the mixture between left- and right-handed states it
does not go to zero for large $\mcha{2}$, but goes through zero at
$\mcha{2} \approx 600 \gev$ due to a cancellation of the 
higgsino (suppressed by the Yukawa term) and the small gaugino
contributions 
to the $\cham{2}\nu_\tau\Staue^\dagger$ coupling. 
In \SZ, on the other hand,
$\cha{2}$ changes from the mixed wino/higgsino state to a wino-like
state at large $\mcha{2}$, and the EW coupling (which is flavor
independent) dominates the decay to the
left-handed sleptons, leading to a rise of the decay widths to 
$\sim 3.2 \gev$ in the case of 
$\DecayCmnSl{2}{\mu}{1}, \DecayCmnSl{2}{e}{1}$ and a reduced
value of $2 \gev$ for $\DecayCmnSl{2}{\tau}{1}$, again due to the mixing
in the stau sector. 
The $\br(\DecayCmnSl{2}{\tau}{1})$ in \SE\ is large
only at the smallest $\mcha{2}$ and below $\sim 0.5 \%$ for most 
$\mcha{2}$ values. For the first and second slepton generation we also
find a monotonous decrease, although a bit weaker than for the $\Staue$.
Within \SZ\ the $\br(\DecayCmnSl{2}{\tau}{1})$ is mostly found at 
$\sim 6\%$, whereas 
$\br(\DecayCmnSl{2}{\mu}{1})$ and $\br(\DecayCmnSl{2}{e}{1})$ are 
somewhat larger,
due to the absence of mixing, and reaches $\sim 9.5\%$. The size of the
one-loop corrections to the decay widths and BR's is only substantial
where the decay widths are small, reaching nearly $-12\%$ at 
$\mcha{2} \approx 500 \gev$ in \SE. In the case of a large BR, i.e.\ in
\SZ, the corrections stay at the level of $\sim 1\%$, staying below
the precision anticipated for the ILC(1000).

\begin{figure}[htb!]
\begin{center}
\begin{tabular}{c}
\includegraphics[width=0.49\textwidth,height=7.5cm]{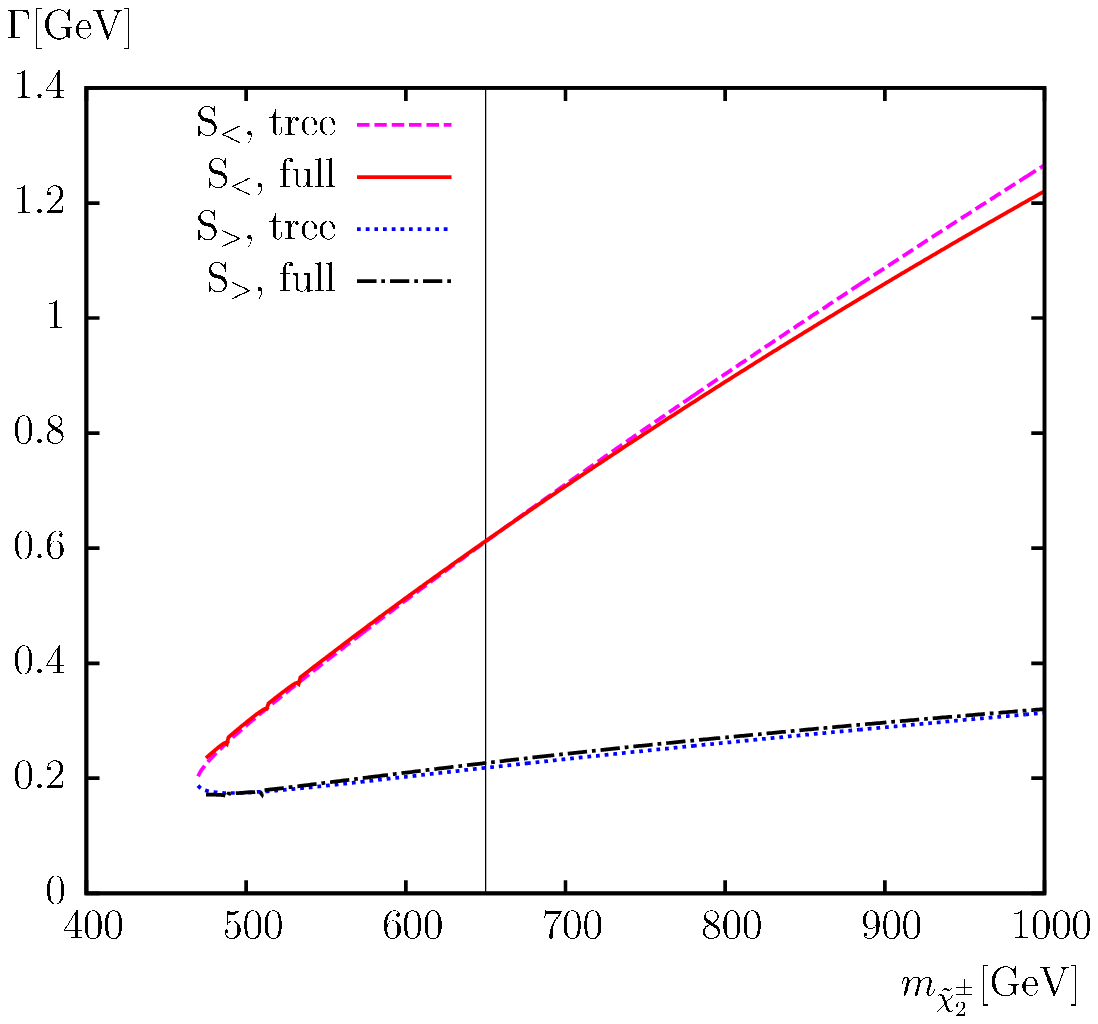}
\hspace{-4mm}
\includegraphics[width=0.49\textwidth,height=7.5cm]{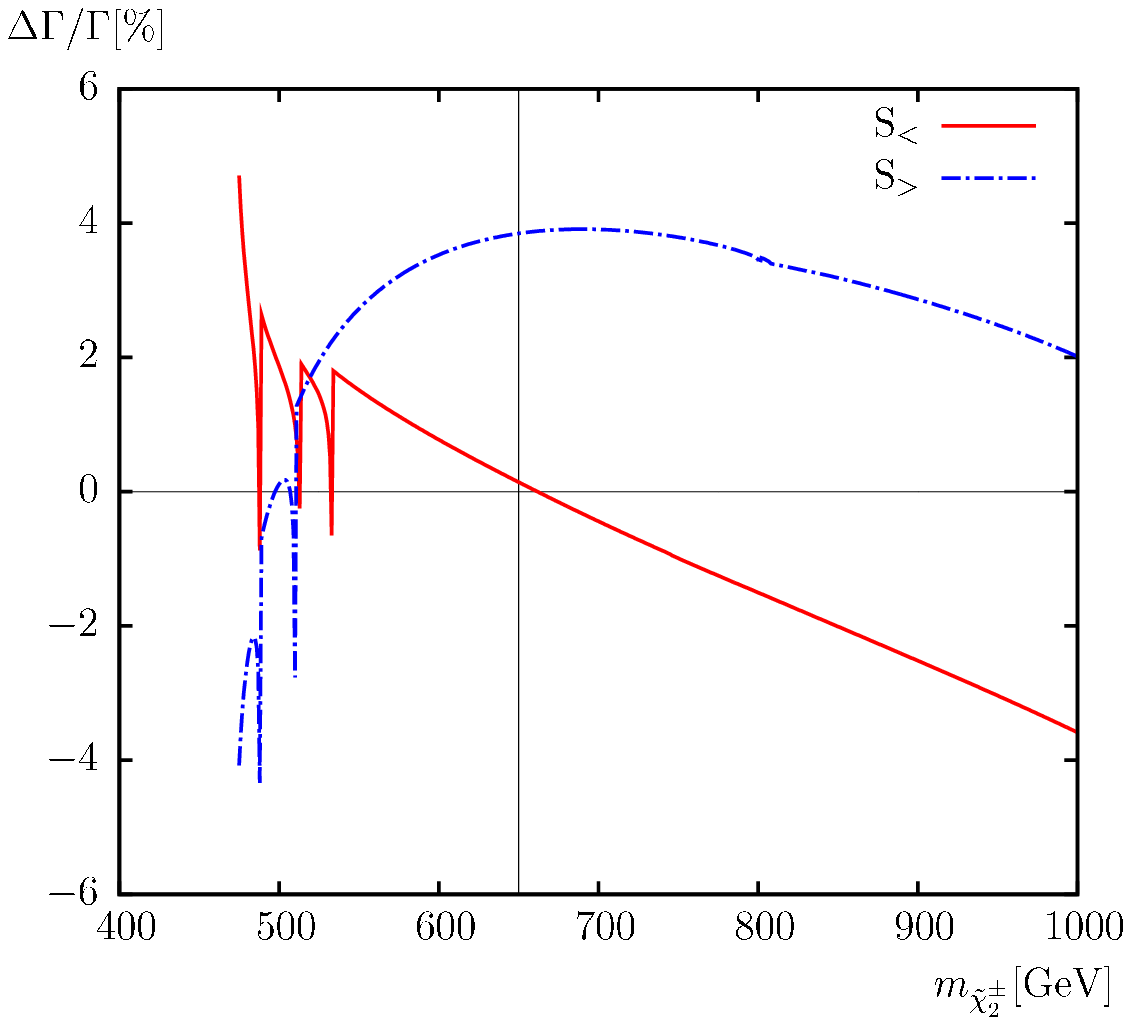} 
\\[4em]
\includegraphics[width=0.49\textwidth,height=7.5cm]{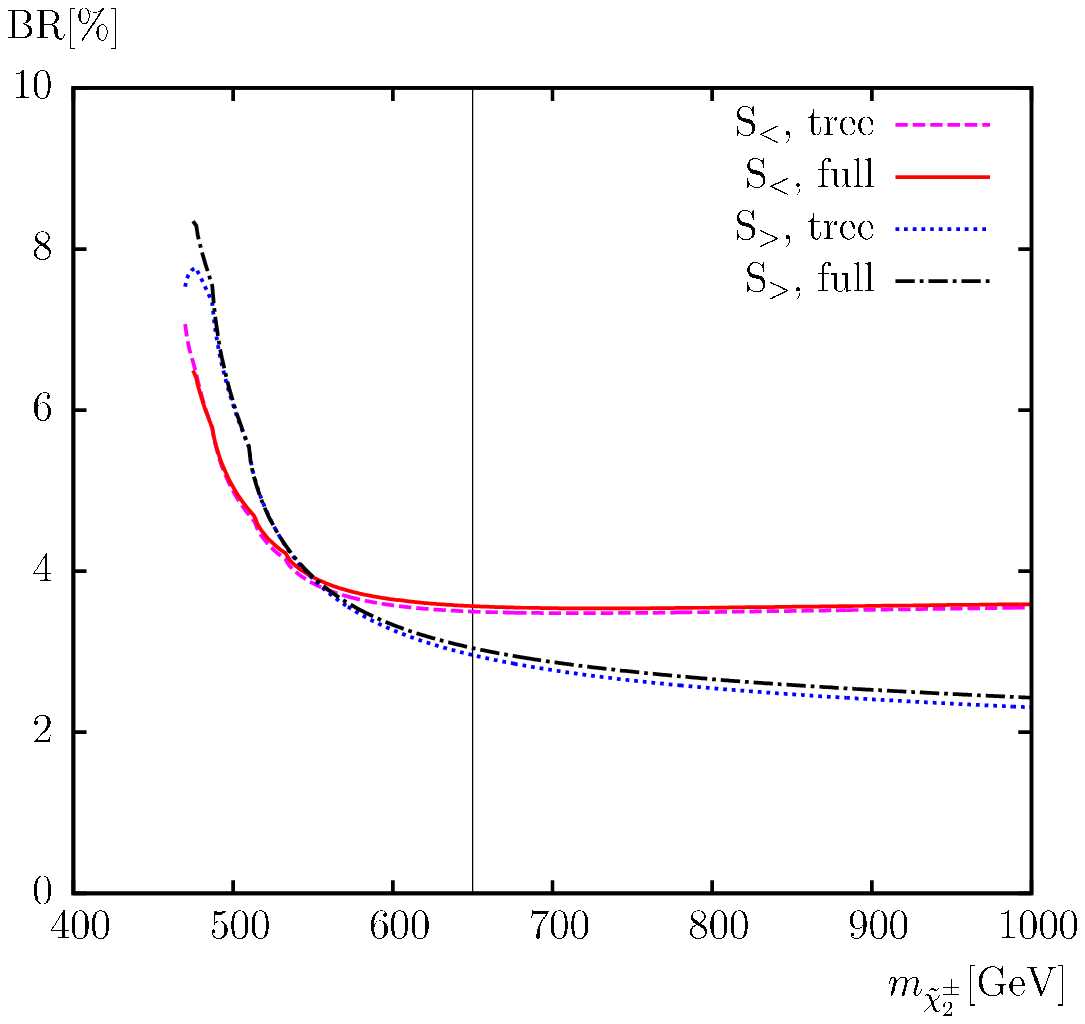}
\hspace{-4mm}
\includegraphics[width=0.49\textwidth,height=7.5cm]{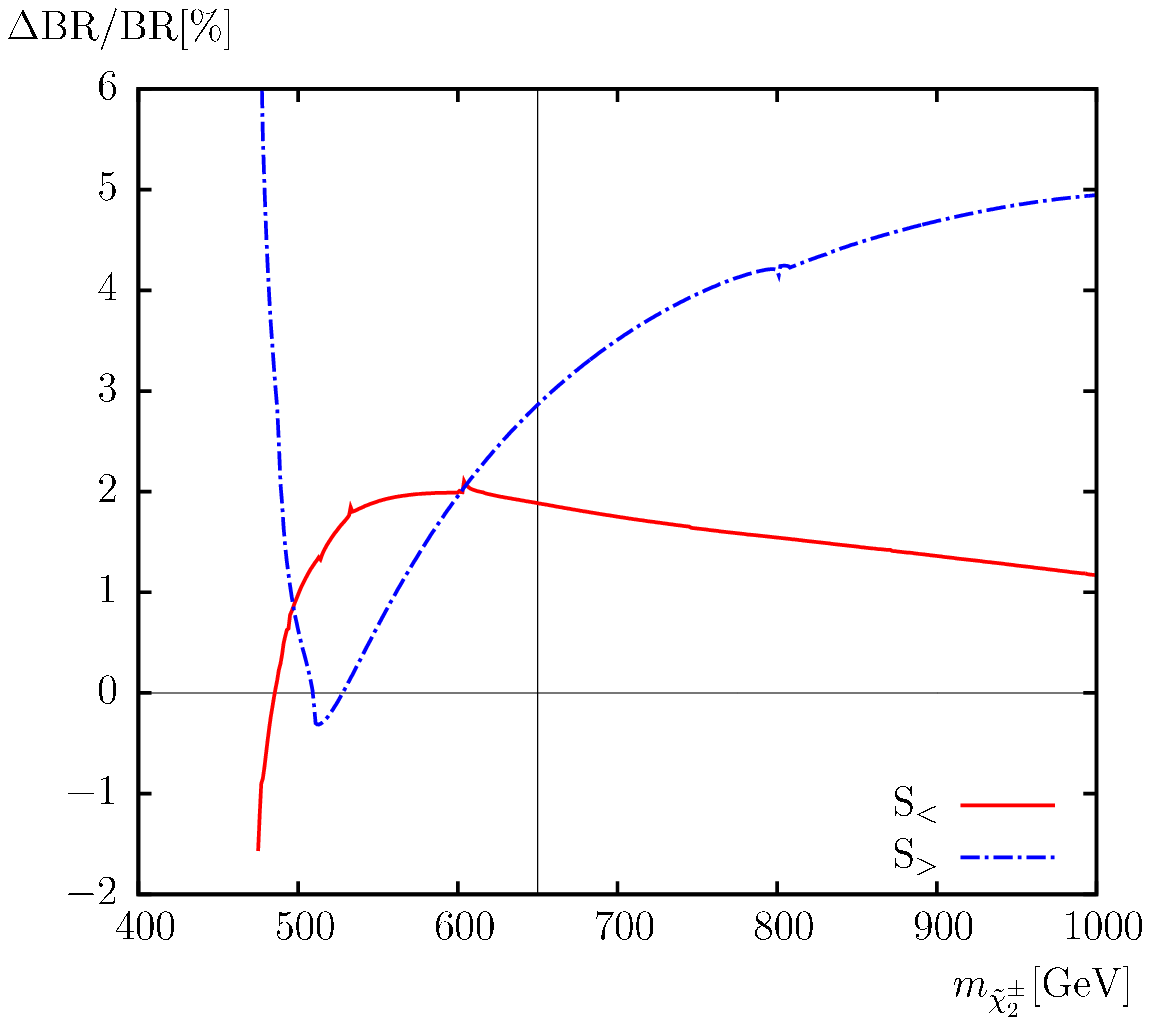}
\end{tabular}
\vspace{2em}
\caption{
  $\Ga(\DecayCmnSl{2}{\tau}{2})$. 
  Tree-level (``tree'') and full one-loop (``full'') corrected 
  decay widths are shown with the parameters chosen according to \SN\
  (see \refta{tab:para}), with $\mcha{2}$ varied.
  The upper left plot shows the decay width, the upper right plot shows 
  the relative size of the corrections.
  The lower left plot shows the BR, the lower right plot shows 
  the relative size of the BR.
  The vertical lines indicate where $\mcha{1} + \mcha{2} = 1000 \gev$, 
  i.e.\ the maximum reach of the ILC(1000).
}
\label{fig:mC2.cha2stau2nu}
\end{center}
\end{figure}

\begin{figure}[htb!]
\begin{center}
\begin{tabular}{c}
\includegraphics[width=0.49\textwidth,height=7.5cm]{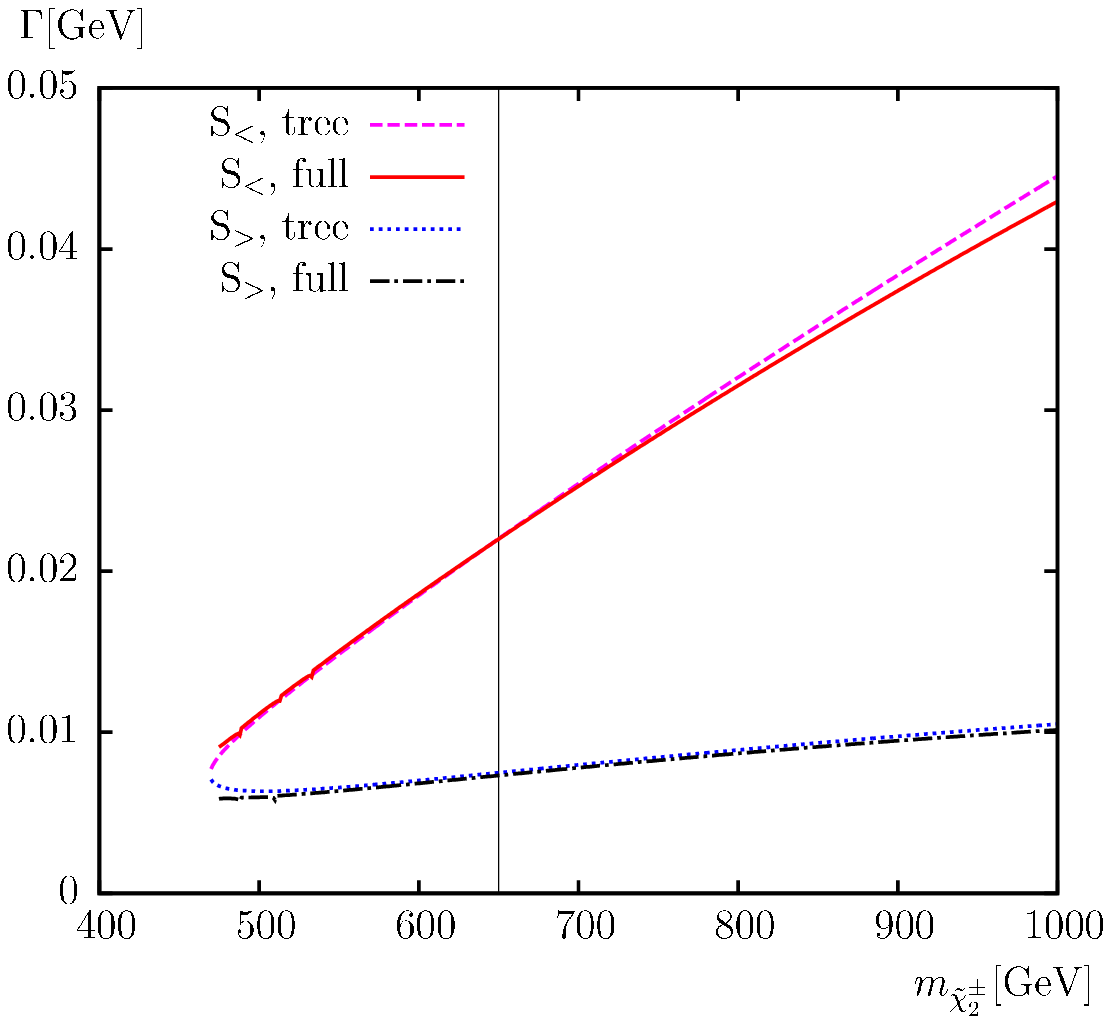}
\hspace{-4mm}
\includegraphics[width=0.49\textwidth,height=7.5cm]{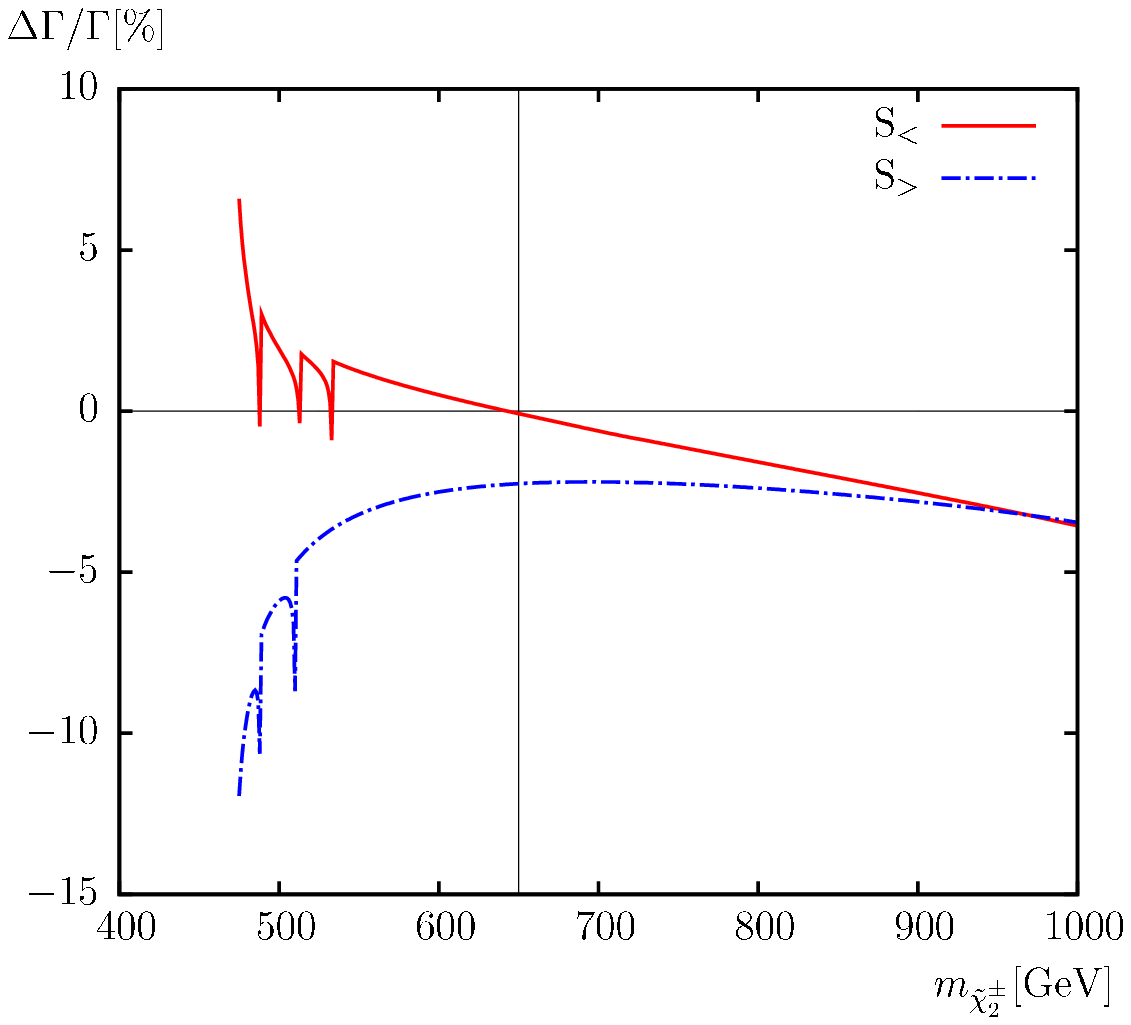} 
\\[4em]
\includegraphics[width=0.49\textwidth,height=7.5cm]{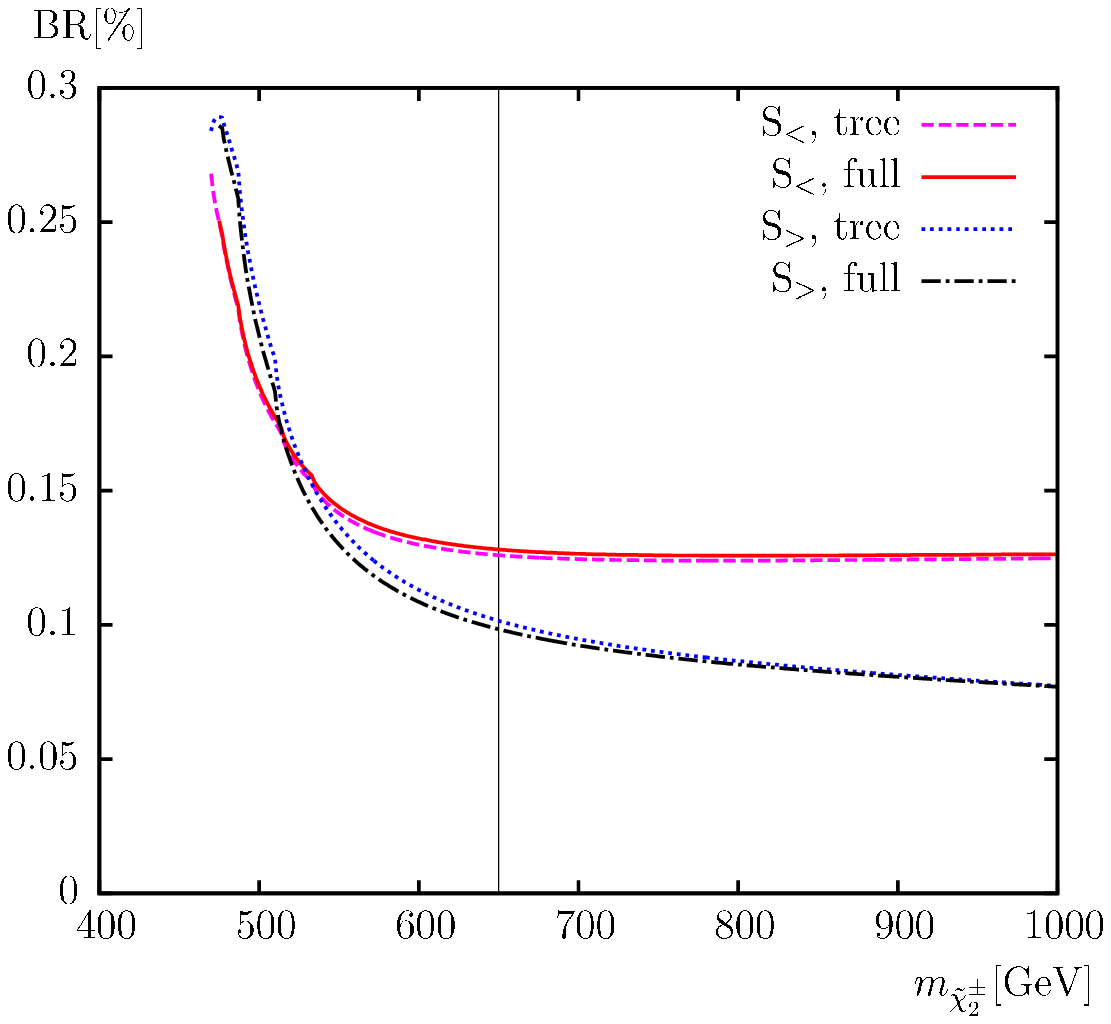}
\hspace{-4mm}
\includegraphics[width=0.49\textwidth,height=7.5cm]{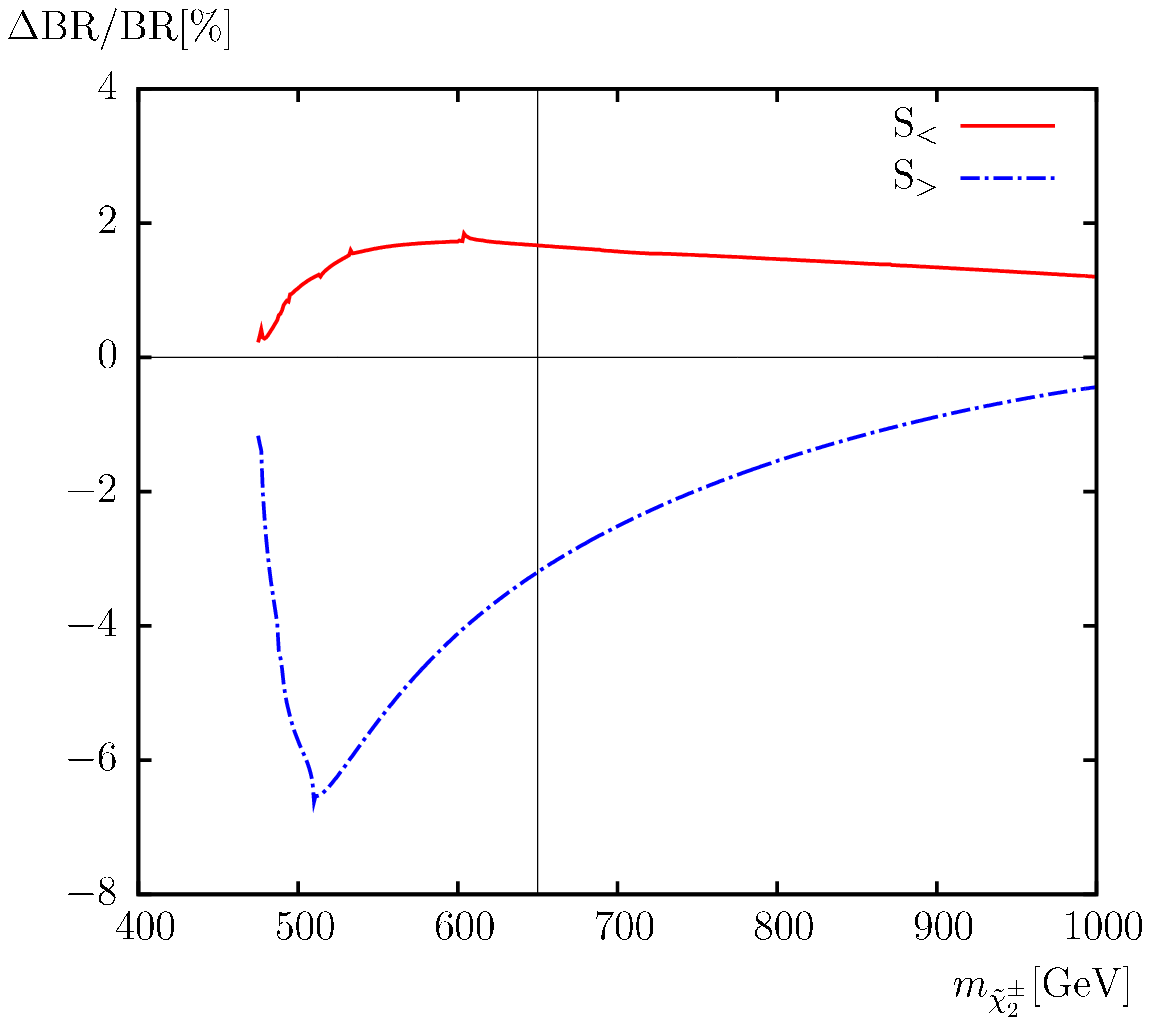}
\end{tabular}
\vspace{2em}
\caption{
  $\Ga(\DecayCmnSl{2}{\mu}{2})$. 
  Tree-level (``tree'') and full one-loop (``full'') corrected 
  decay widths are shown with the parameters chosen according to \SN\
  (see \refta{tab:para}), with $\mcha{2}$ varied.
  The upper left plot shows the decay width, the upper right plot shows 
  the relative size of the corrections.
  The lower left plot shows the BR, the lower right plot shows 
  the relative size of the BR.
  The vertical lines indicate where $\mcha{1} + \mcha{2} = 1000 \gev$, 
  i.e.\ the maximum reach of the ILC(1000).
}
\label{fig:mC2.cha2smu2nu}
\end{center}
\end{figure}

\begin{figure}[htb!]
\begin{center}
\begin{tabular}{c}
\includegraphics[width=0.49\textwidth,height=7.5cm]{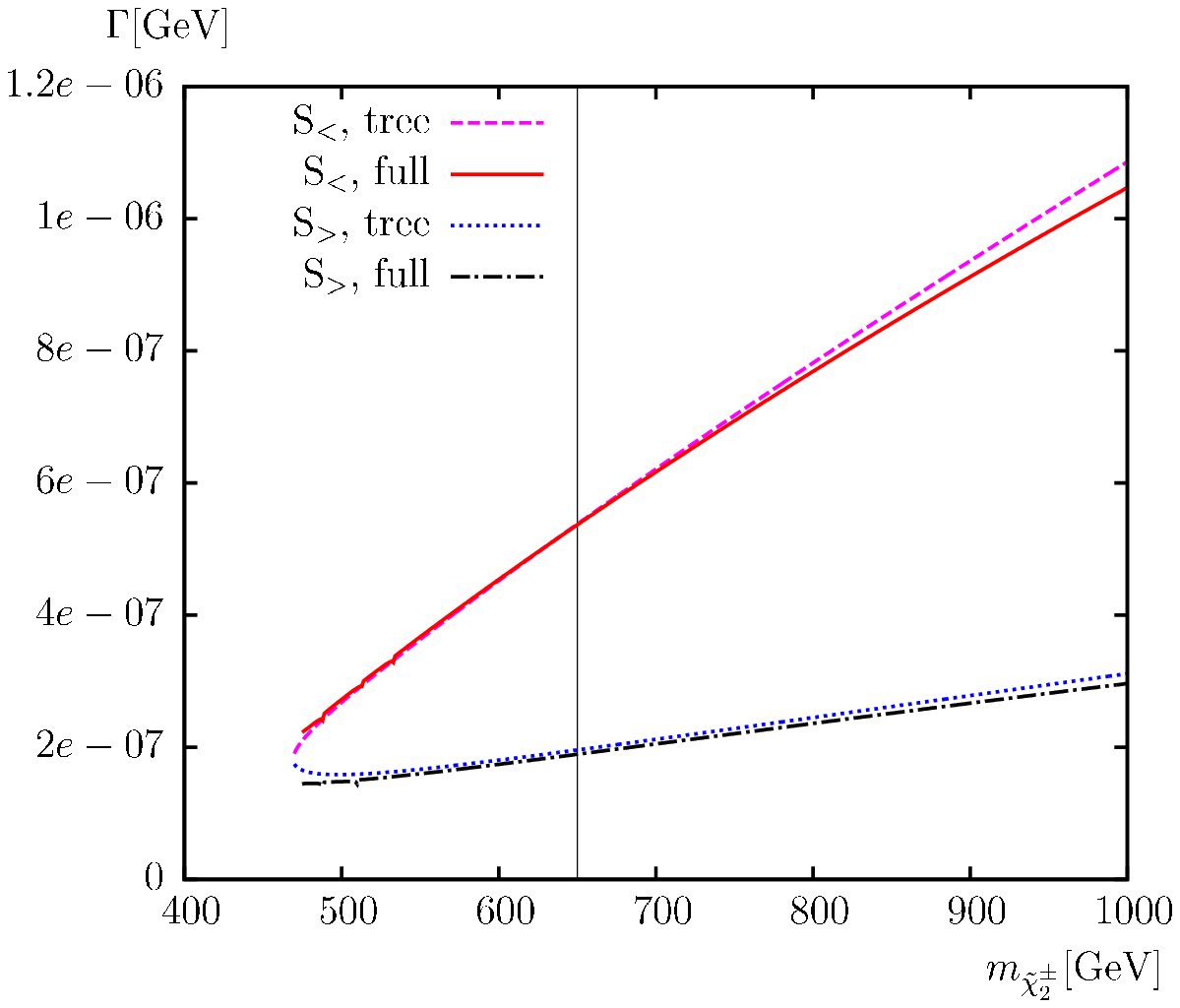}
\hspace{-4mm}
\includegraphics[width=0.49\textwidth,height=7.5cm]{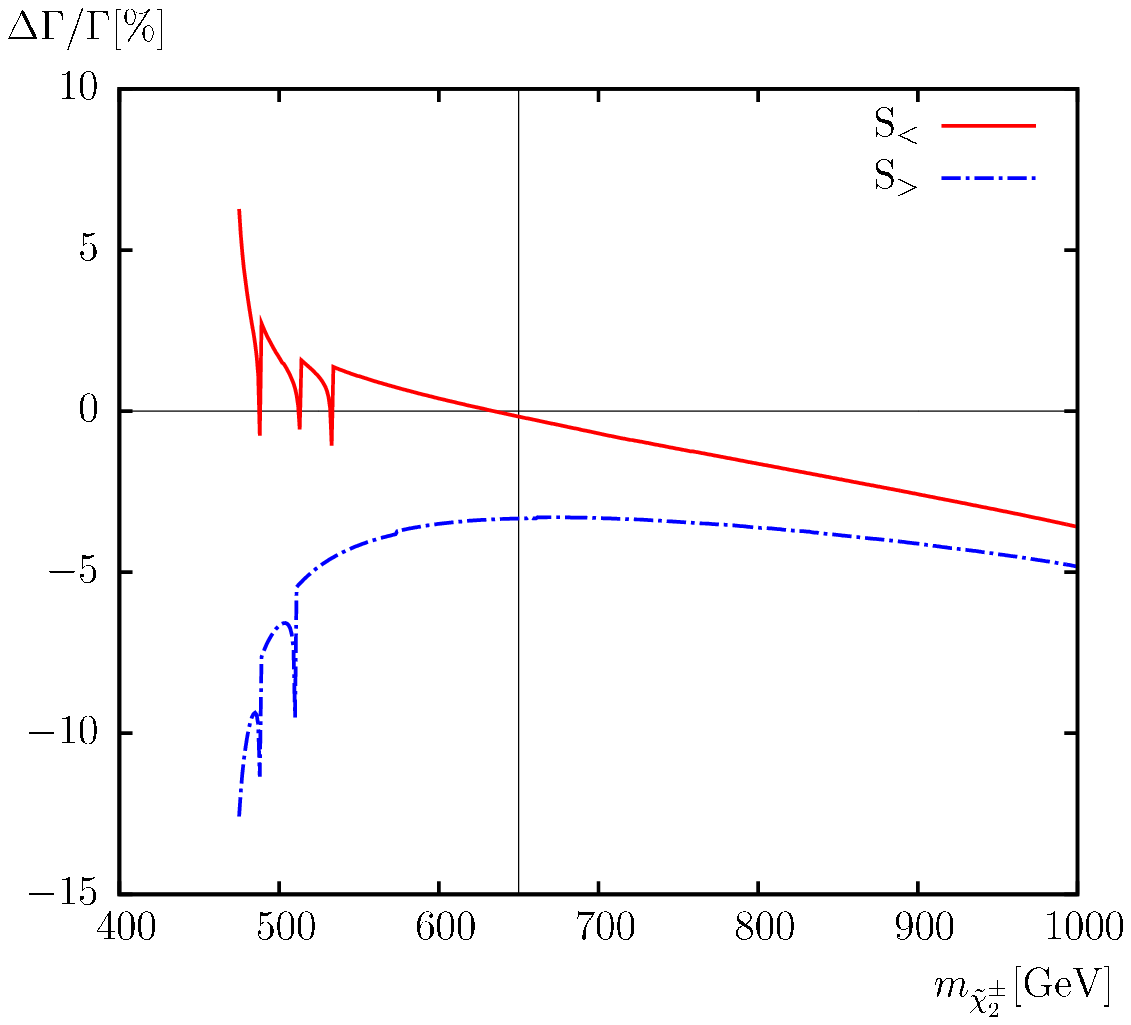} 
\\[4em]
\includegraphics[width=0.49\textwidth,height=7.5cm]{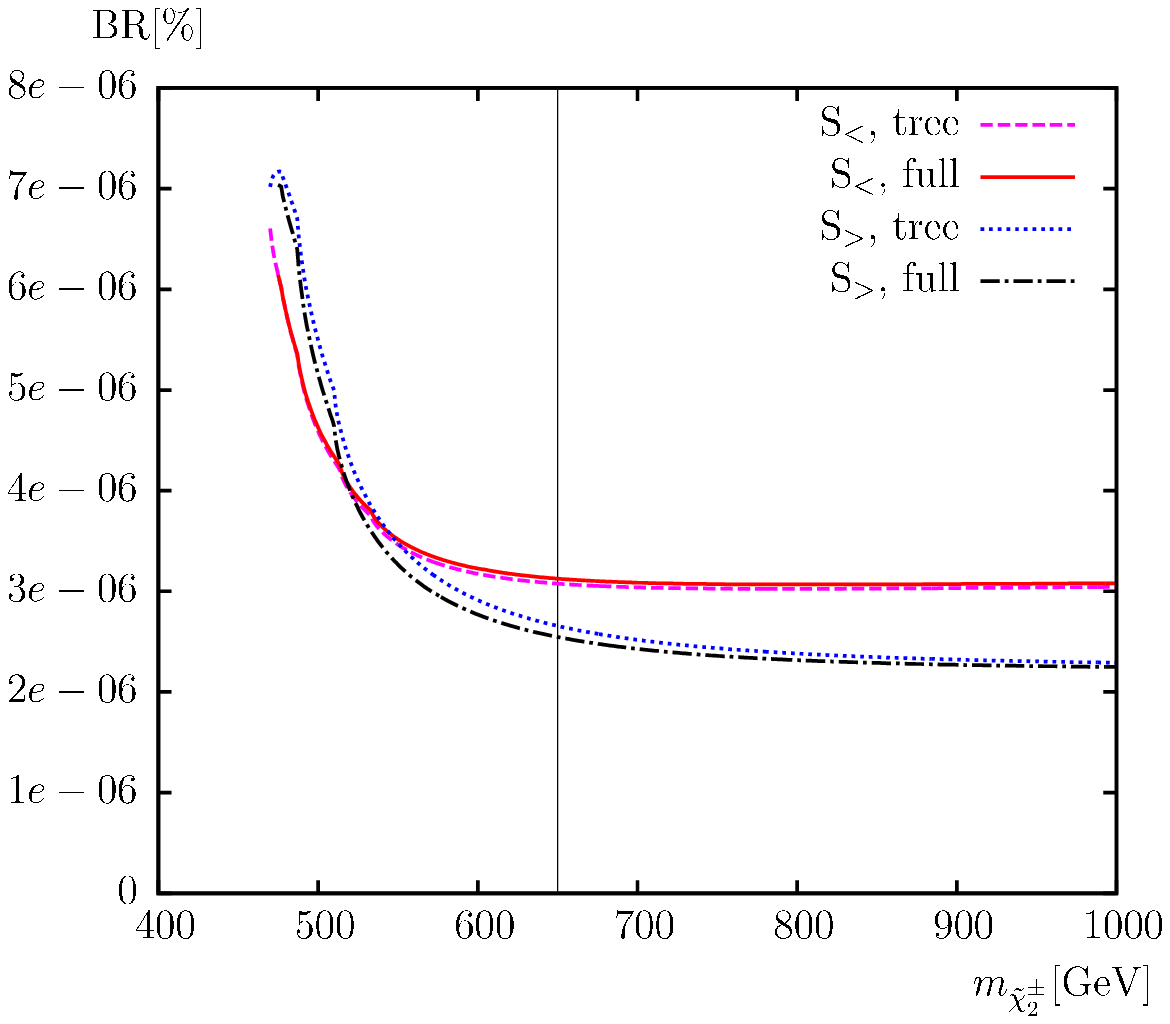}
\hspace{-4mm}
\includegraphics[width=0.49\textwidth,height=7.5cm]{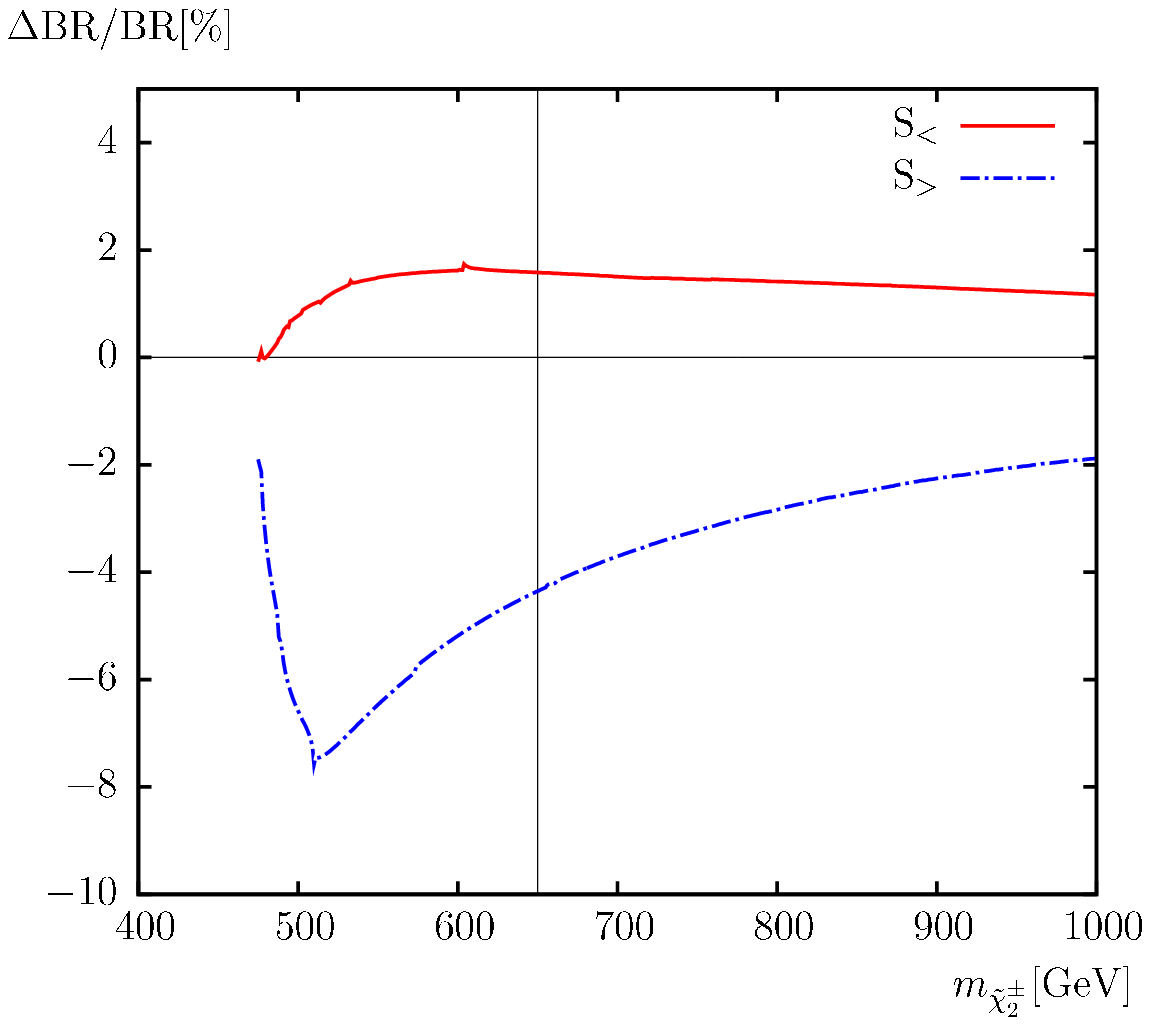}
\end{tabular}
\vspace{2em}
\caption{
  $\Ga(\DecayCmnSl{2}{e}{2})$. 
  Tree-level (``tree'') and full one-loop (``full'') corrected 
  decay widths are shown with the parameters chosen according to \SN\
  (see \refta{tab:para}), with $\mcha{2}$ varied.
  The upper left plot shows the decay width, the upper right plot shows 
  the relative size of the corrections.
  The lower left plot shows the BR, the lower right plot shows 
  the relative size of the BR.
  The vertical lines indicate where $\mcha{1} + \mcha{2} = 1000 \gev$, 
  i.e.\ the maximum reach of the ILC(1000).
}
\label{fig:mC2.cha2sel2nu}
\end{center}
\end{figure}

The decays to the heavier sleptons, 
$\DecayCmnSl{2}{\tau}{2}, \DecayCmnSl{2}{\mu}{2}, \DecayCmnSl{2}{e}{2}$,
shown in \reffis{fig:mC2.cha2stau2nu}, \ref{fig:mC2.cha2smu2nu},
\ref{fig:mC2.cha2sel2nu}, follow a similar pattern. 
The dips visible in
the upper right panels stem from the same thresholds as those observed 
for the decays to the lighter sleptons.
At low $\mcha{2}$ values the mixed higgsino/wino state couples to the
right-handed sleptons through the Yukawa term~$\propto \ml$. At large
$\mcha{2}$ in \SE\ only the higgsino part of $\cha{2}$ survives, leading
to a (Yukawa term) suppressed coupling to the right-handed slepton and 
the corresponding decay widths are very small, below 
$0.3, 0.01, 2 \times 10^{-7} \gev$ for 
$\DecayCmnSl{2}{\tau}{2}, \DecayCmnSl{2}{\mu}{2}, \DecayCmnSl{2}{e}{2}$, 
respectively. In \SZ, on the other hand, the small wino component
couples to the small left-handed admixture of the heavier slepton. 
For $\DecayCmnSl{2}{\tau}{2}$, due to the relatively large mixing, this
still yields a loop-corrected decay width up to $1.2 \gev$ for large
$\mcha{2}$, while for the first and second generation sleptons the
widths stay below $0.045$ and $1.1 \times 10^{-6} \gev$, respectively.
Substantial branching ratios are only found for $\DecayCmnSl{2}{\tau}{2}$,
where values between $\sim 6\%$ and $\sim 2\%$ are realized. In this
case the size of the one-loop corrections varies between $0$~and
$+5\%$. At the high end, this exceeds the anticipated ILC(1000)
accuracy. 

\begin{figure}[htb!]
\begin{center}
\begin{tabular}{c}
\includegraphics[width=0.49\textwidth,height=7.5cm]{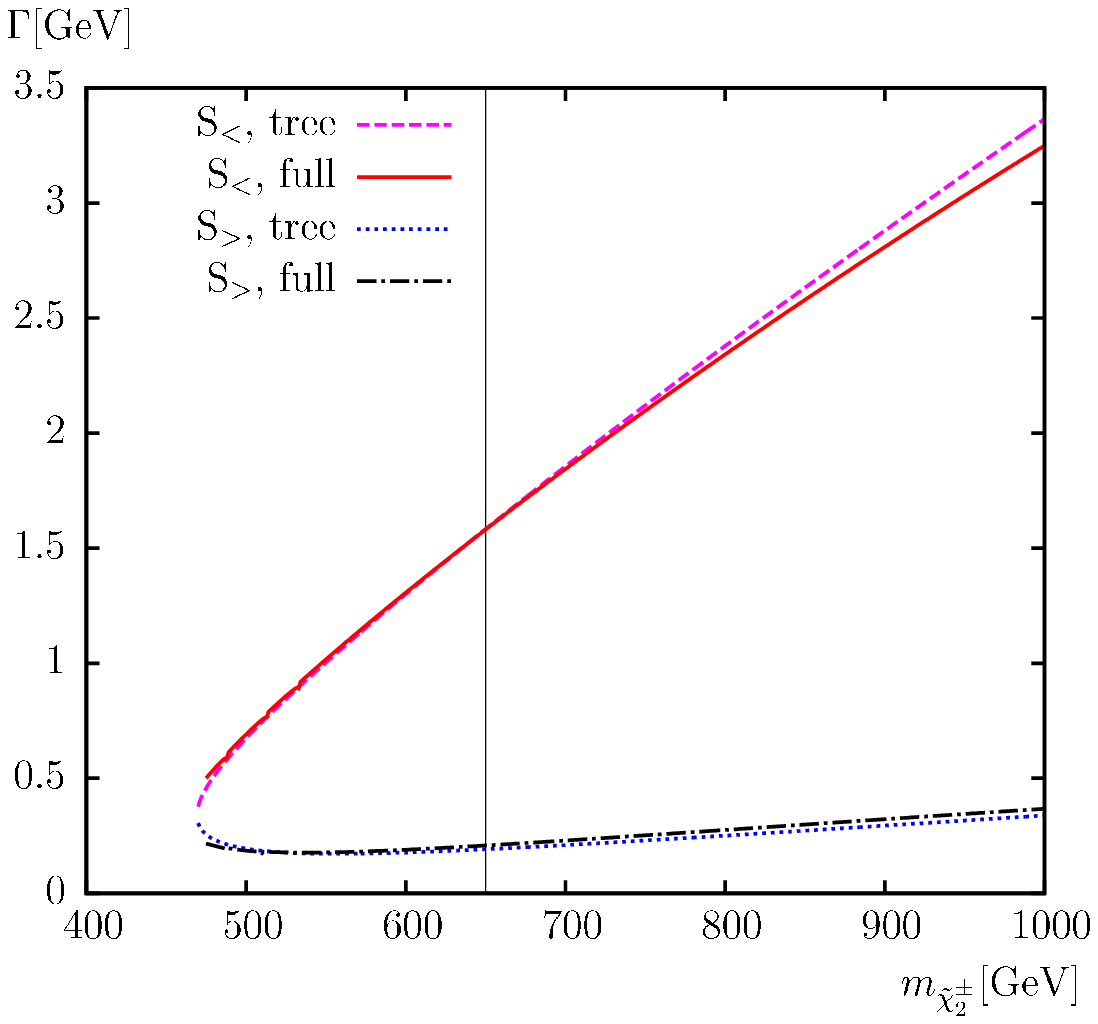}
\hspace{-4mm}
\includegraphics[width=0.49\textwidth,height=7.5cm]{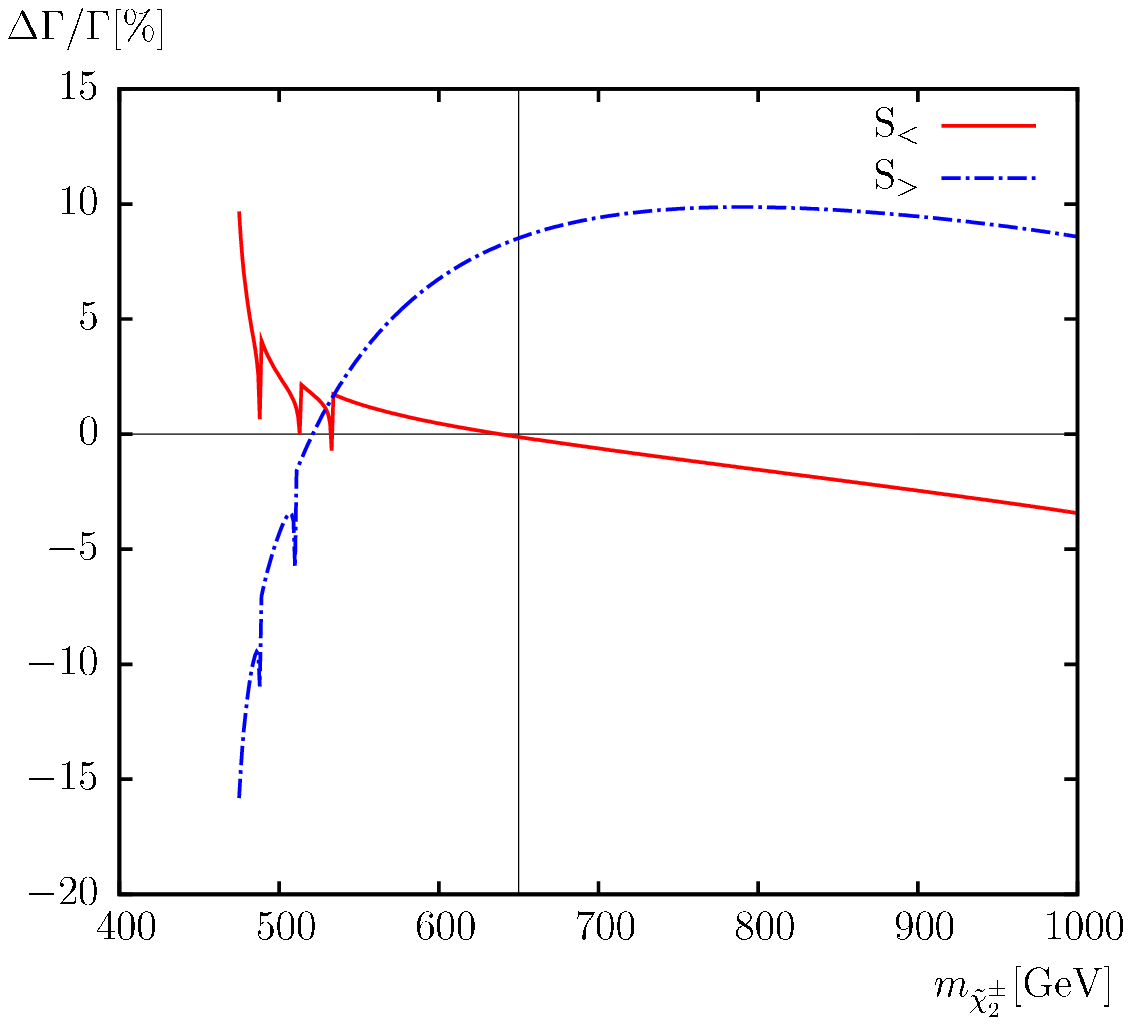} 
\\[4em]
\includegraphics[width=0.49\textwidth,height=7.5cm]{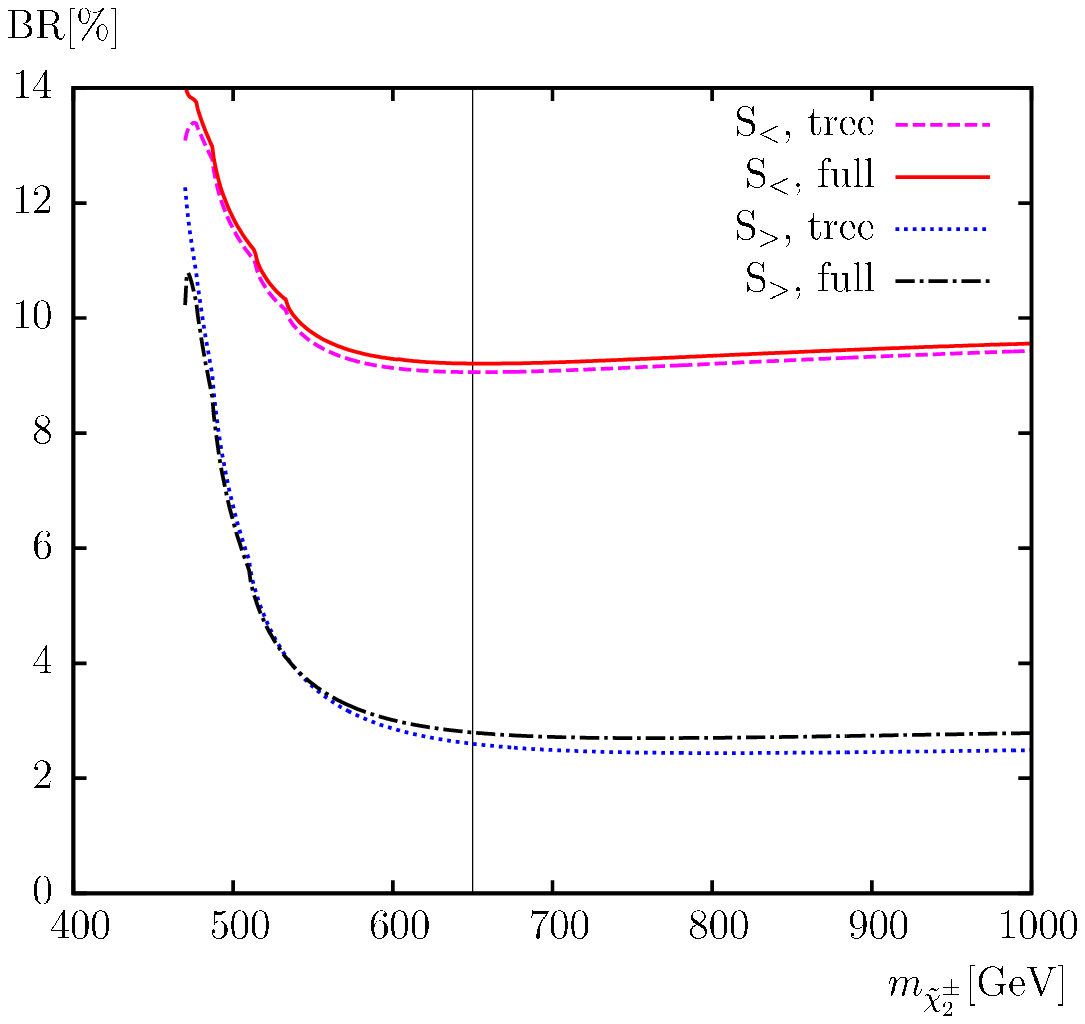}
\hspace{-4mm}
\includegraphics[width=0.49\textwidth,height=7.5cm]{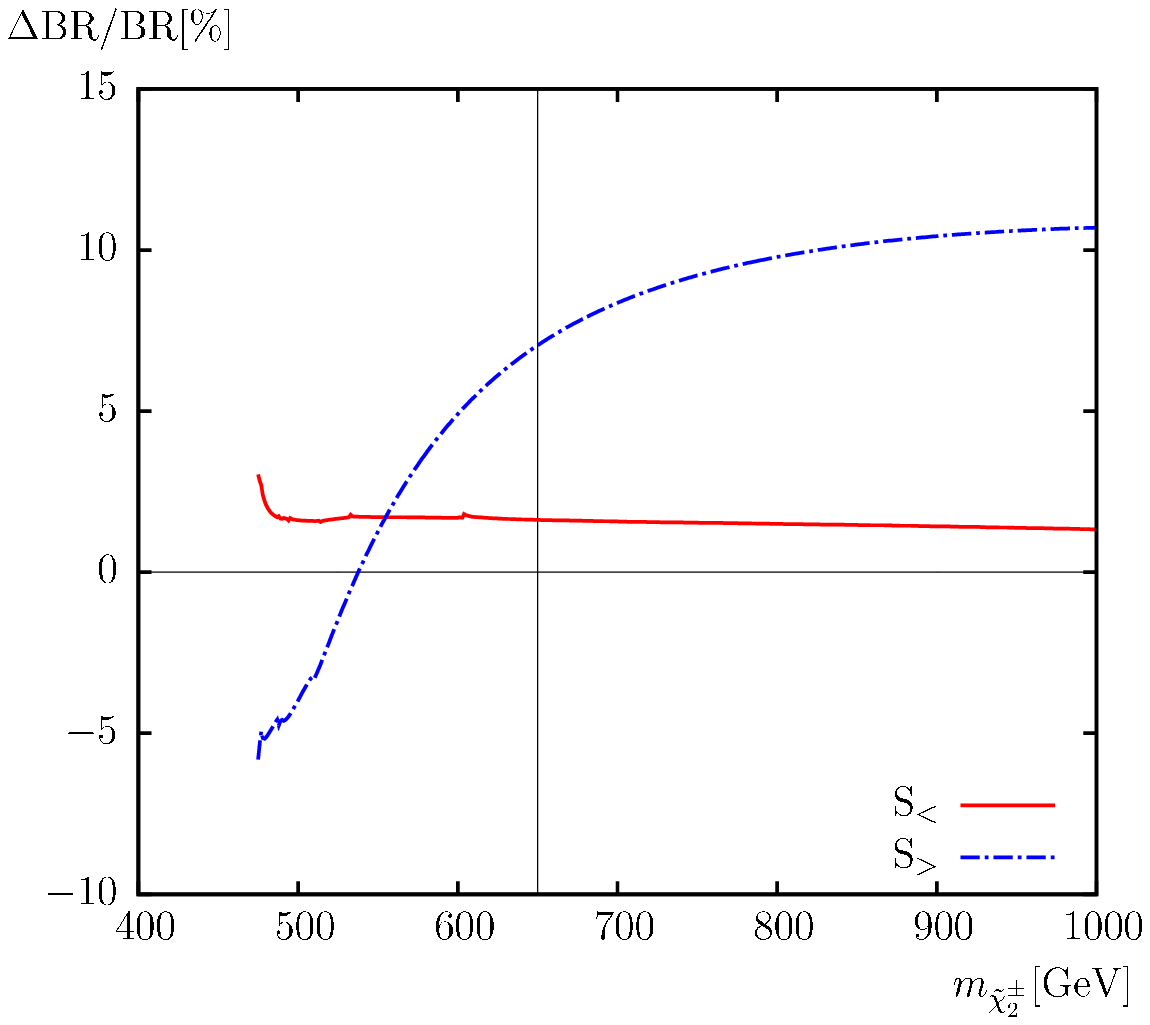}
\end{tabular}
\vspace{2em}
\caption{
  $\Ga(\DecayCmlSn{2}{\tau})$. 
  Tree-level (``tree'') and full one-loop (``full'') corrected 
  decay widths are shown with the parameters chosen according to \SN\
  (see \refta{tab:para}), with $\mcha{2}$ varied.
  The upper left plot shows the decay width, the upper right plot shows 
  the relative size of the corrections.
  The lower left plot shows the BR, the lower right plot shows 
  the relative size of the BR.
  The vertical lines indicate where $\mcha{1} + \mcha{2} = 1000 \gev$, 
  i.e.\ the maximum reach of the ILC(1000).
}
\label{fig:mC2.cha2snutau}
\end{center}
\end{figure}

\begin{figure}[htb!]
\begin{center}
\begin{tabular}{c}
\includegraphics[width=0.49\textwidth,height=7.5cm]{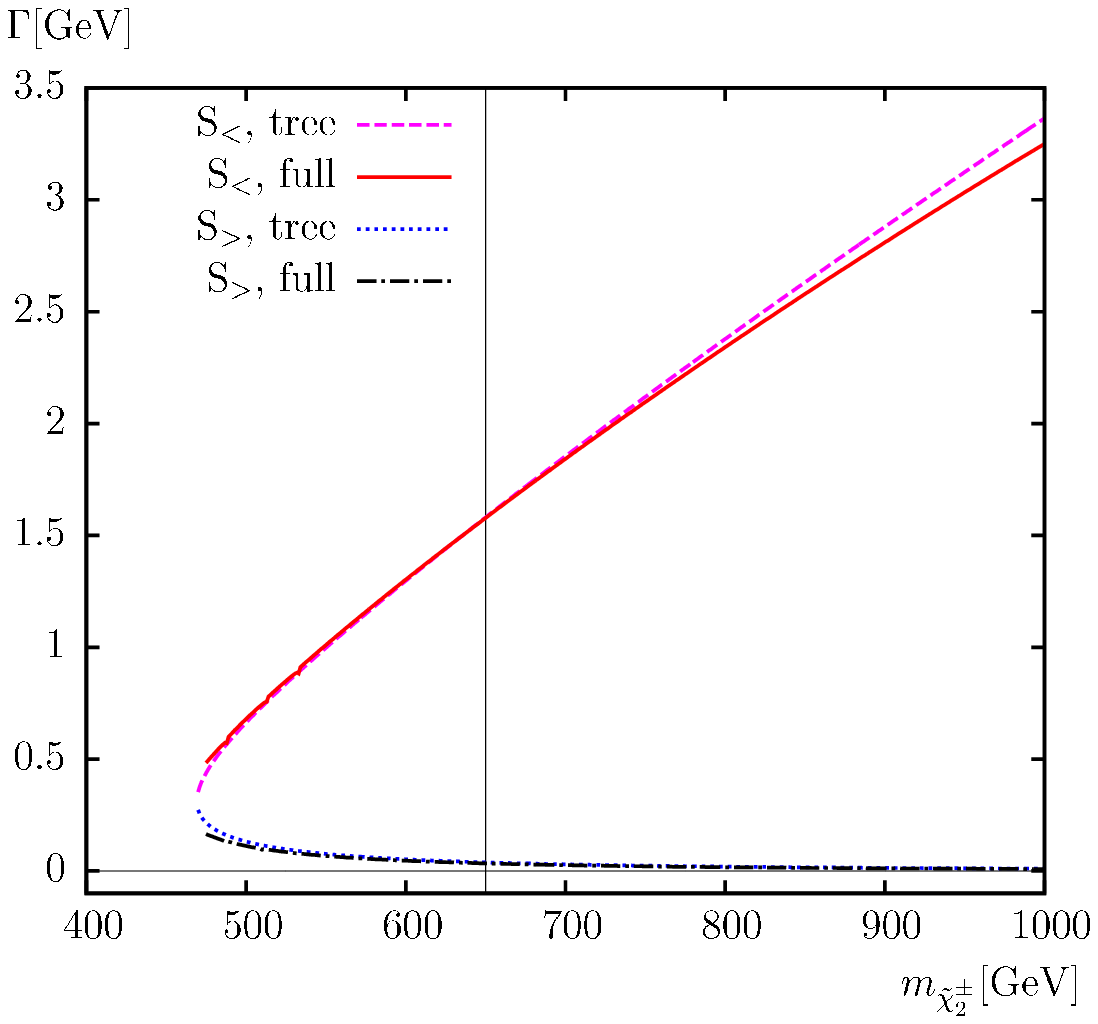}
\hspace{-4mm}
\includegraphics[width=0.49\textwidth,height=7.5cm]{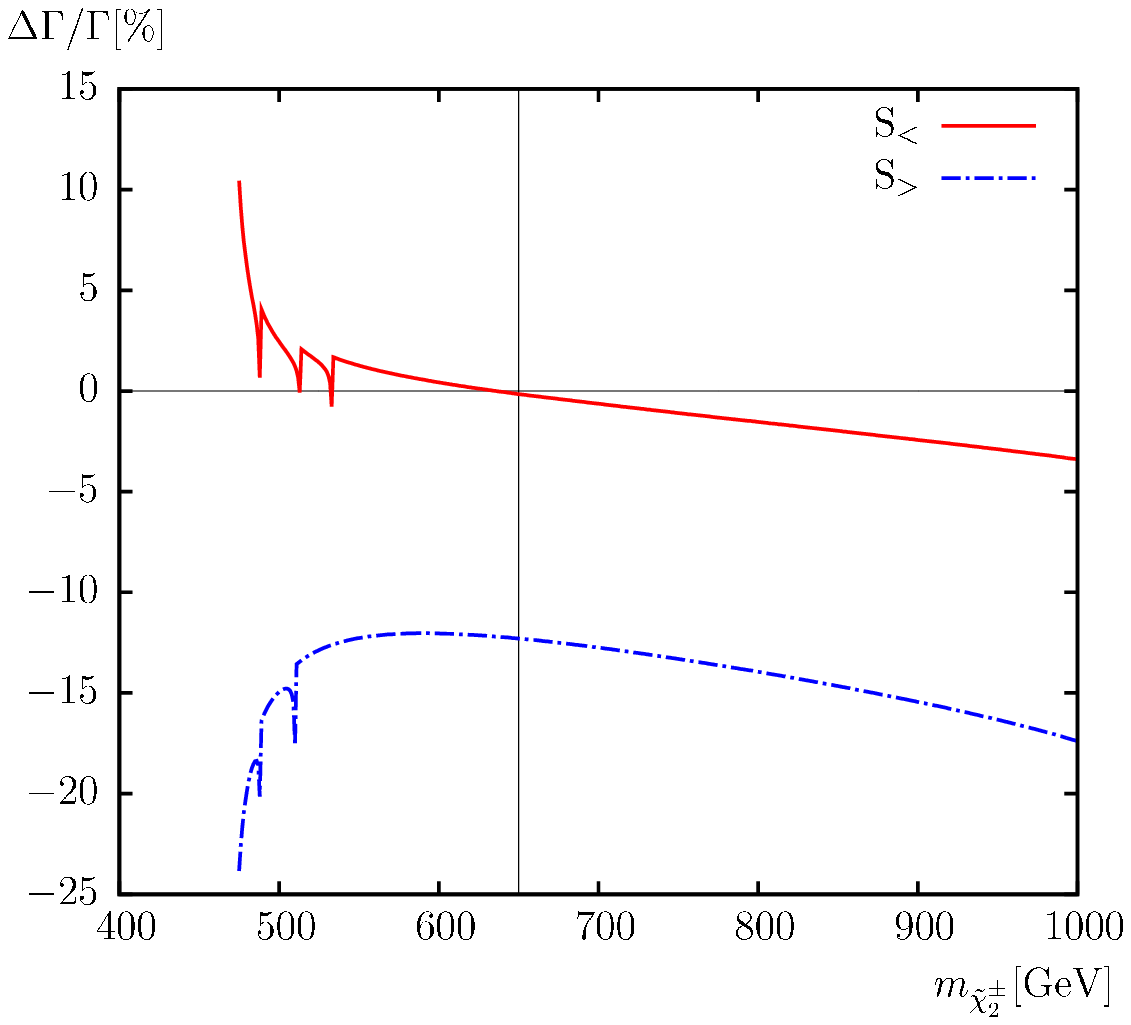} 
\\[4em]
\includegraphics[width=0.49\textwidth,height=7.5cm]{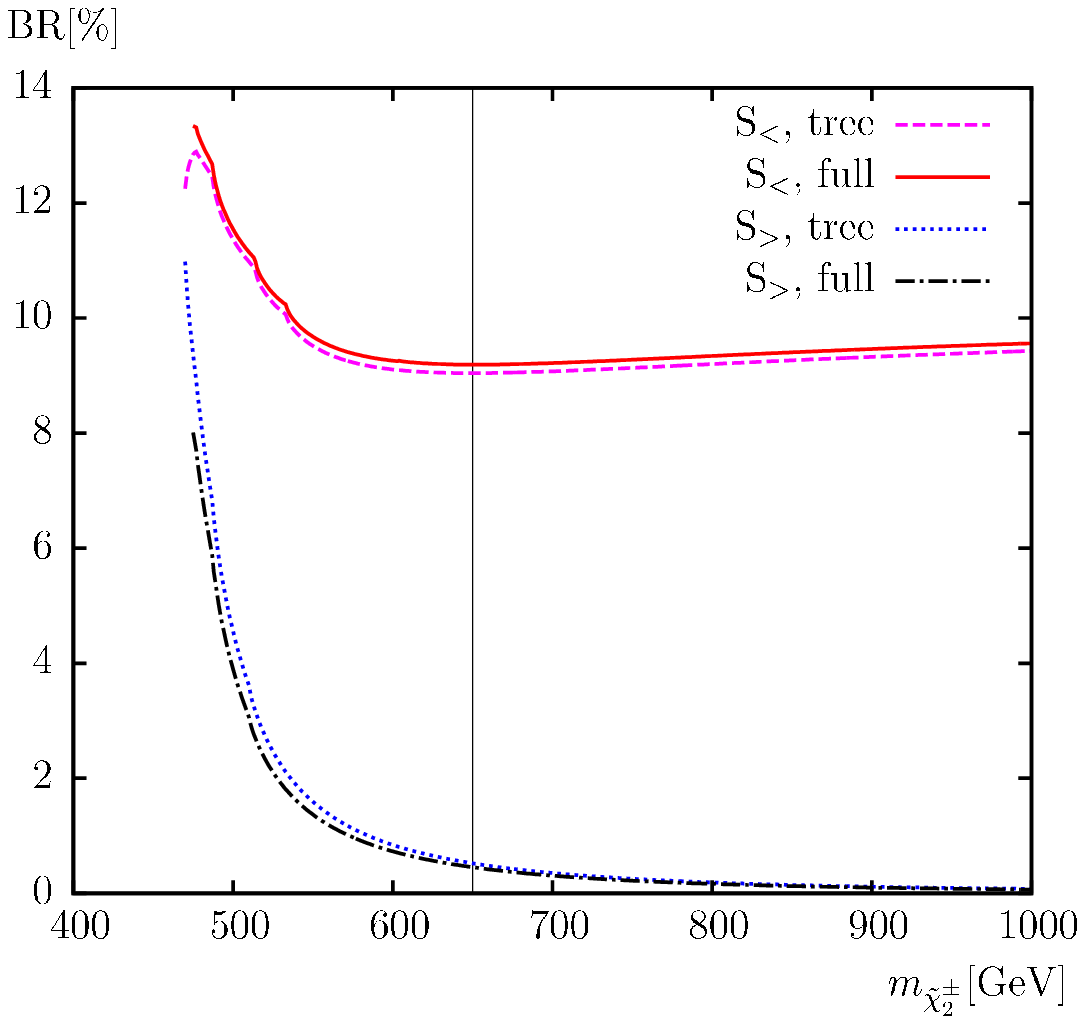}
\hspace{-4mm}
\includegraphics[width=0.49\textwidth,height=7.5cm]{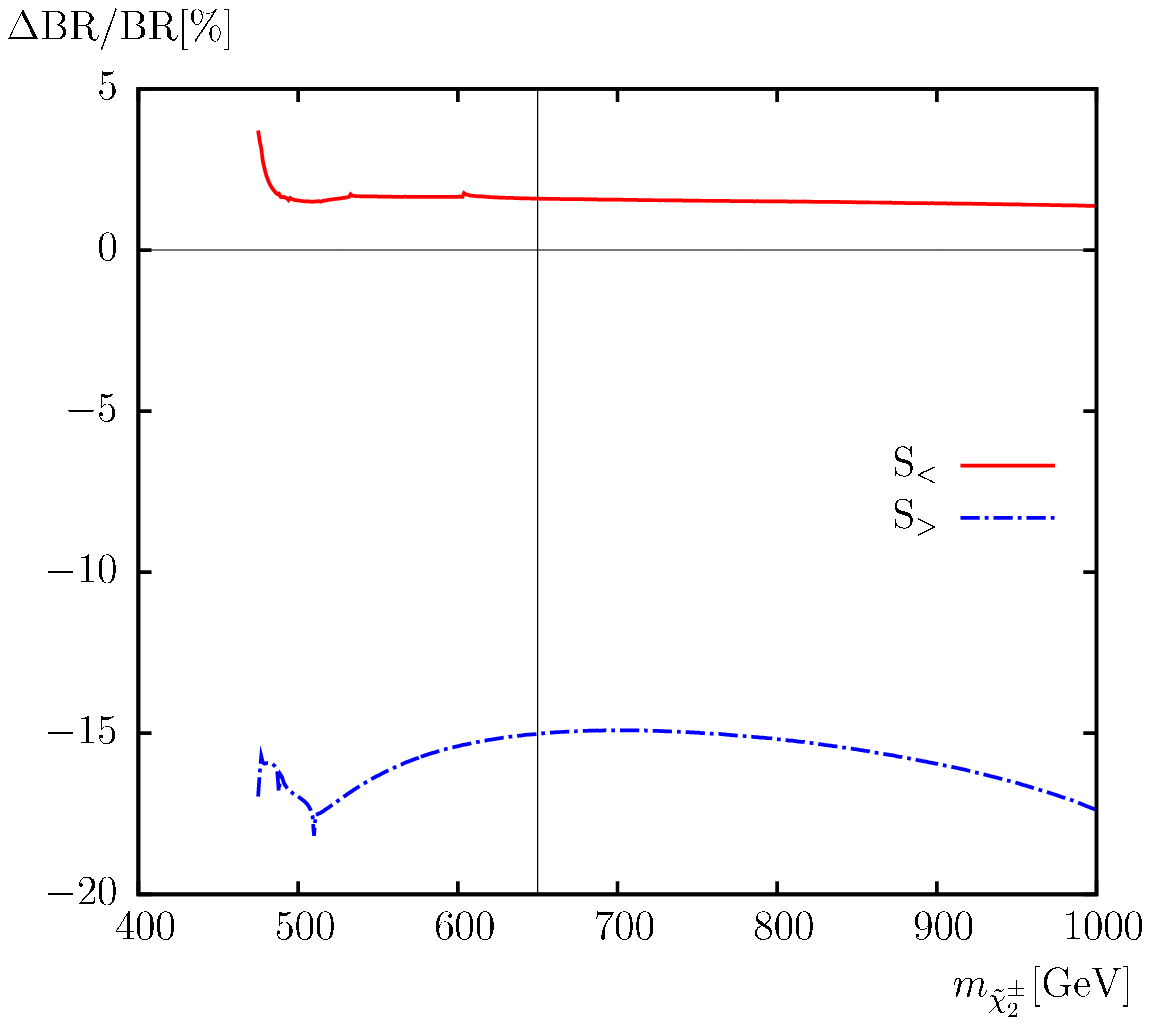}
\end{tabular}
\vspace{2em}
\caption{
  $\Ga(\DecayCmlSn{2}{\mu})$. 
  Tree-level (``tree'') and full one-loop (``full'') corrected 
  decay widths are shown with the parameters chosen according to \SN\
  (see \refta{tab:para}), with $\mcha{2}$ varied.
  The upper left plot shows the decay width, the upper right plot shows 
  the relative size of the corrections.
  The lower left plot shows the BR, the lower right plot shows 
  the relative size of the BR.
  The vertical lines indicate where $\mcha{1} + \mcha{2} = 1000 \gev$, 
  i.e.\ the maximum reach of the ILC(1000).
}
\label{fig:mC2.cha2snumu}
\end{center}
\end{figure}

\begin{figure}[htb!]
\begin{center}
\begin{tabular}{c}
\includegraphics[width=0.49\textwidth,height=7.5cm]{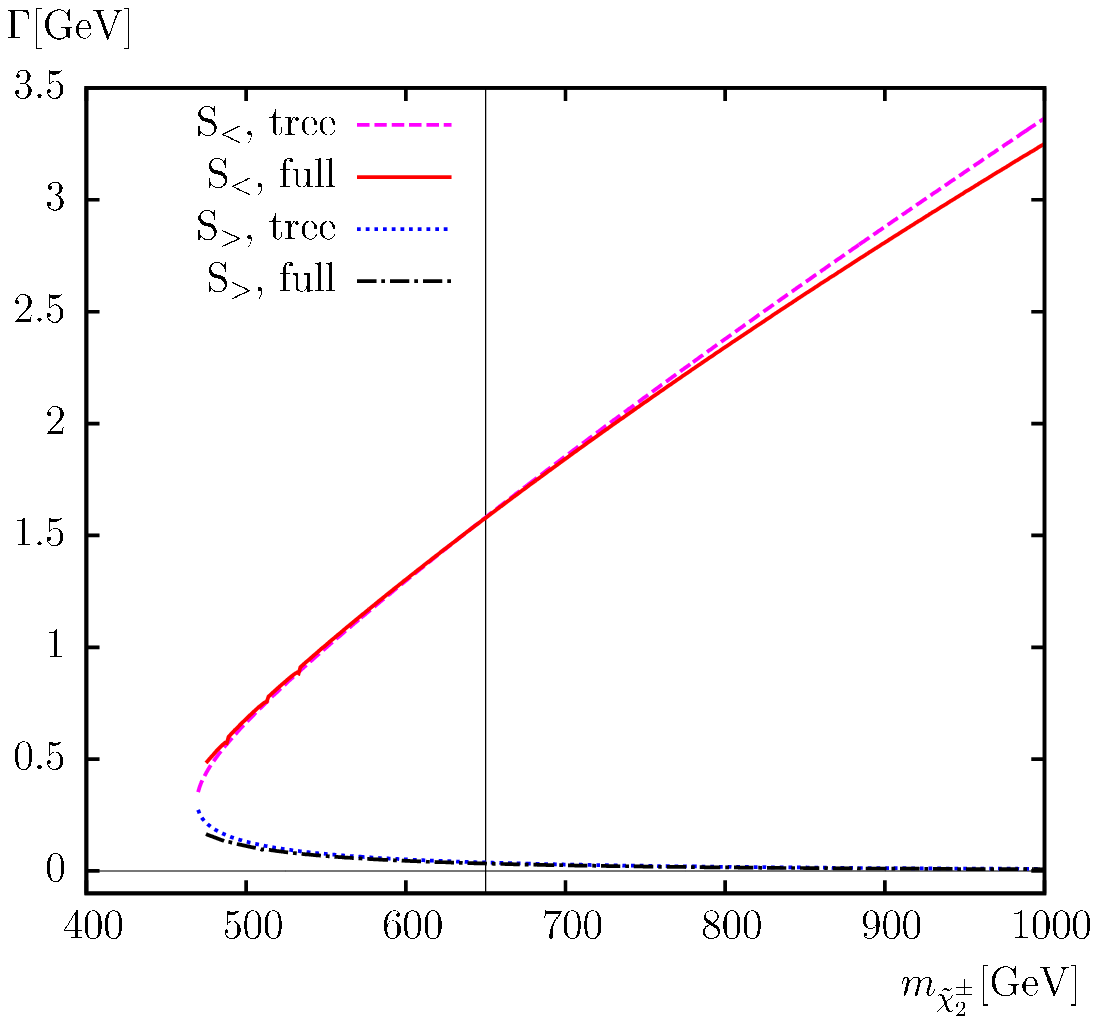}
\hspace{-4mm}
\includegraphics[width=0.49\textwidth,height=7.5cm]{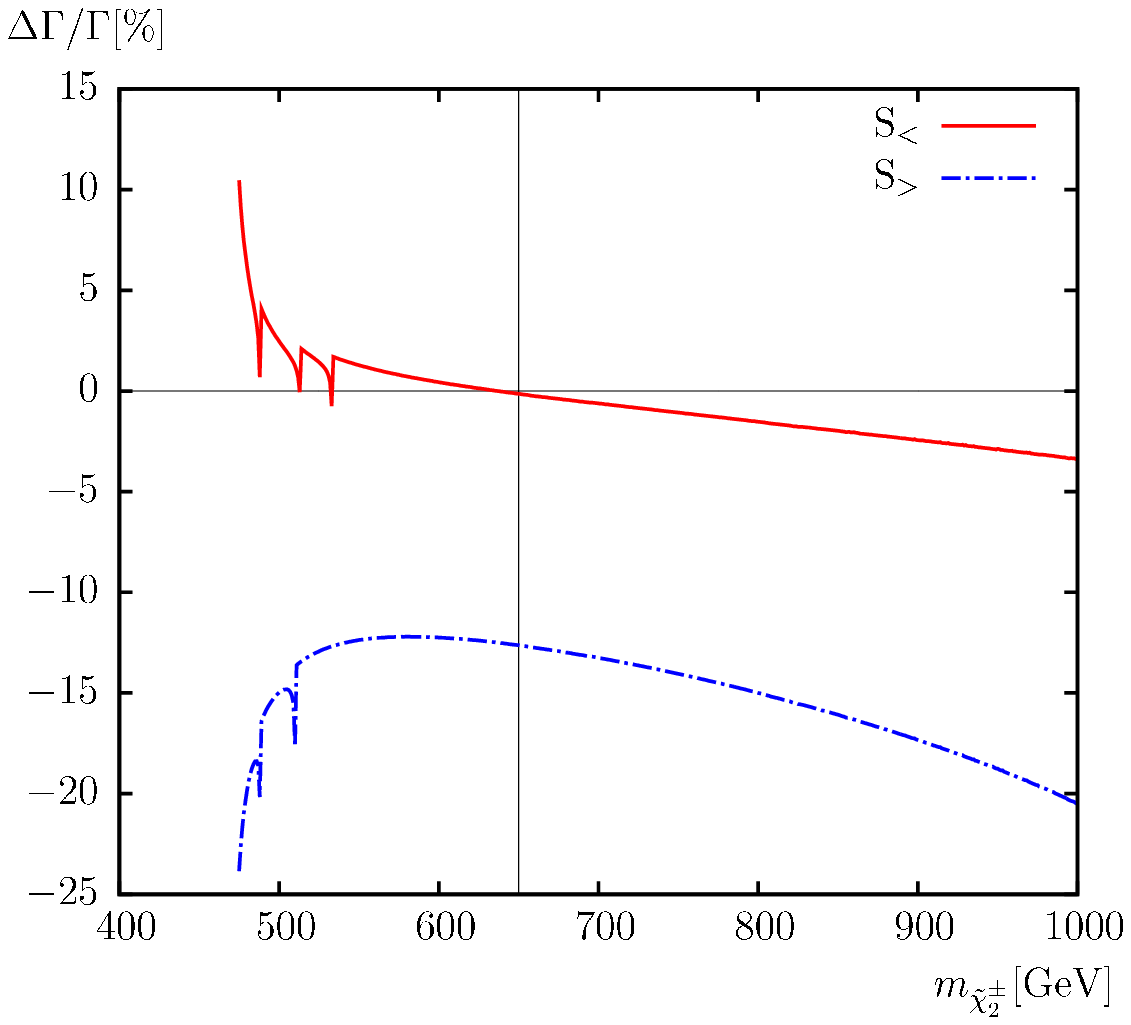} 
\\[4em]
\includegraphics[width=0.49\textwidth,height=7.5cm]{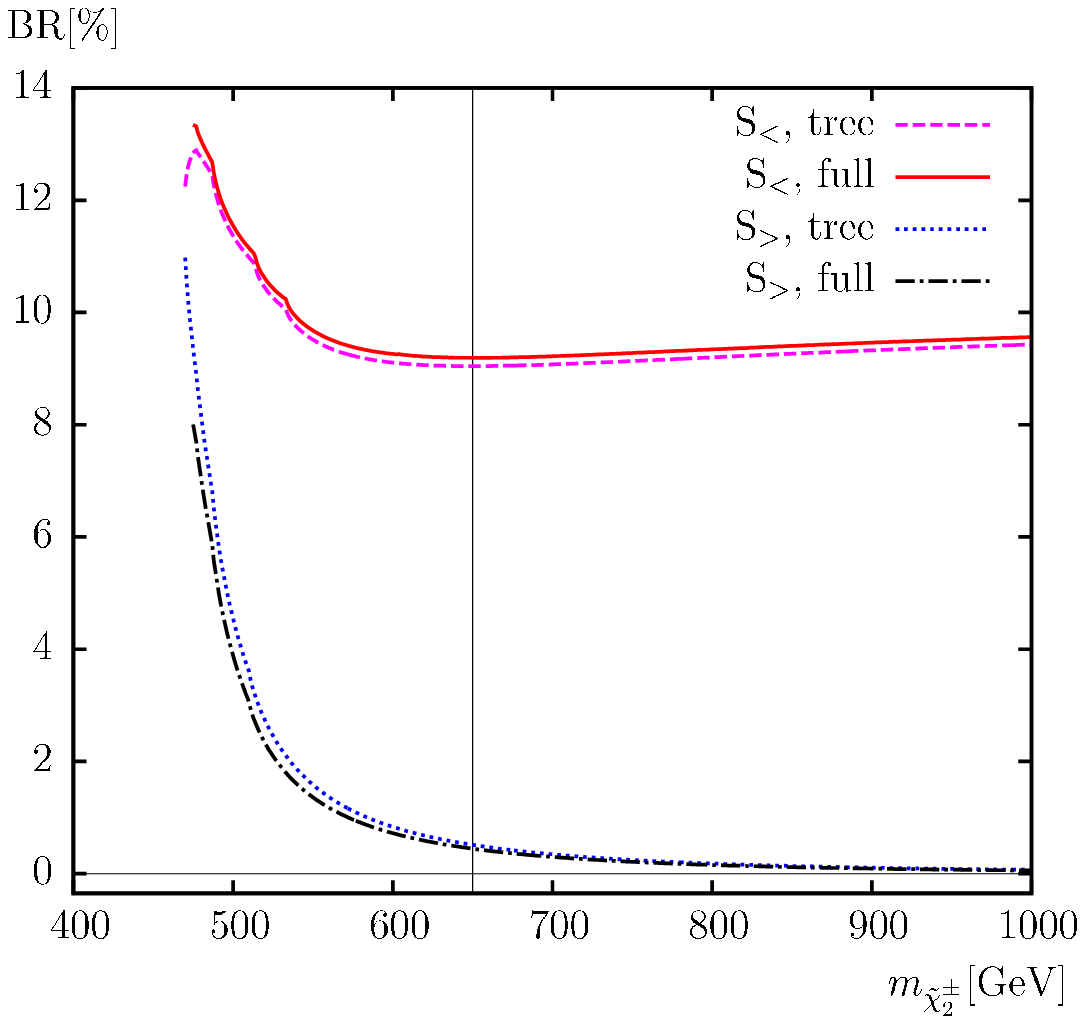}
\hspace{-4mm}
\includegraphics[width=0.49\textwidth,height=7.5cm]{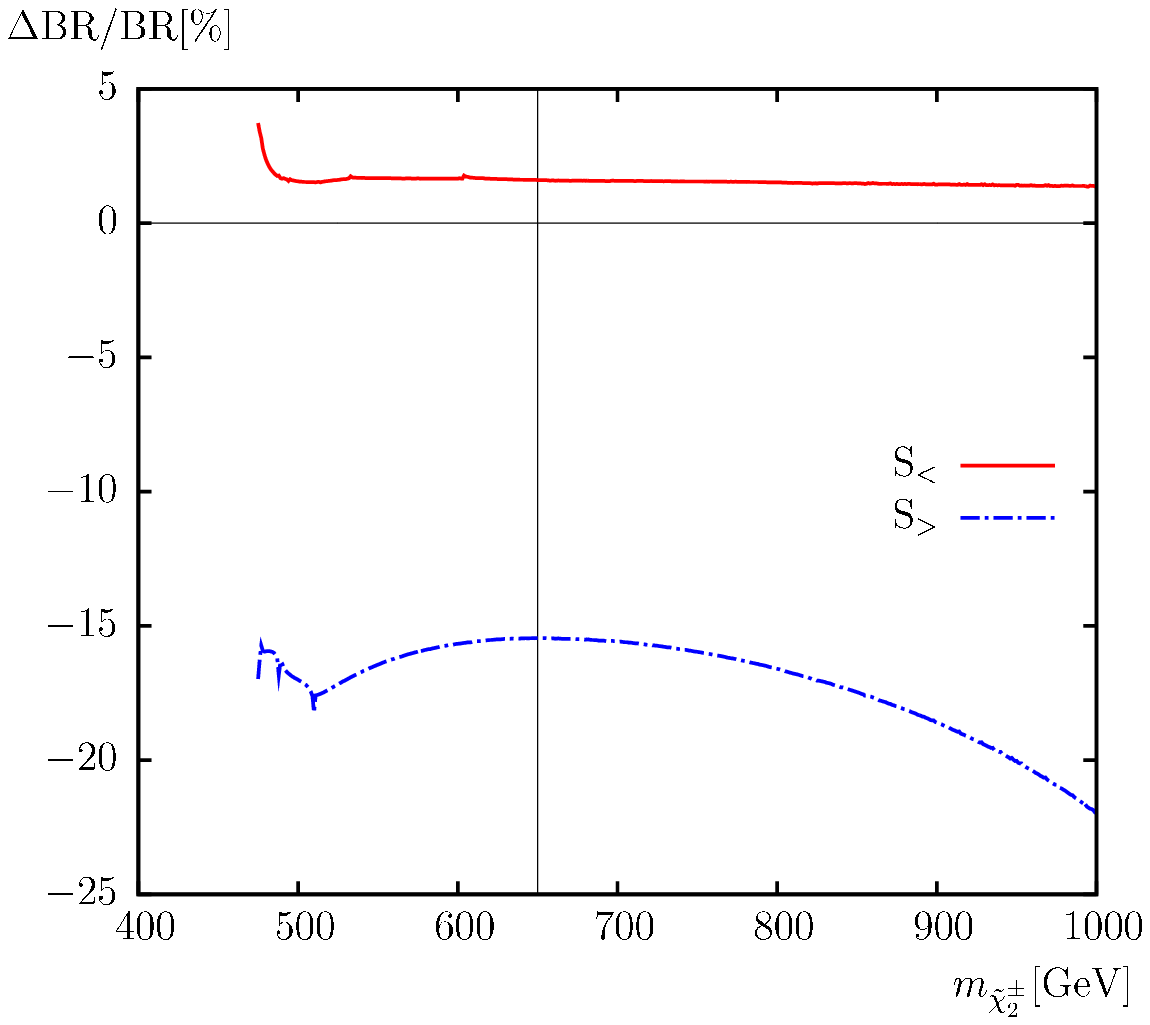}
\end{tabular}
\vspace{2em}
\caption{
  $\Ga(\DecayCmlSn{2}{e})$. 
  Tree-level (``tree'') and full one-loop (``full'') corrected 
  decay widths are shown with the parameters chosen according to \SN\
  (see \refta{tab:para}), with $\mcha{2}$ varied.
  The upper left plot shows the decay width, the upper right plot shows 
  the relative size of the corrections.
  The lower left plot shows the BR, the lower right plot shows 
  the relative size of the BR.
  The vertical lines indicate where $\mcha{1} + \mcha{2} = 1000 \gev$, 
  i.e.\ the maximum reach of the ILC(1000).
}
\label{fig:mC2.cha2snuel}
\end{center}
\end{figure}

The last decays involving scalar leptons are $\DecayCmlSn{2}{l}$ 
($l = \tau, \mu, e$), presented in \reffis{fig:mC2.cha2snutau} --
\ref{fig:mC2.cha2snuel}. 
The dips visible in the upper right panels correspond to the same
thresholds as in the decays into charged sleptons.
The behavior of the decay widths are understood from
\refeq{eq:Lagr.CSnl}. The higgsino part of $\cha{2}$, dominating in \SE,
couples $\propto \ml$, whereas the wino part, dominating in \SZ, couples
with electroweak strength, which is the same for all three
generations. Consequently, we find very similar results for 
$\Ga(\DecayCmlSn{2}{l})$ for $l = \tau, \mu, e$ in \SZ, while the decay
widths are suppressed with $\ml^{2}$ in \SE. The BR's in \SE\ are at 
(or above) the $10\%$ level, whereas in \SZ\ values above 
$2\%$ are only
reached for $\br(\DecayCmlSn{2}{\tau})$, and tiny BR's are found in the
other two cases. The one-loop effects in \SE\ are at the $2\%$ level in
all three generations, and vary between $-5\%$ and $+11\%$ for 
$\br(\DecayCmlSn{2}{\tau})$ in \SZ, exceeding by far the ILC(1000)
accuracy. 

\bigskip

\clearpage

\subsection{Full one-loop results for varying \boldmath{$\mcha{1}$}}
\label{sec:1Lmcha1}

In this section we analyze the $\cham{1}$~decay widths evaluated as
a function of $\mcha{1}$, starting at $\mcha{1} = 300 \gev$.
For the ``tree'' contributions we show results up to 
$\mcha{1} = 480.2 \gev$, about the highest value for a heavy chargino
mass fixed to $\mcha{2} = 600 \gev$.
The ``full'' results are only shown up to $\mcha{1} = 475 \gev$.
For larger values of $\mcha{1}$ the on-shell renormalization scheme 
adopted here leads to incorrect results,
as $\MTwo$ approaches $\mu$, and the potential problems described in
\refse{sec:chaneu} start to take effect.
The leading production channel of the lightest charginos at the 
ILC(1000), $e^+e^- \to \chap{1}\cham{1}$, 
is open for the full parameter range.

In general the line of argument for a behavior of a certain decay width
is identical to the ones given in detail in
\refse{sec:1Lmcha2}. Consequently, we will be very brief about these
arguments here and mainly discuss the size of the effects. 

\begin{figure}[htb!]
\begin{center}
\begin{tabular}{c}
\includegraphics[width=0.49\textwidth,height=7.5cm]{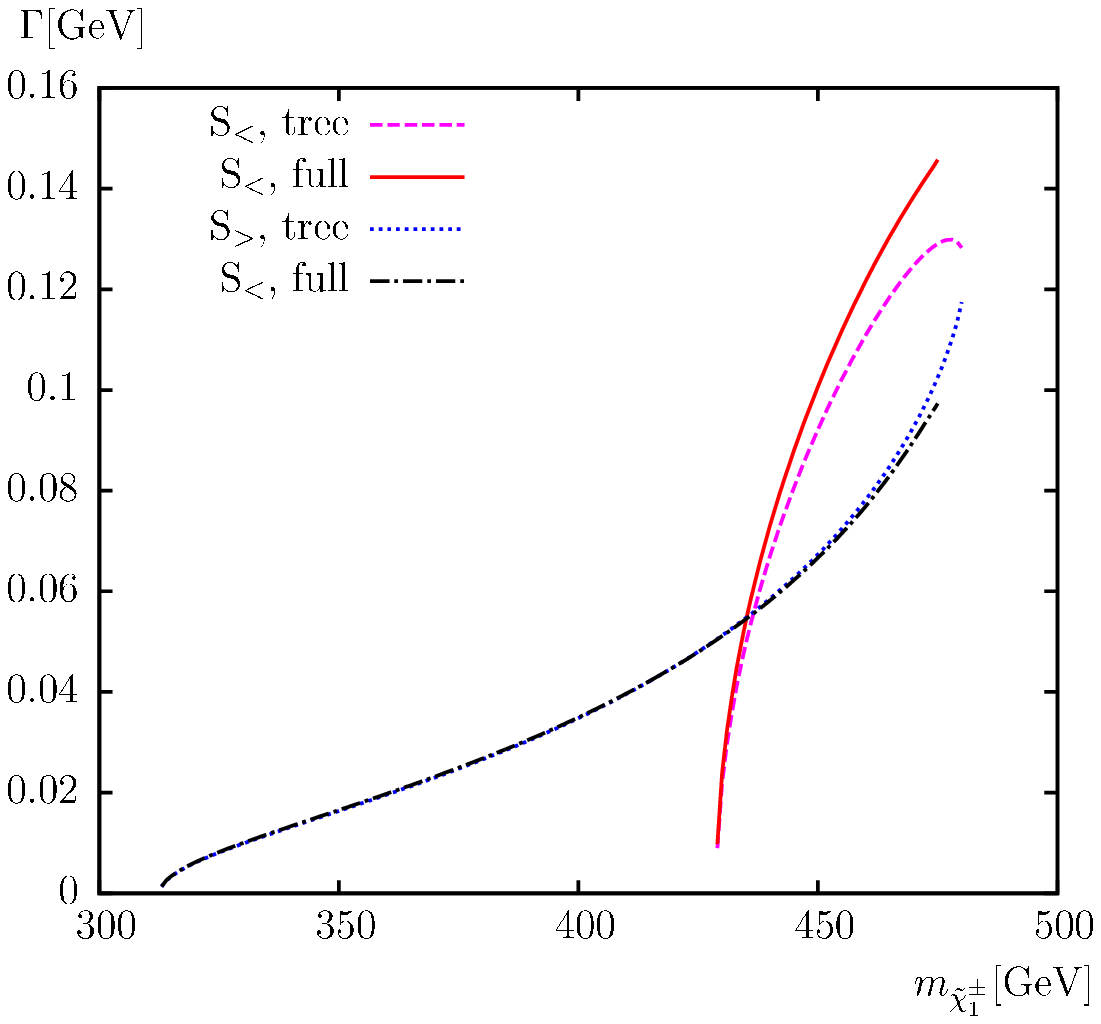}
\hspace{-4mm}
\includegraphics[width=0.49\textwidth,height=7.5cm]{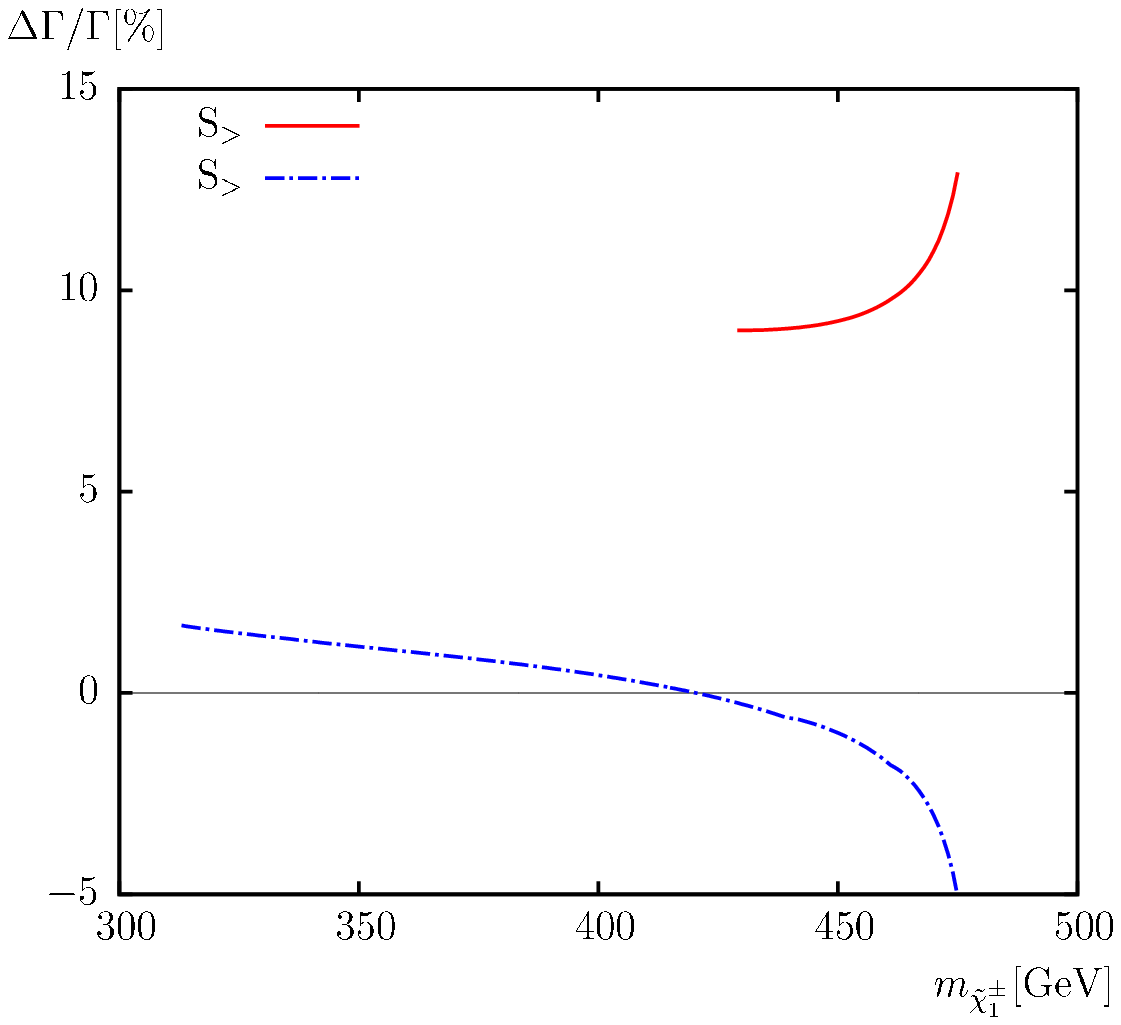} 
\\[5em]
\includegraphics[width=0.49\textwidth,height=7.5cm]{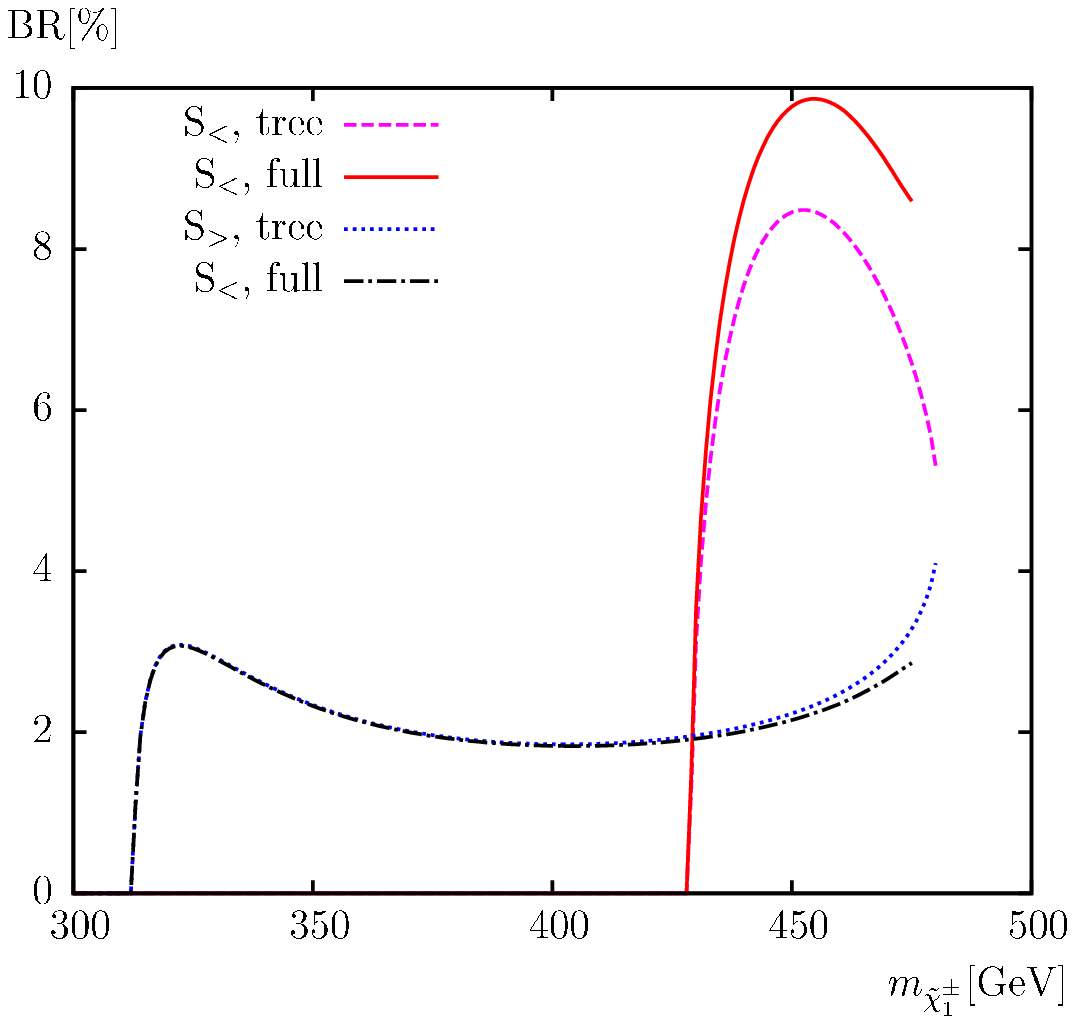}
\hspace{-4mm}
\includegraphics[width=0.49\textwidth,height=7.5cm]{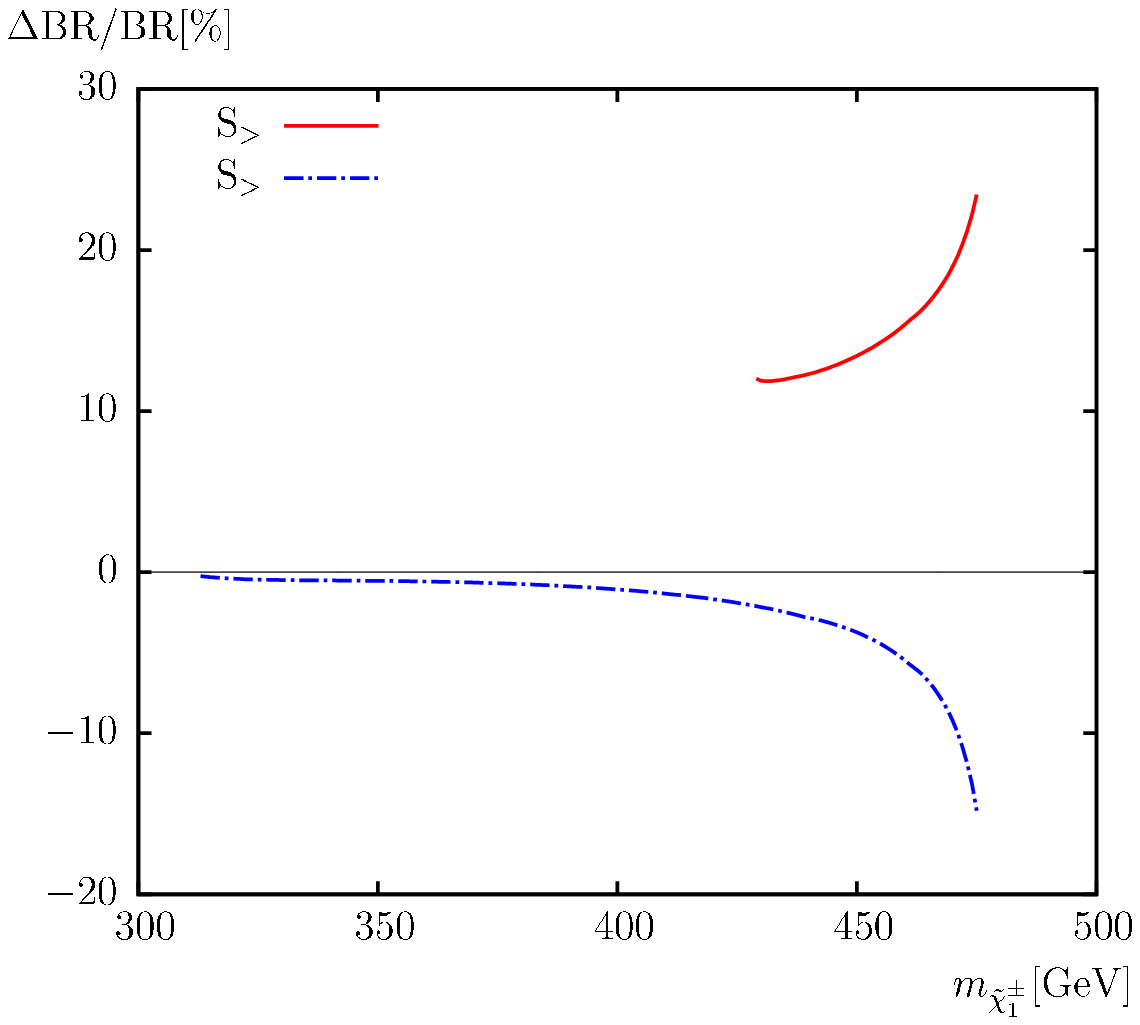}
\end{tabular}
\vspace{2em}
\caption{
  $\Ga(\DecayCmNH{1}{1})$. 
  Tree-level (``tree'') and full one-loop (``full'') corrected 
  decay widths are shown with the parameters chosen according to \SN\
  (see \refta{tab:para}), with $\mcha{1}$ varied.
  The upper left plot shows the decay width, the upper right plot shows 
  the relative size of the corrections.
  The lower left plot shows the BR, the lower right plot shows 
  the relative size of the BR.
}
\label{fig:mC1.cha1neu1hp}
\end{center}
\end{figure}

\medskip
We start with only $\cham{1}$ decay involving Higgs bosons, 
$\DecayCmNH{1}{1}$, shown in \reffi{fig:mC1.cha1neu1hp}. 
The decay widths reach values up to $\sim 0.1\ (0.15) \gev$ in \SE\ (\SZ).
The $\br(\DecayCmNH{1}{1})$ in \SE\ varies around $2\%$,
while in \SZ\ it steeply rises above threshold and goes up to 
nearly $10\%$ at $\mcha{1} \approx 450 \gev$. Within \SE\ the one-loop
effects do hardly exceed $-10\%$, while in \SZ, where the BR is large, a
variation of $\sim 15\%$ is found.

\begin{figure}[htb!]
\begin{center}
\begin{tabular}{c}
\includegraphics[width=0.49\textwidth,height=7.5cm]{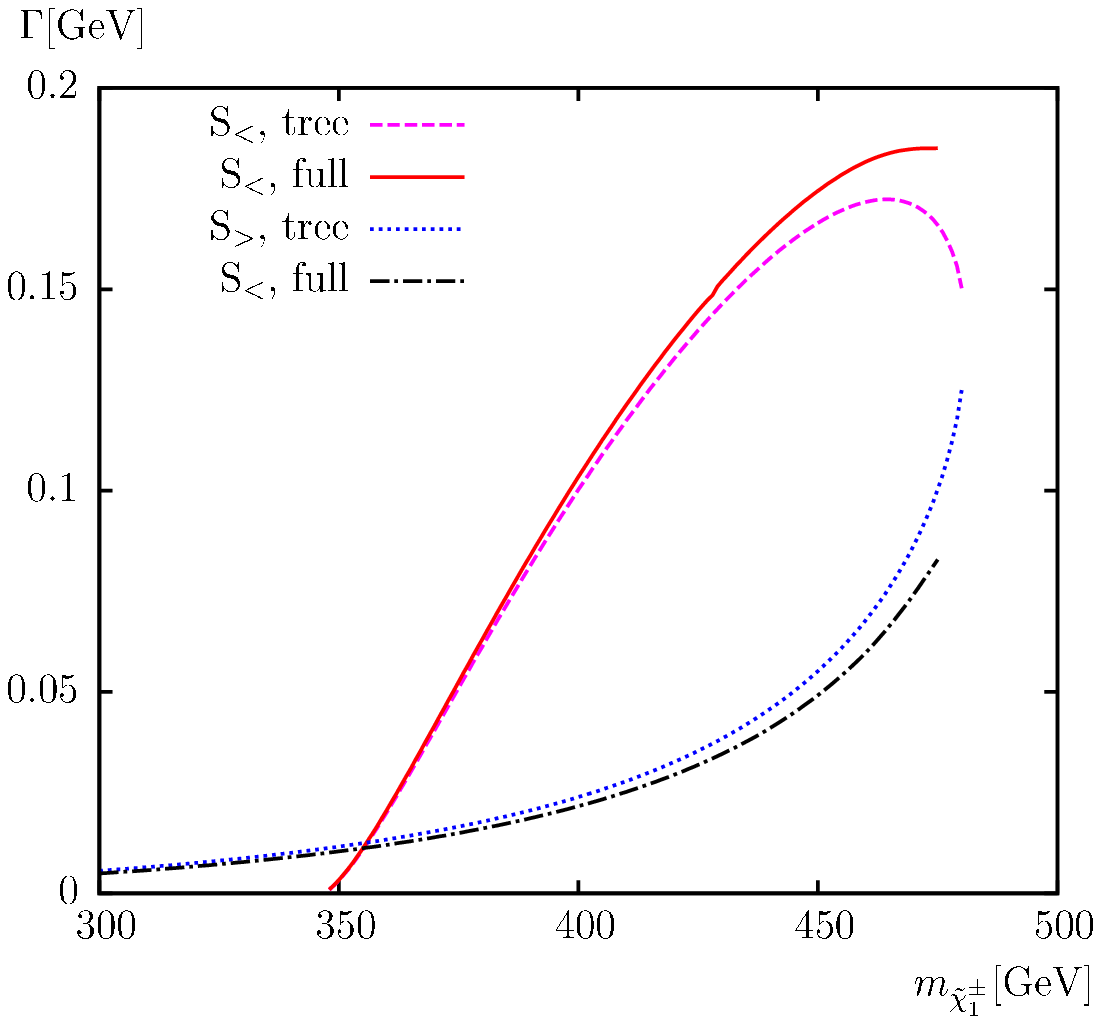}
\hspace{-4mm}
\includegraphics[width=0.49\textwidth,height=7.5cm]{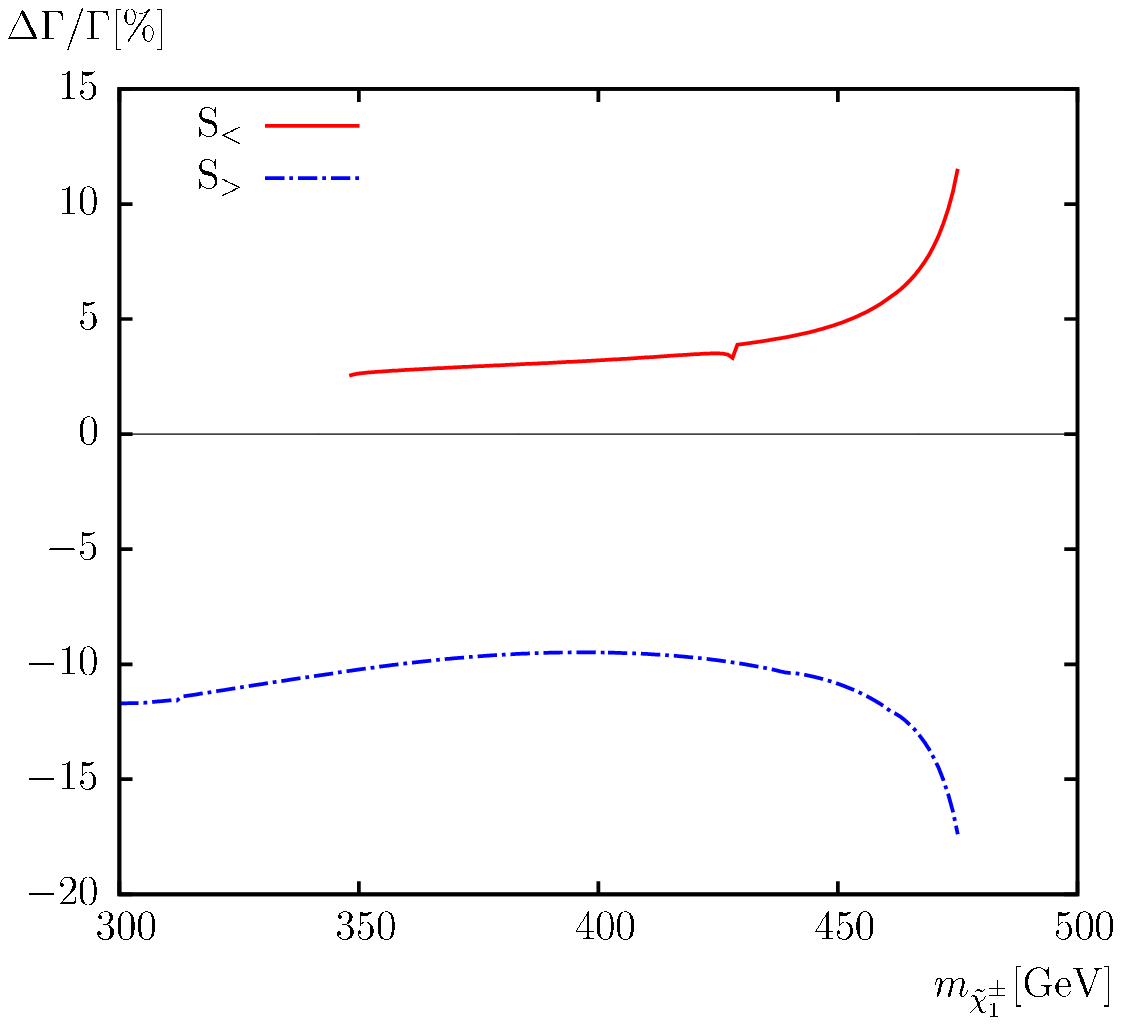} 
\\[5em]
\includegraphics[width=0.49\textwidth,height=7.5cm]{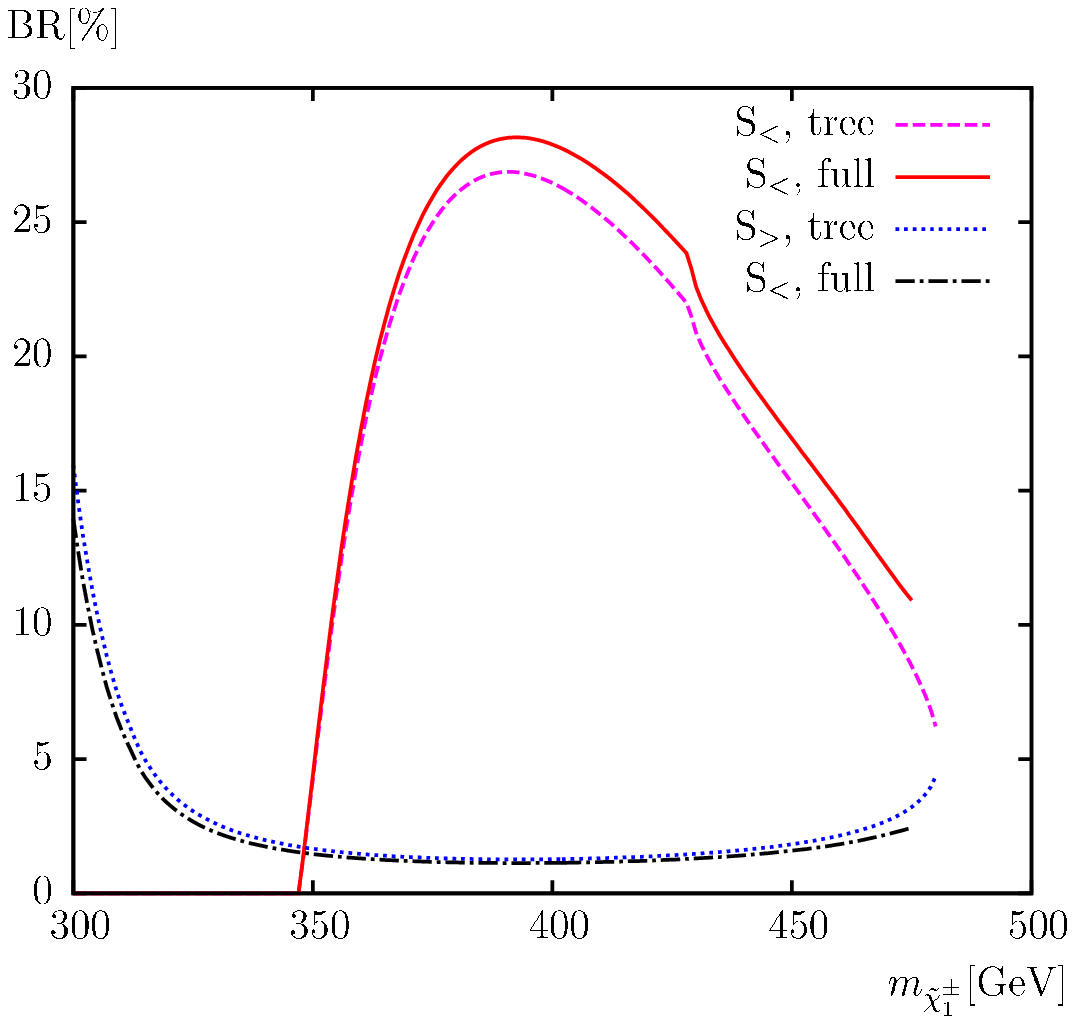}
\hspace{-4mm}
\includegraphics[width=0.49\textwidth,height=7.5cm]{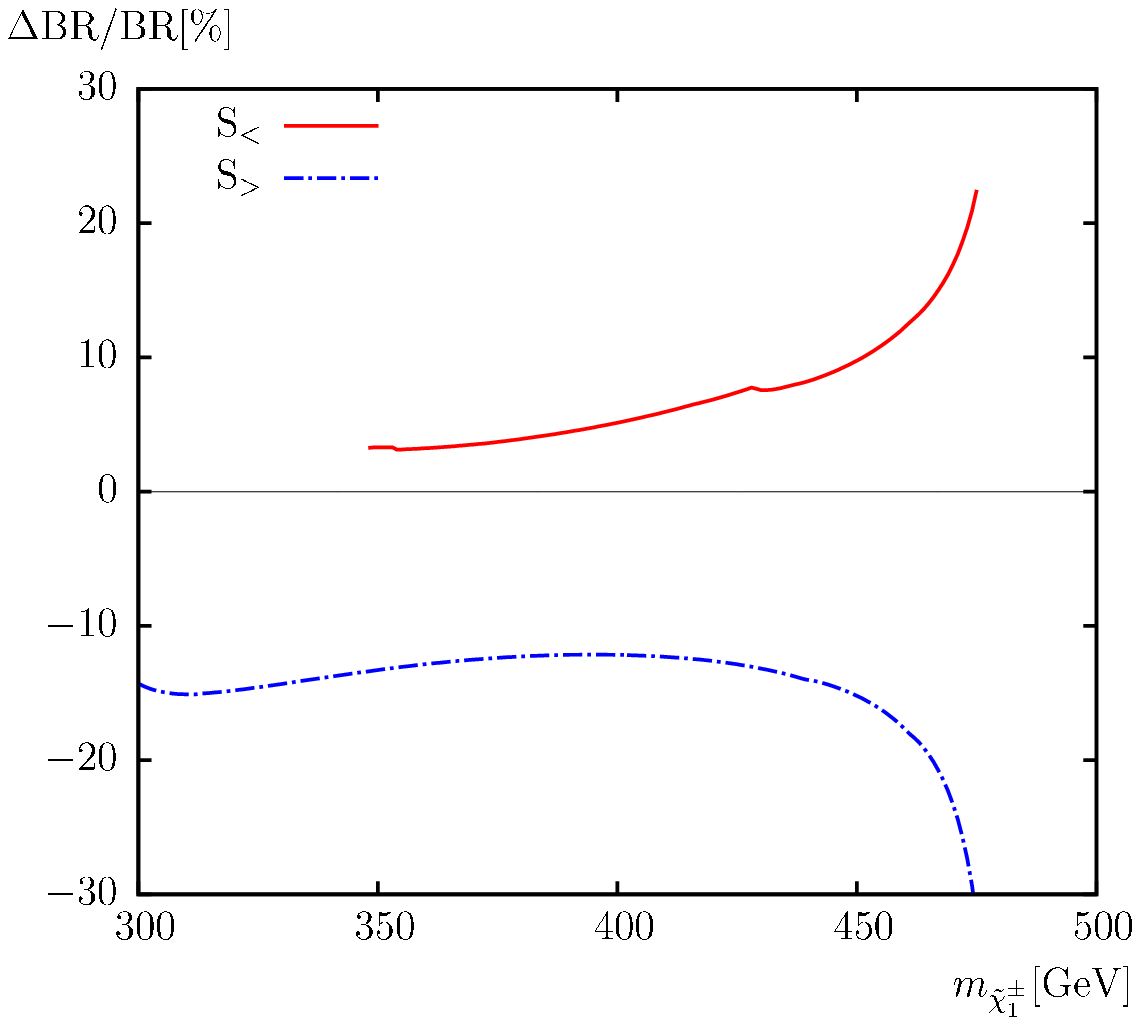}
\end{tabular}
\vspace{2em}
\caption{
  $\Ga(\DecayCmNW{1}{1})$. 
  Tree-level (``tree'') and full one-loop (``full'') corrected 
  decay widths are shown with the parameters chosen according to \SN\
  (see \refta{tab:para}), with $\mcha{1}$ varied.
  The upper left plot shows the decay width, the upper right plot shows 
  the relative size of the corrections.
  The lower left plot shows the BR, the lower right plot shows 
  the relative size of the BR.
}
\label{fig:mC1.cha1neu1w}
\end{center}
\end{figure}

The corresponding decay involving the $W$~boson, $\DecayCmNW{1}{1}$, is
shown in \reffi{fig:mC1.cha1neu1w}. 
The dip at $\mcha{1}=428\gev$ visible in \SZ\ in the upper right panel
is due to the $\DecayCmNH{1}{1}$ threshold. The two decay 
widths in \SE\ and \SZ\ rise above threshold to reach values between 
$\sim 0.1$ and $\sim 0.15 \gev$. 
The BR's behave somewhat differently: in \SE\ values larger than $10\%$
are reached for small $\mcha{1}$, $\br(\DecayCmNH{1}{1})$ reaches about 
$2\%$ for larger $\mcha{1}$. In \SZ, on the other hand, intermediate
values larger than $25\%$ are reached.
The size of the one-loop effects on the BR's are substantial. 
They vary around $-15\%$ in \SE\ and between $3\%$ and more than $10\%$ 
in \SZ. Consequently, these corrections have to be taken into account
in a reliable ILC analysis.

\begin{figure}[htb!]
\begin{center}
\begin{tabular}{c}
\includegraphics[width=0.49\textwidth,height=7.5cm]{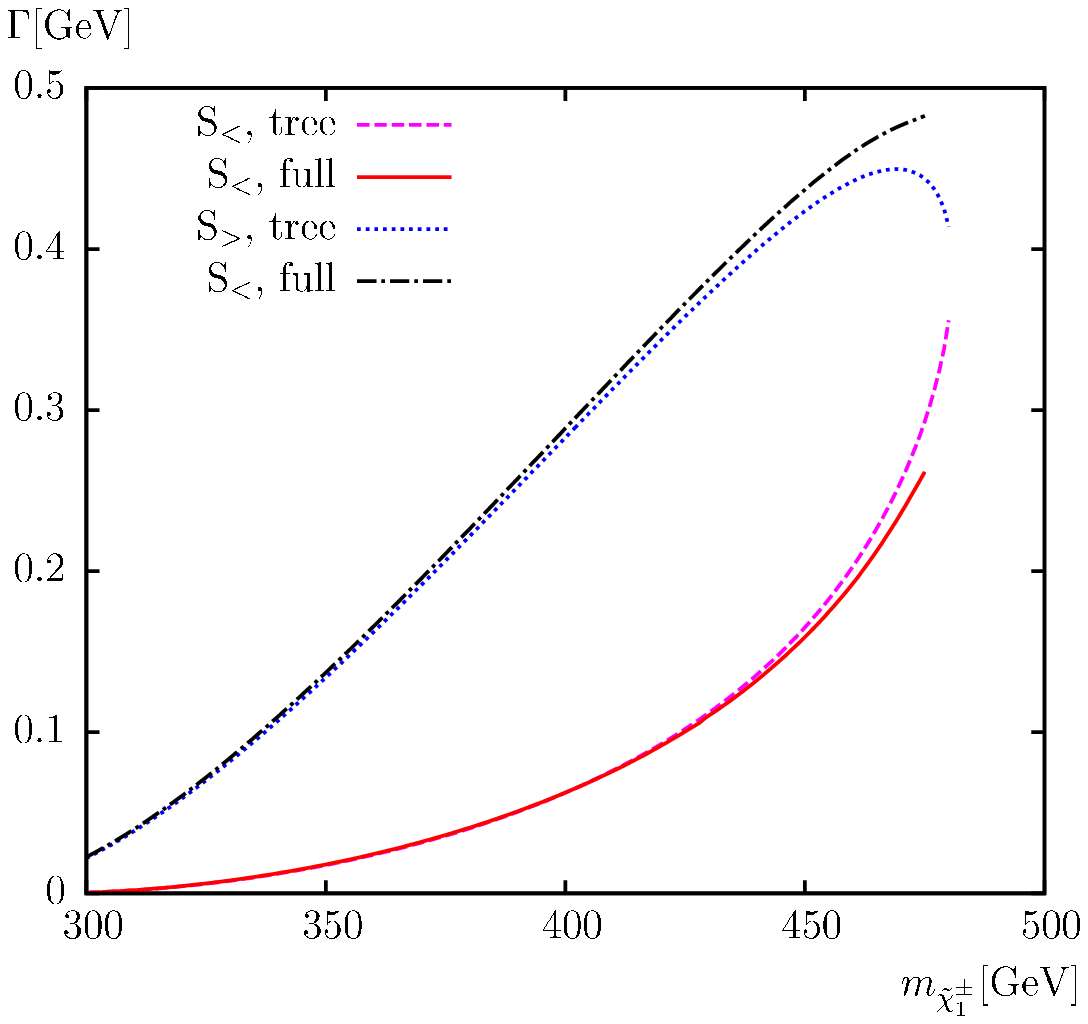}
\hspace{-4mm}
\includegraphics[width=0.49\textwidth,height=7.5cm]{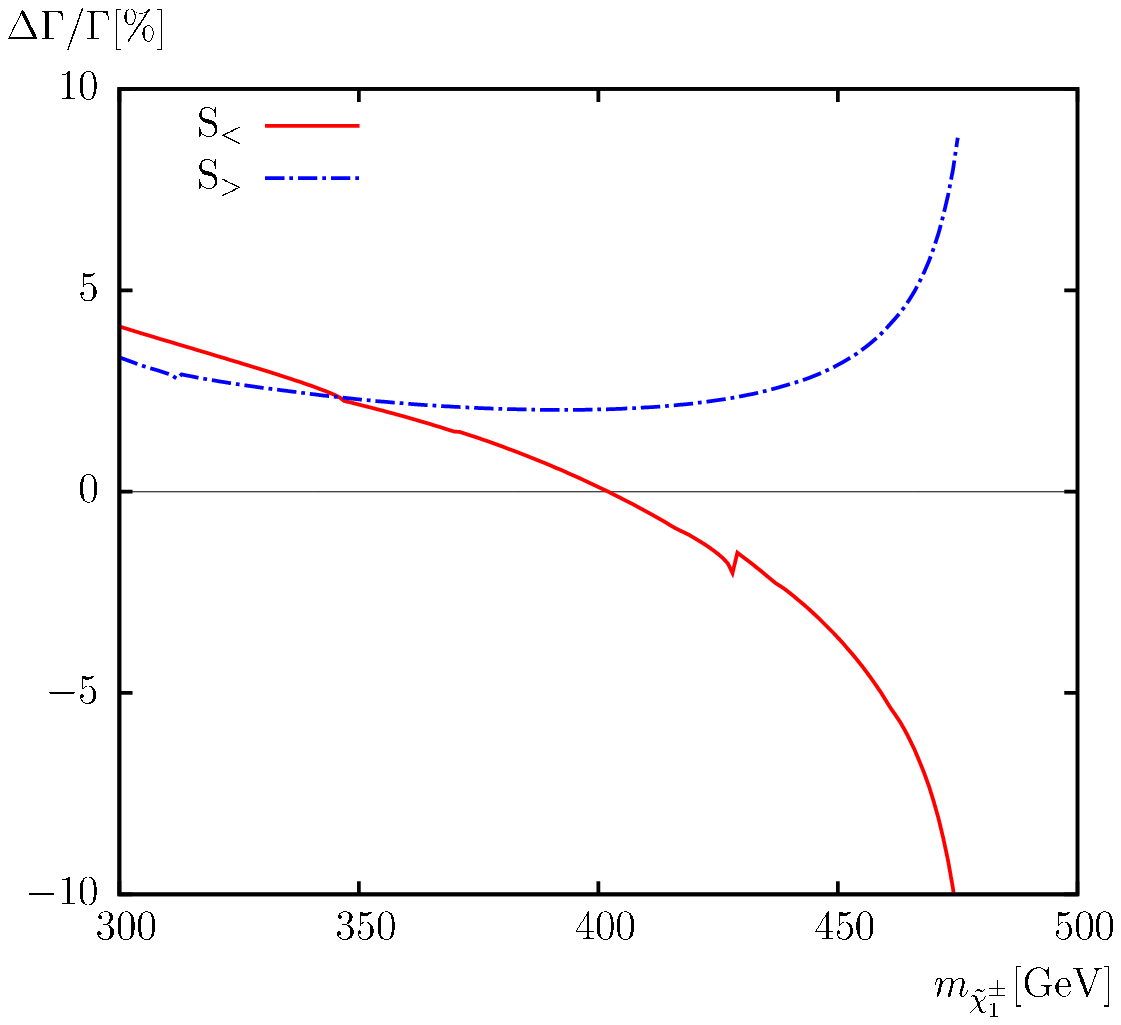} 
\\[5em]
\includegraphics[width=0.49\textwidth,height=7.5cm]{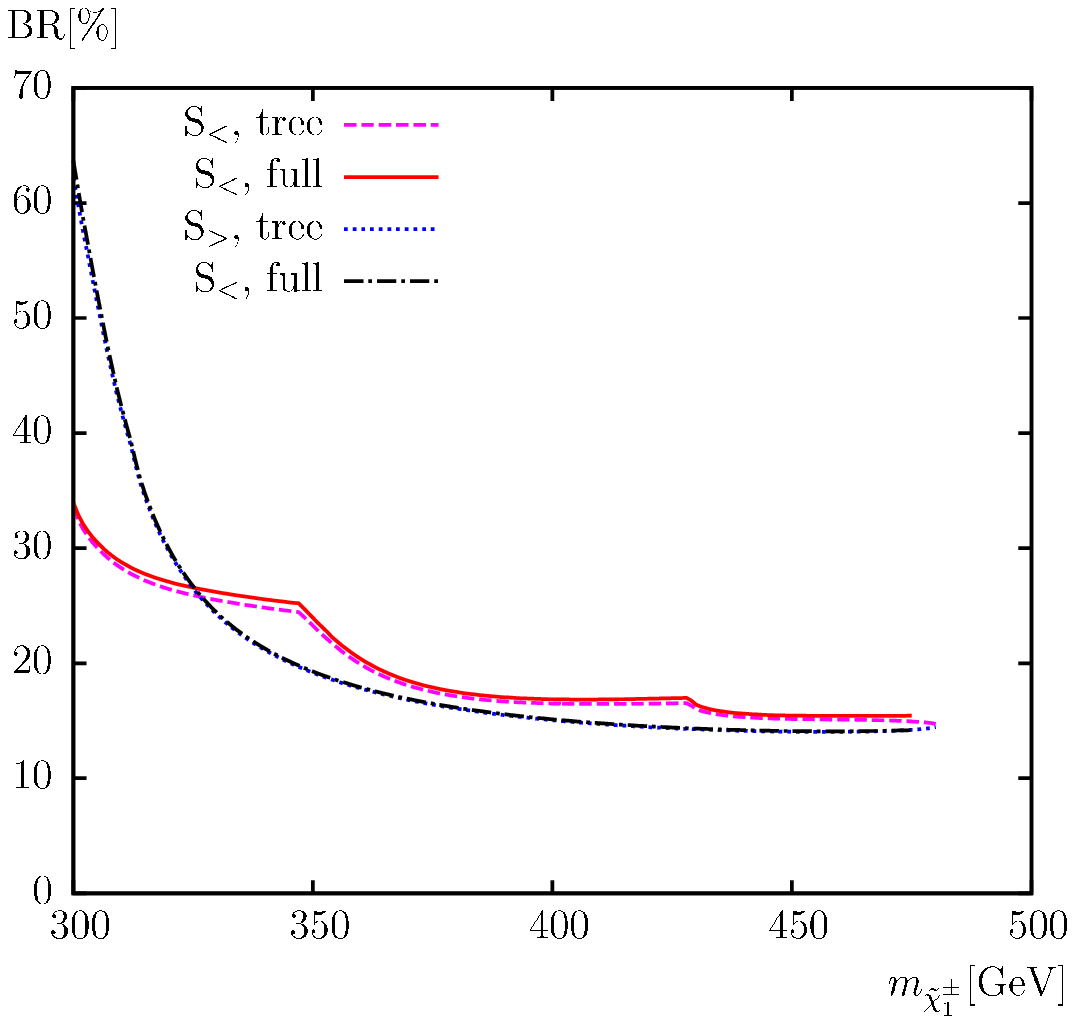}
\hspace{-4mm}
\includegraphics[width=0.49\textwidth,height=7.5cm]{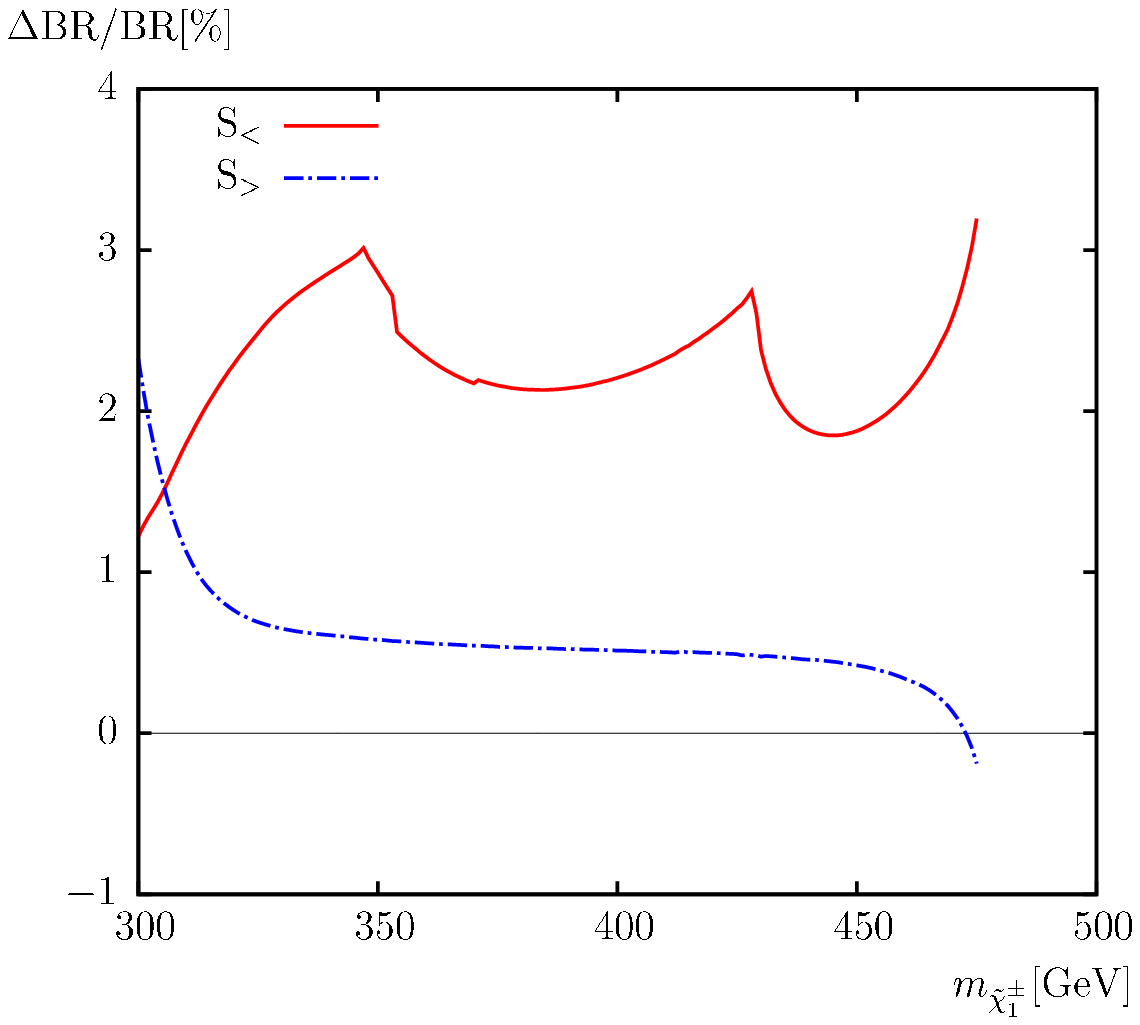}
\end{tabular}
\vspace{2em}
\caption{
  $\Ga(\DecayCmnSl{1}{\tau}{1})$. 
  Tree-level (``tree'') and full one-loop (``full'') corrected 
  decay widths are shown with the parameters chosen according to \SN\
  (see \refta{tab:para}), with $\mcha{1}$ varied.
  The upper left plot shows the decay width, the upper right plot shows 
  the relative size of the corrections.
  The lower left plot shows the BR, the lower right plot shows 
  the relative size of the BR.
}
\label{fig:mC1.cha1stau1nu}
\end{center}
\end{figure}

\begin{figure}[htb!]
\begin{center}
\begin{tabular}{c}
\includegraphics[width=0.49\textwidth,height=7.5cm]{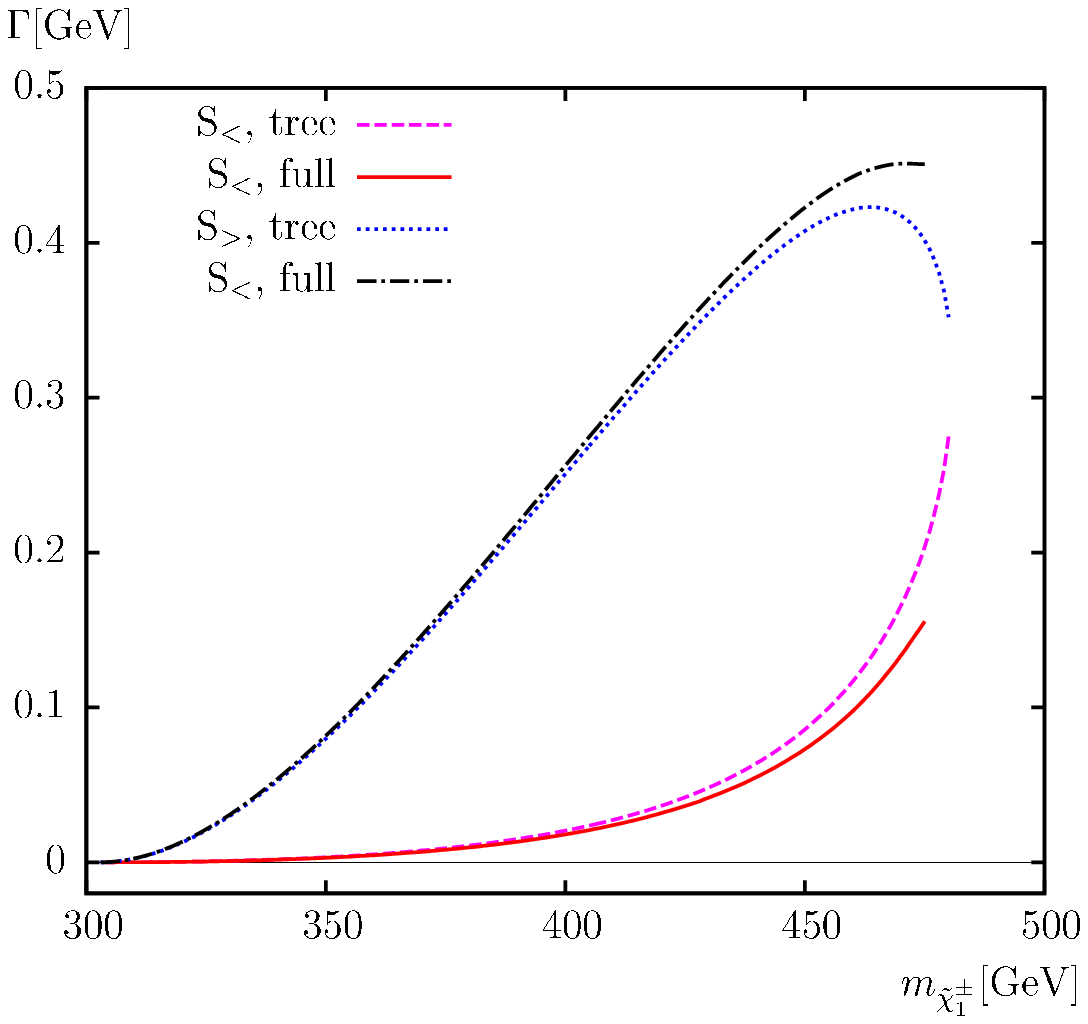}
\hspace{-4mm}
\includegraphics[width=0.49\textwidth,height=7.5cm]{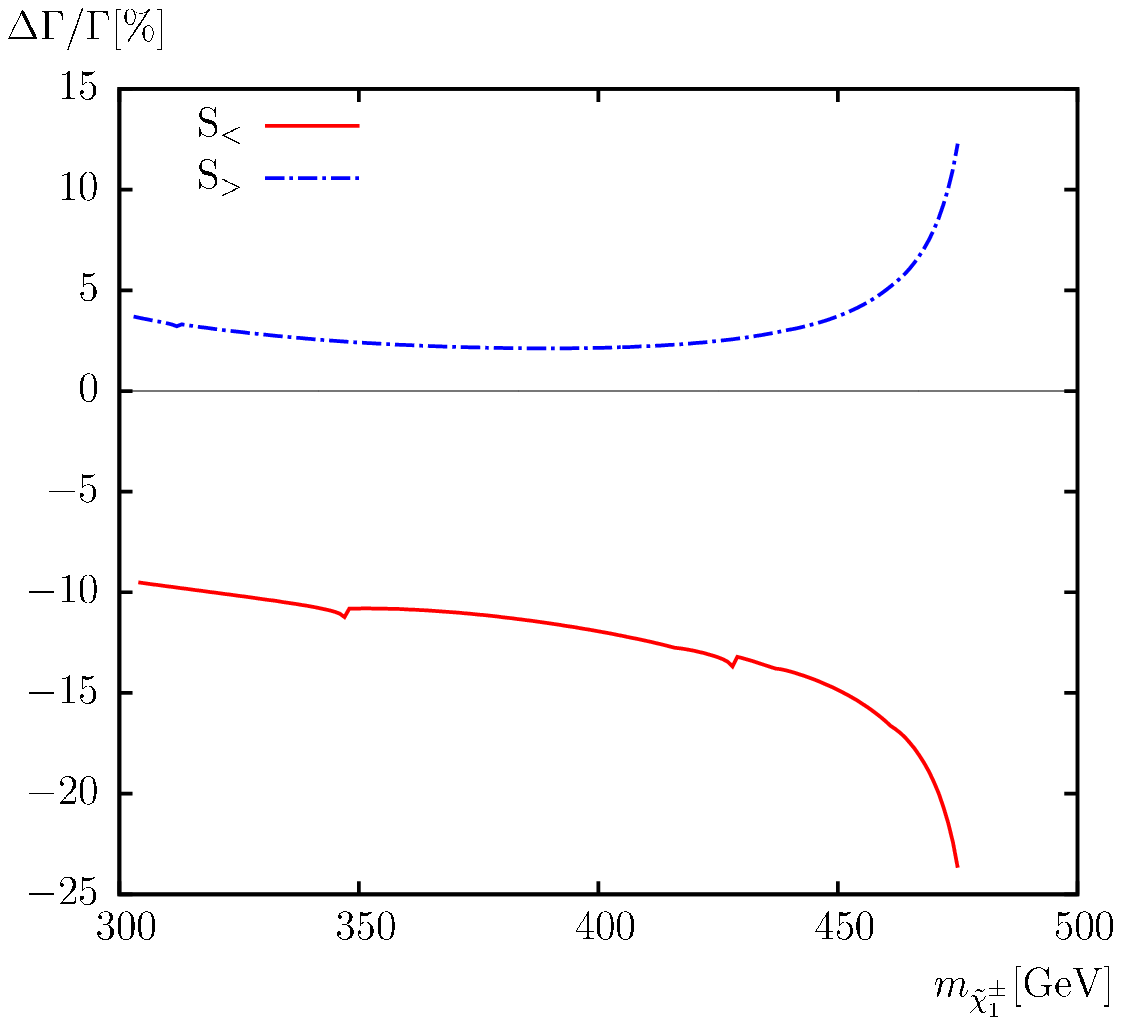} 
\\[5em]
\includegraphics[width=0.49\textwidth,height=7.5cm]{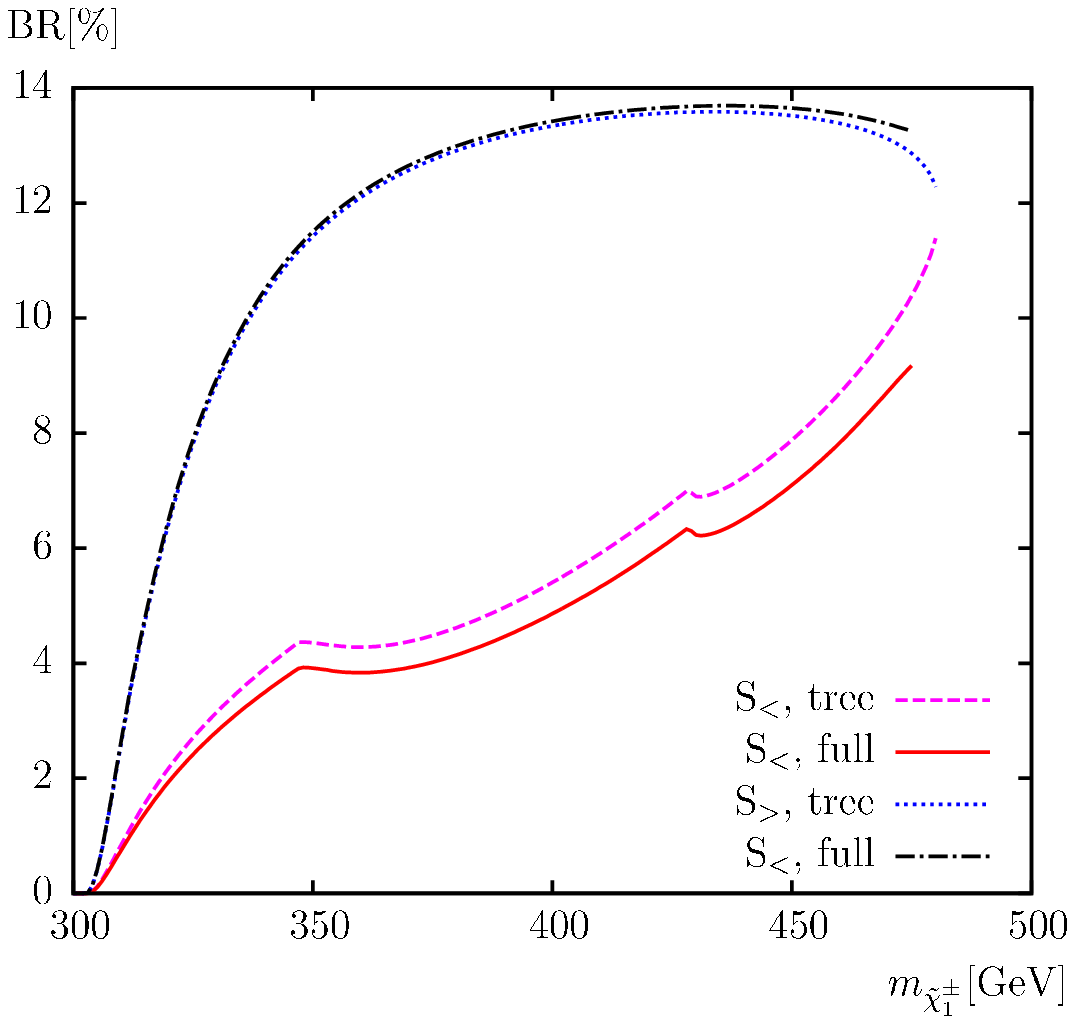}
\hspace{-4mm}
\includegraphics[width=0.49\textwidth,height=7.5cm]{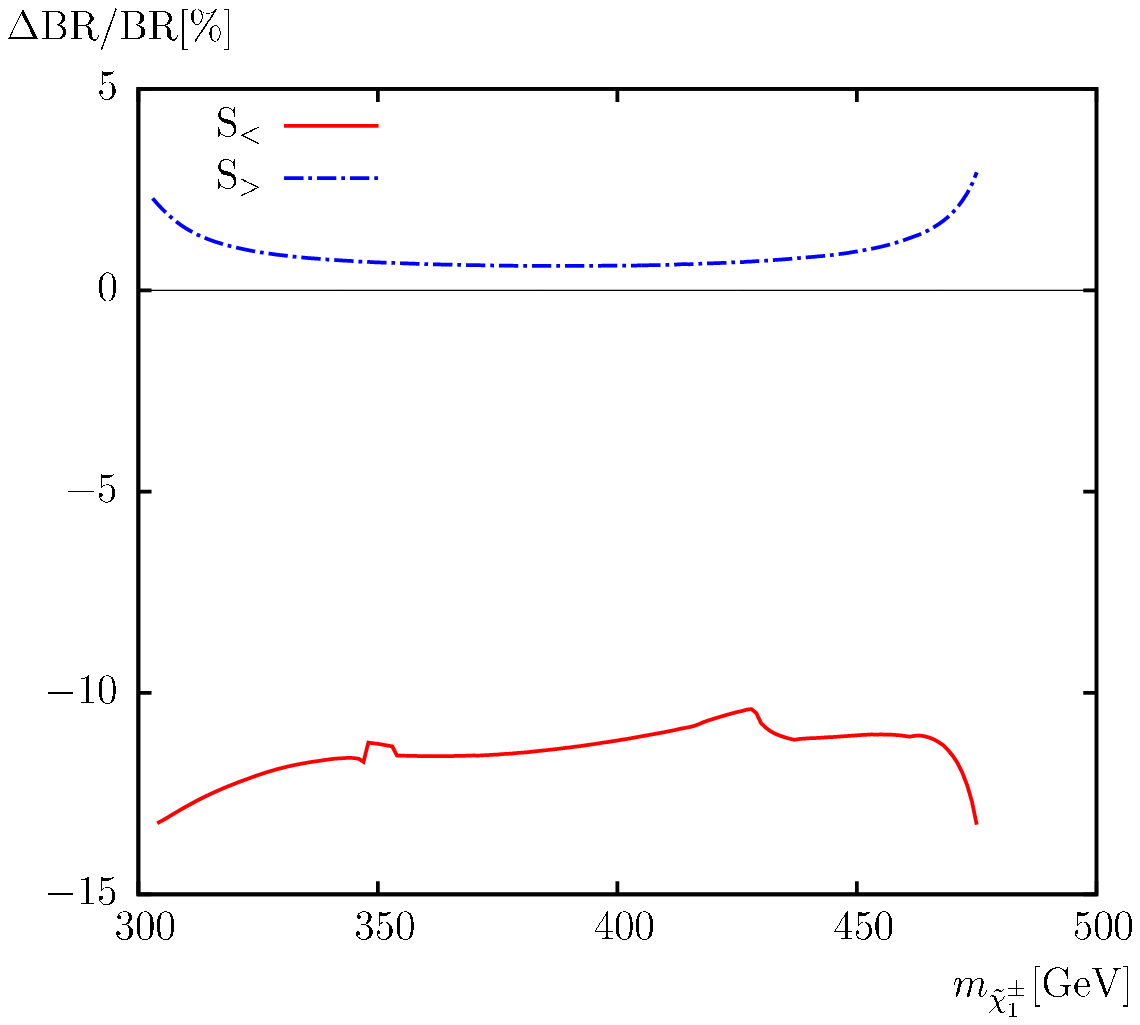}
\end{tabular}
\vspace{2em}
\caption{
  $\Ga(\DecayCmnSl{1}{\mu}{1})$. 
  Tree-level (``tree'') and full one-loop (``full'') corrected 
  decay widths are shown with the parameters chosen according to \SN\
  (see \refta{tab:para}), with $\mcha{1}$ varied.
  The upper left plot shows the decay width, the upper right plot shows 
  the relative size of the corrections.
  The lower left plot shows the BR, the lower right plot shows 
  the relative size of the BR.
}
\label{fig:mC1.cha1smu1nu}
\end{center}
\end{figure}

\begin{figure}[htb!]
\begin{center}
\begin{tabular}{c}
\includegraphics[width=0.49\textwidth,height=7.5cm]{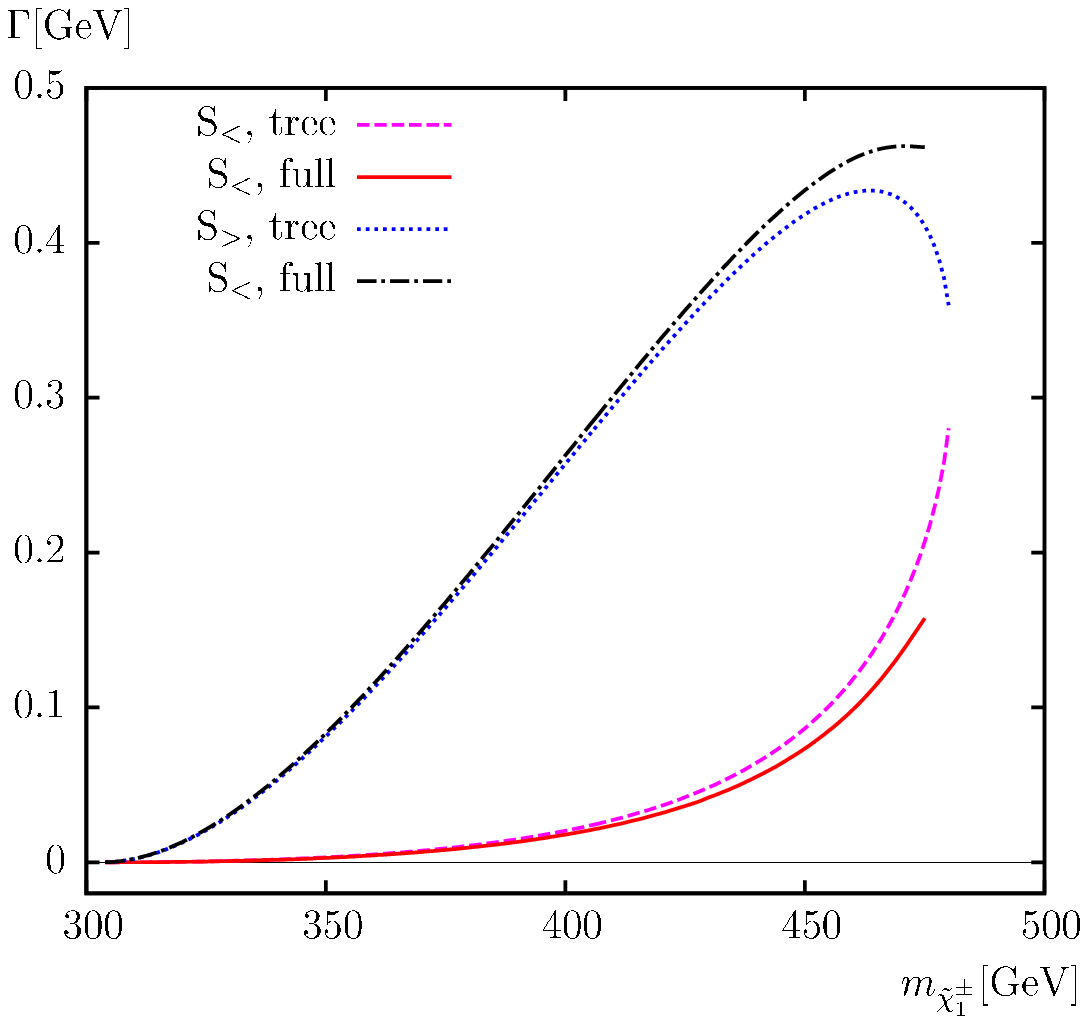}
\hspace{-4mm}
\includegraphics[width=0.49\textwidth,height=7.5cm]{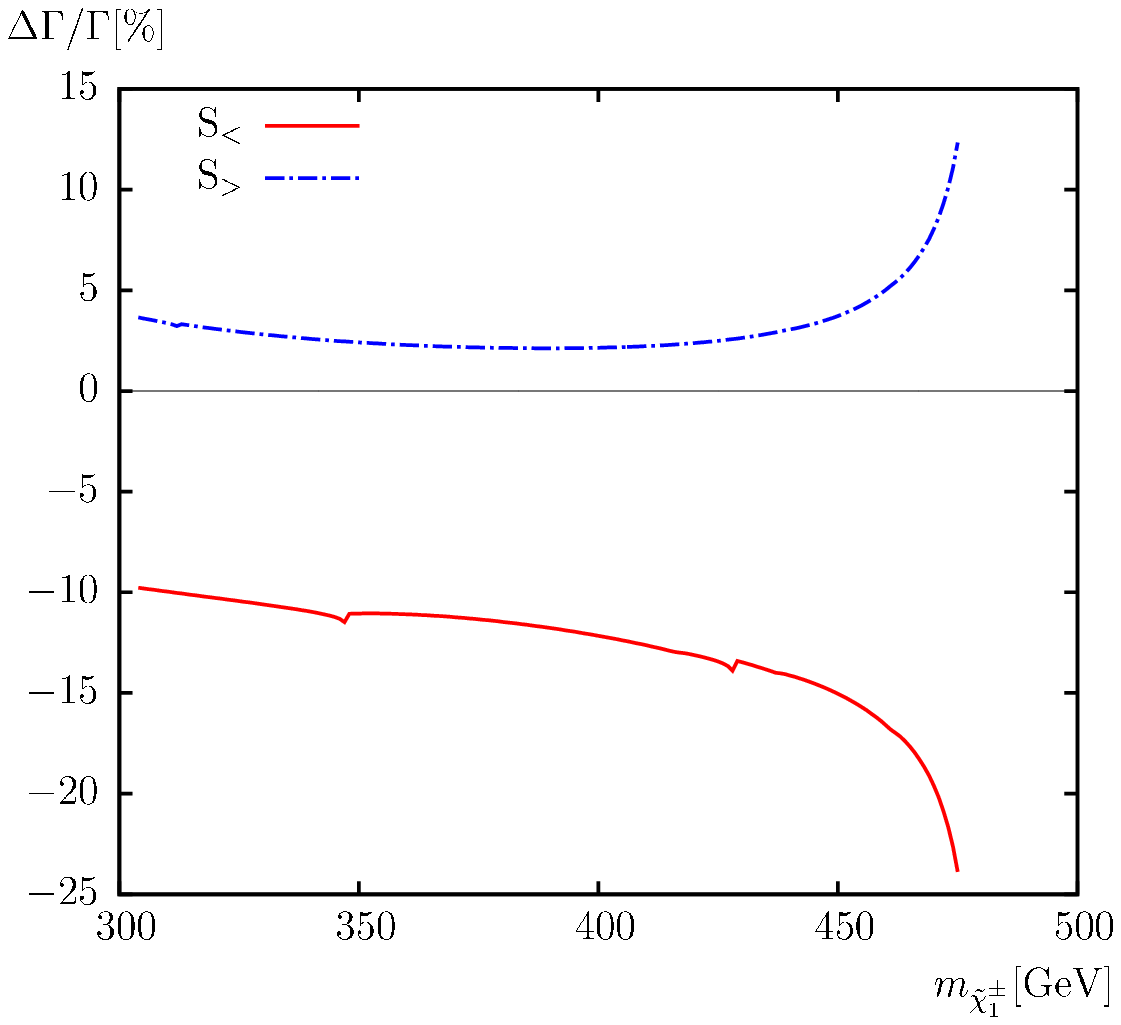} 
\\[5em]
\includegraphics[width=0.49\textwidth,height=7.5cm]{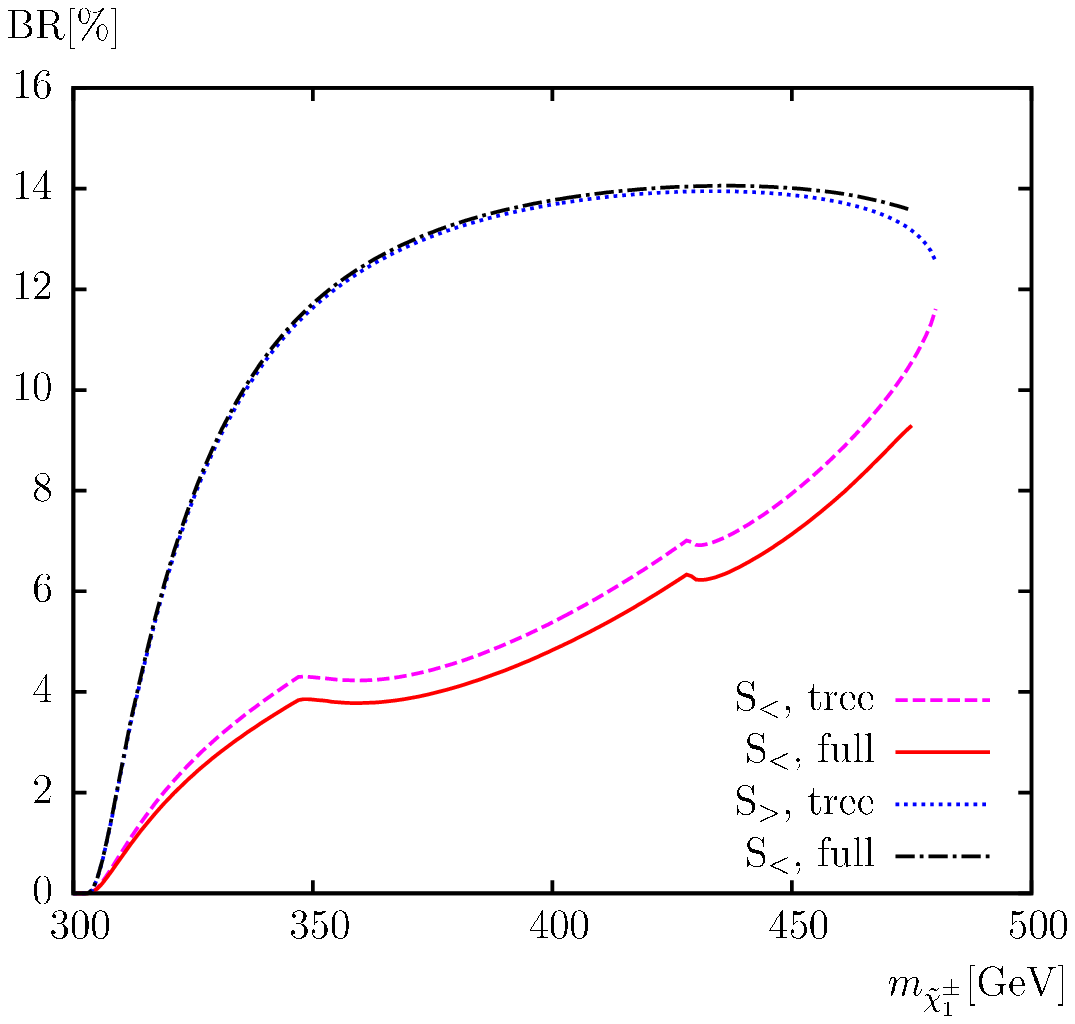}
\hspace{-4mm}
\includegraphics[width=0.49\textwidth,height=7.5cm]{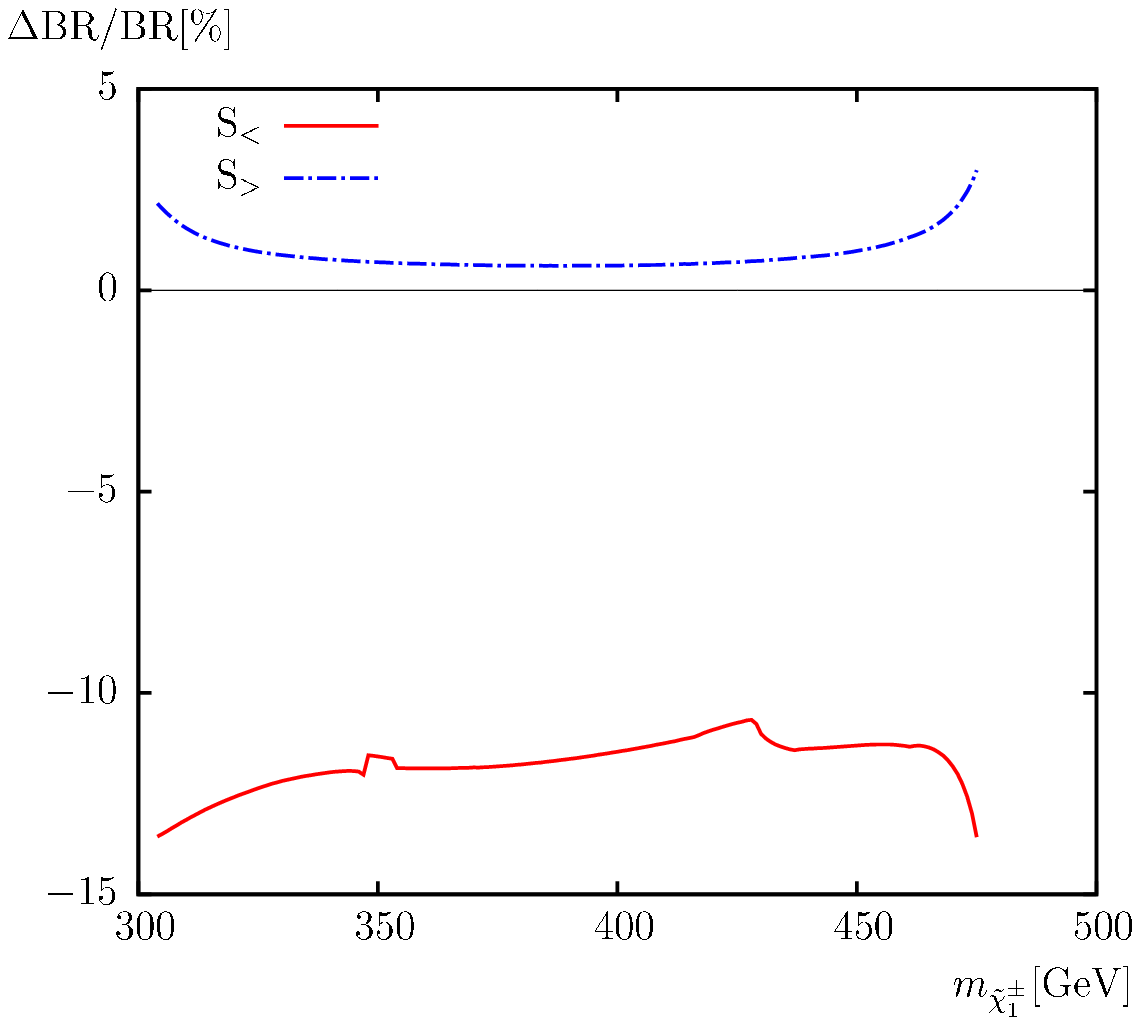}
\end{tabular}
\vspace{2em}
\caption{
  $\Ga(\DecayCmnSl{1}{e}{1})$. 
  Tree-level (``tree'') and full one-loop (``full'') corrected 
  decay widths are shown with the parameters chosen according to \SN\
  (see \refta{tab:para}), with $\mcha{1}$ varied.
  The upper left plot shows the decay width, the upper right plot shows 
  the relative size of the corrections.
  The lower left plot shows the BR, the lower right plot shows 
  the relative size of the BR.
}
\label{fig:mC1.cha1sel1nu}
\end{center}
\end{figure}

\medskip
Next we discuss the $\cha{1}$ decays into scalar leptons. The results for 
$\DecayCmnSl{1}{\tau}{1}$, $\DecayCmnSl{1}{\mu}{1}$, $\DecayCmnSl{1}{e}{1}$ 
are shown in \reffis{fig:mC1.cha1stau1nu}, \ref{fig:mC1.cha1smu1nu},
\ref{fig:mC1.cha1sel1nu}, respectively. 
The dips at $\mcha{1}=347, 428 \gev$, visible best in the upper right panels, 
are due to the $\DecayCmNW{1}{1}$, $\neu{1}\Hm$ thresholds. 
The decay widths grow monotonously up to values of $\sim 0.5 \gev$ 
in \SE\ for all three decays. Within \SZ\ they
rise up to $\sim 0.25 \gev$ for $\DecayCmnSl{1}{\tau}{1}$ due to the
non-vanishing mixing in the scalar tau sector. For the second and first
generation values of $\sim 0.15 \gev$ are
reached. 
The $\br(\DecayCmnSl{1}{\tau}{1})$ is found between $50\%$ ($30\%$)
and $\sim 15\%$ in \SE\ (\SZ). The size of the one-loop effects exceeds
$1\%$ only in \SZ, where it varies between $\sim 2\%$ and
$3\%$, which is roughly at the level of the anticipated ILC(1000)
accuracy. 

\begin{figure}[htb!]
\begin{center}
\begin{tabular}{c}
\includegraphics[width=0.49\textwidth,height=7.5cm]{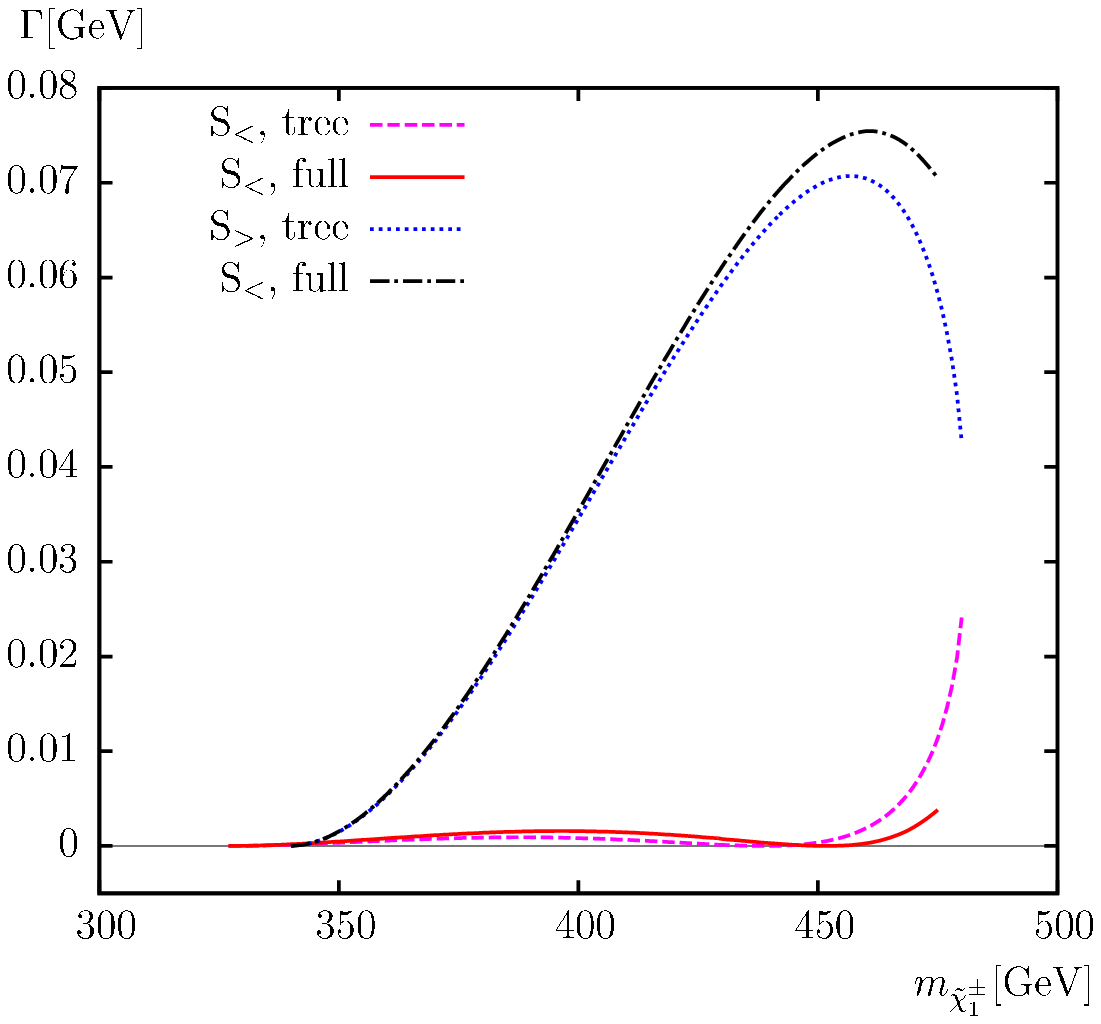}
\hspace{-4mm}
\includegraphics[width=0.49\textwidth,height=7.5cm]{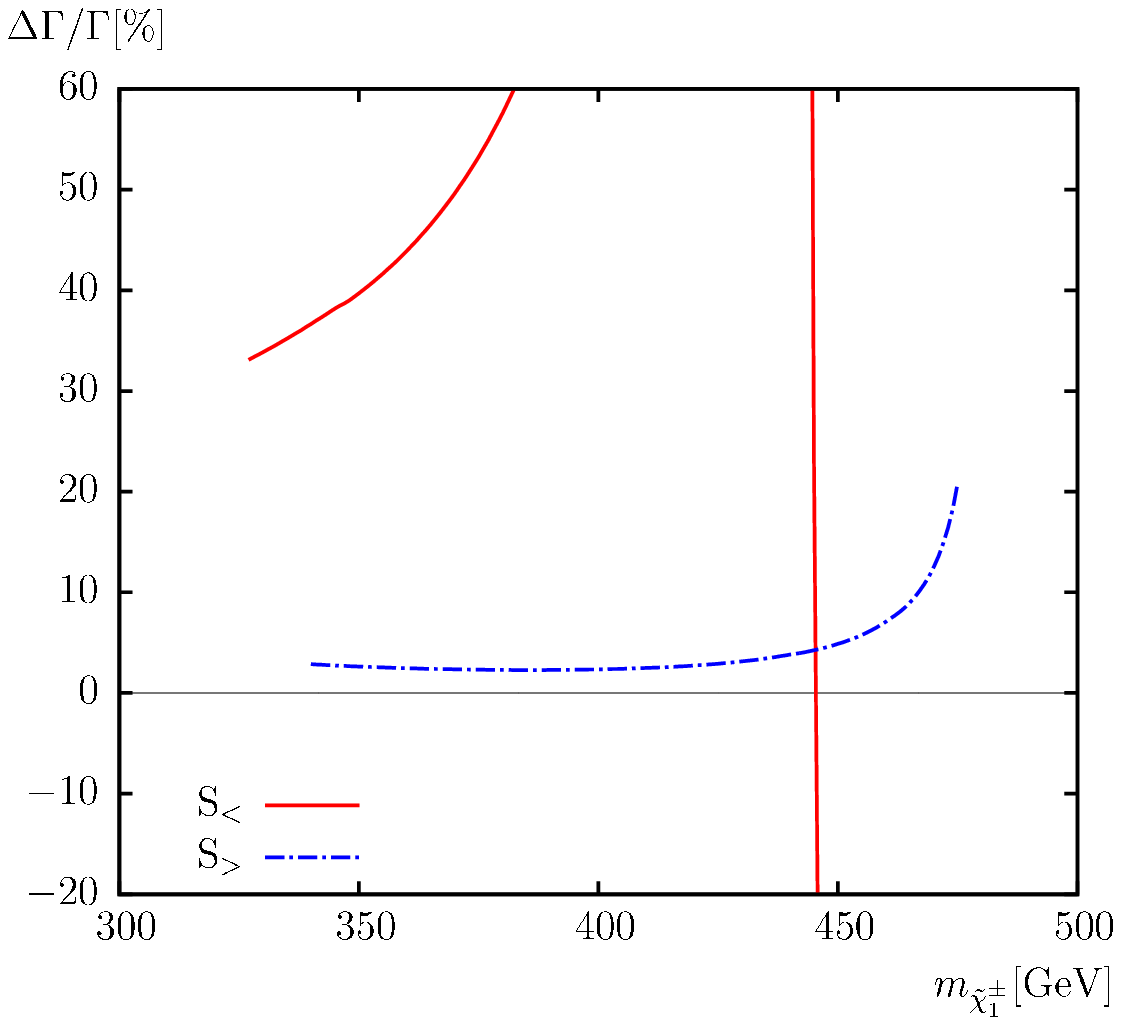} 
\\[5em]
\includegraphics[width=0.49\textwidth,height=7.5cm]{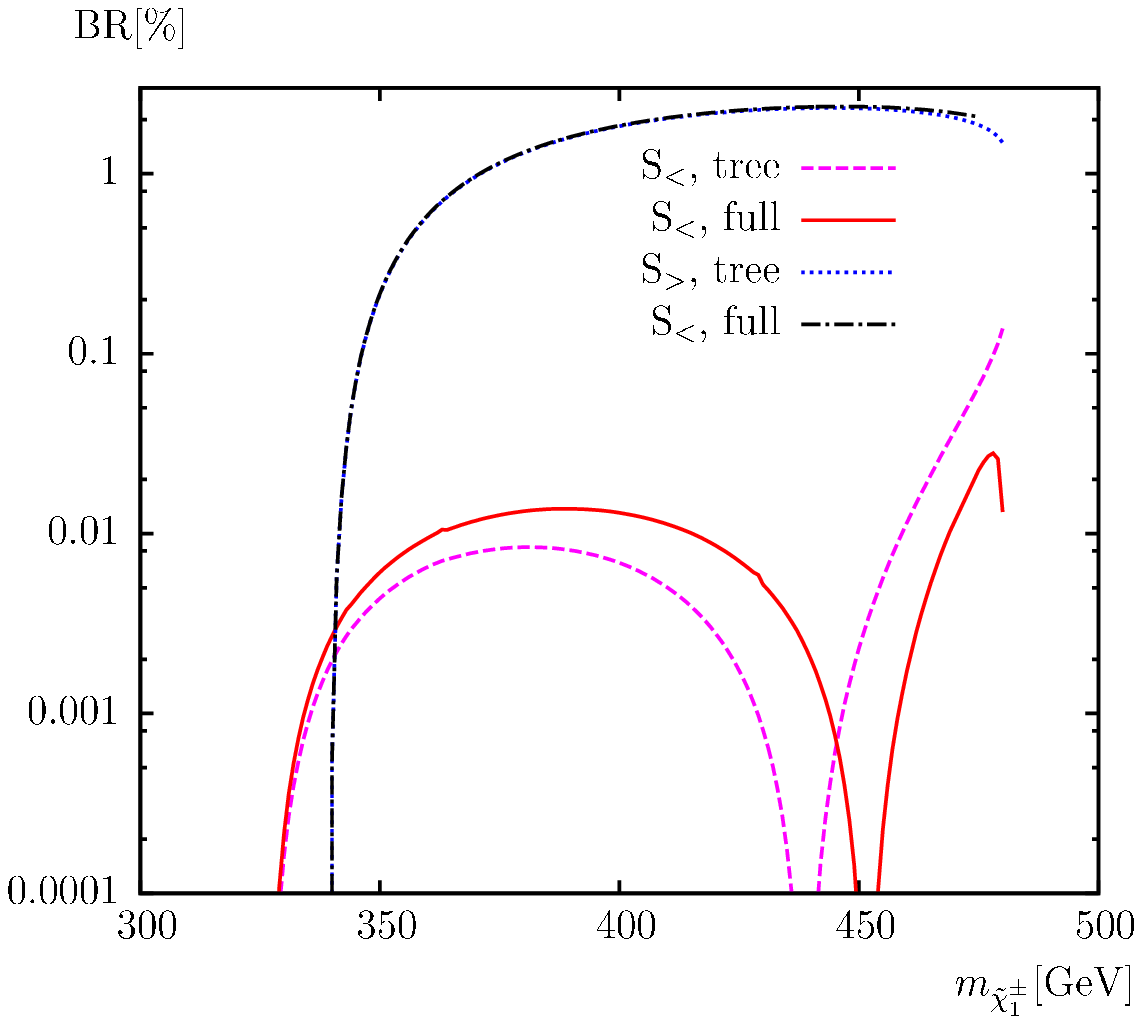}
\hspace{-4mm}
\includegraphics[width=0.49\textwidth,height=7.5cm]{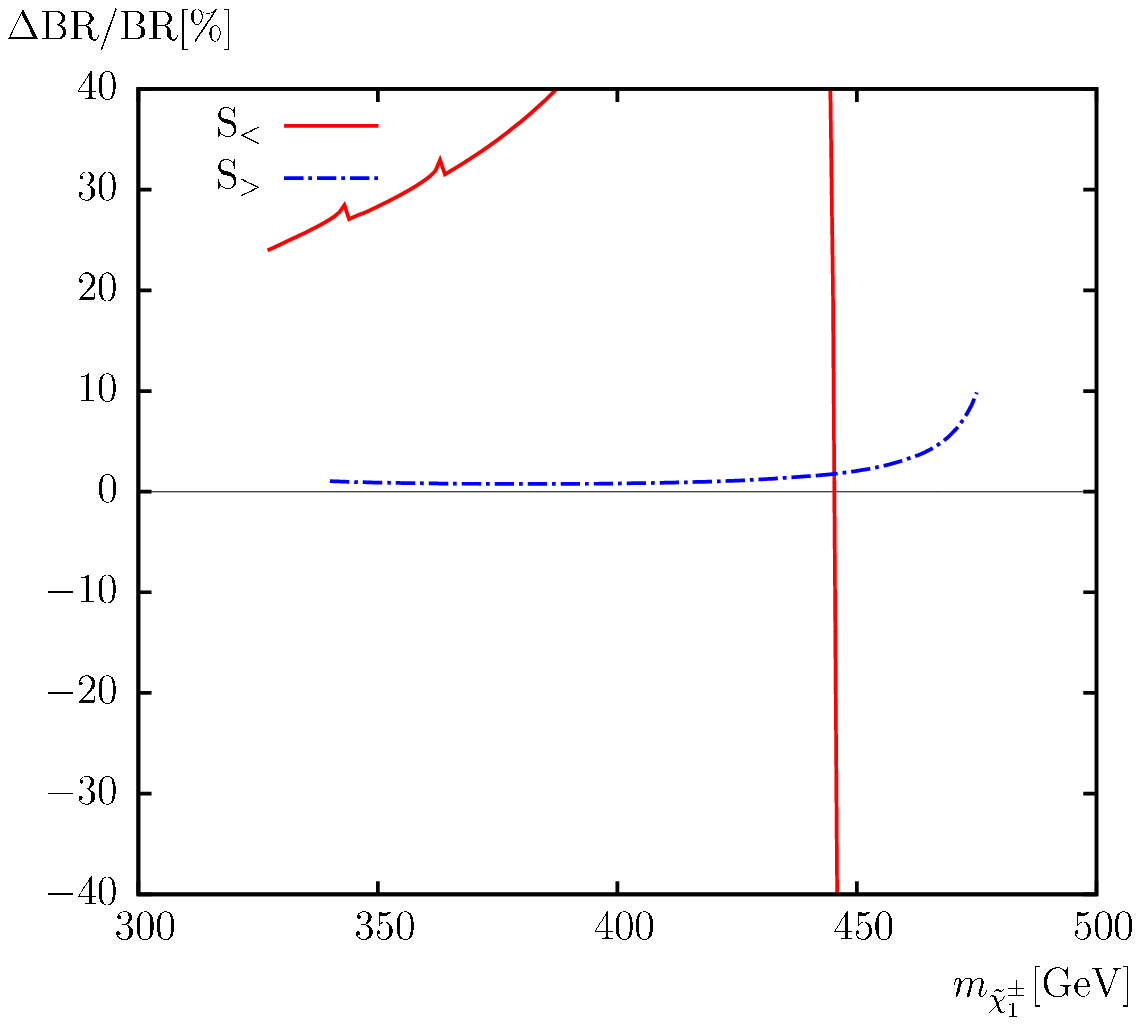}
\end{tabular}
\vspace{2em}
\caption{
  $\Ga(\DecayCmnSl{1}{\tau}{2})$. 
  Tree-level (``tree'') and full one-loop (``full'') corrected 
  decay widths are shown with the parameters chosen according to \SN\
  (see \refta{tab:para}), with $\mcha{1}$ varied.
  The upper left plot shows the decay width, the upper right plot shows 
  the relative size of the corrections.
  The lower left plot shows the BR, the lower right plot shows 
  the relative size of the BR.
}
\label{fig:mC1.cha1stau2nu}
\end{center}
\end{figure}

\begin{figure}[htb!]
\begin{center}
\begin{tabular}{c}
\includegraphics[width=0.49\textwidth,height=7.5cm]{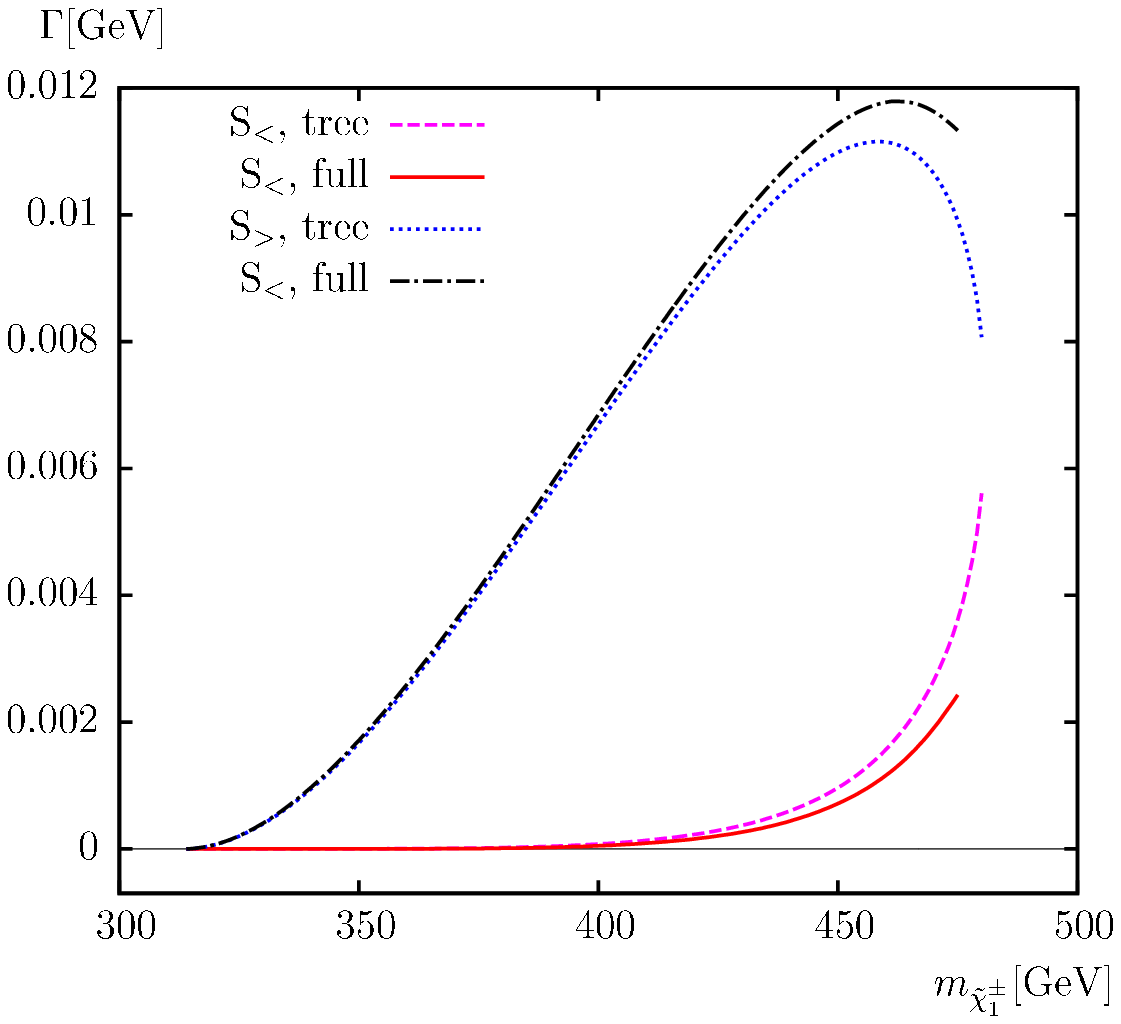}
\hspace{-4mm}
\includegraphics[width=0.49\textwidth,height=7.5cm]{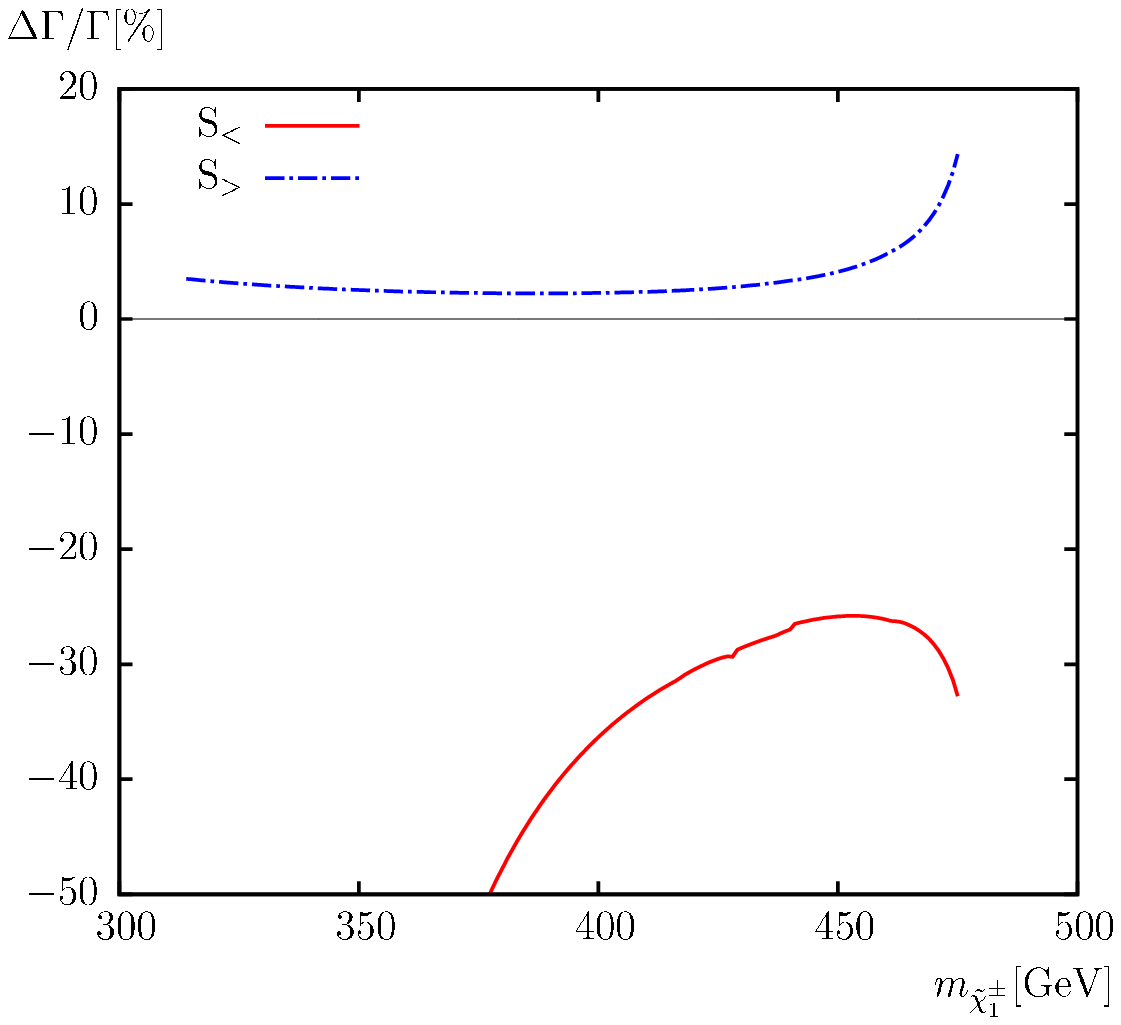} 
\\[5em]
\includegraphics[width=0.49\textwidth,height=7.5cm]{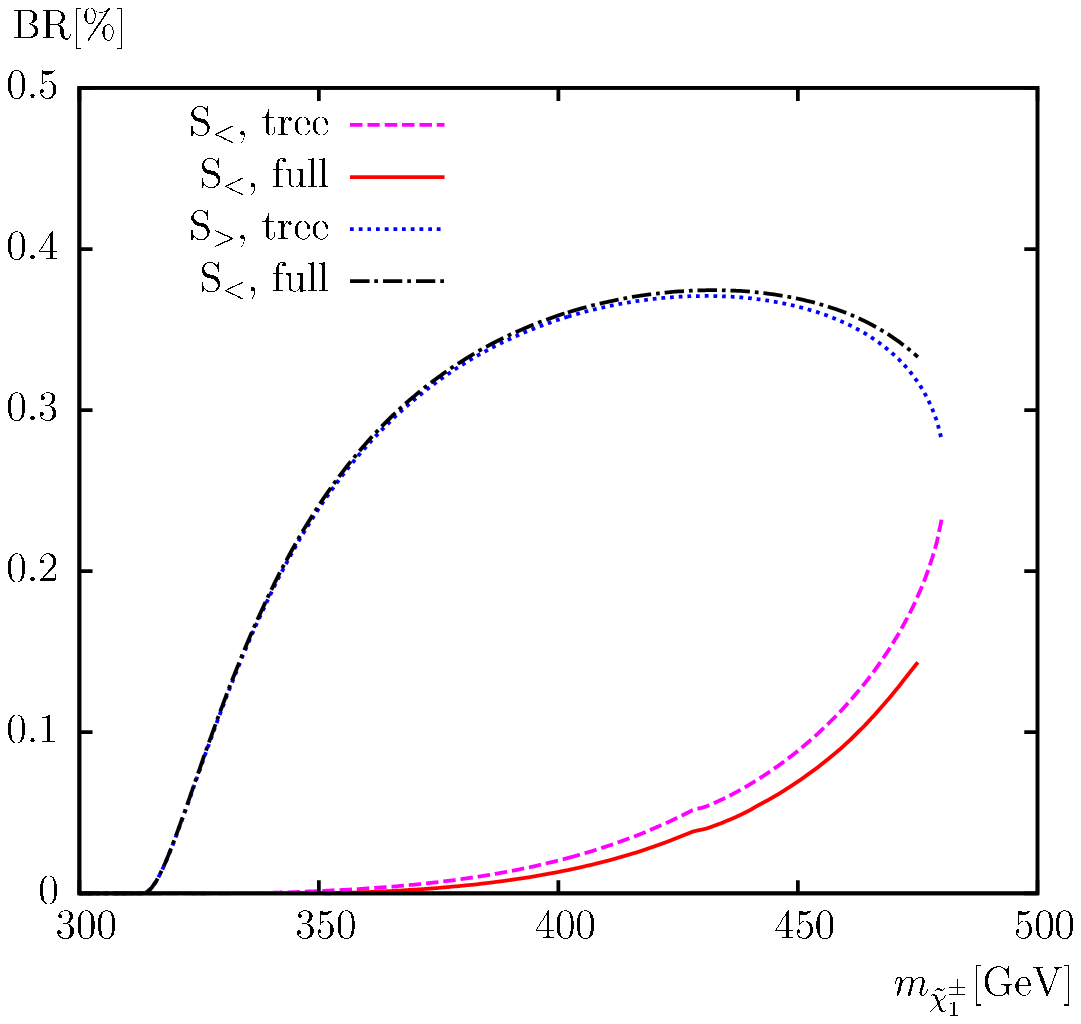}
\hspace{-4mm}
\includegraphics[width=0.49\textwidth,height=7.5cm]{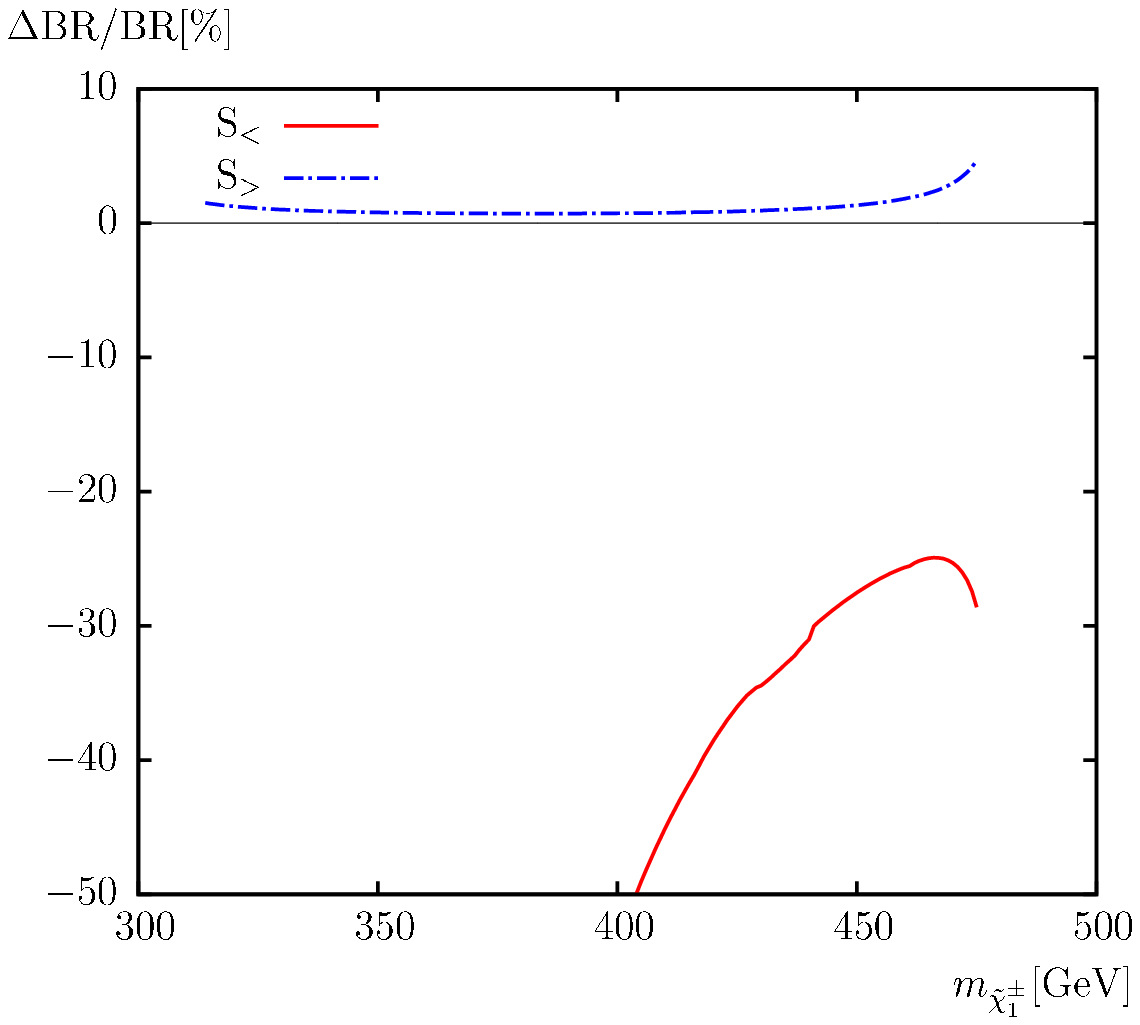}
\end{tabular}
\vspace{2em}
\caption{
  $\Ga(\DecayCmnSl{1}{\mu}{2})$. 
  Tree-level (``tree'') and full one-loop (``full'') corrected 
  decay widths are shown with the parameters chosen according to \SN\
  (see \refta{tab:para}), with $\mcha{1}$ varied.
  The upper left plot shows the decay width, the upper right plot shows 
  the relative size of the corrections.
  The lower left plot shows the BR, the lower right plot shows 
  the relative size of the BR.
}
\label{fig:mC1.cha1smu2nu}
\end{center}
\end{figure}

\begin{figure}[htb!]
\begin{center}
\begin{tabular}{c}
\includegraphics[width=0.49\textwidth,height=7.5cm]{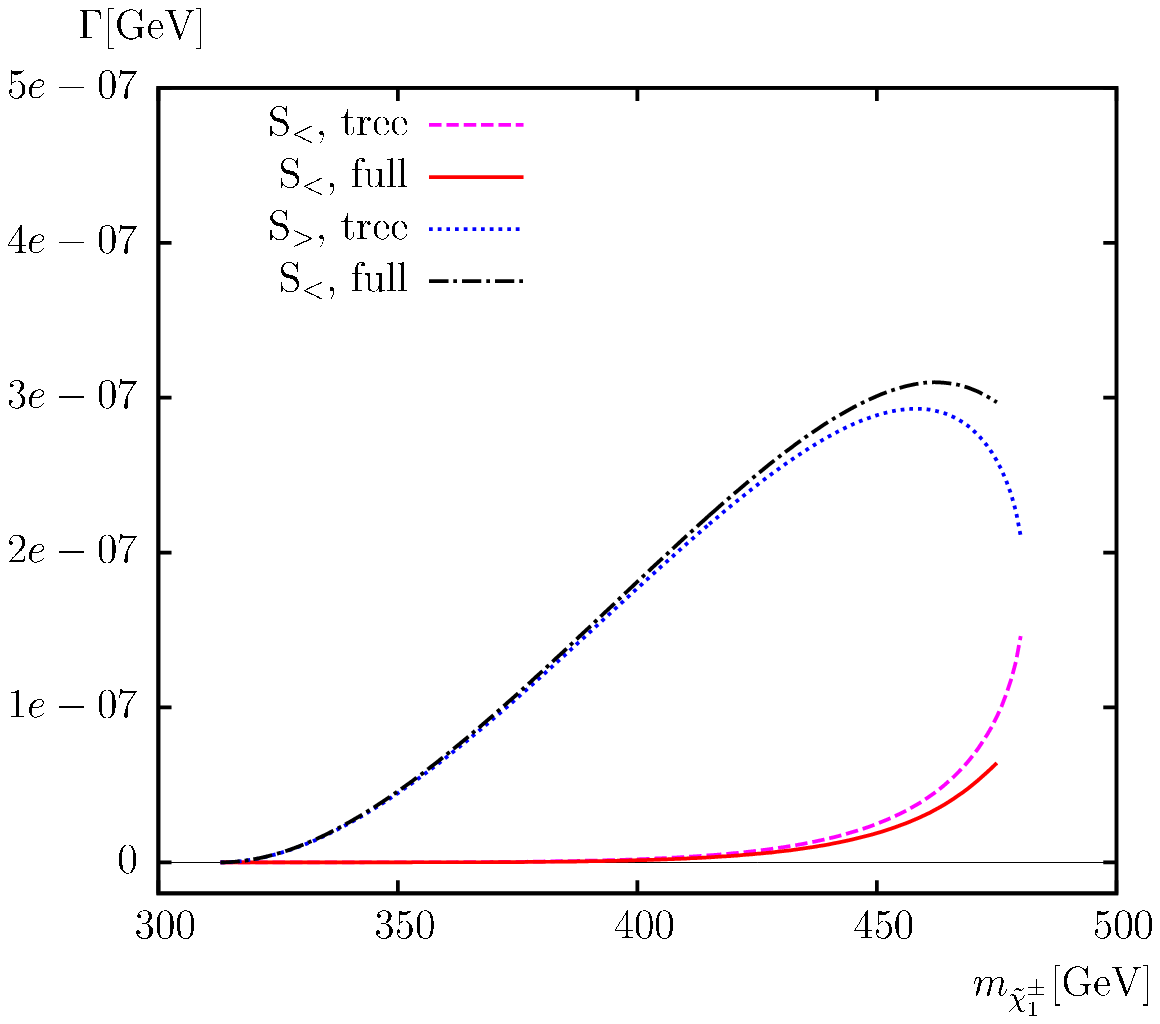}
\hspace{-4mm}
\includegraphics[width=0.49\textwidth,height=7.5cm]{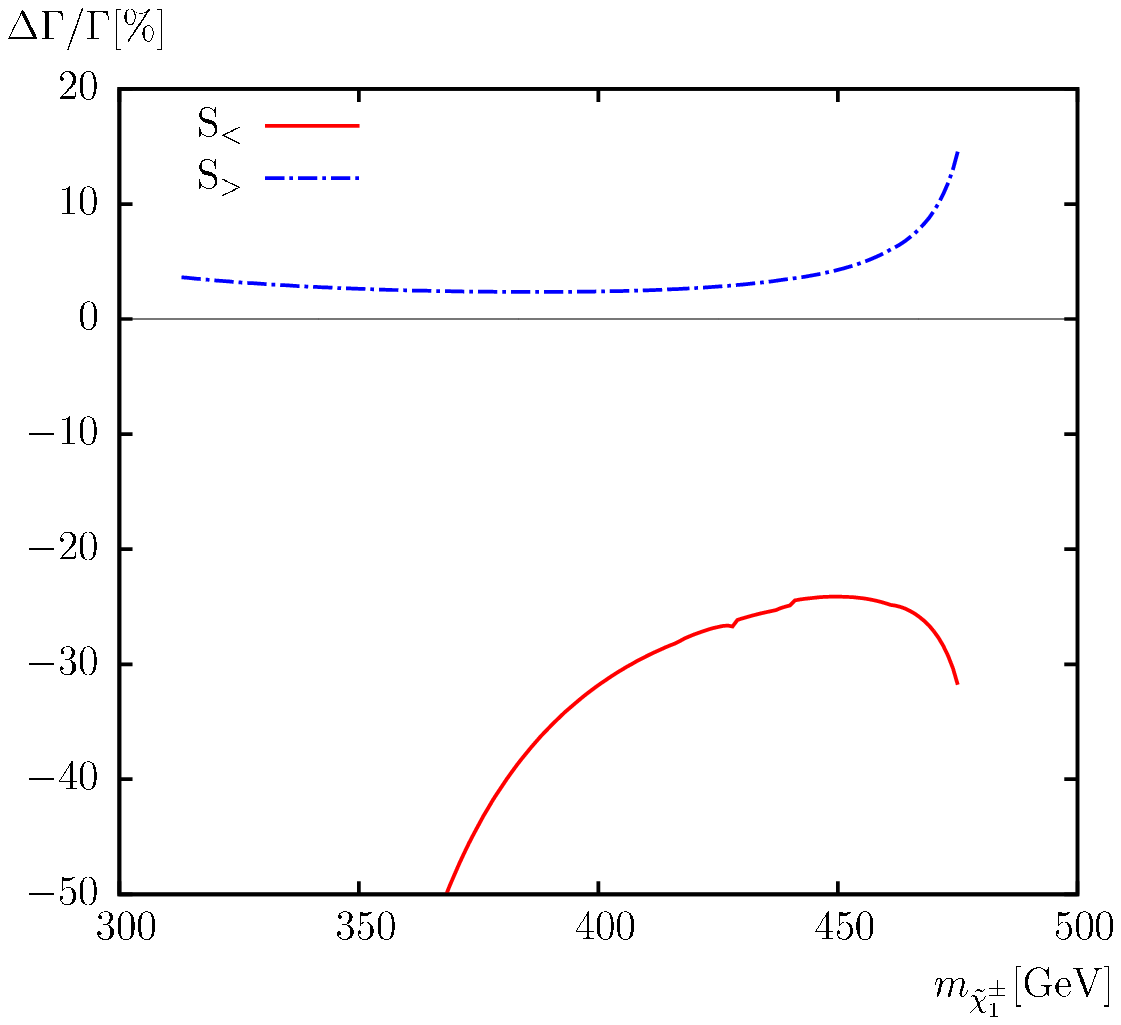} 
\\[5em]
\includegraphics[width=0.49\textwidth,height=7.5cm]{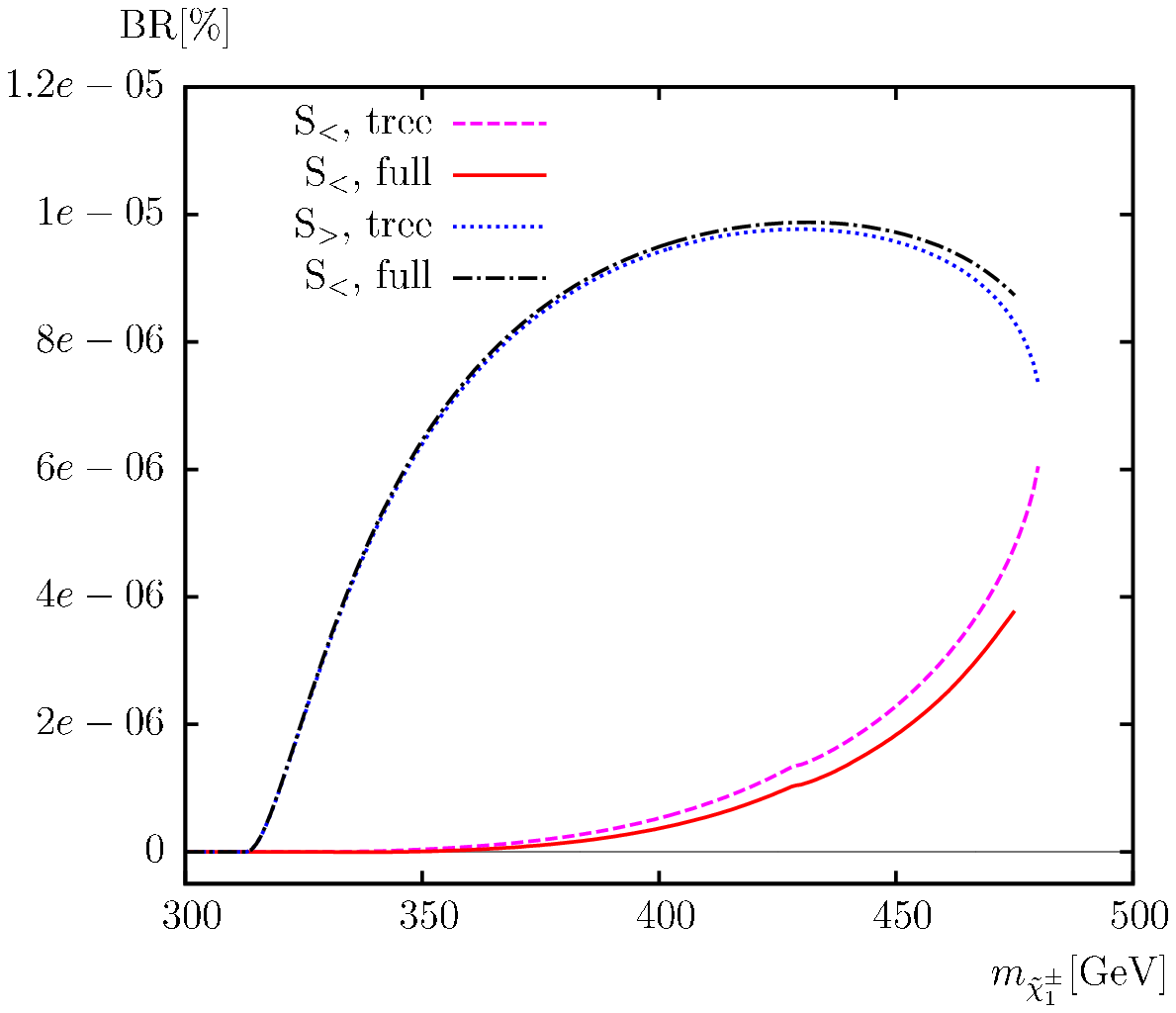}
\hspace{-4mm}

\includegraphics[width=0.49\textwidth,height=7.5cm]{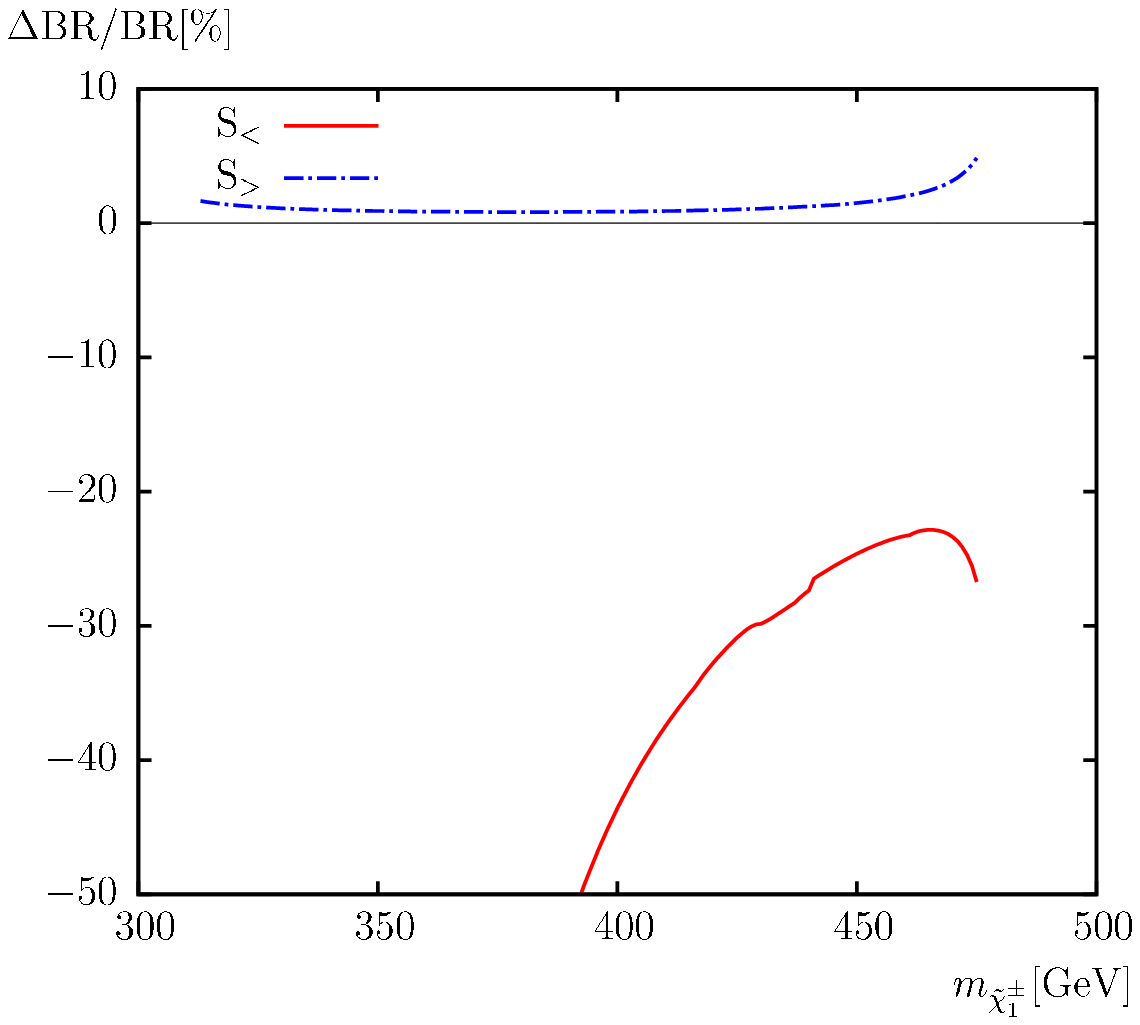}
\end{tabular}
\vspace{2em}
\caption{
  $\Ga(\DecayCmnSl{1}{e}{2})$. 
  Tree-level (``tree'') and full one-loop (``full'') corrected 
  decay widths are shown with the parameters chosen according to \SN\
  (see \refta{tab:para}), with $\mcha{1}$ varied.
  The upper left plot shows the decay width, the upper right plot shows 
  the relative size of the corrections.
  The lower left plot shows the BR, the lower right plot shows 
  the relative size of the BR.
}
\label{fig:mC1.cha1sel2nu}
\end{center}
\end{figure}

The decays to the heavier scalar leptons are, as for the $\cha{2}$
decays, determined by the size of the Yukawa couplings $\propto \ml$. 
The results for 
$\DecayCmnSl{1}{\tau}{2}, \DecayCmnSl{1}{\mu}{2}, \DecayCmnSl{1}{e}{2}$ 
are shown in \reffis{fig:mC1.cha1stau2nu}, \ref{fig:mC1.cha1smu2nu}, 
\ref{fig:mC1.cha1sel2nu}. In the case of the scalar tau the highest
values reached are $\Ga(\DecayCmnSl{1}{\tau}{2}) \lsim 0.075 \gev$ in
\SE, corresponding to a branching ratio below $\sim 2\%$. All other
decays have a very small decay width and a correspondingly small BR. 

\begin{figure}[htb!]
\begin{center}
\begin{tabular}{c}
\includegraphics[width=0.49\textwidth,height=7.5cm]{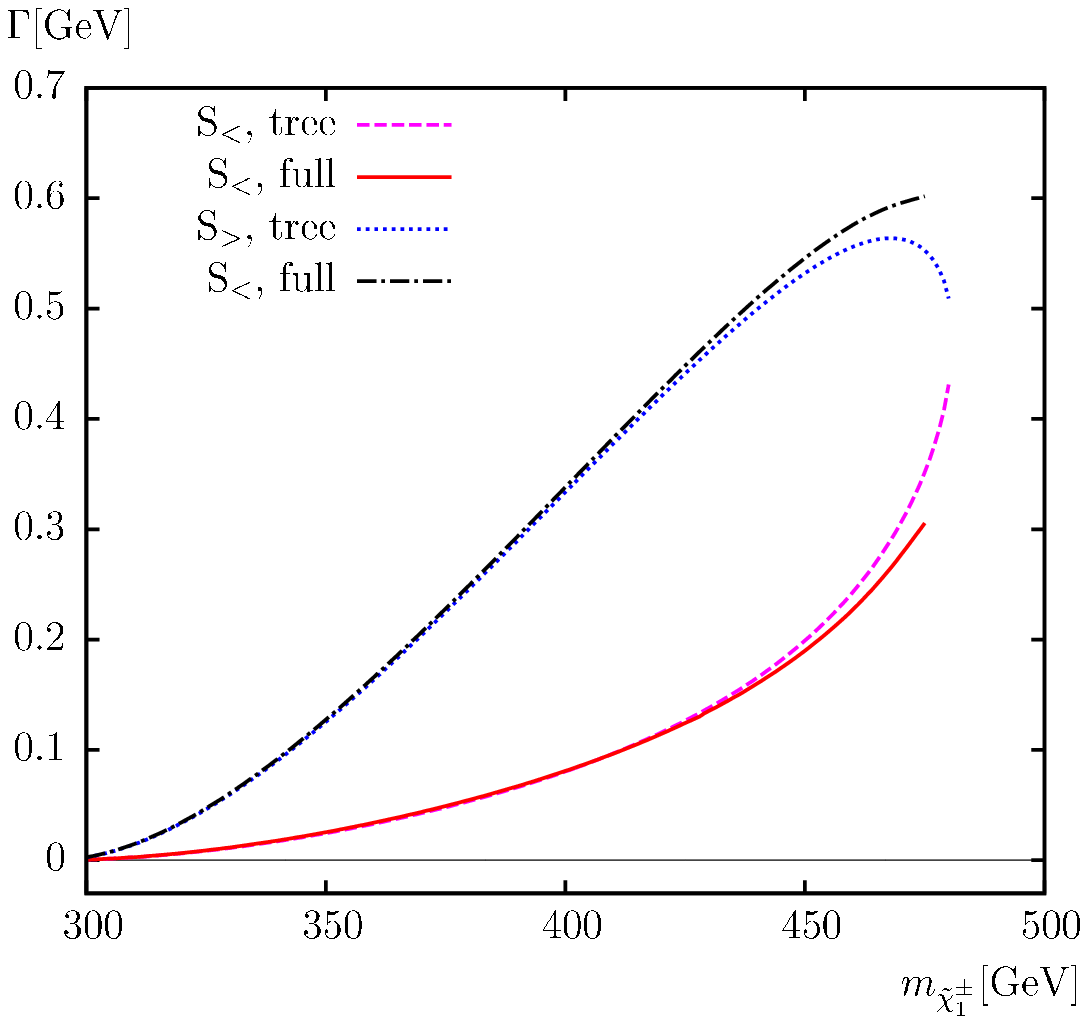}
\hspace{-4mm}
\includegraphics[width=0.49\textwidth,height=7.5cm]{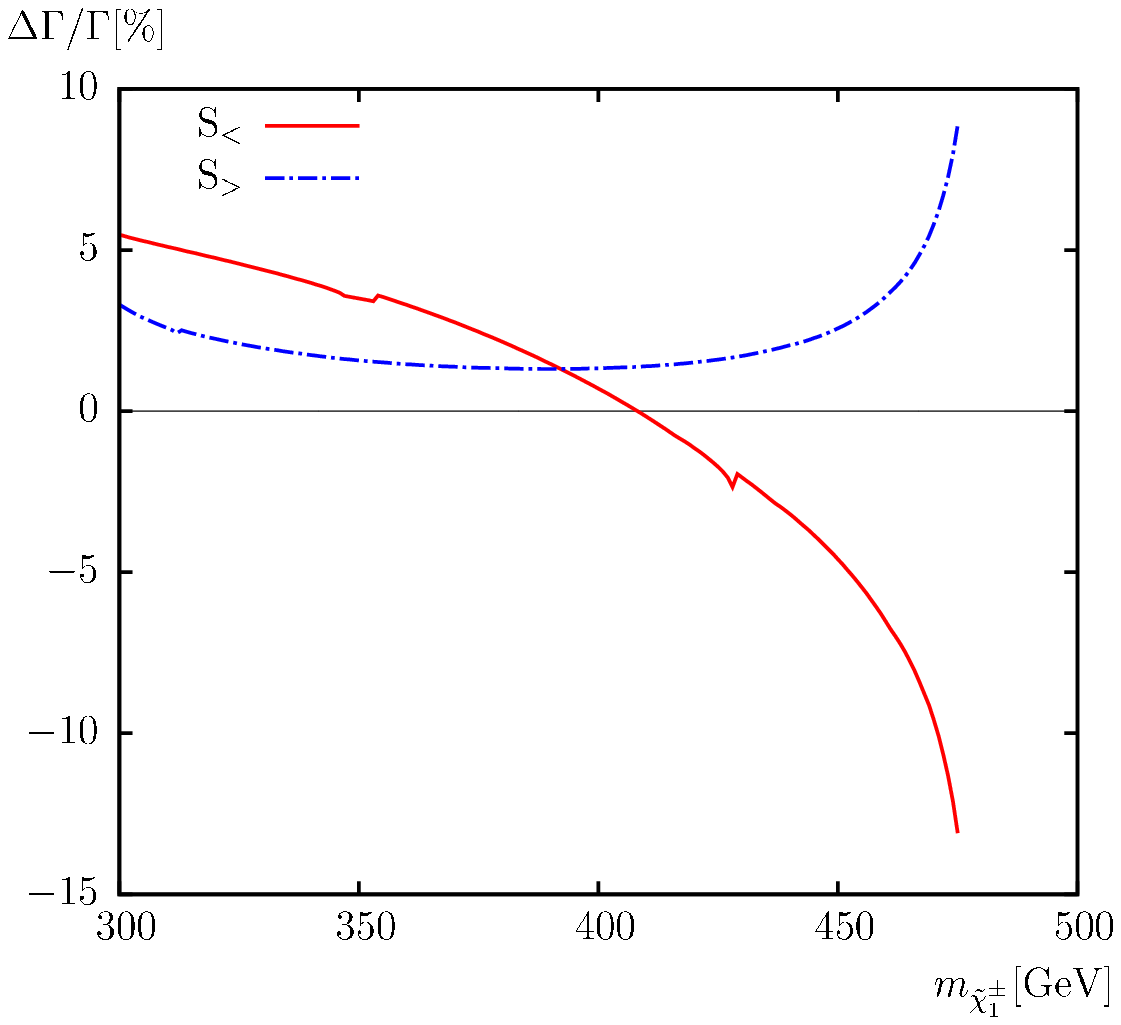} 
\\[5em]
\includegraphics[width=0.49\textwidth,height=7.5cm]{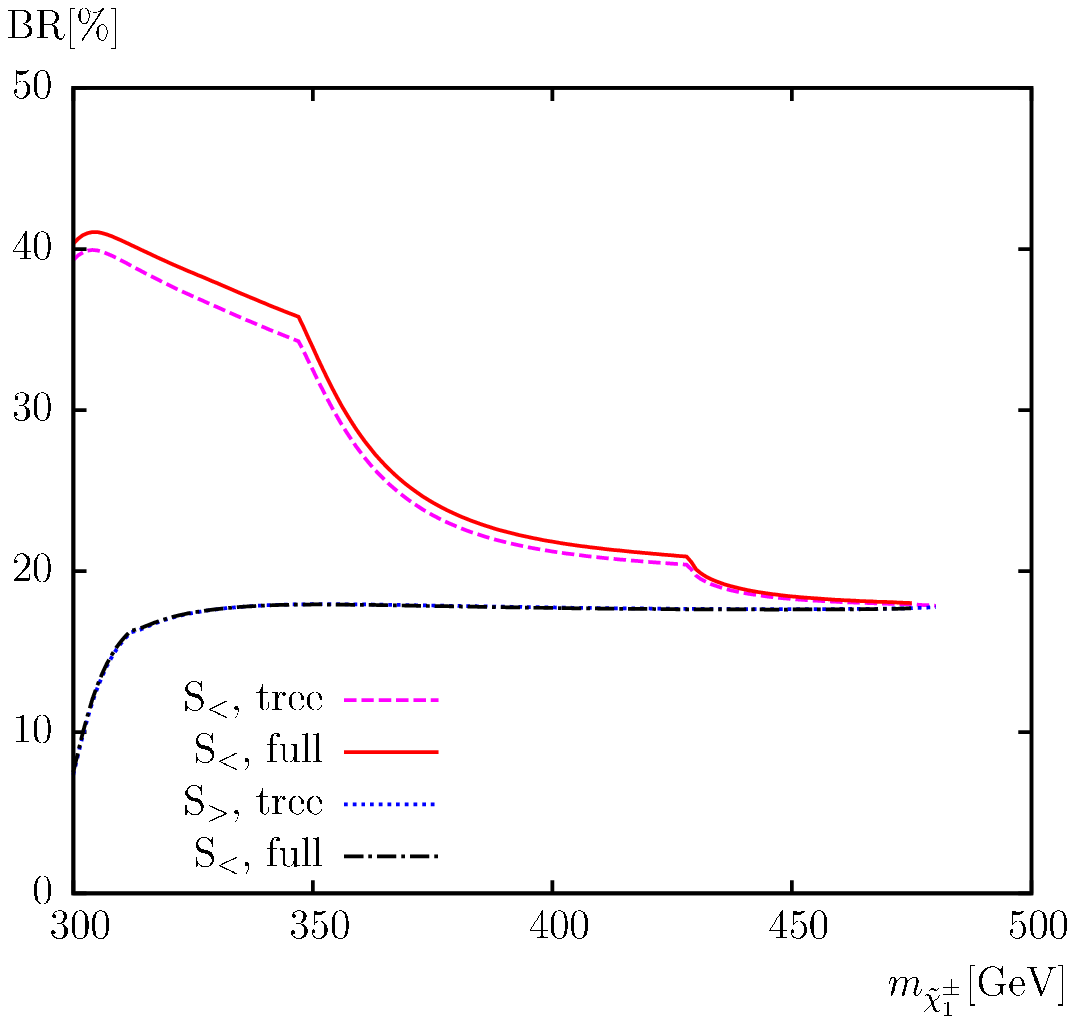}
\hspace{-4mm}
\includegraphics[width=0.49\textwidth,height=7.5cm]{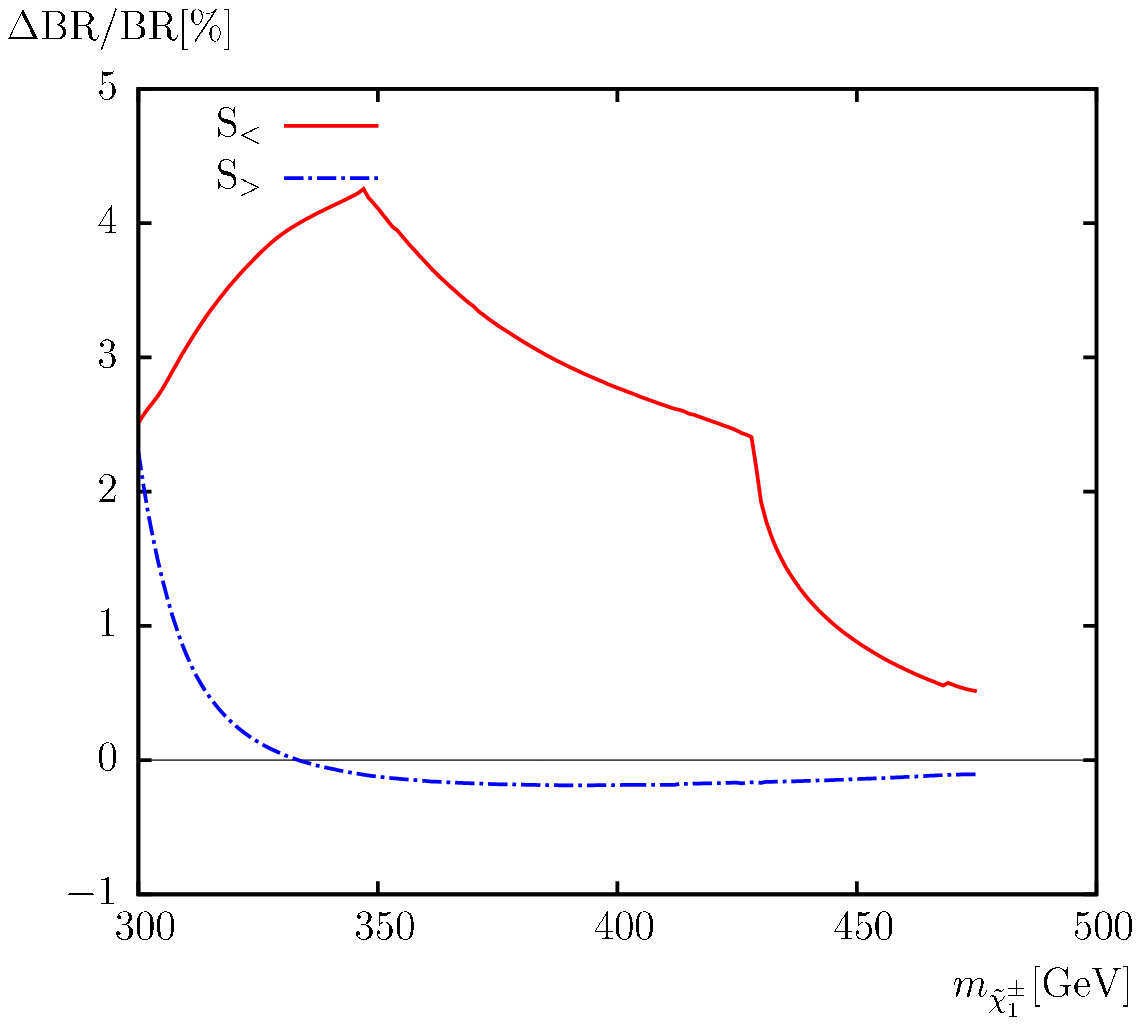}
\end{tabular}
\vspace{2em}
\caption{
  $\Ga(\DecayCmlSn{1}{\tau})$. 
  Tree-level (``tree'') and full one-loop (``full'') corrected 
  decay widths are shown with the parameters chosen according to \SN\
  (see \refta{tab:para}), with $\mcha{1}$ varied.
  The upper left plot shows the decay width, the upper right plot shows 
  the relative size of the corrections.
  The lower left plot shows the BR, the lower right plot shows 
  the relative size of the BR.
}
\label{fig:mC1.cha1snutau}
\end{center}
\end{figure}

\begin{figure}[htb!]
\begin{center}
\begin{tabular}{c}
\includegraphics[width=0.49\textwidth,height=7.5cm]{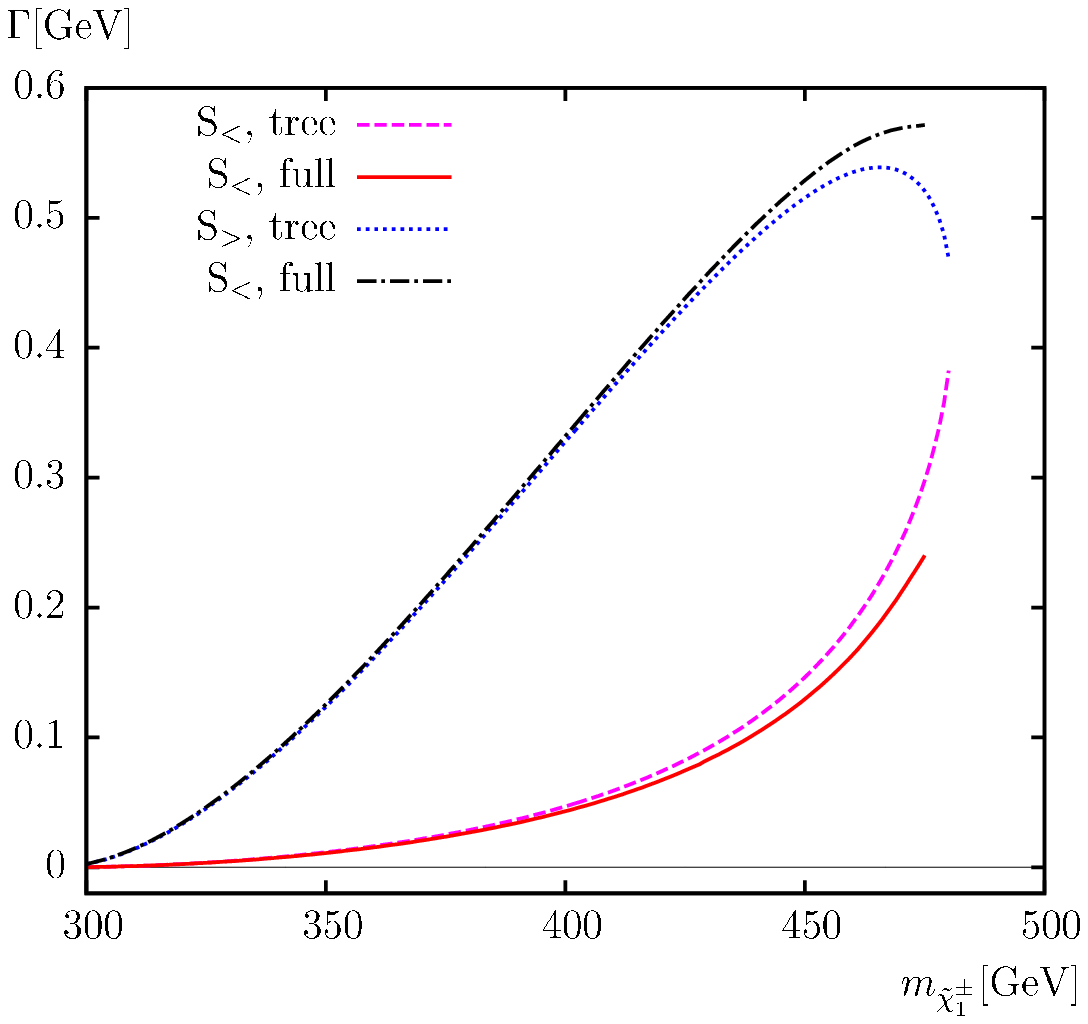}
\hspace{-4mm}
\includegraphics[width=0.49\textwidth,height=7.5cm]{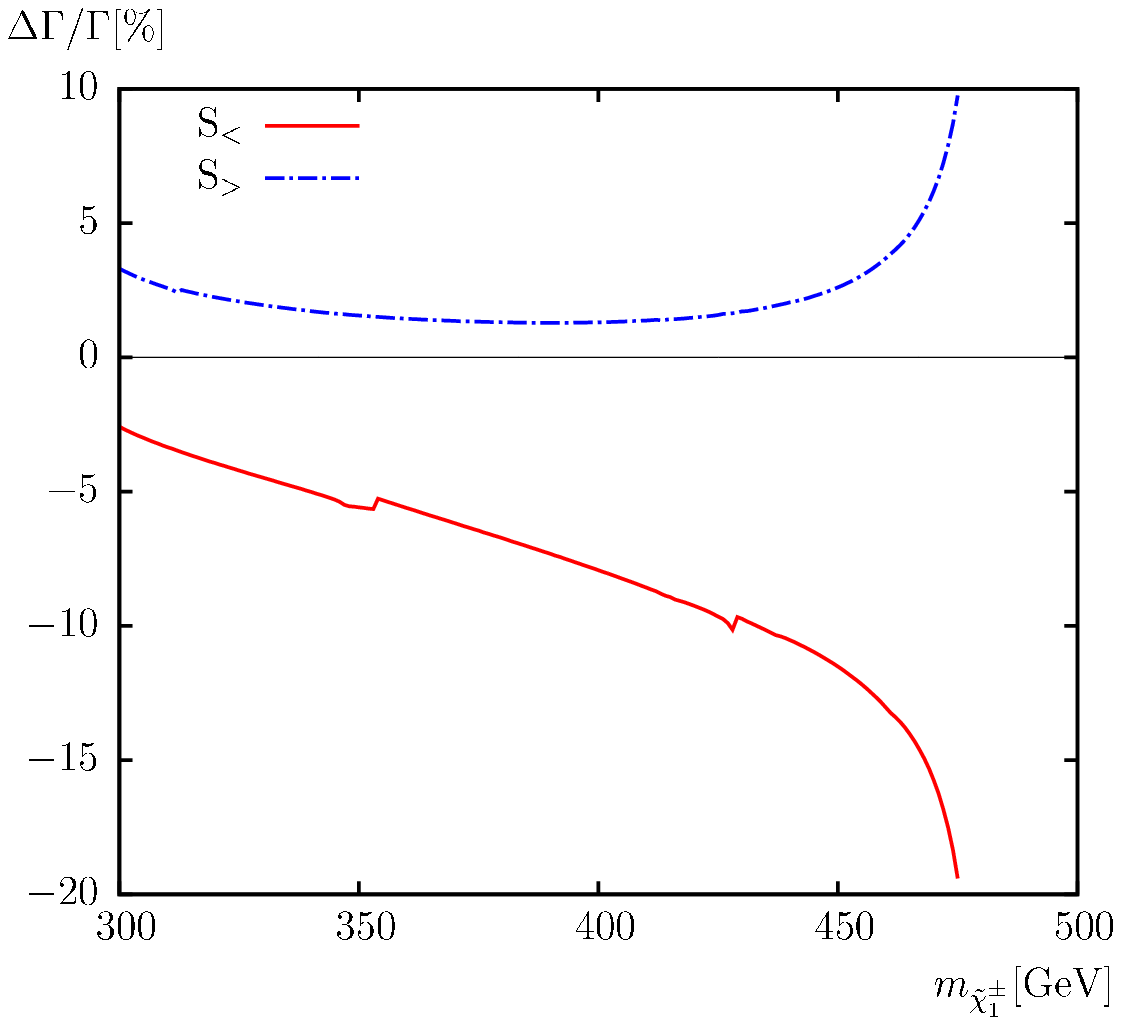} 
\\[5em]
\includegraphics[width=0.49\textwidth,height=7.5cm]{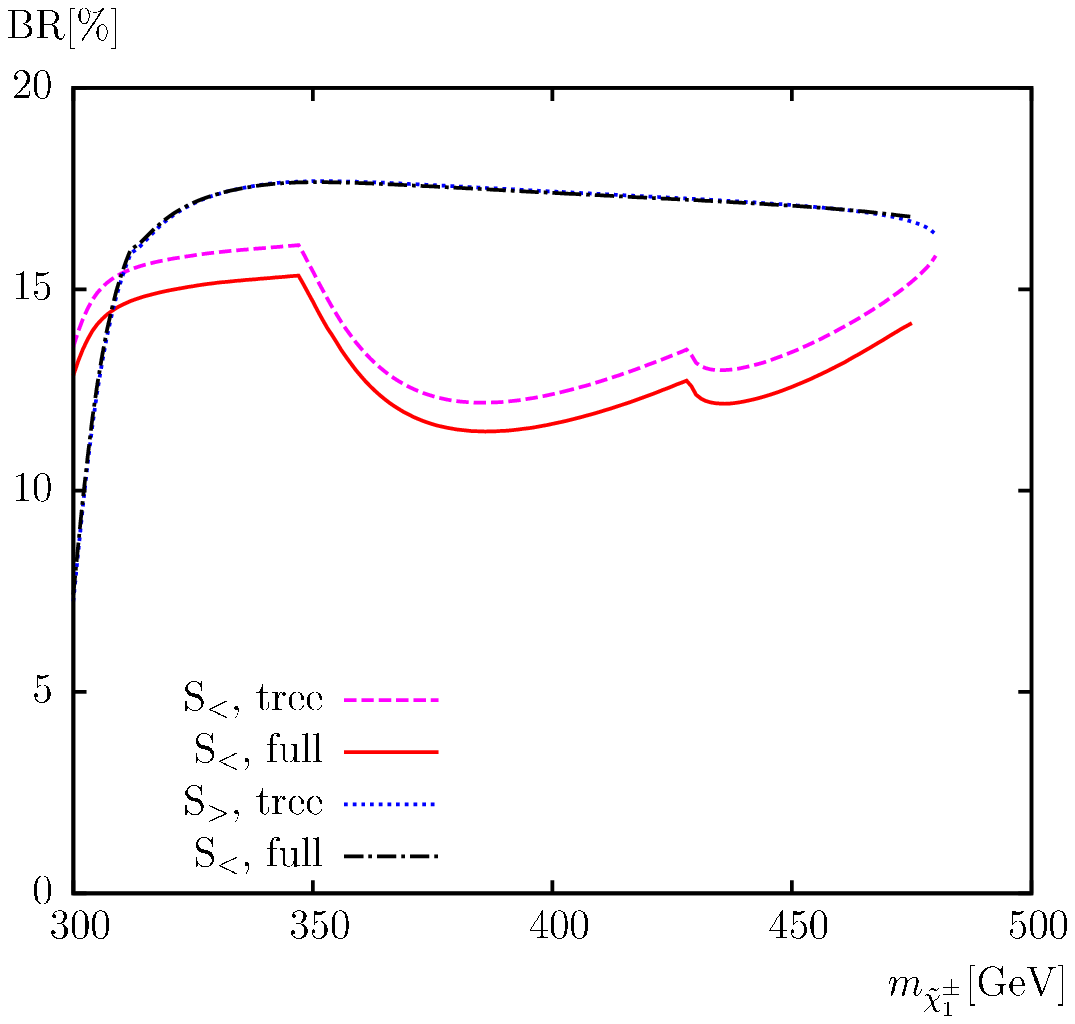}
\hspace{-4mm}
\includegraphics[width=0.49\textwidth,height=7.5cm]{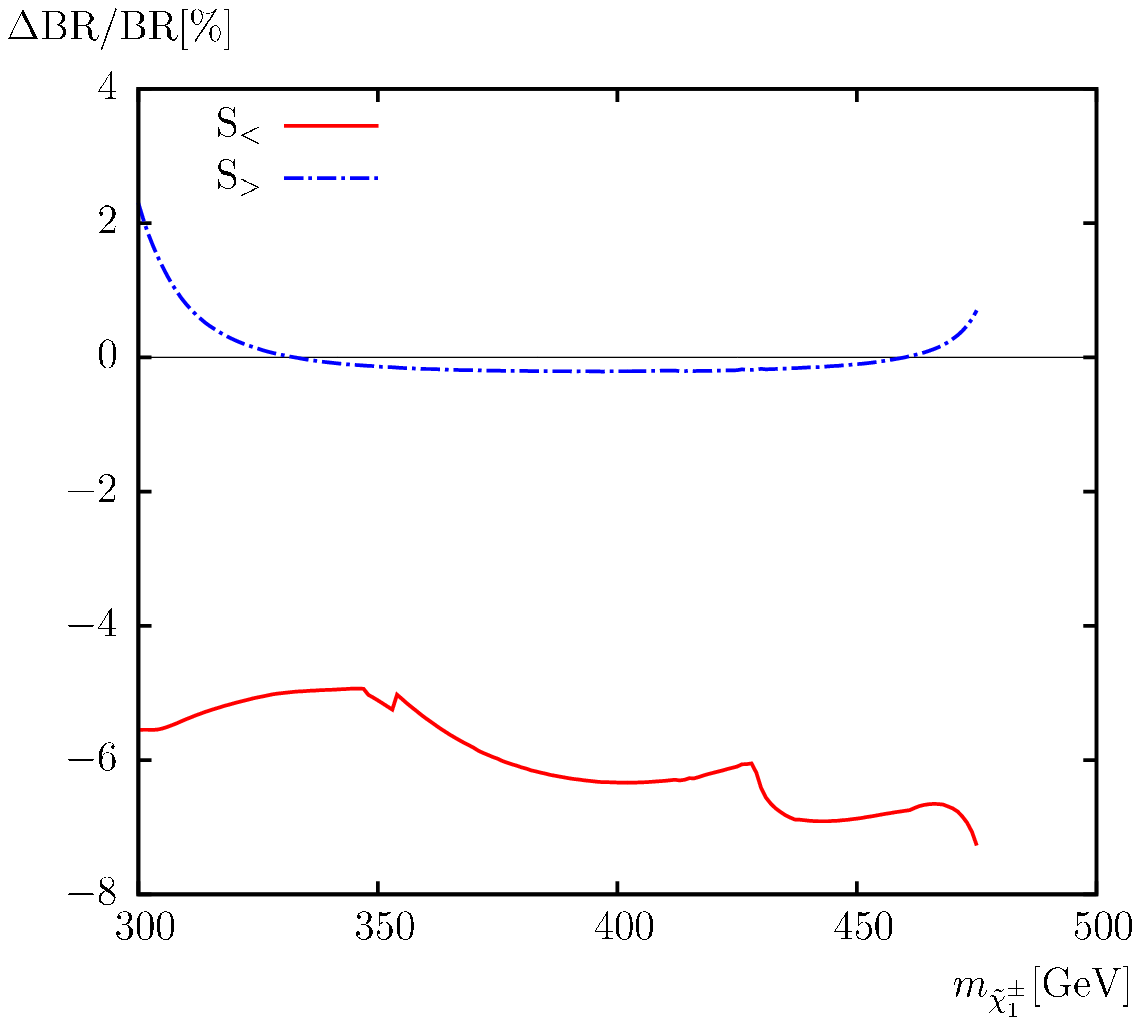}
\end{tabular}
\vspace{2em}
\caption{
  $\Ga(\DecayCmlSn{1}{\mu})$. 
  Tree-level (``tree'') and full one-loop (``full'') corrected 
  decay widths are shown with the parameters chosen according to \SN\
  (see \refta{tab:para}), with $\mcha{1}$ varied.
  The upper left plot shows the decay width, the upper right plot shows 
  the relative size of the corrections.
  The lower left plot shows the BR, the lower right plot shows 
  the relative size of the BR.
}
\label{fig:mC1.cha1snumu}
\end{center}
\end{figure}

\begin{figure}[htb!]
\begin{center}
\begin{tabular}{c}
\includegraphics[width=0.49\textwidth,height=7.5cm]{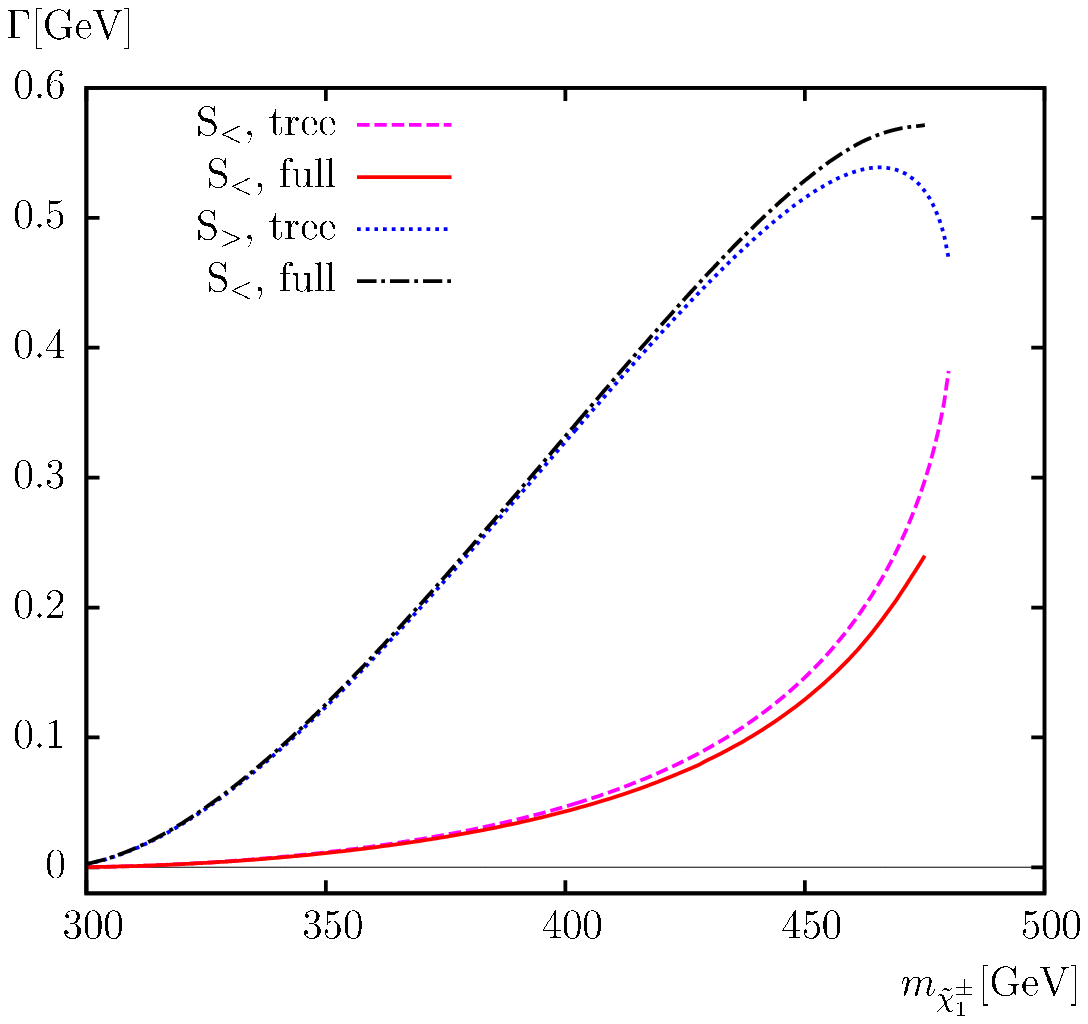}
\hspace{-4mm}
\includegraphics[width=0.49\textwidth,height=7.5cm]{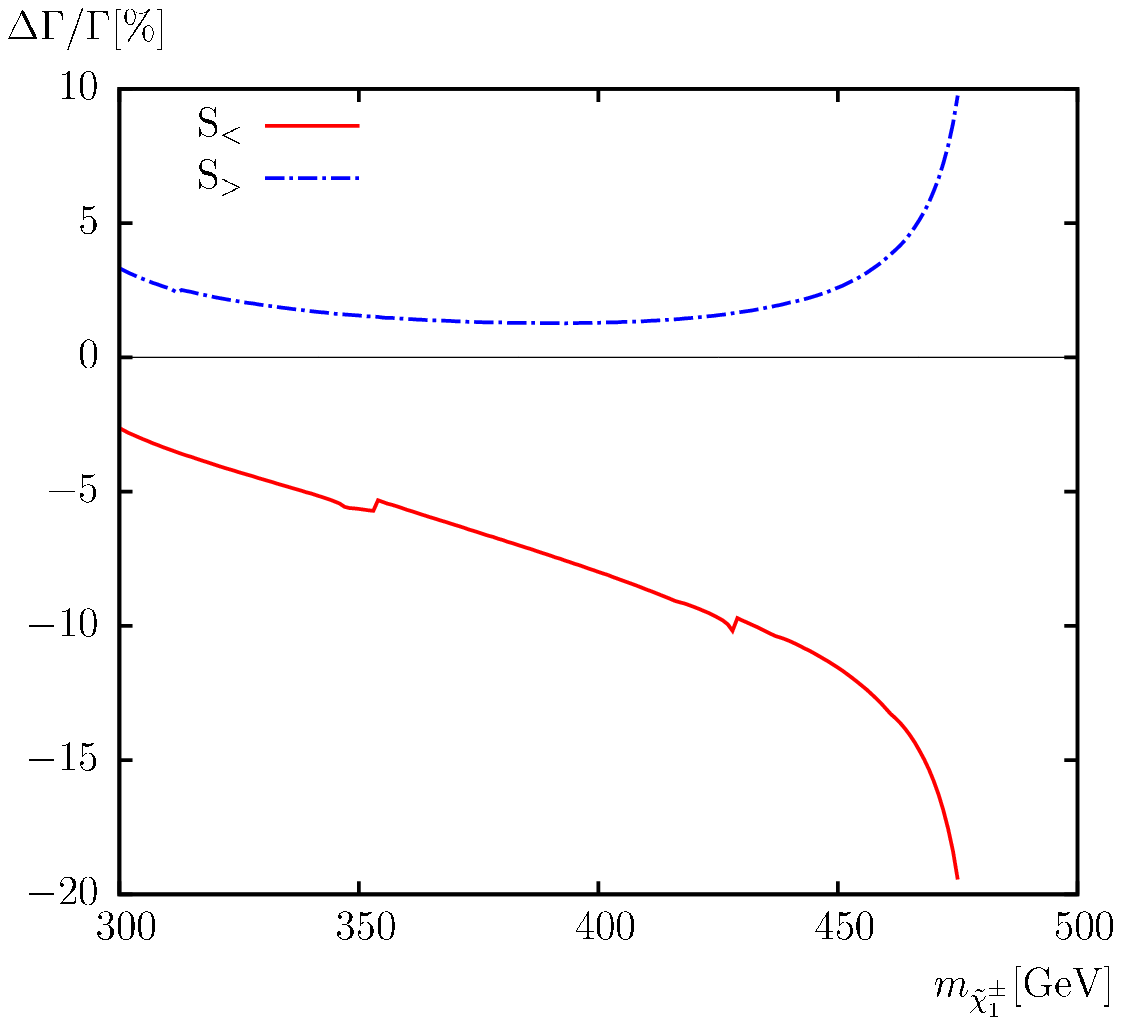} 
\\[5em]
\includegraphics[width=0.49\textwidth,height=7.5cm]{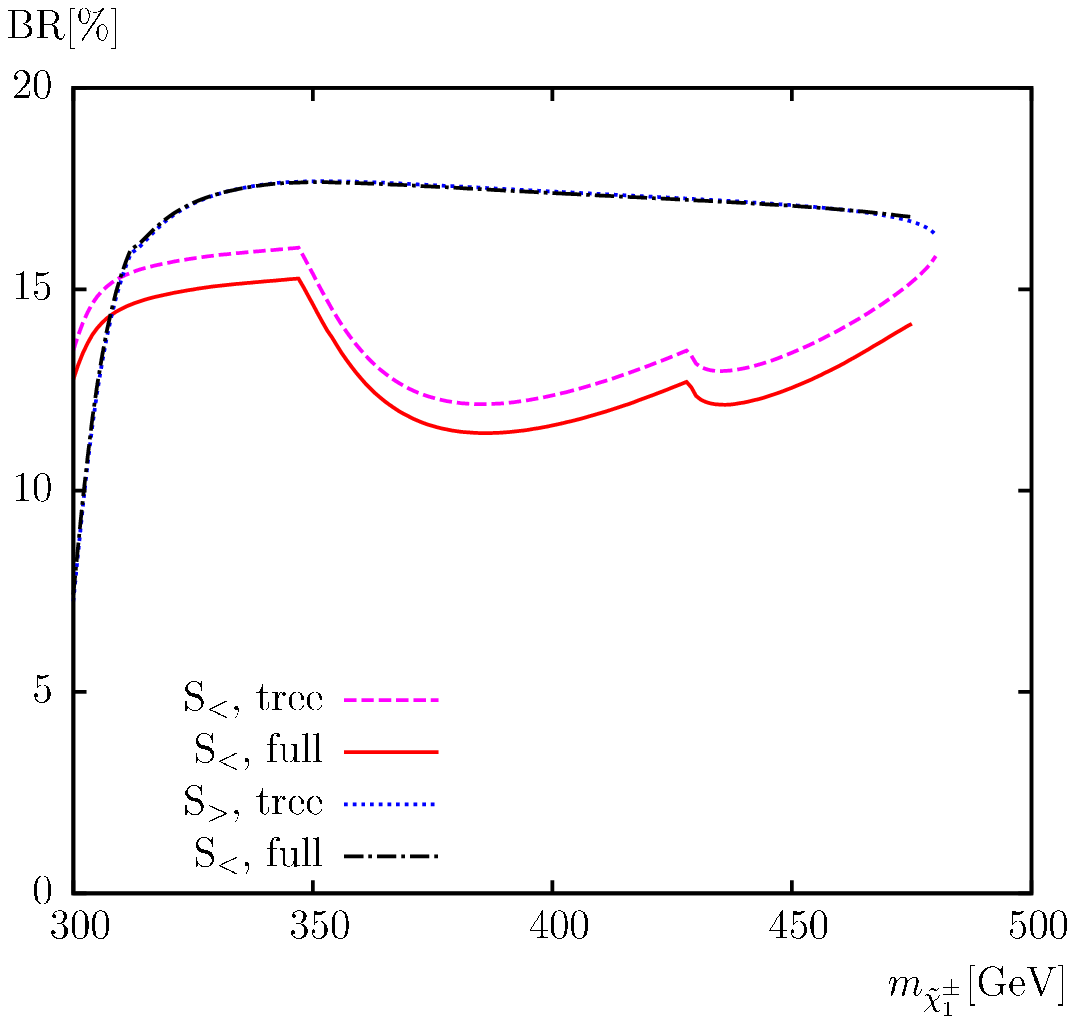}
\hspace{-4mm}
\includegraphics[width=0.49\textwidth,height=7.5cm]{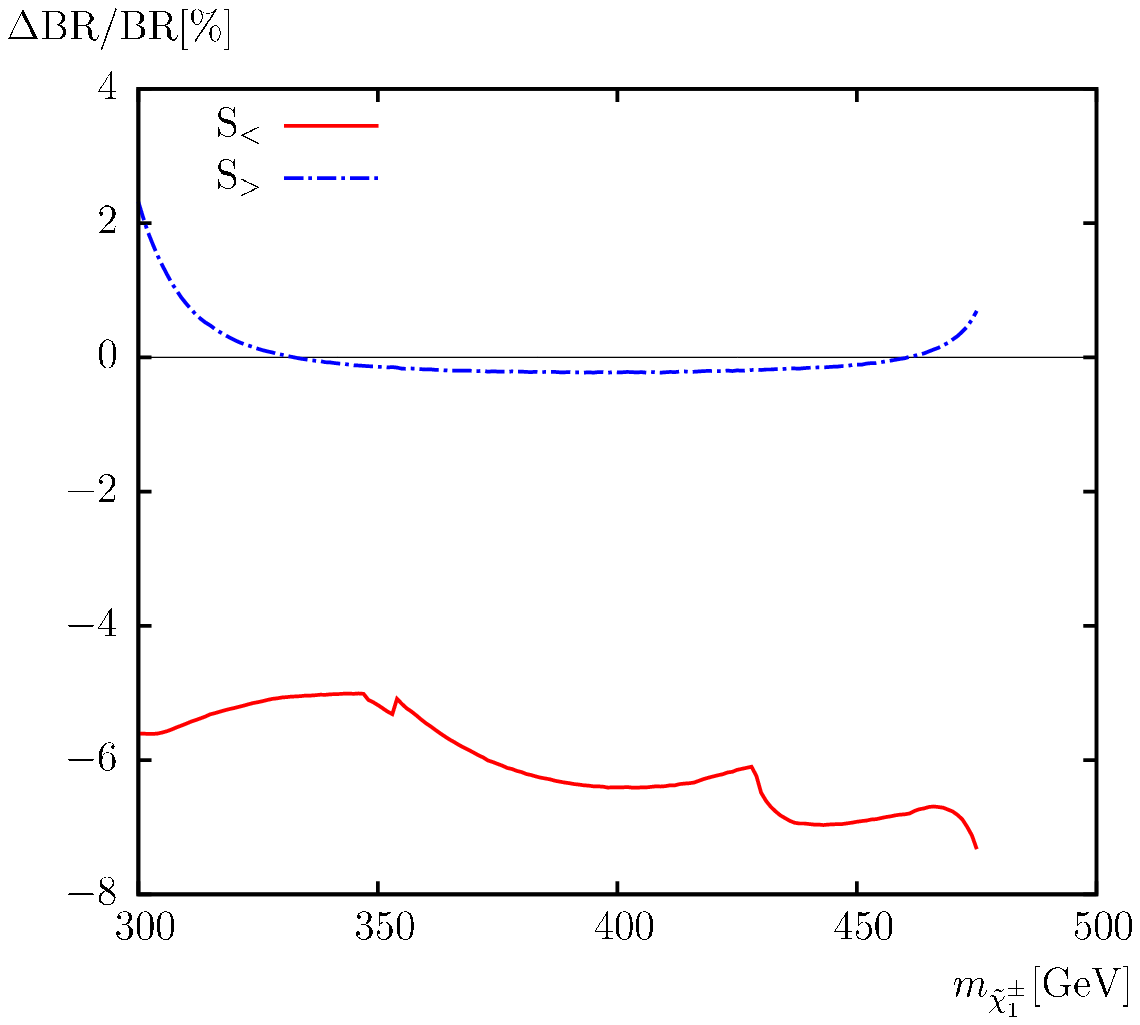}
\end{tabular}
\vspace{2em}
\caption{
  $\Ga(\DecayCmlSn{1}{e})$. 
  Tree-level (``tree'') and full one-loop (``full'') corrected 
  decay widths are shown with the parameters chosen according to \SN\
  (see \refta{tab:para}), with $\mcha{1}$ varied.
  The upper left plot shows the decay width, the upper right plot shows 
  the relative size of the corrections.
  The lower left plot shows the BR, the lower right plot shows 
  the relative size of the BR.
}
\label{fig:mC1.cha1snuel}
\end{center}
\end{figure}

Finally, the decays $\DecayCmlSn{1}{l}$ ($l = \tau, \mu, e$)
are presented in
\reffis{fig:mC1.cha1snutau} -- \ref{fig:mC1.cha1snuel}. These decays
proceed mainly with electroweak strength and thus are very similar for
the three generations, and can indeed be substantial. The dips 
are due to the $\neu{1}W^-$ and $\neu{1}\Hm$ thresholds.
The size of the decay widths reaches about $0.6 \gev$ in \SZ\ and 
$\sim 0.3 \gev$ in \SE. The $\br(\DecayCmlSn{1}{\tau})$ is nearly $18\%$
for most $\mcha{1}$ values, while it drops from $40\%$ at small
$\mcha{1}$ to about $18\%$ at large $\mcha{1}$. The size of the one-loop
effects in the latter case varies between $\sim 3\%$ and $\sim 0.5\%$. 
For $\br(\DecayCmlSn{1}{l})$ ($l = \mu, e$) 
values between $\sim 18\%$ in \SE\ and
$\sim 12\%$ in \SZ\ are found. In the latter case the one-loop 
corrections can be sizable around $-6\%$, which are relevant 
for a reliable ILC analysis.

\subsection{Full one-loop results for varying \boldmath{$\phiMe$}}
\label{sec:1LphiMe}

As in the previous sections, the results shown in this 
subsection consist of ``tree'', which denotes the tree-level 
value and of ``full'', which is the decay width including {\em all} one-loop 
corrections as described in \refse{sec:calc}.
We also show the result leaving out the contributions from absorptive 
parts of the one-loop self-energy corrections as discussed in
\refse{sec:cMSSM}, labelled as ``full R''. 
We concentrate on the dependence on $\phiMe$ for the decays with a final
neutralino. For all other decays with no external neutralinos, the
neutralinos appear only as virtual particles in the loops, resulting in a
negligible dependence on $\phiMe$.
It should be noted, however, that all decay channels must be computed to
obtain the correct branching ratios. 
The parameters are chosen according to \refta{tab:para}, and
consequently the full parameter range is accessible at the ILC(1000).\\

\begin{figure}[htb!]
\begin{center}
\begin{tabular}{c}
\includegraphics[width=0.49\textwidth,height=7.5cm]{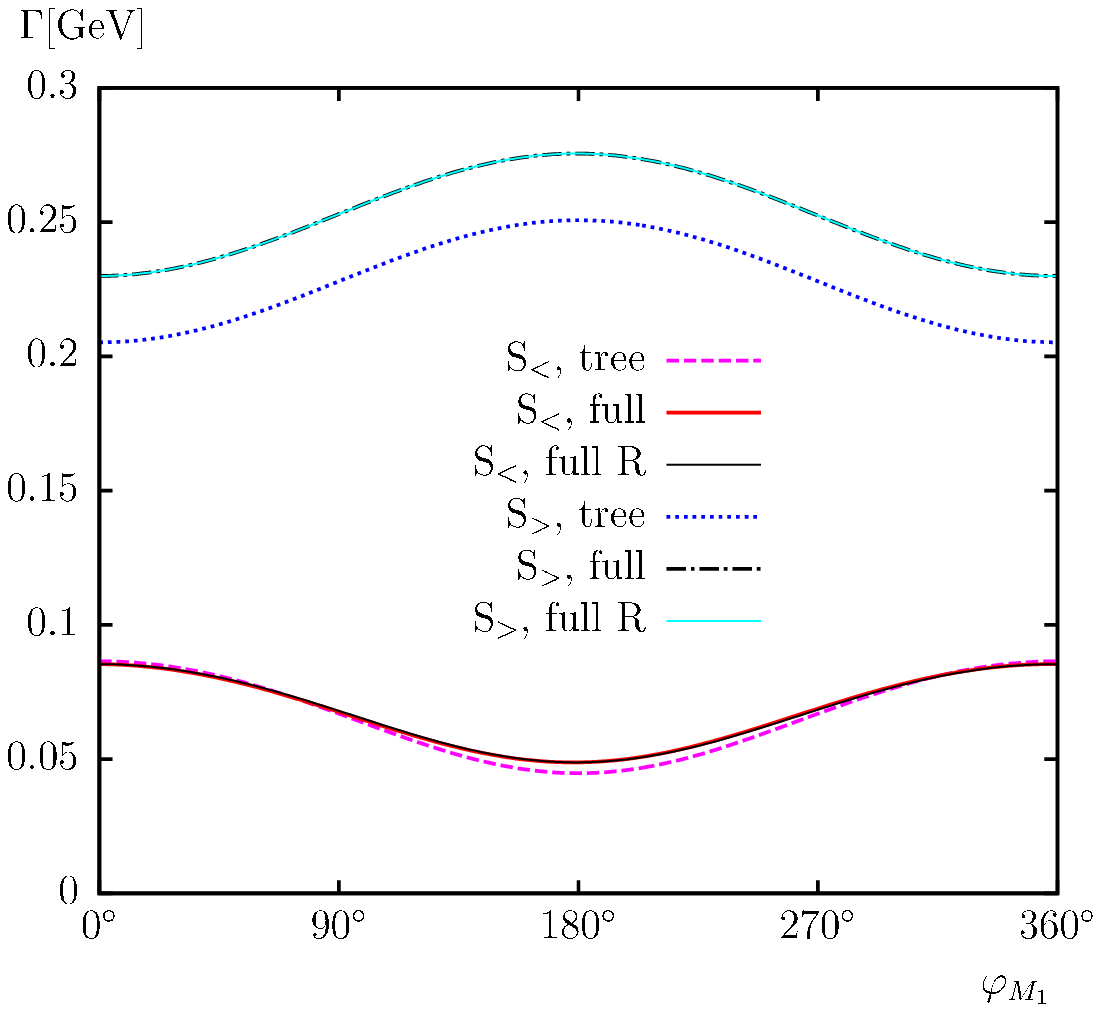}
\hspace{-4mm}
\includegraphics[width=0.49\textwidth,height=7.5cm]{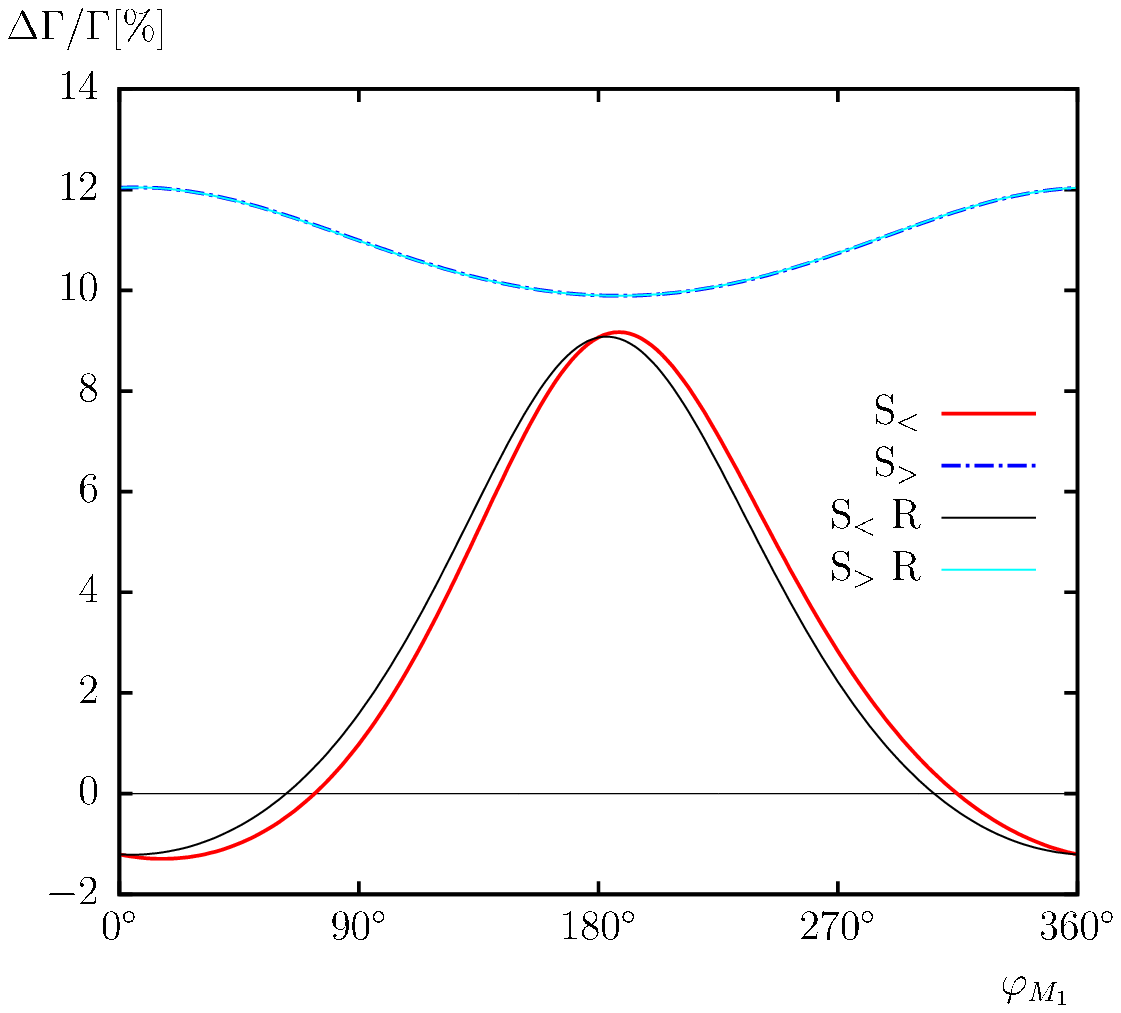} 
\\[4em]
\includegraphics[width=0.49\textwidth,height=7.5cm]{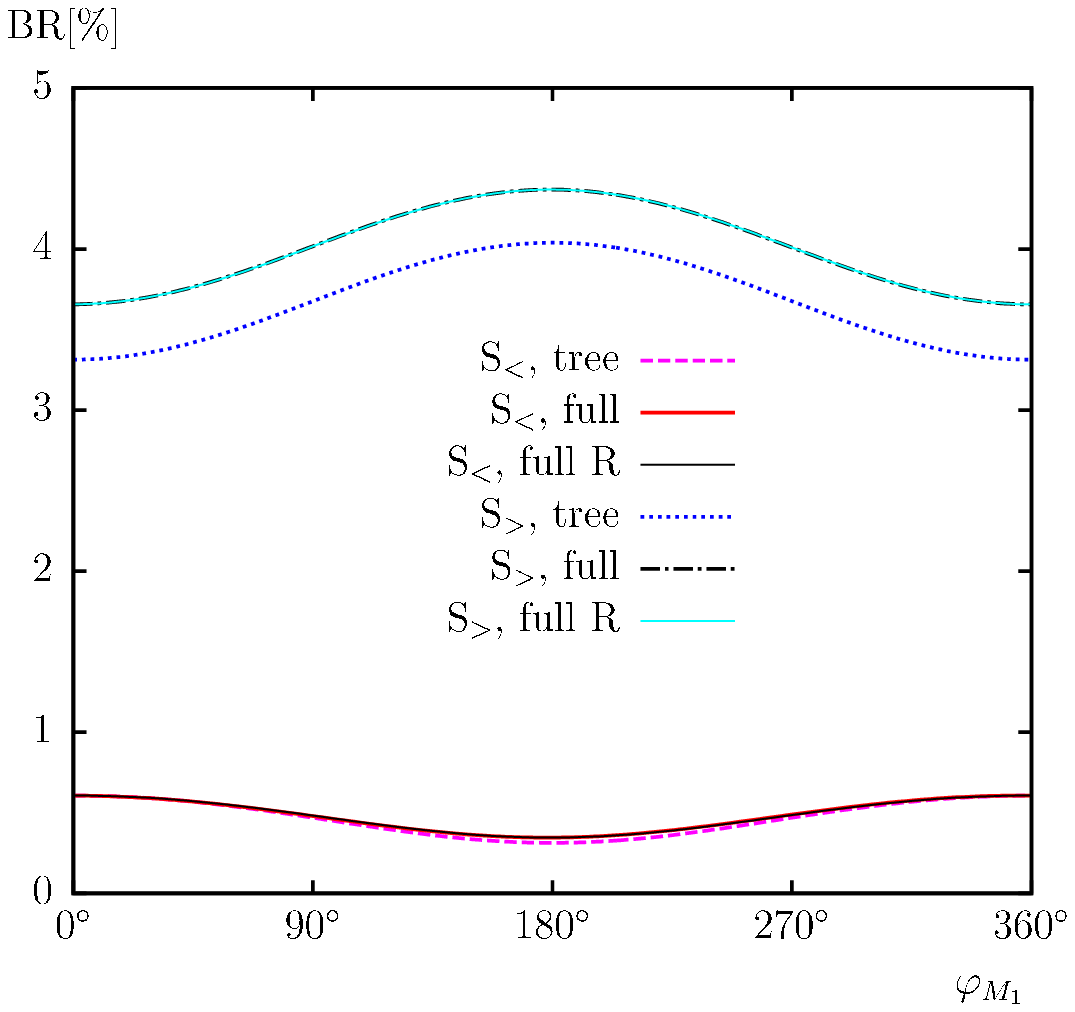}
\hspace{-4mm}
\includegraphics[width=0.49\textwidth,height=7.5cm]{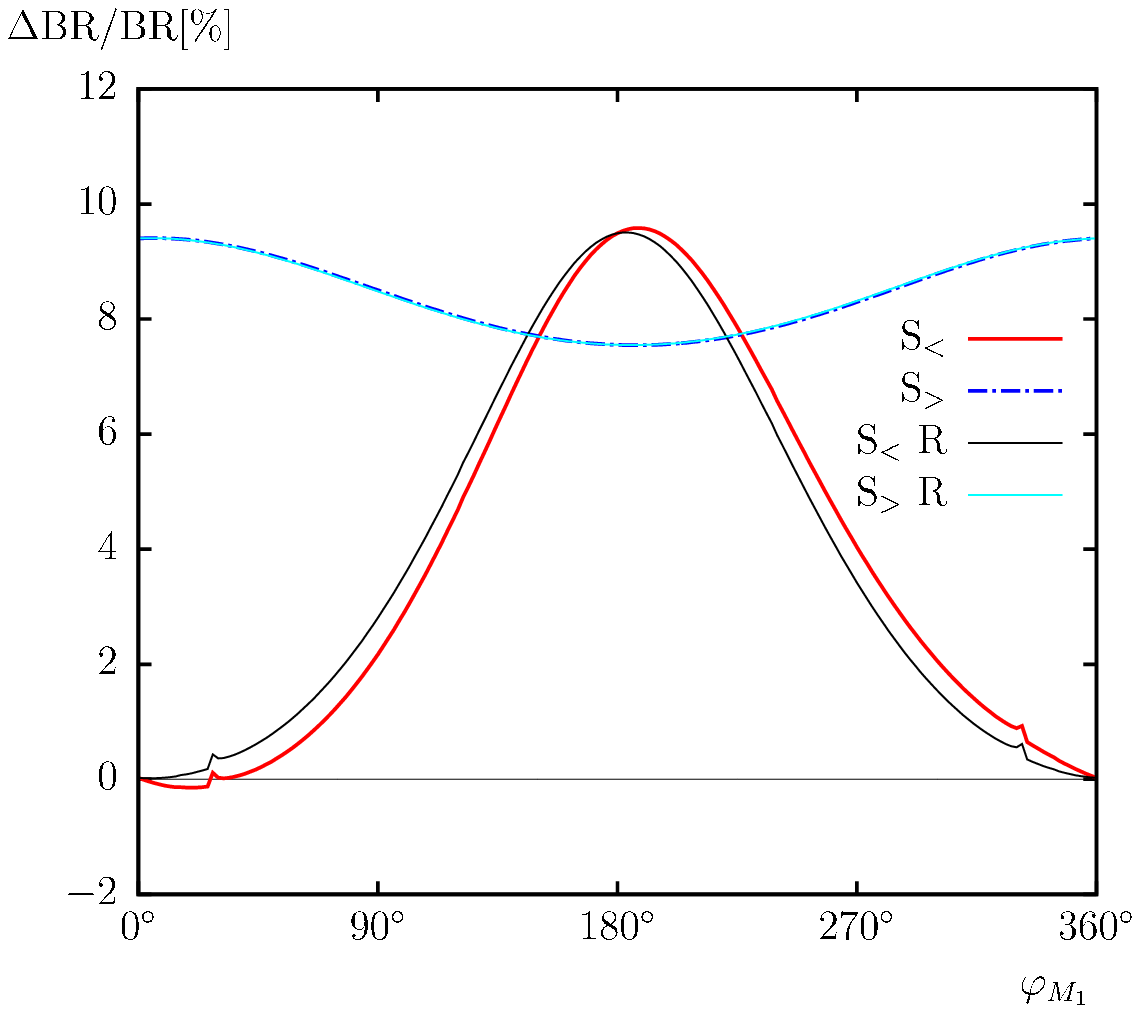}
\end{tabular}
\vspace{2em}
\caption{
  $\Ga(\DecayCmNH{2}{1})$. 
  Tree-level (``tree'') and full one-loop (``full'') corrected 
  decay widths are shown with the parameters chosen according to \SN\
  (see \refta{tab:para}), with $\phi_{\MOne}$ varied.
  Also shown are the full one-loop corrected decay widths omitting
  the absorptive contributions (``full R'').
  The upper left plot shows the decay width, the upper right plot shows 
  the relative size of the corrections.
  The lower left plot shows the BR, the lower right plot shows 
  the relative size of the BR.
}
\label{fig:PhiM1.cha2neu1hp}
\end{center}
\end{figure}

\begin{figure}[htb!]
\begin{center}
\begin{tabular}{c}
\includegraphics[width=0.49\textwidth,height=7.5cm]{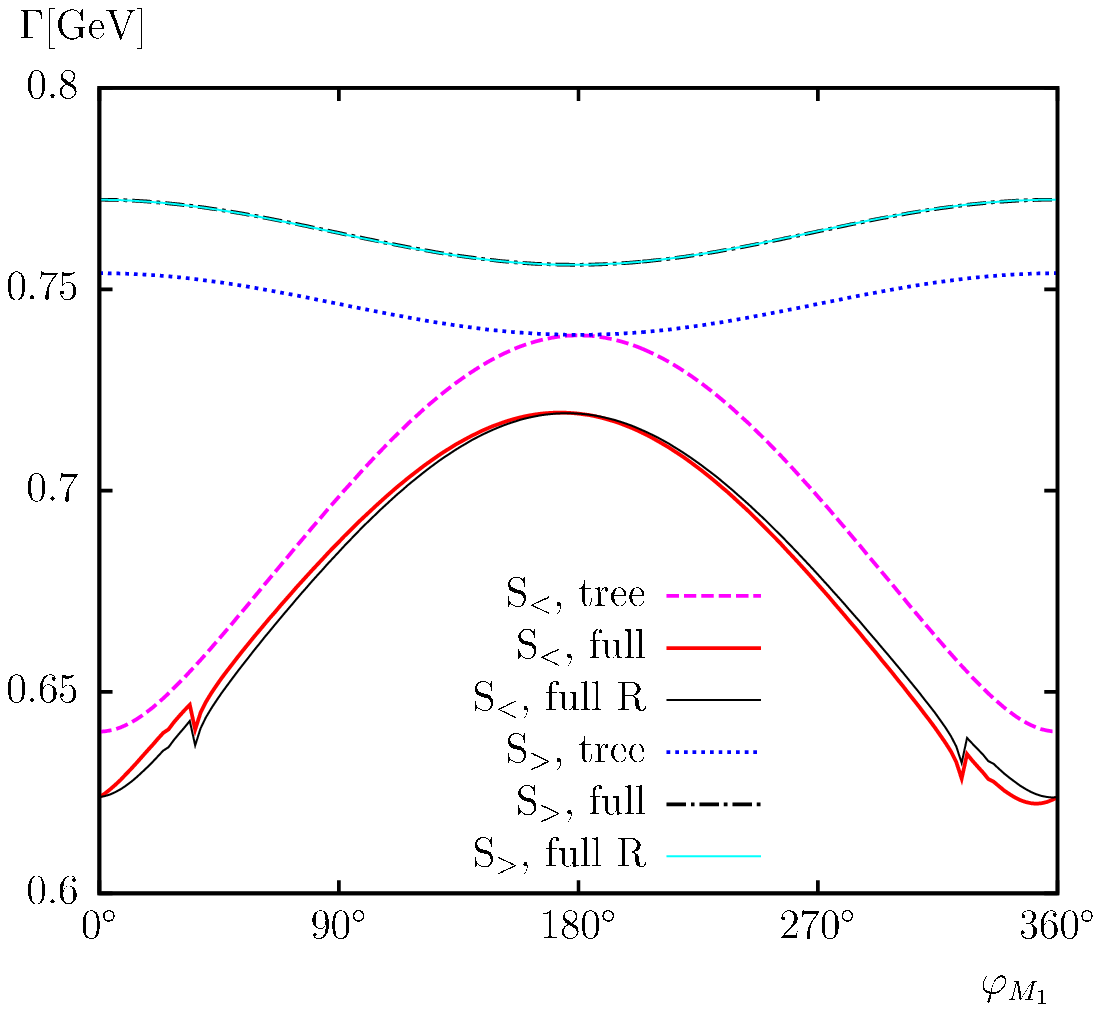}
\hspace{-4mm}
\includegraphics[width=0.49\textwidth,height=7.5cm]{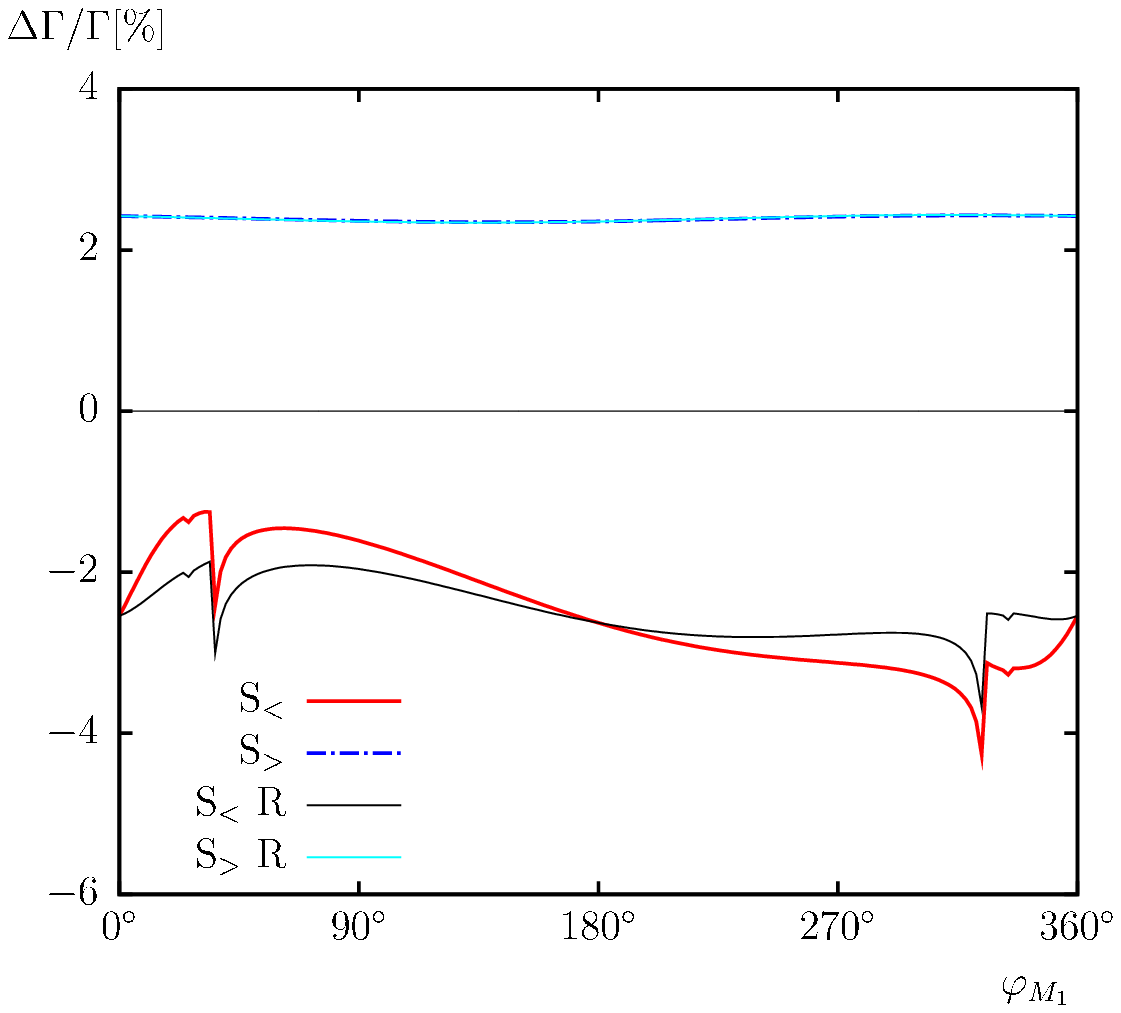}
\\[4em]
\includegraphics[width=0.49\textwidth,height=7.5cm]{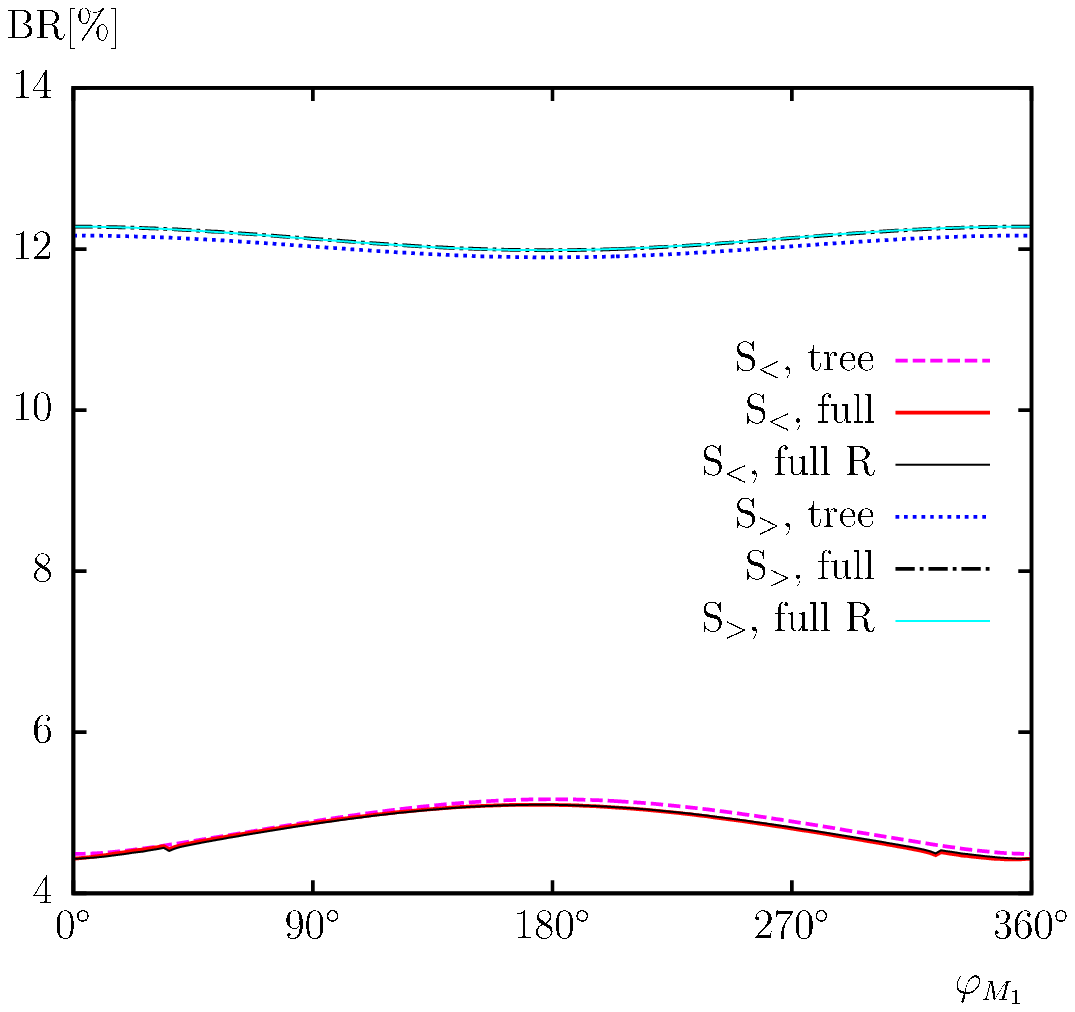}
\hspace{-4mm}
\includegraphics[width=0.49\textwidth,height=7.5cm]{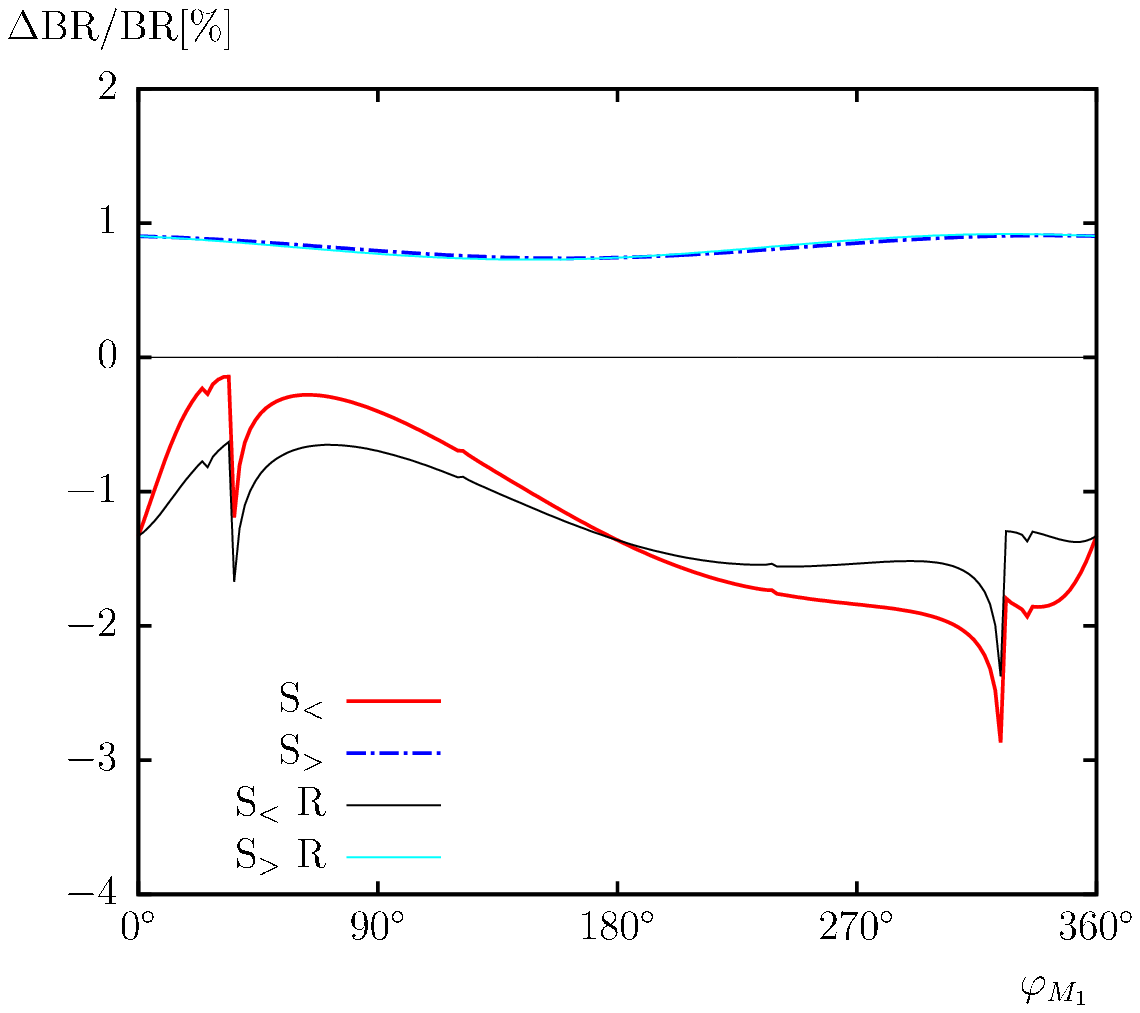}
\end{tabular}
\vspace{2em}
\caption{
  $\Ga(\DecayCmNH{2}{2})$. 
  Tree-level (``tree'') and full one-loop (``full'') corrected 
  decay widths are shown with the parameters chosen according to \SN\
  (see \refta{tab:para}), with $\phi_{\MOne}$ varied.
  Also shown are the full one-loop corrected decay widths omitting
  the absorptive contributions (``full R'').
  The upper left plot shows the decay width, the upper right plot shows 
  the relative size of the corrections.
  The lower left plot shows the BR, the lower right plot shows 
  the relative size of the BR.
}
\label{fig:PhiM1.cha2neu2hp}
\end{center}
\end{figure}

\begin{figure}[htb!]
\begin{center}
\begin{tabular}{c}
\includegraphics[width=0.49\textwidth,height=7.5cm]{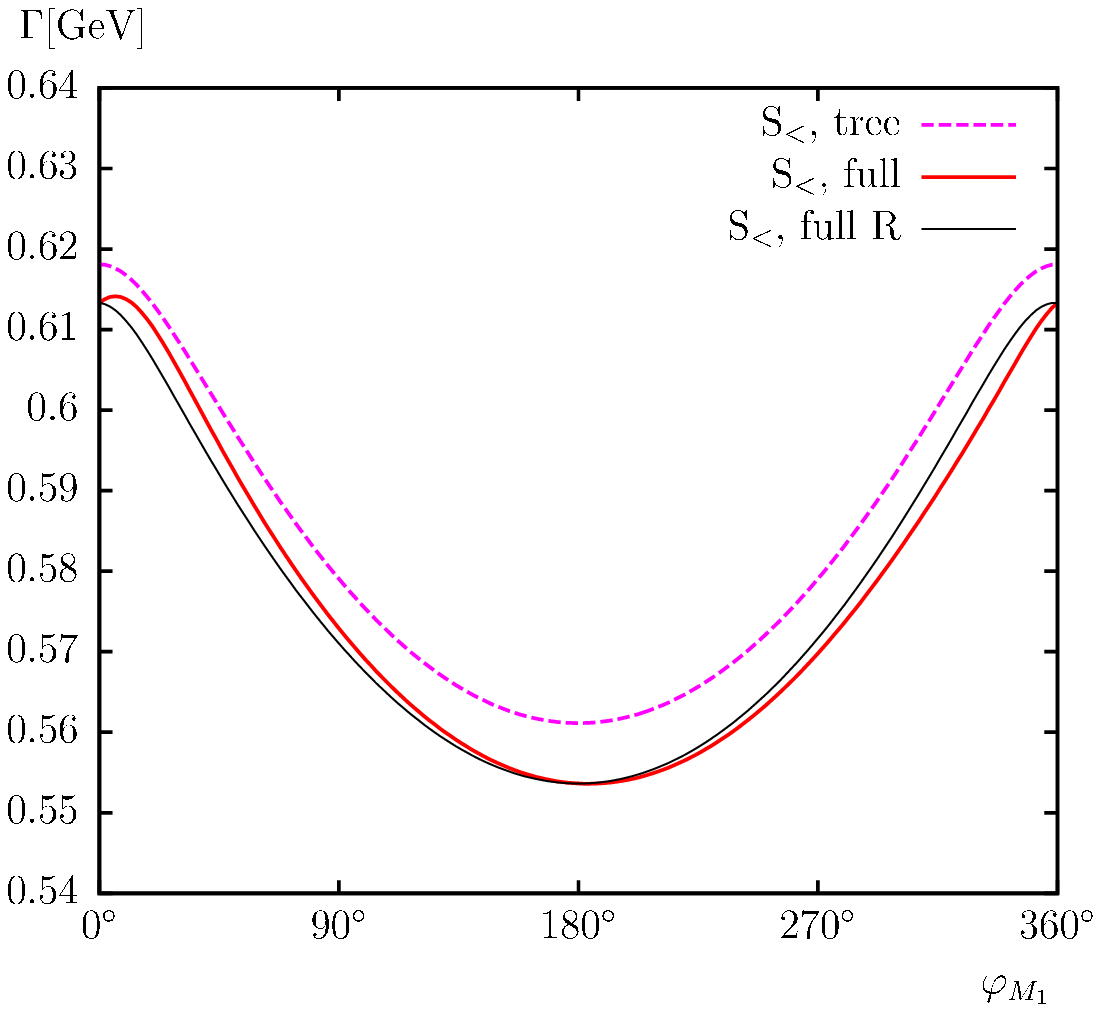}
\hspace{-4mm}
\includegraphics[width=0.49\textwidth,height=7.5cm]{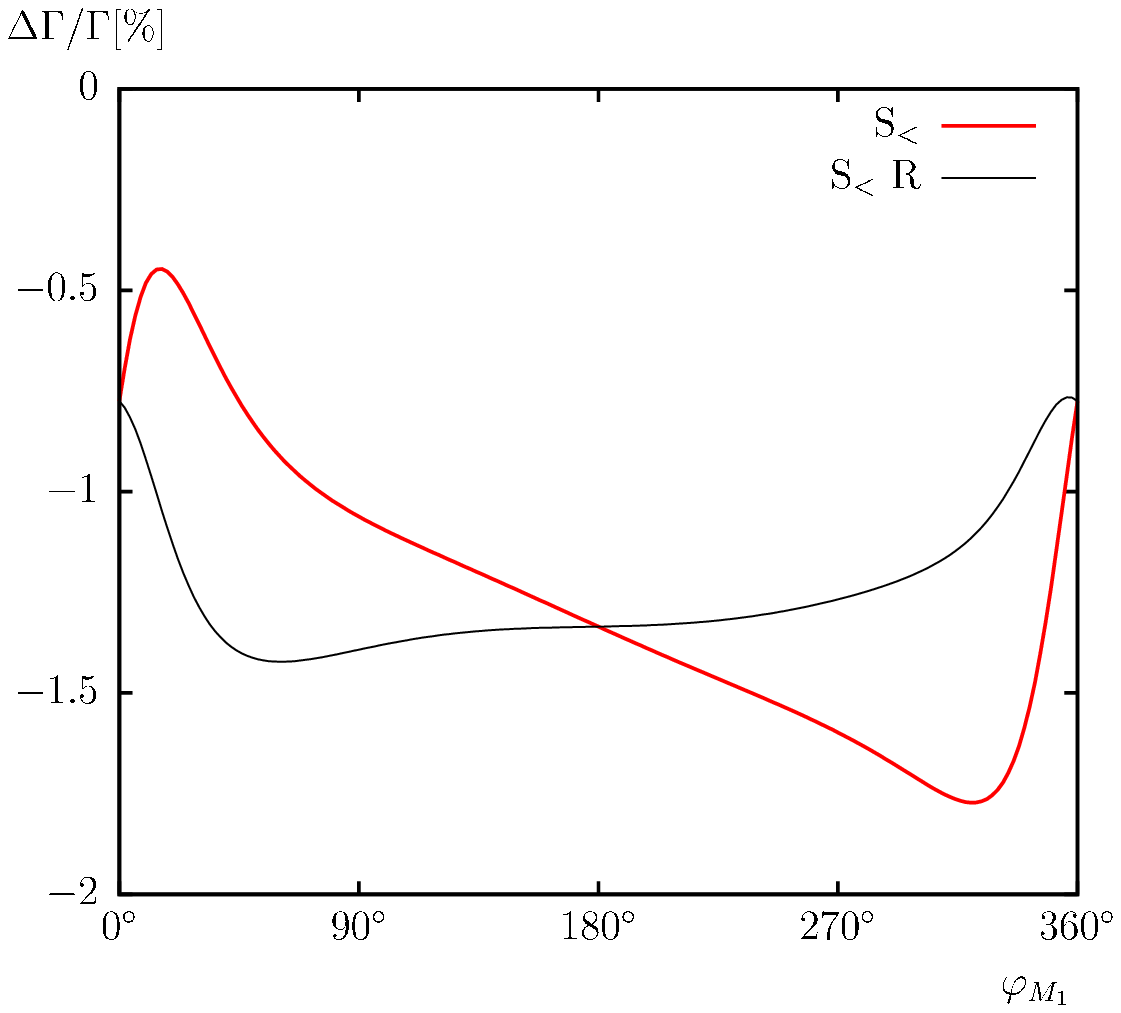}
\\[4em]
\includegraphics[width=0.49\textwidth,height=7.5cm]{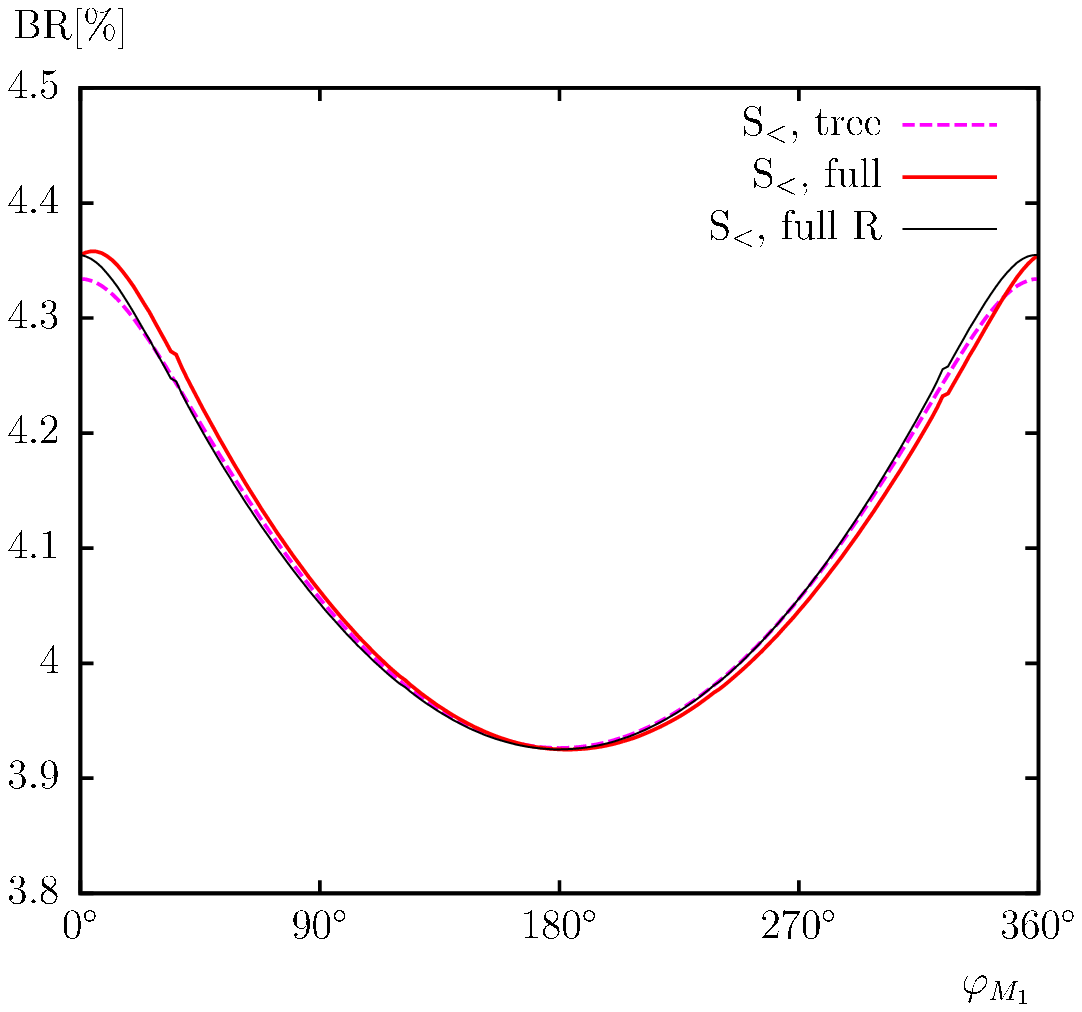}
\hspace{-4mm}
\includegraphics[width=0.49\textwidth,height=7.5cm]{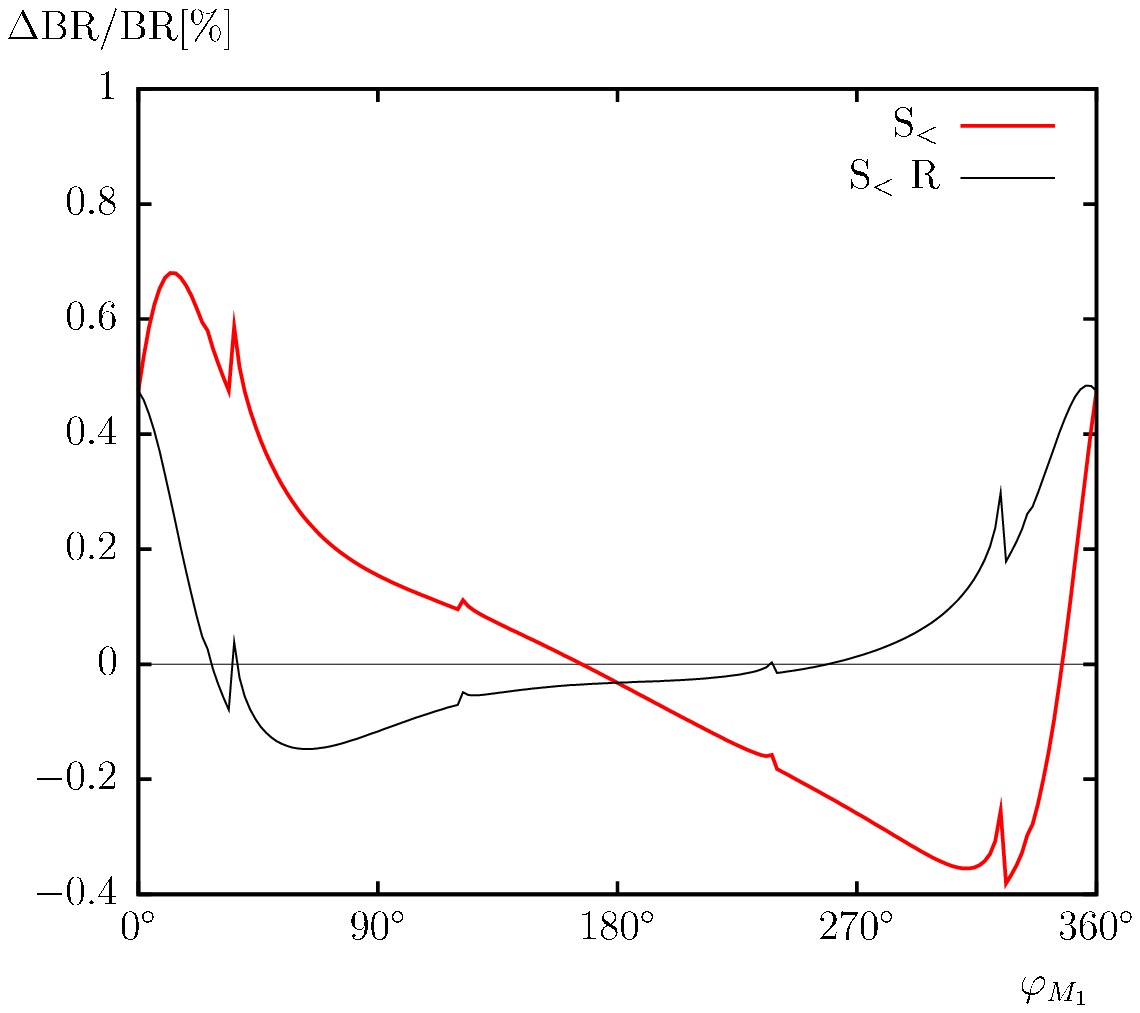}
\end{tabular}
\vspace{2em}
\caption{
  $\Ga(\DecayCmNH{2}{3})$. 
  Tree-level (``tree'') and full one-loop (``full'') corrected 
  decay widths are shown with the parameters chosen according to \SN\
  (see \refta{tab:para}), with $\phi_{\MOne}$ varied.
  Also shown are the full one-loop corrected decay widths omitting
  the absorptive contributions (``full R'').
  The upper left plot shows the decay width, the upper right plot shows 
  the relative size of the corrections.
  The lower left plot shows the BR, the lower right plot shows 
  the relative size of the BR.
}
\label{fig:PhiM1.cha2neu3hp}
\end{center}
\end{figure}

In \reffis{fig:PhiM1.cha2neu1hp} -- \ref{fig:PhiM1.cha2neu3hp}
we present the results for the decays involving the charged Higgs boson, 
$\DecayCmNH{2}{1,2,3}$. The decay to the lightest neutralino 
(see \reffi{fig:PhiM1.cha2neu1hp}) reaches 
decay widths around $0.25\ (0.07) \gev$ in \SE\ (\SZ), where the
dependence on $\phiMe$ in the latter is substantial, varying between
$-1\%$ and $+9\%$. 
The inclusion of the absorptive self-energy parts, on
the other hand, yields only a small effect for the parameter chosen
here. The $\br(\DecayCmNH{2}{1})$ stays below $1\%$ in \SZ\ 
and reaches around $4\%$ in \SE. Here also the
one-loop effects on the BR are substantial around $+9\%$. 
Consequently, an analysis of $\phiMe$ at the ILC(1000) requires the
inclusion of the full one-loop corrections.

The case of $\DecayCmNH{2}{2}$ is shown in \reffi{fig:PhiM1.cha2neu2hp}. 
The dips, best visible again in the upper right panel are due to 
the $\DecayNNZ{2}{1}$ threshold, at $\phiMe=15^\circ$ 
and the $\DecayNNh{2}{1}{1}$ threshold, at $\phiMe=28^\circ$. 
Due to ${\cal CPT}$-invariance the masses are invariant under
$\phiMe\to -\phiMe$ and  
mirrored dips are observed at 
$\phiMe=332^\circ$ and $\phiMe=345^\circ$.
Notice that at tree-level the lightest Higgs boson $\He$ and the $Z$~boson 
are typically almost degenerate.
The widths reach values around $\sim 0.7 \gev$ in both scenarios,
leading to branching rations at the level of $12\%$ in \SE\ and $4.5\%$
in \SZ\ with a small variation due to $\phiMe$. Again the effects of the
absorptive self-energy contributions are small.
The relative effect of the one-loop corrections is also small at the
level of $\pm 1\%$, roughly at the level of the anticipated
ILC(1000) precision.

Since the decay $\DecayCmNH{2}{3}$ (\reffi{fig:PhiM1.cha2neu3hp}) 
is kinematically forbidden in \SE\ we
show it only for \SZ. We find $\br(\DecayCmNH{2}{3}) \sim 4\%$, again
with a small variation with $\phiMe$. The effects of the absorptive
self-energy contributions is visible at the level of $\sim 1\%$ in the
corrections to $\Ga(\DecayCmNH{2}{3})$, whereas the one-loop effects on
the BR is below $\pm 1\%$. 

\begin{figure}[htb!]
\begin{center}
\begin{tabular}{c}
\includegraphics[width=0.49\textwidth,height=7.5cm]{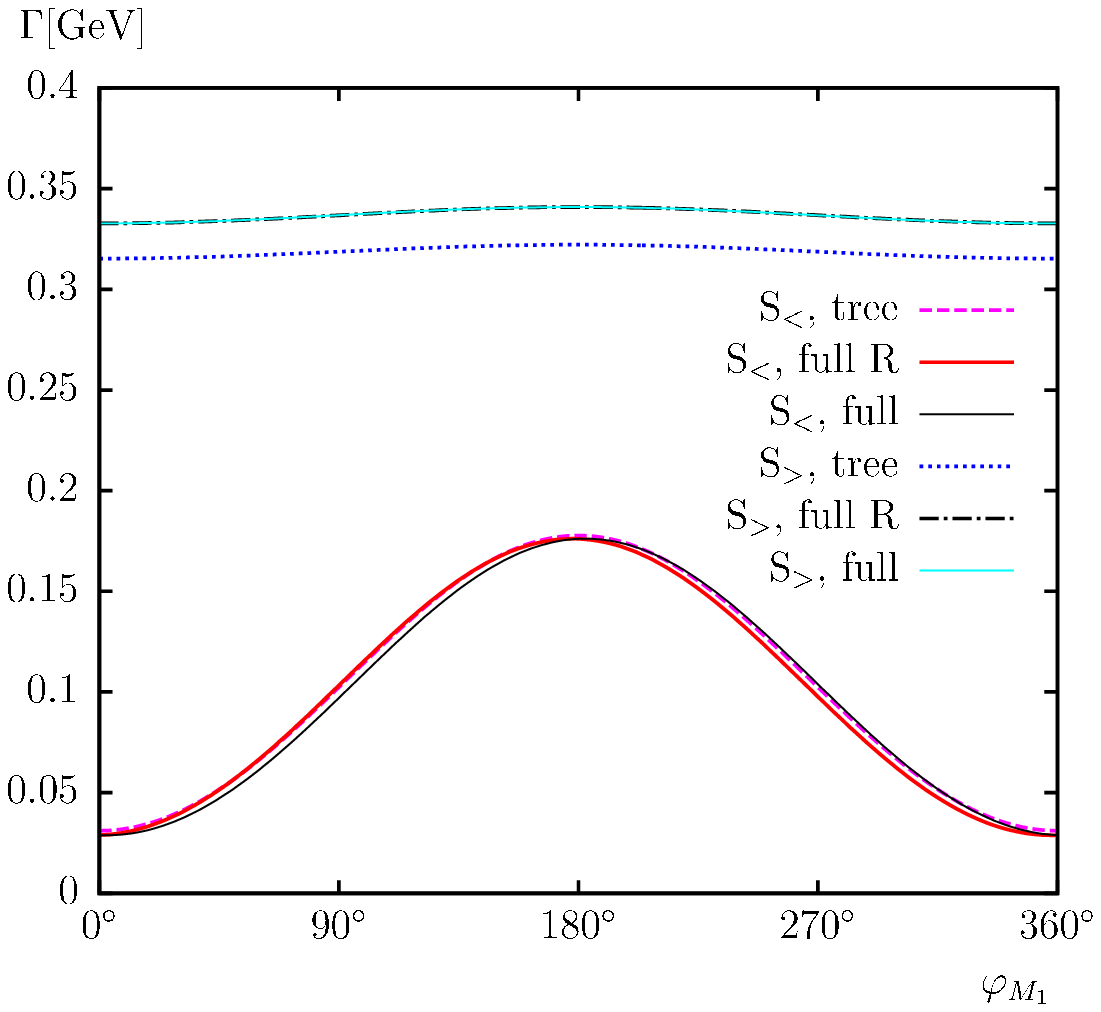}
\hspace{-4mm}
\includegraphics[width=0.49\textwidth,height=7.5cm]{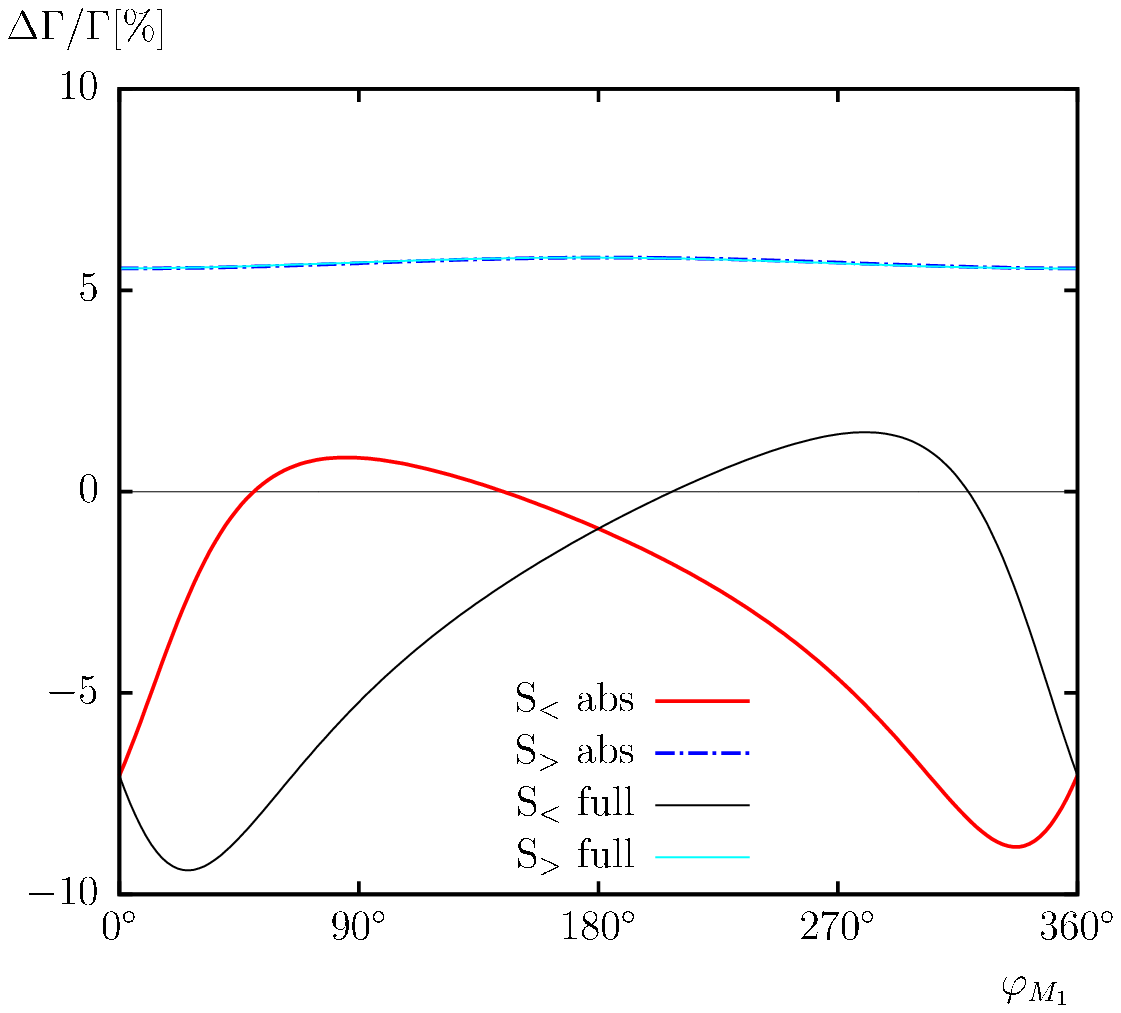} 
\\[4em]
\includegraphics[width=0.49\textwidth,height=7.5cm]{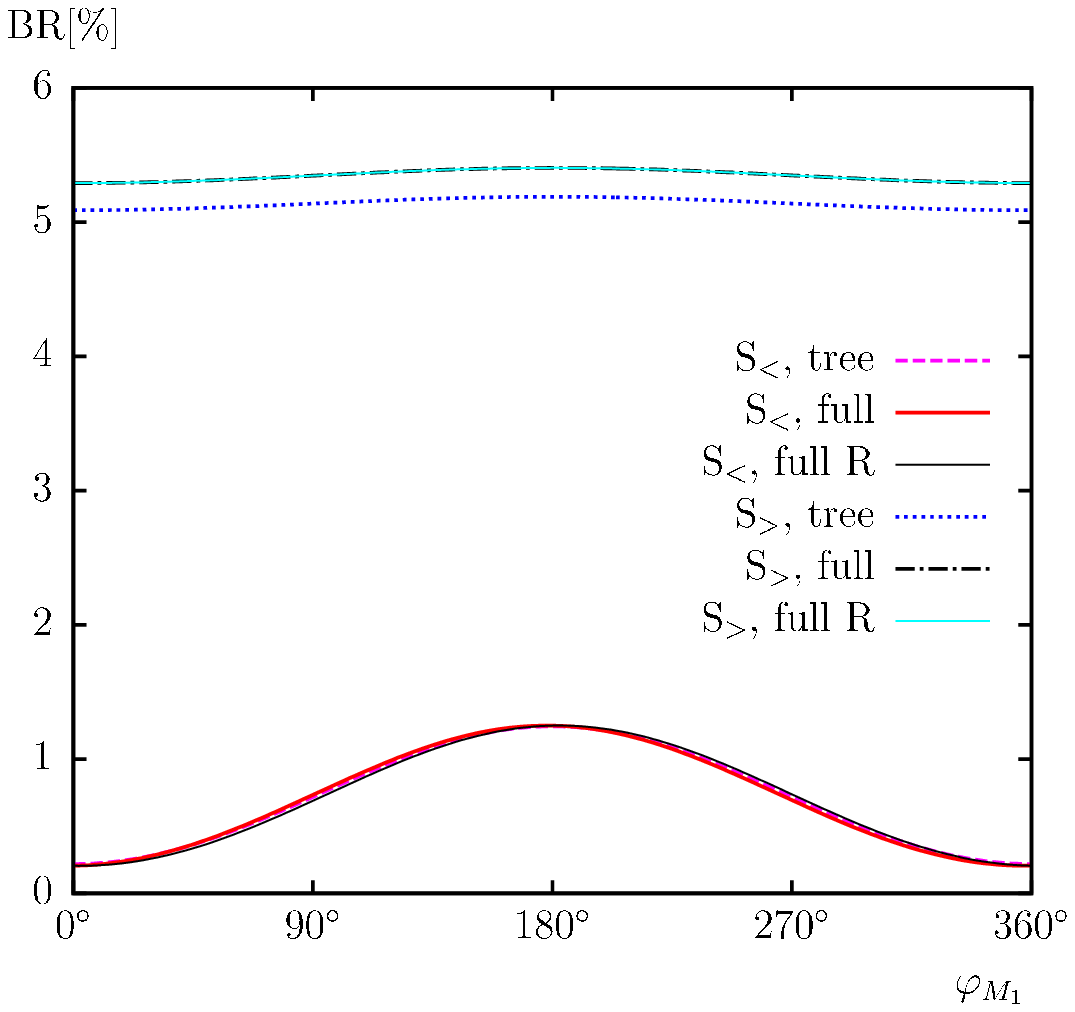}
\hspace{-4mm}
\includegraphics[width=0.49\textwidth,height=7.5cm]{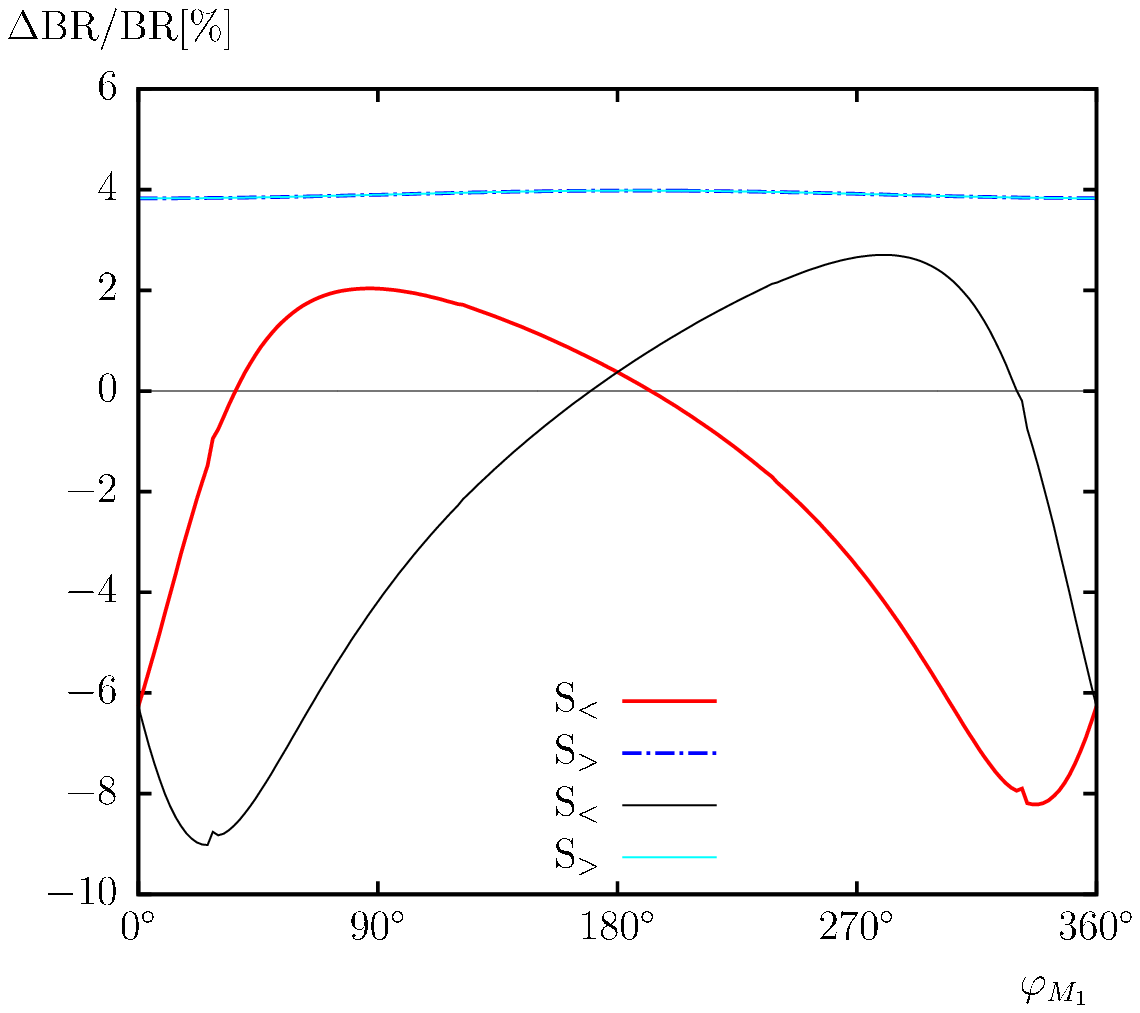}
\end{tabular}
\vspace{2em}
\caption{
  $\Ga(\DecayCmNW{2}{1})$. 
  Tree-level (``tree'') and full one-loop (``full'') corrected 
  decay widths are shown with the parameters chosen according to \SN\
  (see \refta{tab:para}), with $\phi_{\MOne}$ varied.
  Also shown are the full one-loop corrected decay widths omitting
  the absorptive contributions (``full R'').
  The upper left plot shows the decay width, the upper right plot shows 
  the relative size of the corrections.
  The lower left plot shows the BR, the lower right plot shows 
  the relative size of the BR.
}
\label{fig:PhiM1.cha2neu1w}
\end{center}
\end{figure}

\begin{figure}[htb!]
\begin{center}
\begin{tabular}{c}
\includegraphics[width=0.49\textwidth,height=7.5cm]{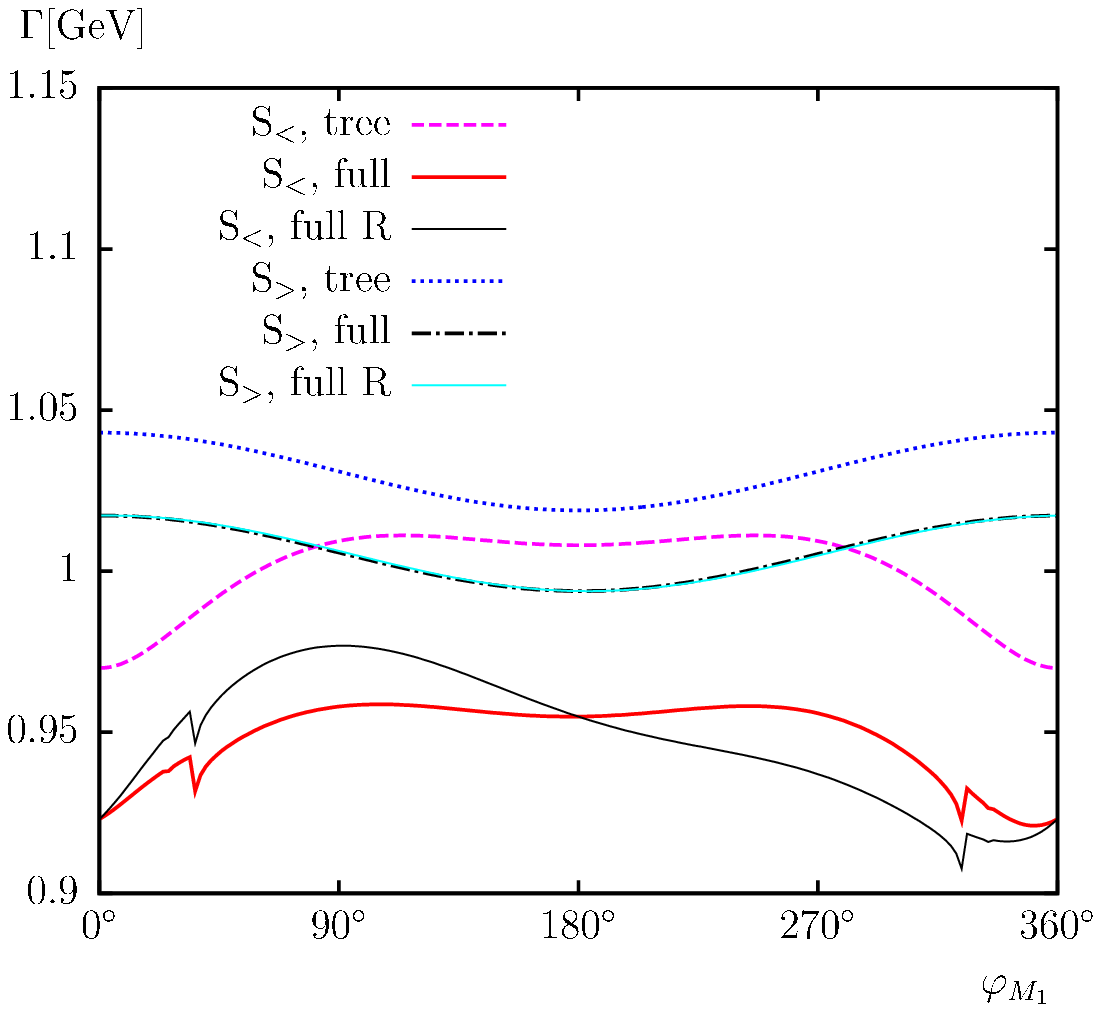}
\hspace{-4mm}
\includegraphics[width=0.49\textwidth,height=7.5cm]{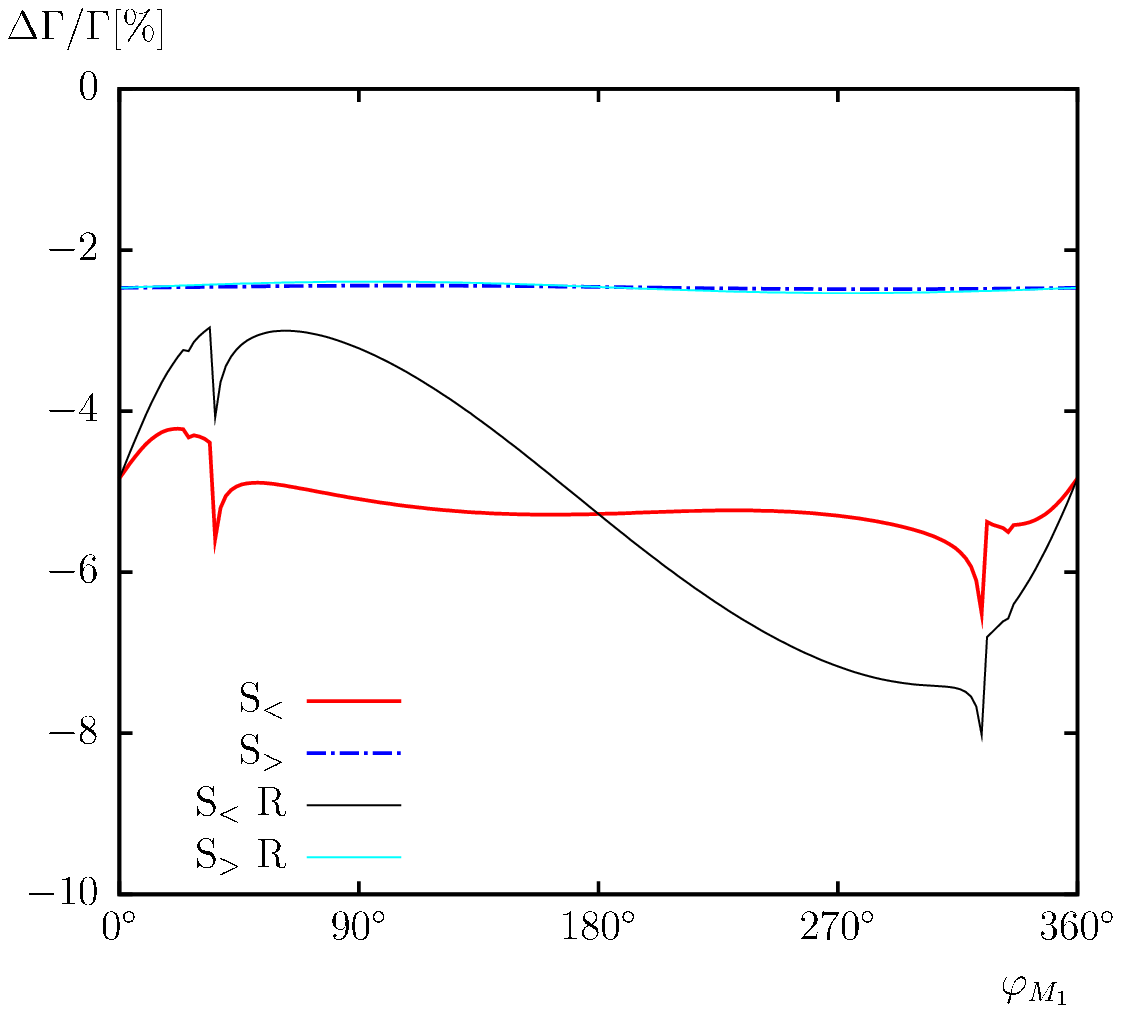} 
\\[4em]
\includegraphics[width=0.49\textwidth,height=7.5cm]{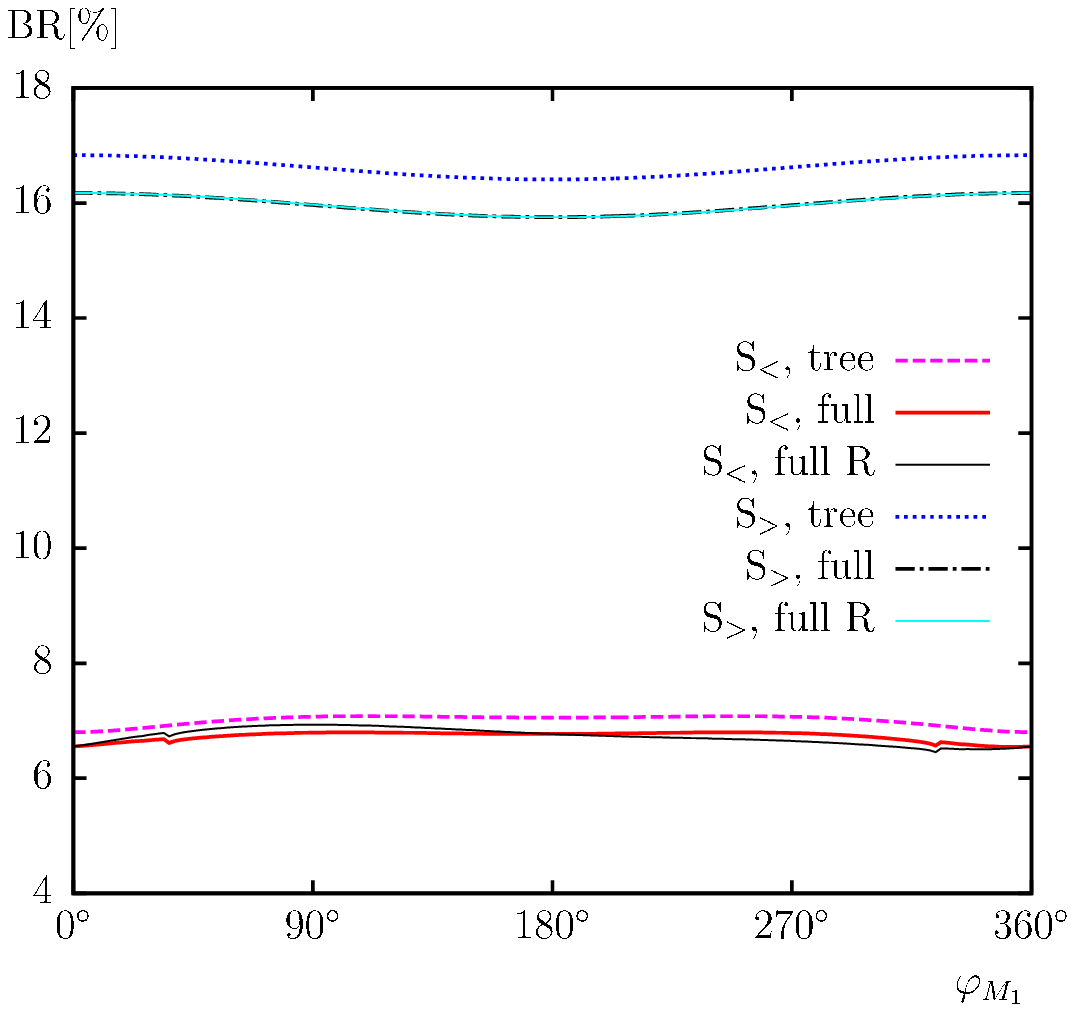}
\hspace{-4mm}
\includegraphics[width=0.49\textwidth,height=7.5cm]{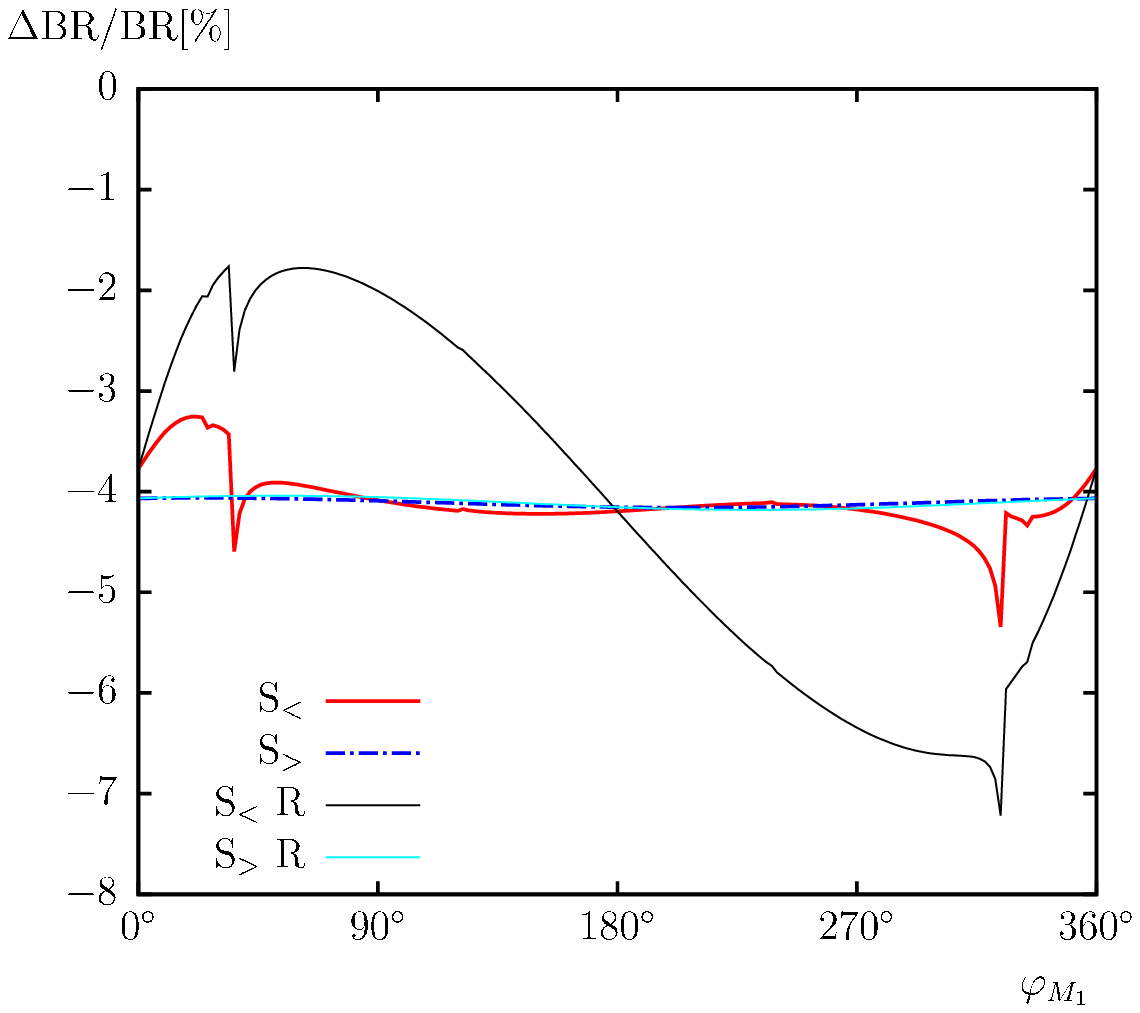}
\end{tabular}
\vspace{2em}
\caption{
  $\Ga(\DecayCmNW{2}{2})$. 
  Tree-level (``tree'') and full one-loop (``full'') corrected 
  decay widths are shown with the parameters chosen according to \SN\
  (see \refta{tab:para}), with $\phi_{\MOne}$ varied.
  Also shown are the full one-loop corrected decay widths omitting
  the absorptive contributions (``full R'').
  The upper left plot shows the decay width, the upper right plot shows 
  the relative size of the corrections.
  The lower left plot shows the BR, the lower right plot shows 
  the relative size of the BR.
}
\label{fig:PhiM1.cha2neu2w}
\end{center}
\end{figure}

\begin{figure}[htb!]
\begin{center}
\begin{tabular}{c}
\includegraphics[width=0.49\textwidth,height=7.5cm]{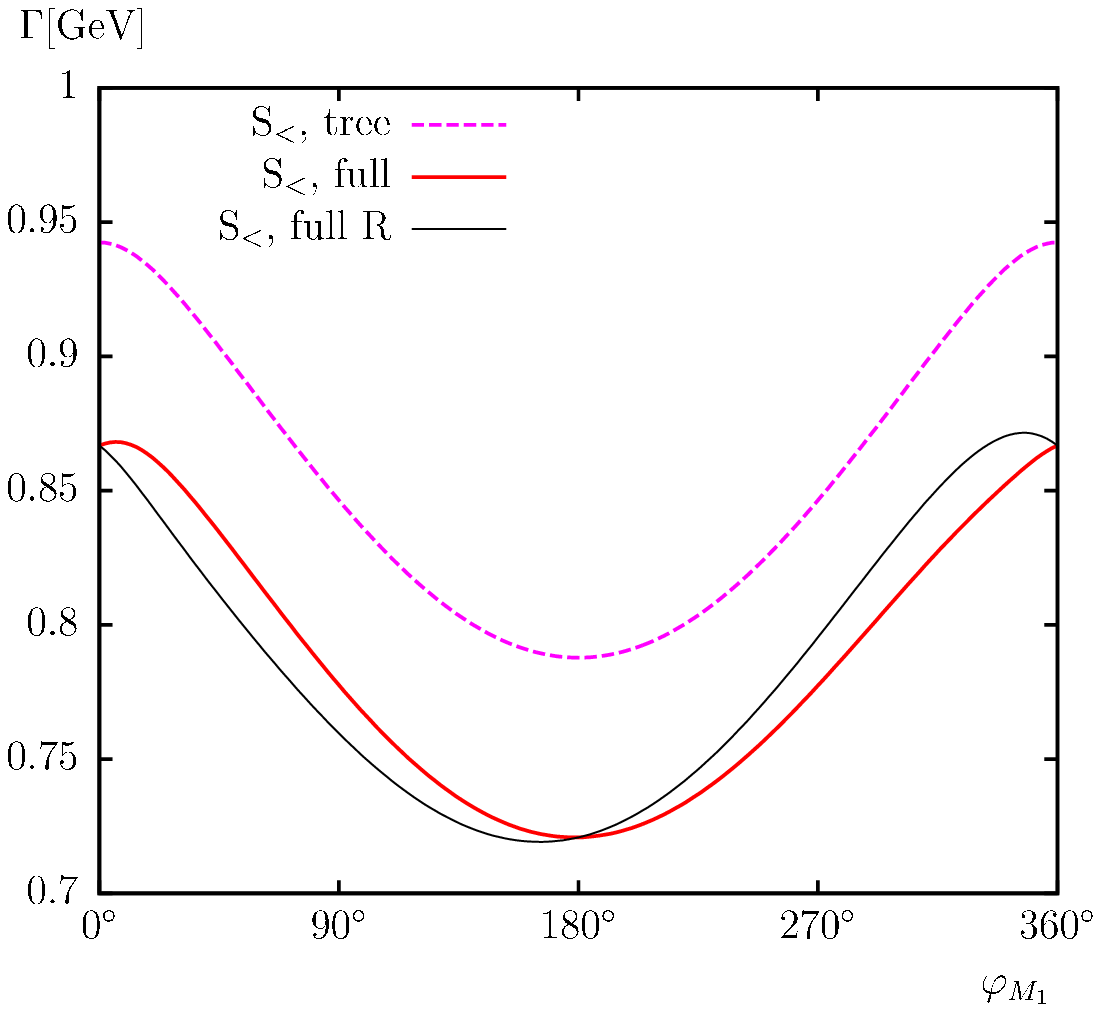}
\hspace{-4mm}
\includegraphics[width=0.49\textwidth,height=7.5cm]{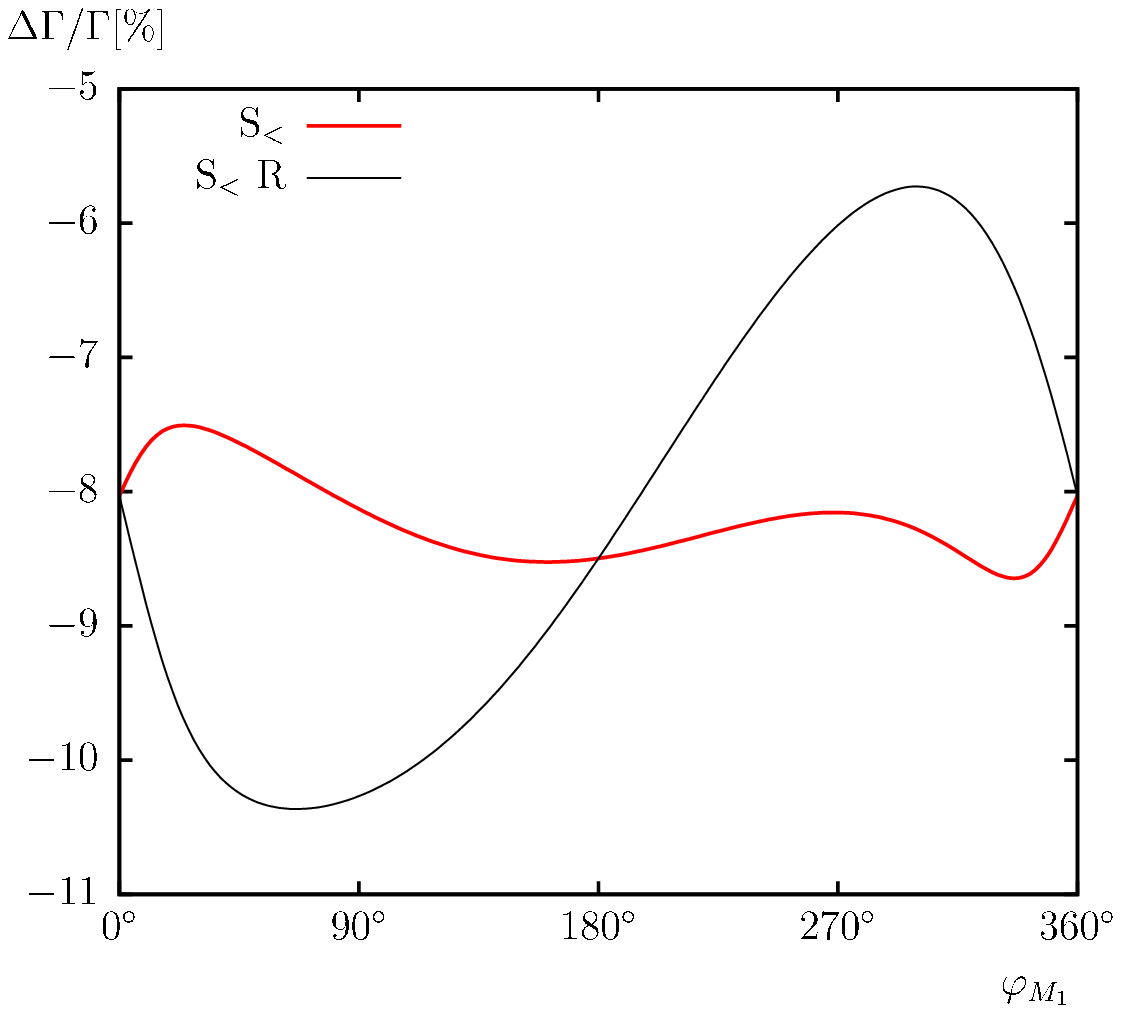} 
\\[4em]
\includegraphics[width=0.49\textwidth,height=7.5cm]{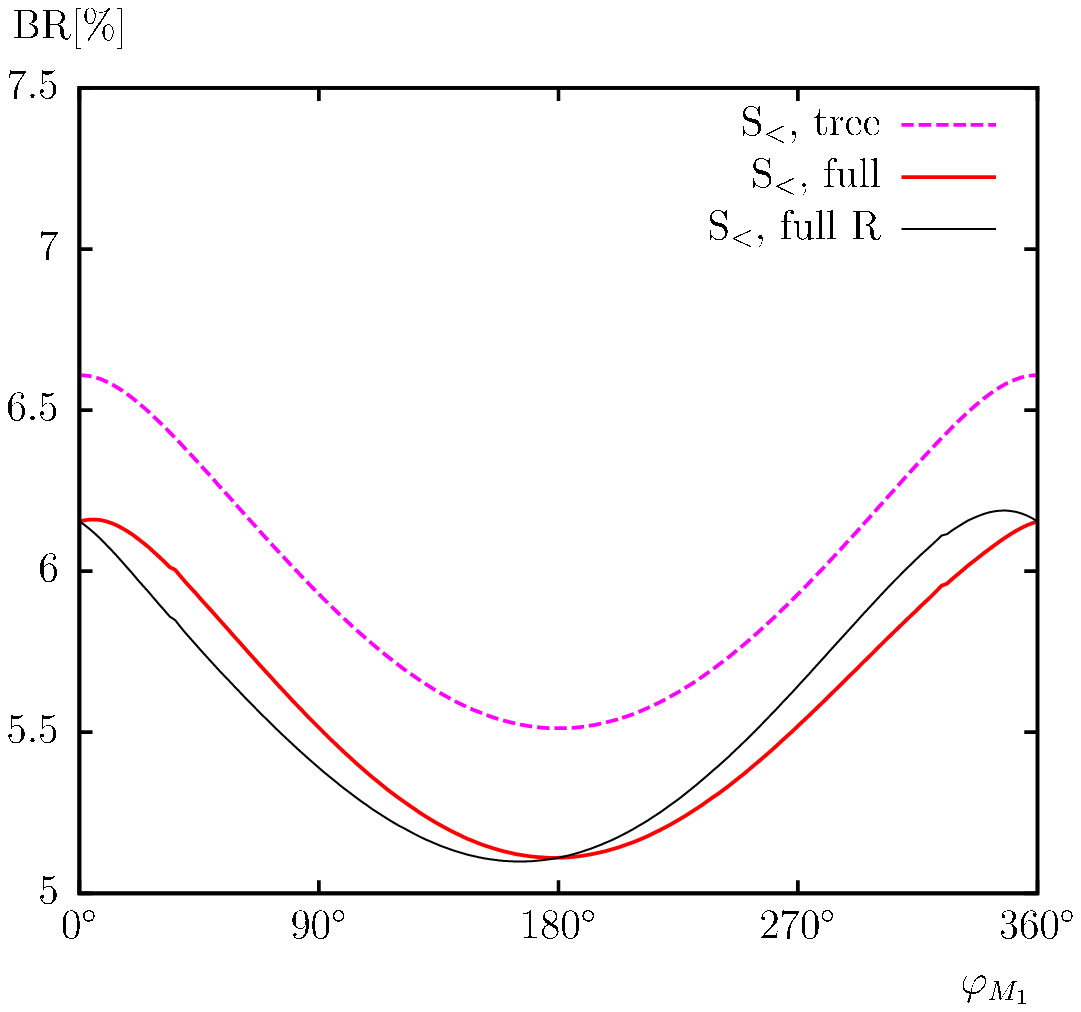}
\hspace{-4mm}
\includegraphics[width=0.49\textwidth,height=7.5cm]{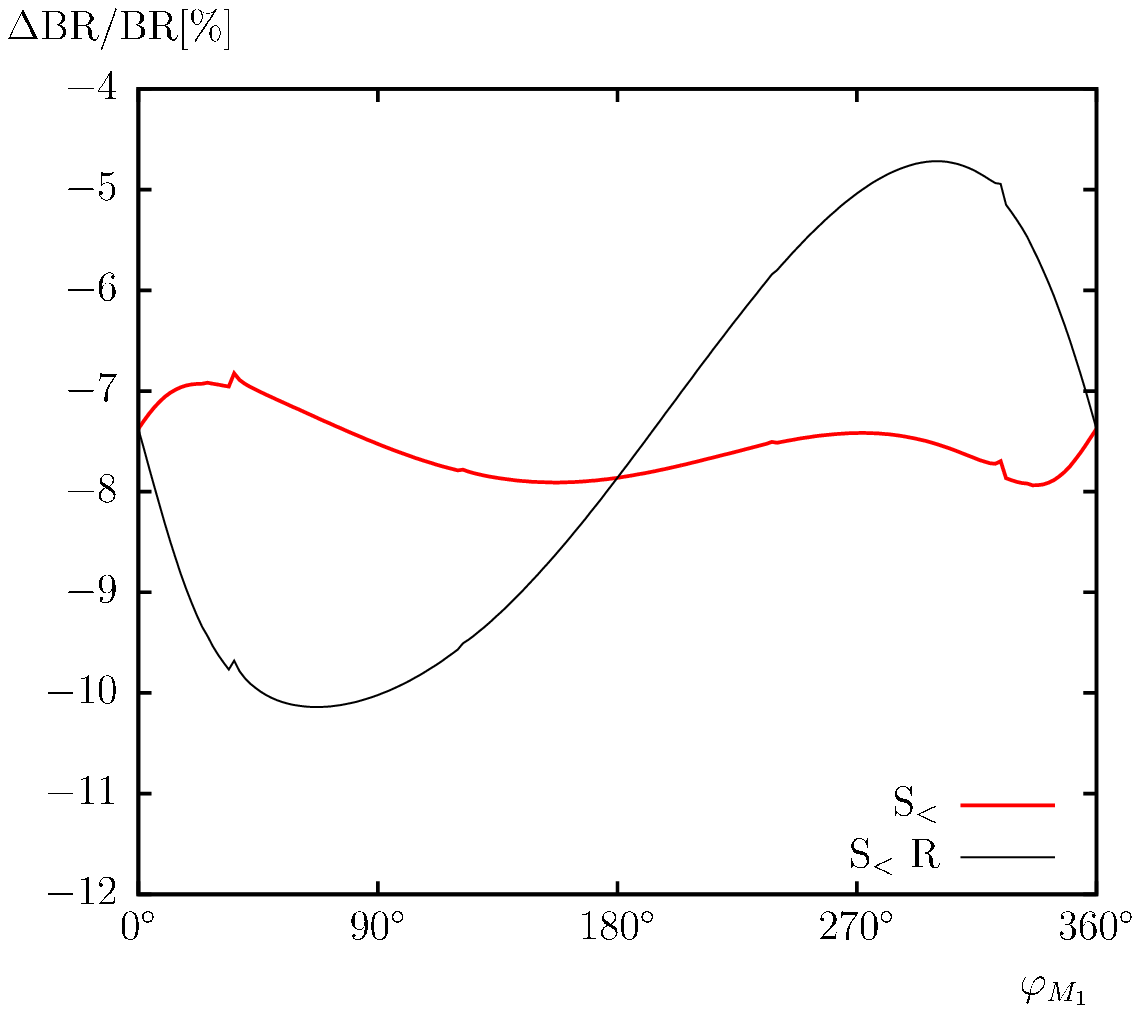}
\end{tabular}
\vspace{2em}
\caption{
  $\Ga(\DecayCmNW{2}{3})$. 
  Tree-level (``tree'') and full one-loop (``full'') corrected 
  decay widths are shown with the parameters chosen according to \SN\
  (see \refta{tab:para}), with $\phi_{\MOne}$ varied.
  Also shown are the full one-loop corrected decay widths omitting
  the absorptive contributions (``full R'').
  The upper left plot shows the decay width, the upper right plot shows 
  the relative size of the corrections.
  The lower left plot shows the BR, the lower right plot shows 
  the relative size of the BR.
}
\label{fig:PhiM1.cha2neu3w}
\end{center}
\end{figure}

\medskip
Now we turn to the decays involving a $W$~boson, $\DecayCmNW{2}{1,2,3}$,
as shown in \reffis{fig:PhiM1.cha2neu1w} -- \ref{fig:PhiM1.cha2neu3w}. 
For $\Ga(\DecayCmNW{2}{1})$ we find values of $\sim 0.34 \gev$ in
\SE\ with a small dependence on $\phiMe$ and values between $0.03 \gev$
and $0.18 \gev$ with a large dependence on $\phiMe$. The one-loop
effects appear at the level of $+5\%$ and $-5\%$, respectively, where in
the latter case a sizable effect of the absorptive self-energy
contributions can be observed. The $\br(\DecayCmNW{2}{1})$ yield 
$\sim 5\%$ in \SE\ and $\sim 1\%$ in \SZ, where the one-loop effects 
are found to be $\sim 4\%$ (\SE) and between $-9\%$ and $+2\%$
(\SZ). Especially in the latter case a reliable ILC analysis
requires the inclusion of the full one-loop calculation.

The decay $\DecayCmNW{2}{2}$ (see \reffi{fig:PhiM1.cha2neu2w})
yields decay widths around $\sim 1 \gev$ in
both scenarios, corresponding to BR's of $\sim 16\%$ in \SE\ and 
$\sim 7\%$ in \SZ, with a small dependence on $\phiMe$. 
The same dips as in $\DecayCmNH{2}{2}$ can be observed.
The one-loop
effects on the BR's is found to be 
$\sim -4\%$ and the variation with $\phiMe$
is small {\em after} the inclusion of the absorptive self-energy
contributions, as can be seen in the lower
right panel. However, the size of the corrections still exceed the
anticipated ILC(1000) accuracy.

The last decay, $\DecayCmNW{2}{3}$ (\reffi{fig:PhiM1.cha2neu3w}), 
which is again only realized in \SZ, gives a BR around $\sim 6\%$, 
where the one-loop corrections can be substantial at the level of $-7\%$. 
Again, the variation with $\phiMe$ becomes small {\em after} the
inclusion of the absorptive self-energy contributions. 

\begin{figure}[htb!]
\begin{center}
\begin{tabular}{c}
\includegraphics[width=0.49\textwidth,height=7.5cm]{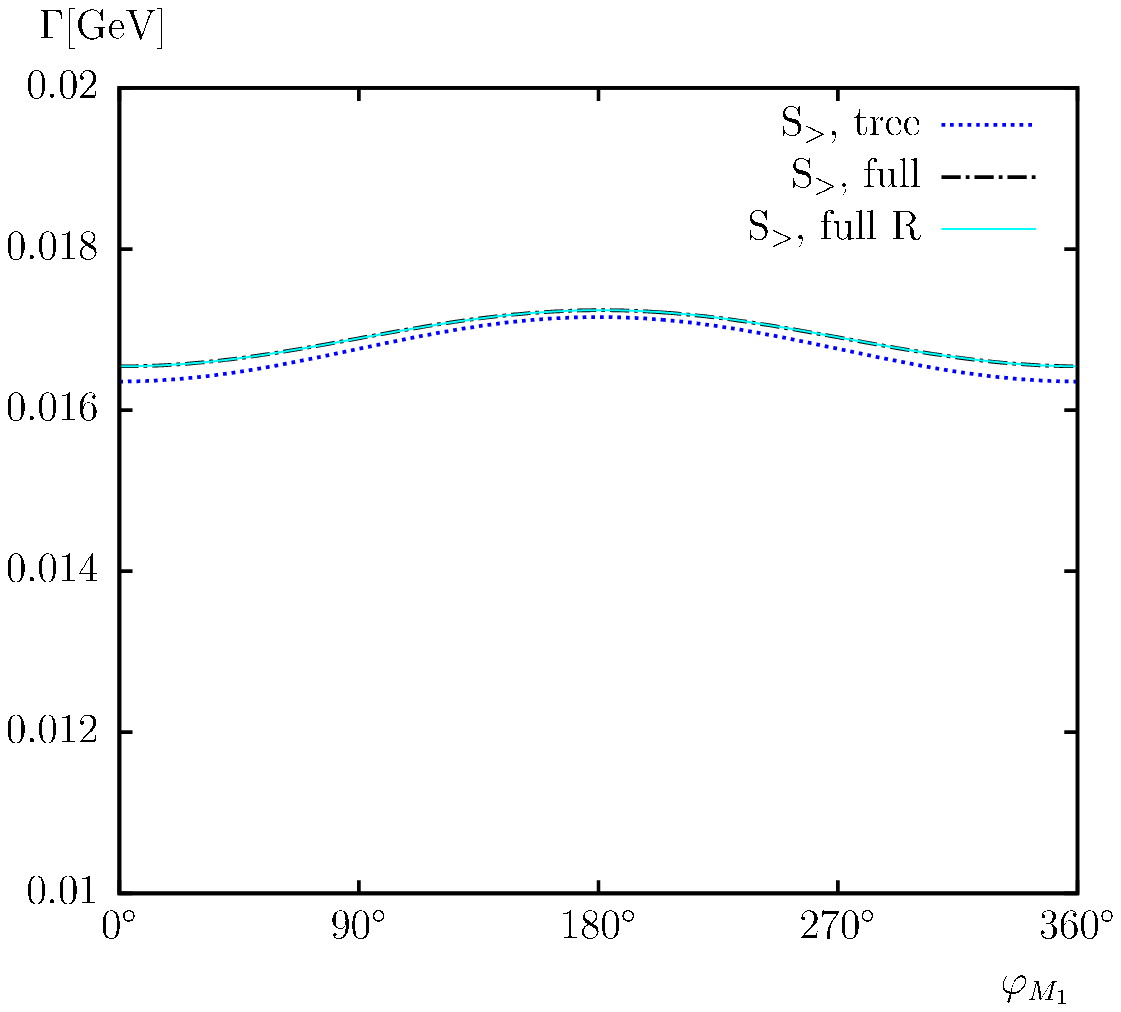}
\hspace{-4mm}
\includegraphics[width=0.49\textwidth,height=7.5cm]{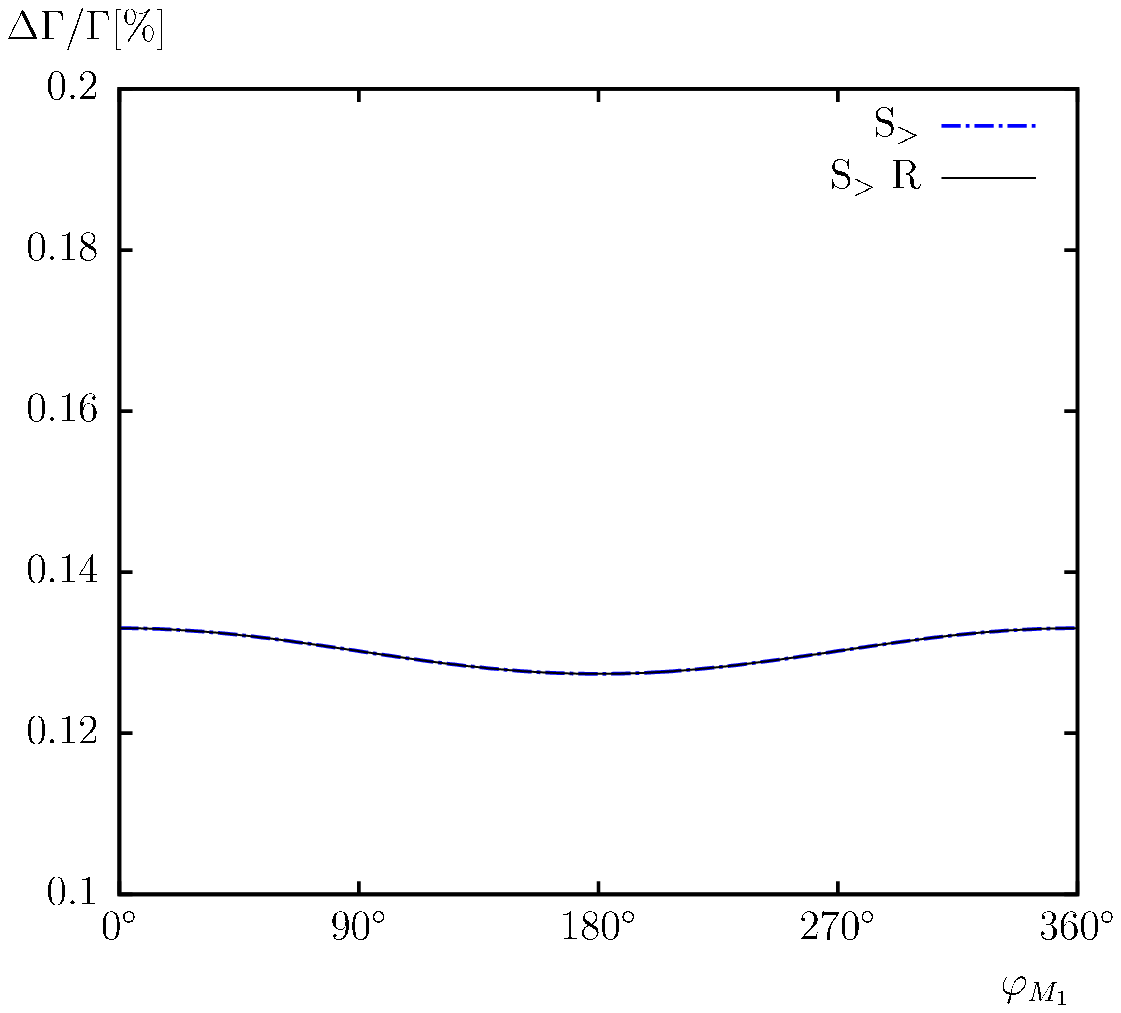} 
\\[4em]
\includegraphics[width=0.49\textwidth,height=7.5cm]{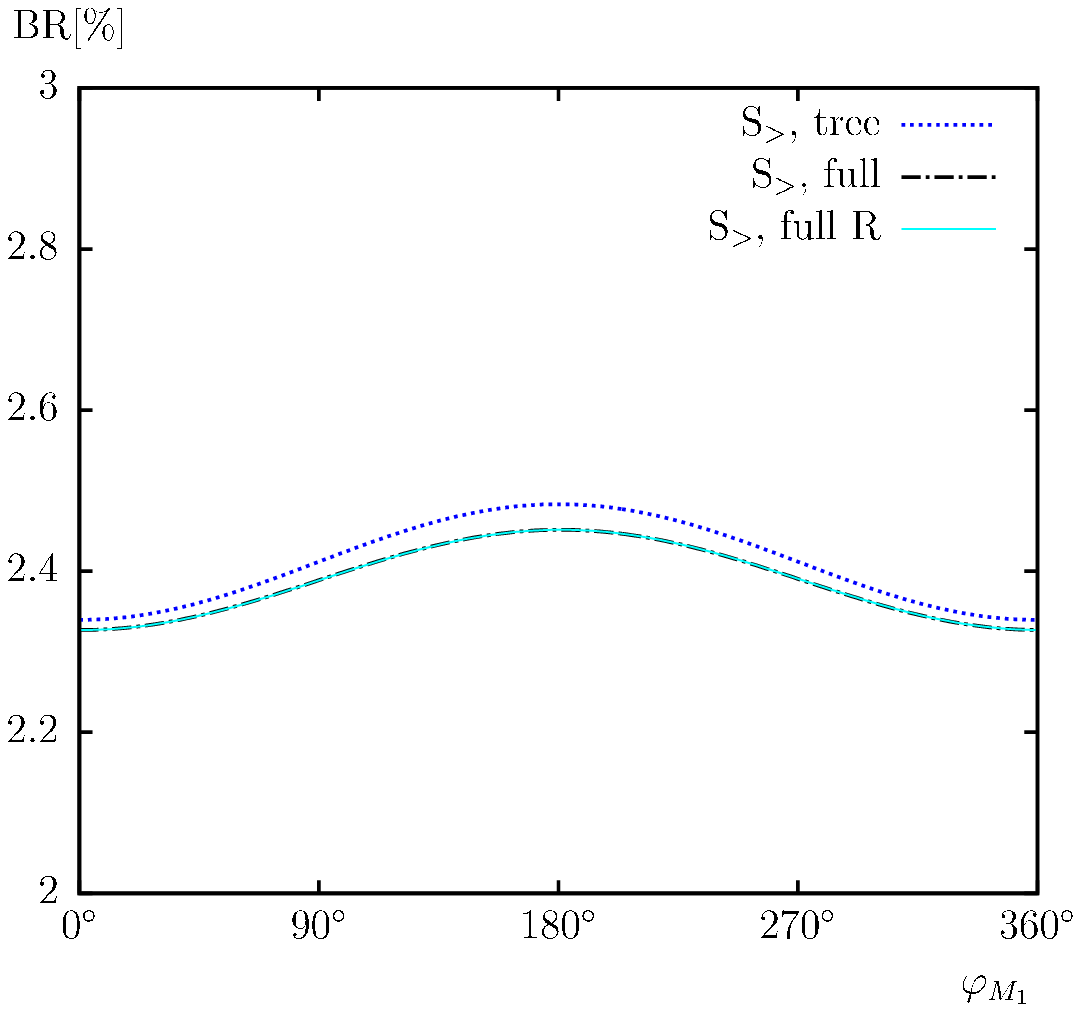}
\hspace{-4mm}
\includegraphics[width=0.49\textwidth,height=7.5cm]{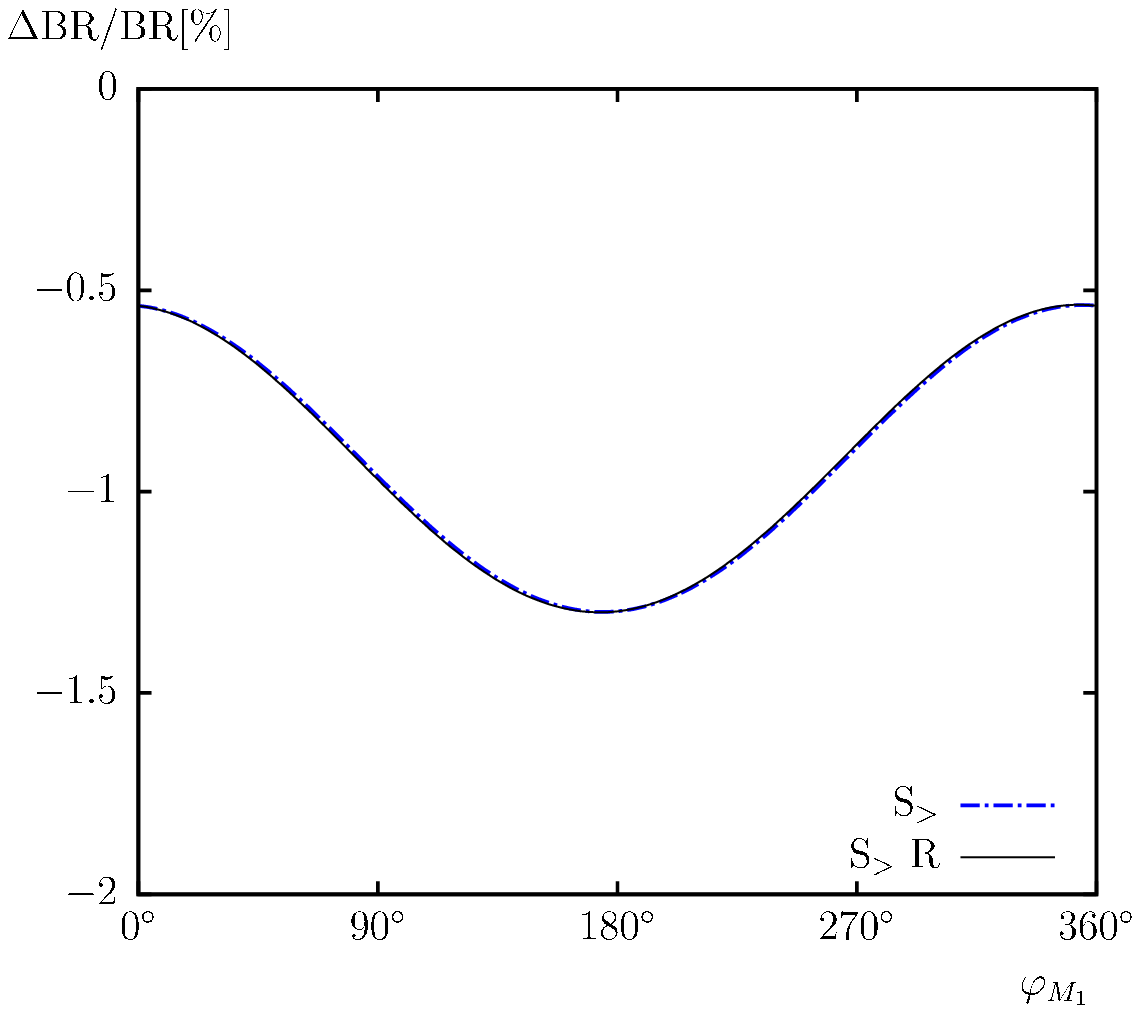}
\end{tabular}
\vspace{2em}
\caption{
  $\Ga(\DecayCmNH{1}{1})$. 
  Tree-level (``tree'') and full one-loop (``full'') corrected 
  decay widths are shown with the parameters chosen according to \SN\
  (see \refta{tab:para}), with $\phi_{\MOne}$ varied.
  Also shown are the full one-loop corrected decay widths omitting
  the absorptive contributions (``full R'').
  The upper left plot shows the decay width, the upper right plot shows 
  the relative size of the corrections.
  The lower left plot shows the BR, the lower right plot shows 
  the relative size of the BR.
}
\label{fig:PhiM1.cha1neu1hp}
\end{center}
\end{figure}

\medskip
Finally we discuss the two relevant $\cham{1}$ decays. 
In \reffi{fig:PhiM1.cha1neu1hp} we present the results for
$\DecayCmNH{1}{1}$. The decay is kinematically allowed only in
\SE\ (with the other parameters chosen according to \refta{tab:para}). 
The decay width is small, not exceeding $0.017 \gev$, 
corresponding to a $\br(\DecayCmNH{1}{1})$ below $2.5\%$. The variation
with $\phiMe$ is small.
The effect of the absorptive self-energy contributions is negligible in
both scenarios. 
In view of the anticipated ILC(1000) accuracy at the per-cent level
these corrections should still be taken into account for a reliable
analysis.

\begin{figure}[htb!]
\begin{center}
\begin{tabular}{c}
\includegraphics[width=0.49\textwidth,height=7.5cm]{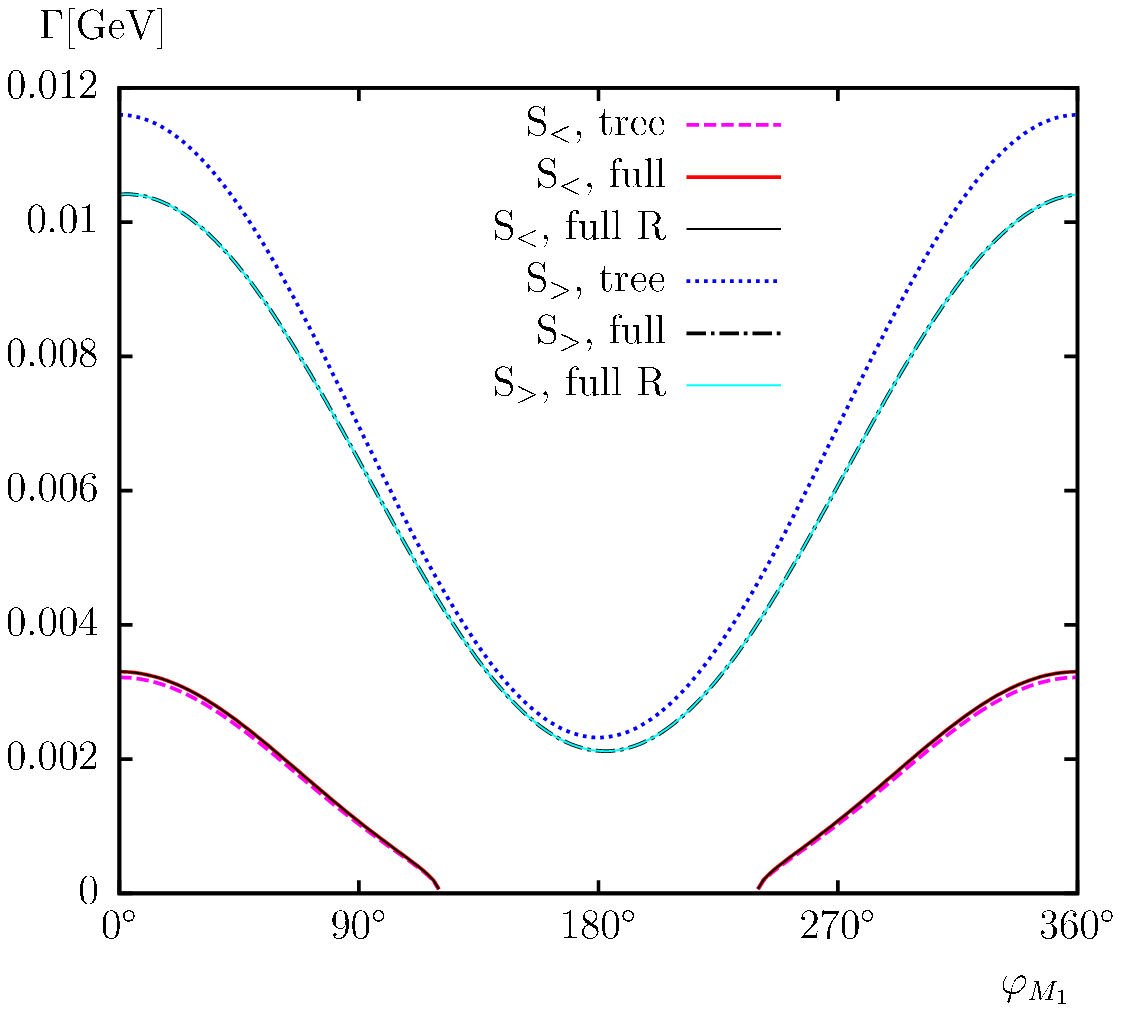}
\hspace{-4mm}
\includegraphics[width=0.49\textwidth,height=7.5cm]{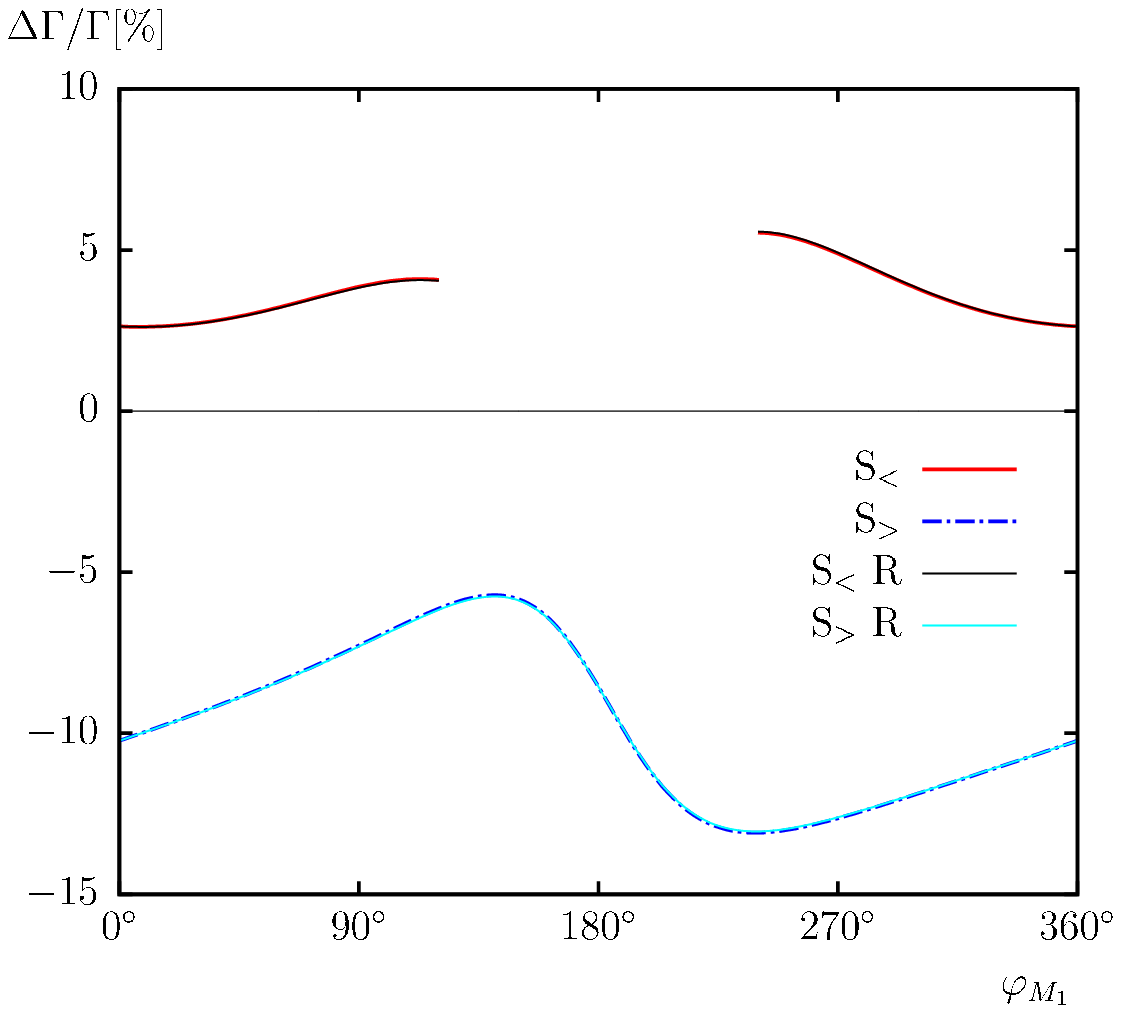} 
\\[4em]
\includegraphics[width=0.49\textwidth,height=7.5cm]{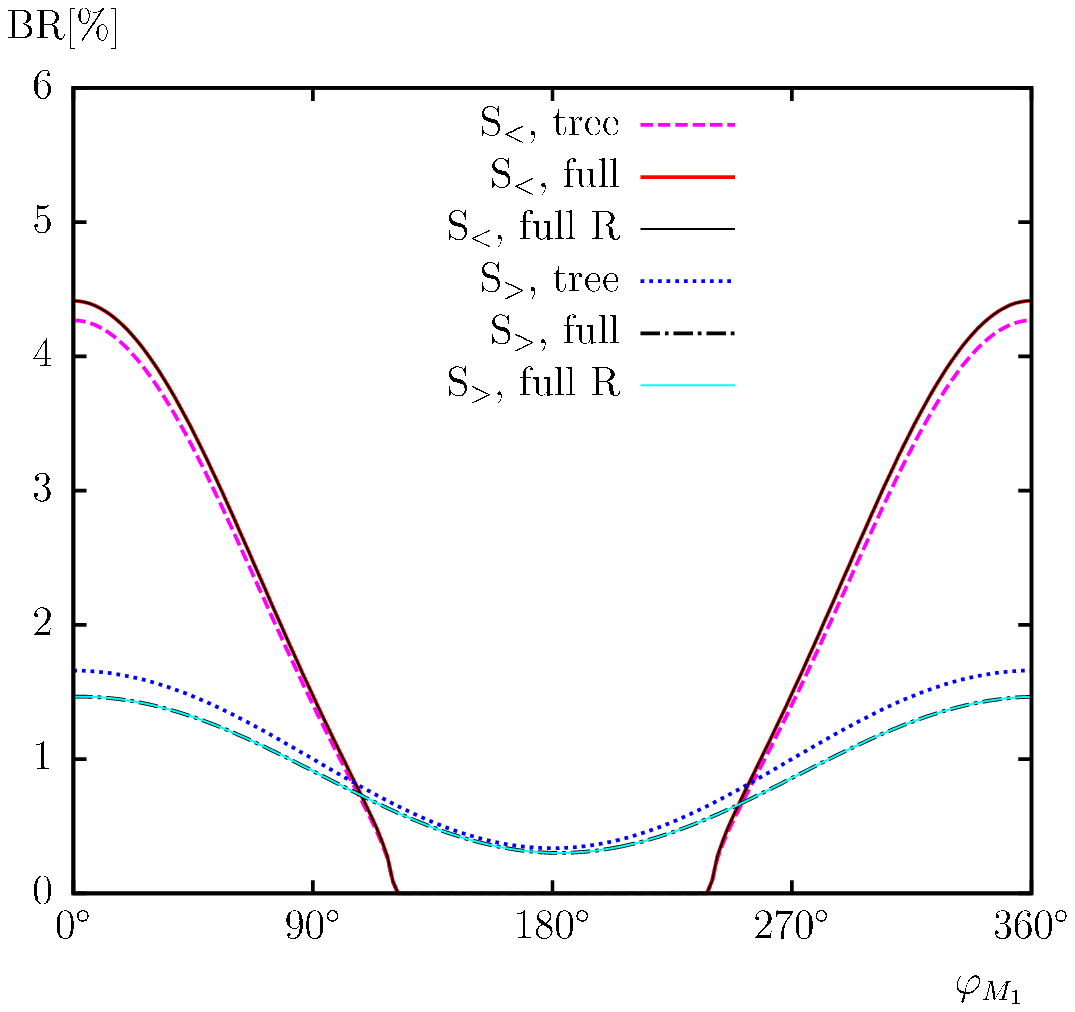}
\hspace{-4mm}
\includegraphics[width=0.49\textwidth,height=7.5cm]{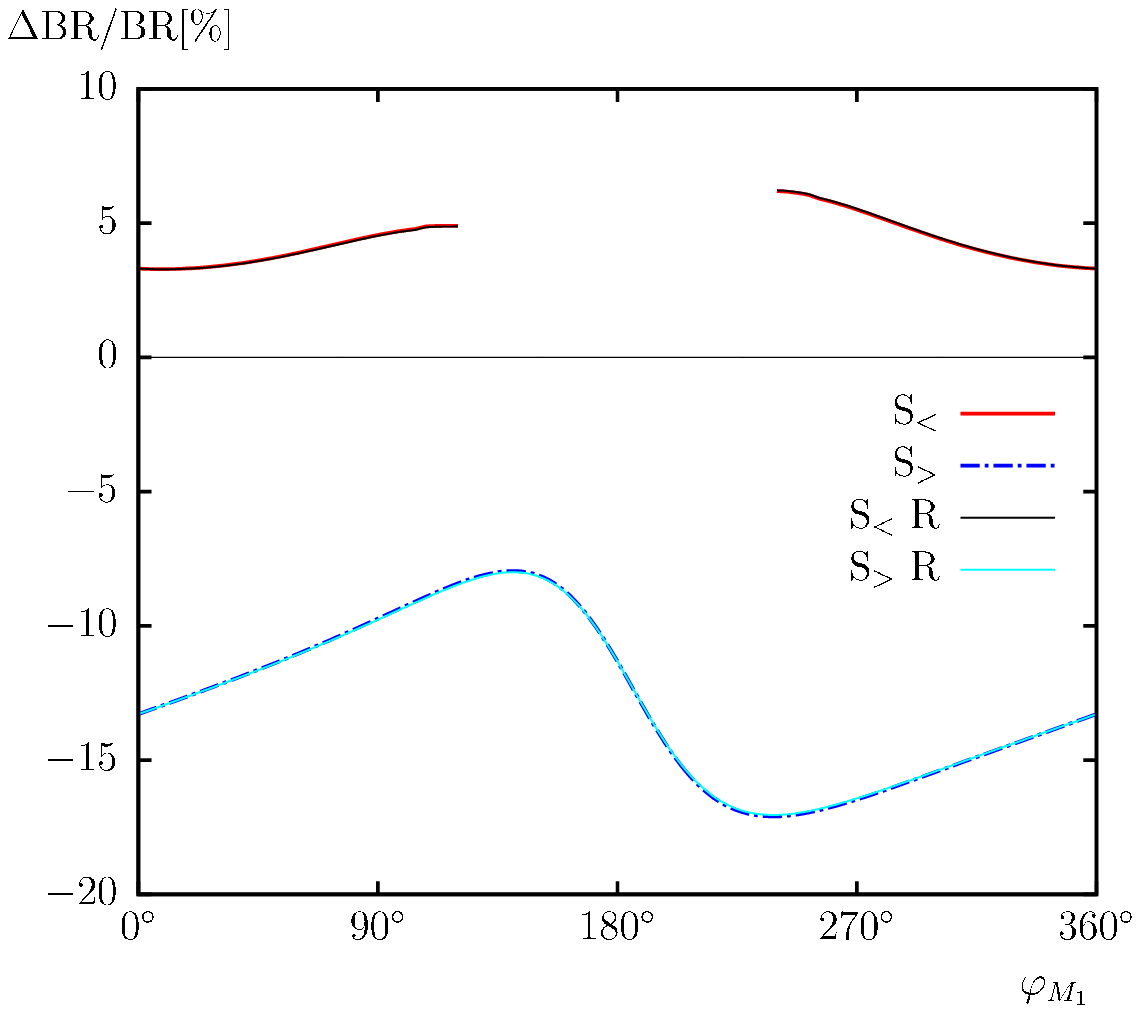}
\end{tabular}
\vspace{2em}
\caption{
  $\Ga(\DecayCmNW{1}{1})$. 
  Tree-level (``tree'') and full one-loop (``full'') corrected 
  decay widths are shown with the parameters chosen according to \SN\
  (see \refta{tab:para}), with $\phi_{\MOne}$ varied.
  Also shown are the full one-loop corrected decay widths omitting
  the absorptive contributions (``full R'').
  The upper left plot shows the decay width, the upper right plot shows 
  the relative size of the corrections.
  The lower left plot shows the BR, the lower right plot shows 
  the relative size of the BR.
}
\label{fig:PhiM1.cha1neu1w}
\end{center}
\end{figure}

The last decay is $\DecayCmNW{1}{1}$ shown in
\reffi{fig:PhiM1.cha1neu1w}. While in \SE\ the decay is kinematically
allowed for the full range of $\phiMe$ only in \SE, while in \SZ\ it
remains kinematically forbidden for $120^\circ \le \phiMe \le 240^\circ$.
The decay widths are below $0.012\ (0.004) \gev$ in \SE\ (\SZ),
corresponding to BR's at the level of $\sim 1\%$ (\SE) and between 
$\sim 4\%$ and $0\%$ (\SZ). Here a strong dependence on $\phiMe$ is
visible. The size of the one-loop effects exceed $-15\%$ in \SE\ and are
around $+4\%$ in \SZ. 
The effect of the absorptive self-energy contributions is negligible.
As before, the corrections are potentially relevant for an
ILC(1000) analysis.

\bigskip
We have also analyzed the effects of a variation with $\phiatau$, but
found negligible effects for the parameters in \refta{tab:para}. The
situation would be different if the off-diagonal element in the scalar
tau mass matrix, $X_\tau = \Atau - \mu\tb$ were depending strongly on
$\Atau$, i.e.\ for small $\mu$ and/or $\tb$. However, a detailed
analysis of these effects is beyond the scope of our paper.

\clearpage
\newpage

\subsection{Full one-loop results: total decay widths}
\label{sec:gatot}

\begin{figure}[htb!]
\begin{center}
\begin{tabular}{c}
\includegraphics[width=0.49\textwidth,height=7.5cm]{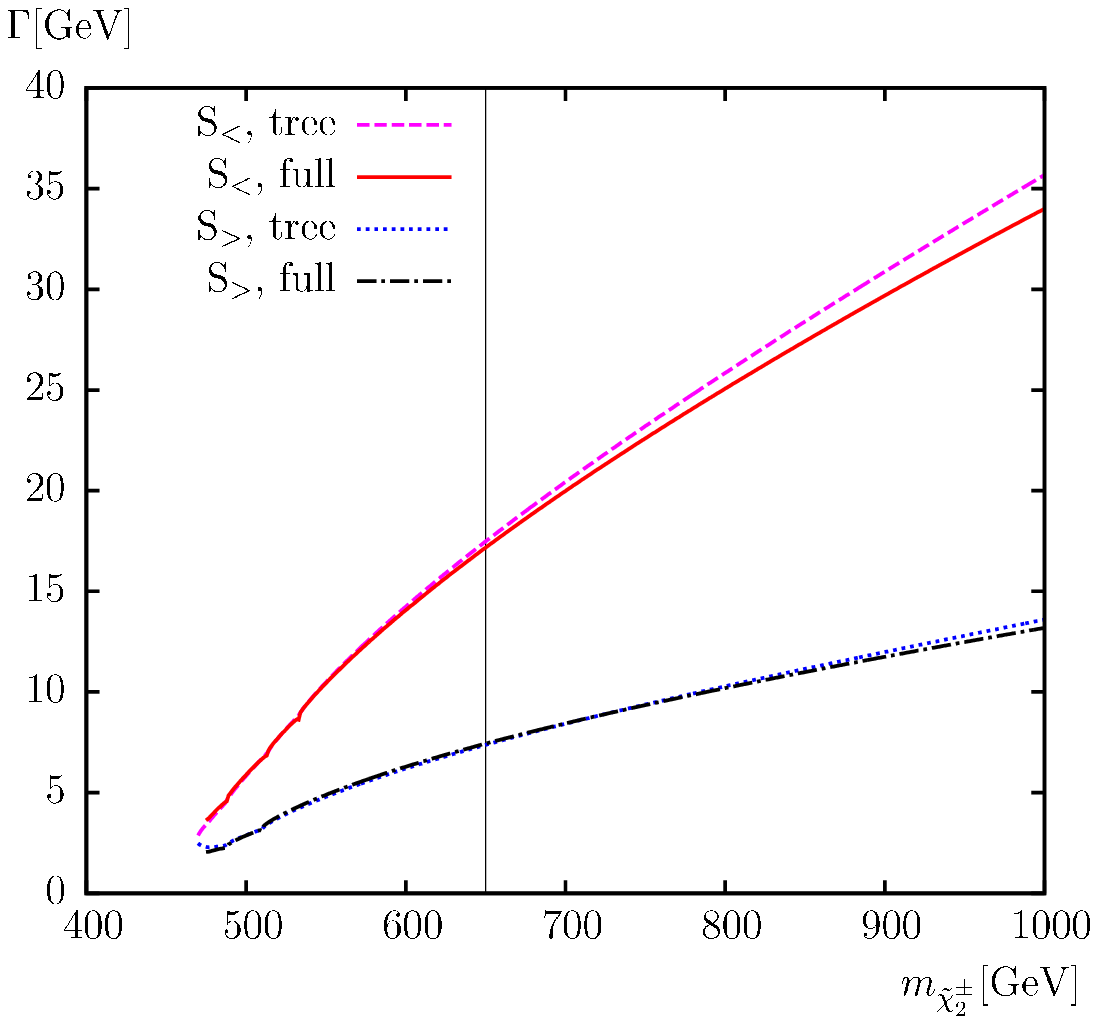}
\hspace{-4mm}
\includegraphics[width=0.49\textwidth,height=7.5cm]{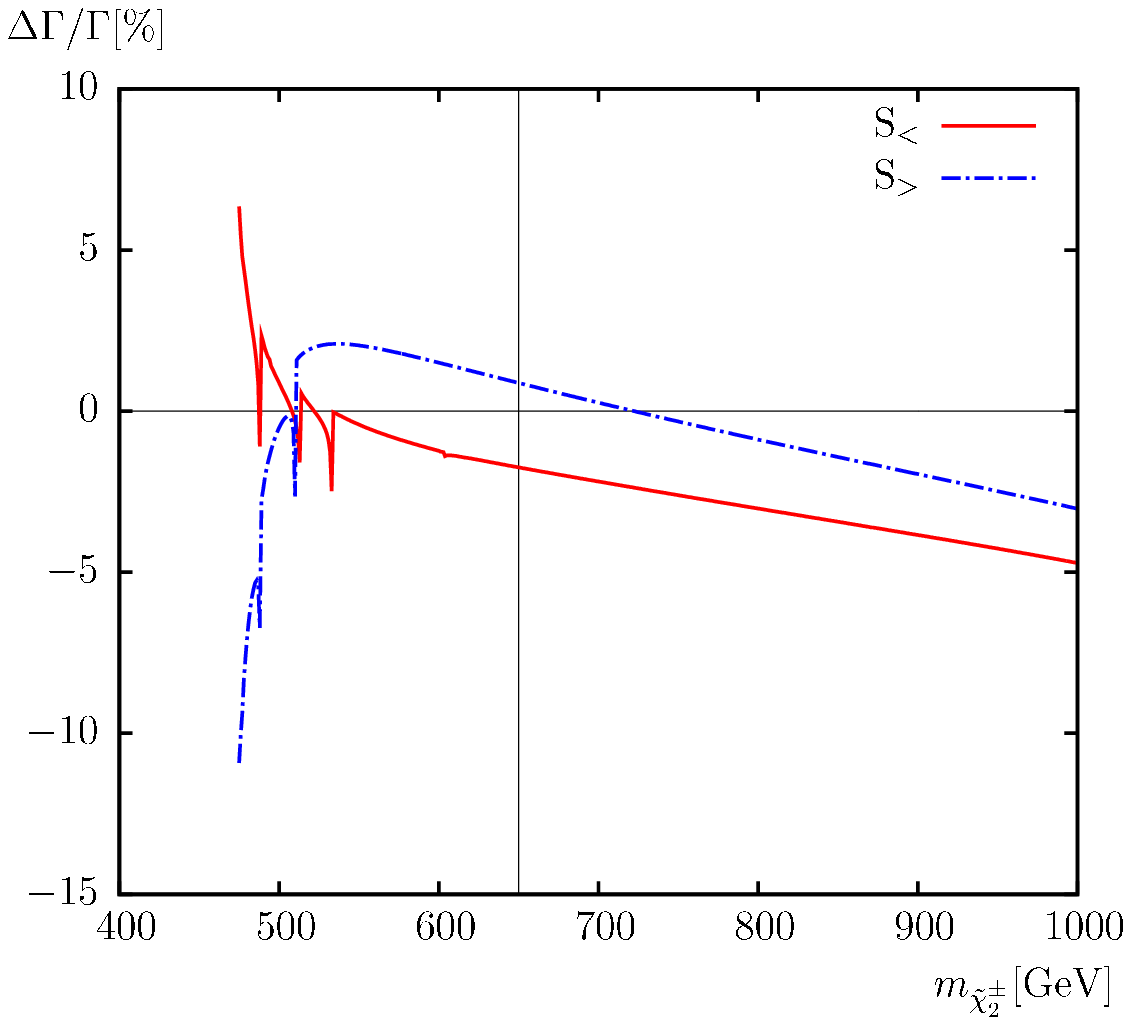} 
\\[2em]
\includegraphics[width=0.49\textwidth,height=7.5cm]{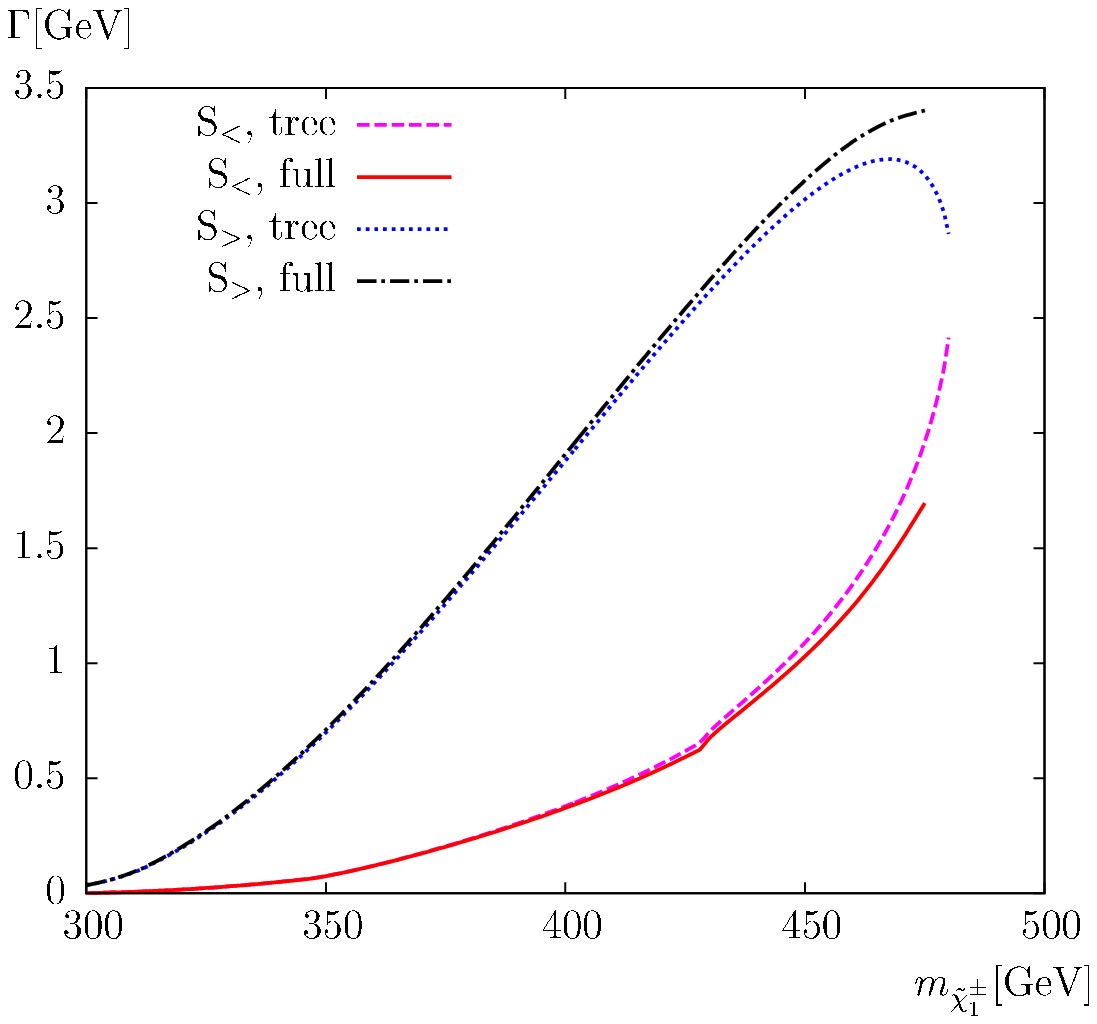}
\hspace{-4mm}
\includegraphics[width=0.49\textwidth,height=7.5cm]{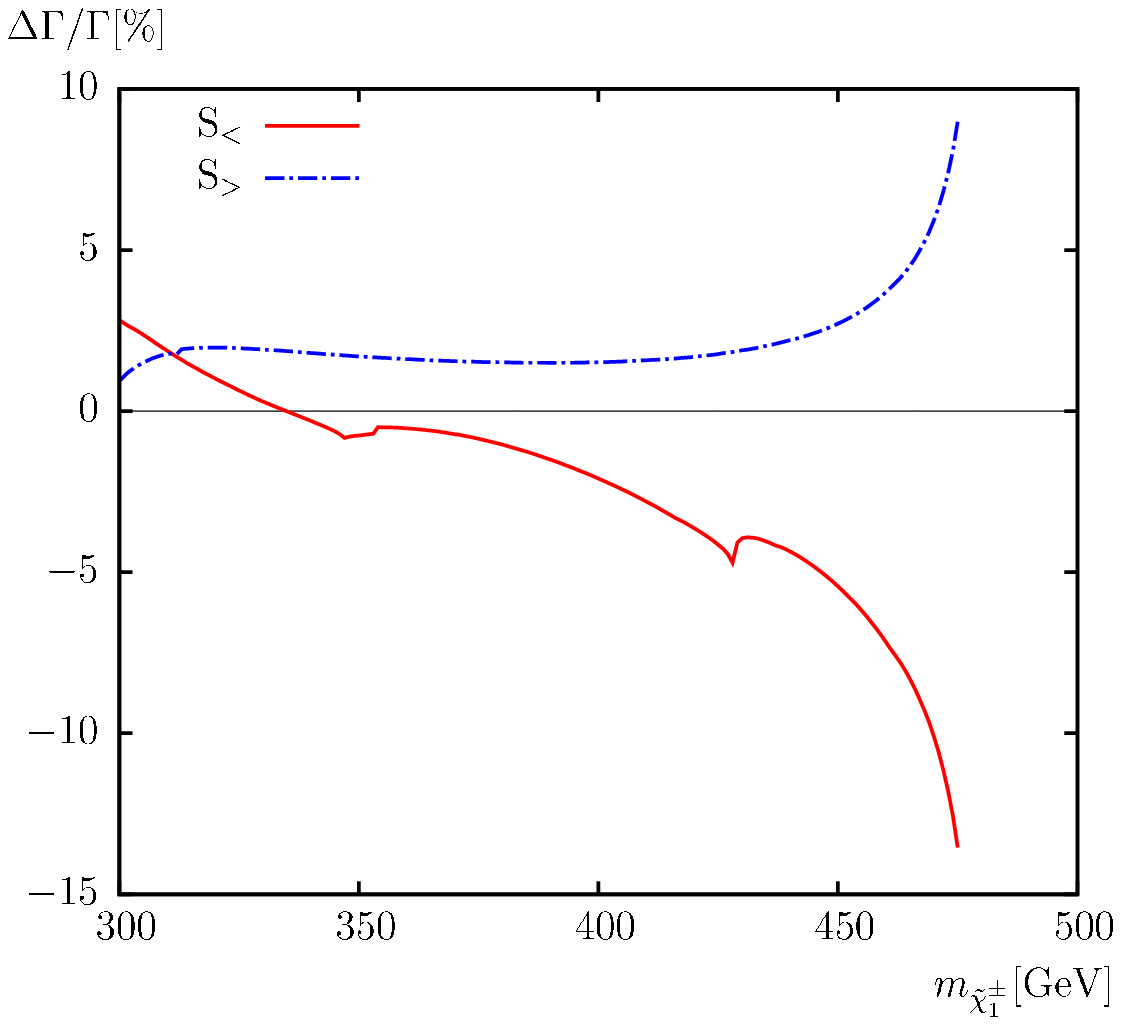}
\end{tabular}
\vspace{2em}
\caption{
  $\Ga(\cham{i} \to all)$, $i=1,2$.
  Tree-level (``tree'') and full one-loop (``full'') corrected total 
  decay widths are shown with the parameters chosen according to \SN\ 
  (see \refta{tab:para}), with $\mcha{i}$ varied.
  The upper left plot shows the decay width of $\cham{2}$ and 
  the upper right plot shows the corresponding relative size of the 
  corrections, both as a function of its mass.
  The lower left and right plots show the same observables for $\cham{1}$.
  The vertical lines indicate where $\mcha{1} + \mcha{2} = 1000 \gev$, 
  i.e.\ the maximum reach of the ILC(1000) 
  for $\cha{1}\champ{2}$ pair production.
}
\vspace{2em}
\label{fig:mCi.chaitotal}
\end{center}
\end{figure}

In this final subsection we briefly show the results for the total decay
widths in \reffi{fig:mCi.chaitotal}. The results for $\cha{2}$ are shown
in the upper row. The total width rises from its lowest values at
$\mcha{2} = 475 \gev$ to about 
$13 \gev$ at $\mcha{2} = 1000 \gev$ in \SE. In \SZ\ the width goes up to 
$34 \gev$, once loop corrections are included. The overall size of the
one-loop corrections varies strongly with $\mcha{2}$ as can be seen in
the upper right plot. Values of $\pm 5\%$ can easily be
reached. 

In the lower row of \reffi{fig:mCi.chaitotal} we show the total width of
the lighter chargino. It rises only up to about $3.3 \gev$ in \SE\ and 
$\sim 1.5 \gev$ in \SZ. Again the size of the one-loop corrections
vary with $\mcha{1}$, where away from threshold we find corrections at
the level of $+2\%$ in \SE\ and between $+2\%$ and $\sim -10\%$ in \SZ.

\section{Conclusions}
\label{sec:conclusions}

We have evaluated two-body decay widths of charginos in 
the Minimal Supersymmetric Standard Model with complex parameters
(cMSSM). Assuming heavy scalar quarks we take into account all decay channels
involving charginos, neutralinos, (scalar) leptons, Higgs bosons and SM
gauge bosons. 
The decay modes are given in \refeqs{CNH} -- (\ref{CSnl}). 
The evaluation of the decay widths is based on a full one-loop calculation 
including hard and soft QED radiation. 
Such a calculation is necessary to derive a reliable
prediction of any two-body branching ratio.
Three-body decay modes can become sizable only if all the two-body channels
are kinematically (nearly) closed and have thus been neglected
throughout the paper. The same applies to two-body decay modes that
appear only at the one-loop level. 

We first reviewed the one-loop renormalization of the cMSSM, which is
relevant for our calculation. 
We have given details for the lepton/slepton sector, whereas the details
for the chargino/neutralino and the Higgs boson sector can be found in
\citere{Stop2decay}. 

We have discussed the calculation of the one-loop diagrams, the
treatment of UV- and IR-divergences that are canceled by the inclusion
of soft QED radiation. 
Our calculation set-up can easily be extended to other two-body decays
involving (scalar) quarks. 

For the numerical analysis we have chosen a parameter set that allows
simultaneously {\em all} two-body decay modes under investigation
(but could potentially be in conflict with the most recent SUSY
search results from the LHC).
The masses of the charginos in this scenario are $600$ and $350 \gev$. 
This scenario allows copious 
production of the charginos in SUSY cascades at the LHC.
Furthermore, the production of $\cha{1}\champ{2}$ or $\chap{1}\cham{1}$
at the ILC(1000), i.e.\ with $\sqrt{s} = 1000 \gev$, via 
$e^+e^- \to \cha{1,2}\champ{1}$ will be possible,
with all the subsequent decay modes (\ref{CNH}) -- (\ref{CSnl})
being (in principle) open. The clean environment of the ILC would
then permit a detailed, statistically dominated study of the chargino
decays. Depending on the channel and the polarization, a precision at
the per-cent level seems to be achievable.
Special attention is paid to chargino decays involving the 
Lightest Supersymmetric Particle (LSP), i.e.\ the lightest 
neutralino, or a neutral or charged Higgs boson.

In our numerical analysis we have shown results for varying $\mcha{1,2}$ and
$\phiMe$, the phase of the soft SUSY-breaking parameter~$\MOne$. 
In the results with varied chargino masses only the lighter values allow
$\cha{1}\champ{2}$ production at the ILC(1000), whereas the results with
varied $\phiMe$ have sufficiently light charginos to permit 
$e^+e^- \to \cha{1}\champ{2}$. 
In the numerical analysis we compared the tree-level width with the one-loop
corrected decay width. In the analysis with $\phiMe$ varied we
explicitly took into account contributions from the absorptive parts of
self-energy contributions on external legs. 
We also analyzed the relative change of the width
to demonstrate the size of the loop corrections on each individual
channel. In order to see the effect on the experimentally accessible
quantities we also show the various branching ratios at tree-level (all
channels are evaluated at tree-level) and at the one-loop level (with
all channels evaluated including the full one-loop
contributions). Furthermore we presented the relative change of the BRs
that can directly be compared with the anticipated experimental
accuracy.

We found sizable corrections in many of the decay channels. 
Especially, the higher-order corrections of the chargino decay widths
involving the LSP can easily reach 
a level of about $\pm 10\%$.
Decay modes involving Higgs bosons turn out to have slightly smaller
corrections. The size of the full one-loop corrections to the decay
widths and the branching ratios also depends strongly on $\phiMe$. The
one-loop contributions, again being roughly of \order{5\%}, often vary
by a factor of $2-3$ as a function of $\phiMe$. 
All results on partial decay widths are given in detail in
\refses{sec:1Lmcha2} -- \ref{sec:1LphiMe}, while the total decay widths
are shown in \refse{sec:gatot}. 

The numerical results we have shown are of course dependent on choice of
the SUSY parameters. Nevertheless, they give an idea of the relevance
of the full one-loop corrections. 
For other choices of SUSY masses the 
corrections to the decay widths would stay the same, but the branching 
ratios would look very different. 
Channels (and their respective one-loop corrections) that may look 
unobservable due to the smallness of their BR in our numerical examples
could become important if other channels are kinematically forbidden.

Following our analysis it is evident that the full one-loop corrections
are mandatory for a precise prediction of the various branching ratios.
This applies to LHC analyses, but even more to analyses at the ILC,
where a precision at the per-cent level is anticipated for the
determination of chargino branching ratios (depending on the chargino
masses, the center-of-mass energy and the integrated luminosity).
The results for the chargino decays will be implemented into the
Fortran code {\tt FeynHiggs}.


\subsection*{Acknowledgements}

We thank  
A.~Bharucha,
M.~Drees,
A.~Fowler,
H.~Haber,
T.~Hahn, 
O.~Kittel,
S.~Liebler,
H.~Rzehak
and
G.~Weiglein
for helpful discussions.  
The work of S.H.\ was partially supported by CICYT (grant FPA
2007--66387 and FPA 2010--22163-C02-01).
F.v.d.P.\ was supported by 
the Spanish MICINN's Consolider-Ingenio 2010 Programme under grant MultiDark CSD2009-00064.



\begin{thebibliography}{99}
\bibitem{feynarts-mf} The couplings can be found in the files
                      {\tt MSSM.ps.gz}, {\tt MSSMQCD.ps.gz} and 
                      {\tt HMix.ps.gz} as part of the
                      \fa~package~\cite{feynarts}.
\end{thebibliography}
\end{document}